\documentclass[notitlepage,showpacs,aps,floatfix,prd] {revtex4-1}
\input revstyle.tex
\input epsf
\usepackage{graphicx}
\begin{document}\title{Warped    modes in flux  compactification 
of type II \ b  supergravity  on the  conifold}
\author{M. Chemtob} 
\email{marc.chemtob@ipht.fr} 
\affiliation{Institut  de  Physique Th\'eorique, CEA-Saclay F-91191
Gif-sur-Yvette Cedex FRANCE}  \thanks {\it Supported by 
Direction  de la Recherche Fondamentale et aux  Energies  Alternatives
Saclay } \date{\today} 
\begin{abstract} 
We examine the Kaluza-Klein theory for warped flux compactifications of type
$II\ b $ string theory on  a   Minkowski spacetime $ M_4$ times  
a conic  Calabi-Yau orientifold  $X_6$.   The  
region  glued along the internal  space directions to  the bulk of
$X_6$ is modeled by the  warped undeformed conifold  $\calc _6$.  The
resulting   classical vacum solution of  Klebanov-Tseytlin solution, 
valid at  large radial distances from the conic singularity,
describes a spacetime  asymptotic to $ AdS_5  $  times 
the conifold compact base   $ T^{1,1}$.    
The metric  tensor, axio-dilaton
and  antisymmetric  fields part of the 10-d supergravity multiplet
bosonic sector decompose into towers of scalar and tensor fields of fixed  mass
associated to  harmonic modes of $T^{1,1}$ and $M_4$.  
The warped throat deformations  by
compactification effects   are described  by means of AdS/CFT
duality.   The applications are  developed within the hard wall
approximation  for a truncated $ AdS_5 $ spacetime
ending on flat  boundaries.   Numerical results are presented for the
mass spectra and interactions of a selected set of singlet and charged
modes   and  for the interactions of graviton modes 
with embedded  branes.  We examine the possibility that an
interacting gas of warped massive modes produced after brane-antibrane
inflation might  reheat the universe and leave  cold dark matter relics.
\end{abstract}  
\maketitle 
\renewcommand{\thefootnote}{alph{footnote}}
\tableofcontents\addtocontents{toc}{\protect\setcounter{tocdepth}{1}}


\section{Introduction}
\label{secintro}

 The   string theory flux compactifications~\cite{polchger96,gvw99}
inspired  by   AdS/CFT  duality~\cite{malda99}  have  opened 
novel perspectives for the construction of   particle physics  models~\cite{malda99}.    Embedding    fluxes and branes    
near isolated singularities of internal 
manifolds yields   solutions  involving 
non-factorizable spacetime  regions with tunable  amount of 
warping. While the viability of warped  compactifications  
was recognized a decade
earlier~\cite{strom86,dewitdass87}, what motivated a renewed activity was the
hope to  provide a stringy basis to Randall-Sundrum~\cite{rs9905,rs9906}
mechanism of mass hierarchies.  Satisfactory vacuum solutions were
indeed found in type $ II \ b $~\cite{verlinde99,gubser99} and
heterotic~\cite{mayr00} string theories and  
in the wider context of M- and     
F-theory~\cite{becker96,dasgupta99,becker00,greene00,becker01,gubserF00}.
One  concern then was in  circumventing  obstacles set by
no-go theorems on   warped backgrounds   free from 
bad   type  singularities~\cite{gubser00,malnez00}.  These efforts
culminated in the identification by Giddings, Kachru and Polchinski
(GKP)~\cite{gkp01} of supersymmetric solutions for
type $ II \ b $ string theory and F-theory with warped spacetime
regions sourced by 3-fluxes.  Useful reviews on flux
compactifications are provided in~\cite{dougkach05,granarev05}.

The  purpose of the present work   is to  construct 
the     Kaluza-Klein theory for compactifications  of type $ II \ b $
supergraviy theory  on singular Calabi-Yau (CY) orientifolds $X_6$
times   a   maximally symmetric  4-d  spacetime 
$M_4$  preserving  $\caln =1 $  supersymmetry. It is possible to 
source warped  geometries  in these  backgrounds  by  embedding   
3- and 5-fluxes and/or  branes  embedded  near 
conic type singularities   in  the  moduli space  of the internal
manifold~\cite{gkp01}.  
Workable  models   are built by gluing a copy of the conifold
near the  singularity along the  internal   directions.  
The  conifold     separates  into  the compact base  manifold
$X_5= T^{1,1} $ and   the  semi-infinite radial
direction~\cite{candelas90}    which 
combines   with the  non-compact spacetime   $M_4$  into a manifold
that asymptotes    $AdS _5 $  spacetime.   Various  generalizations
of this scheme were discussed for  type $ II\b $ and F-theory
compactifications  on manifolds involving    multiple
throats  and   embedding    $D3  /D7 $-branes. 

We  wish to  study  the dimensional reduction of the  10-d
supergravity multiplet in   the large radial distance regime  of     
the warped conifold  for a  background   spacetime asymptotic to 
 the direct product of      equal,   constant  and opposite
(negative and  positive) sign curvature  manifolds $AdS _5 \times T^{1,1} $. 
In parallel   with   the  classical 10-d approach,  
a quantum   4-d field theory  description can  also  be  considered
where  one uses the  interaction superpotentials from
fluxes~\cite{gvw99} and  instantons~\cite{witteninst96}
to   stabilize  the    closed  string (axio-dilaton  and
 geometric)  moduli as well as  the  open  string moduli
 of embedded  branes. Both the 10- and 4-d descriptions  were        
employed  in     
applications ranging from astroparticle physics~\cite{dimo01} to brane
inflation~\cite{kklt03,kklmmt03,baumann06,baumann08,baumann208,dewolfe08,baumann10}
to mechanisms of spontaneous supersymmetry
breaking~\cite{kachpearson01,dewolfelinde04,browndW09,benini09} and to
Standard Model  extensions~\cite{cascales04}. 

The gravitational infrared shift effect in warped throats gives the
ability to  bind      strings  whose  mass spectra  are  
set  by the local  string   scale, as in     Randall-Sundrum
(two- and one-brane) models~\cite{rs9905,rs9906}.
The massive string states  above zero modes  are naturally 
likened to  the   massive  particle modes
descending   from   dimensional reduction of the 10-d
supergravity theory.  
Their  description can then  be developed by means of the  widely documented 
Kaluza-Klein  approach on  which there  exist  reviews of general
 character~\cite{salamdee81,appel87,overduin97} and ones    
 specialized to  supergravity~\cite{vannieuw84,duffpope86,bailin87}.  
It should  be  noted, however,  that  additional layers of complexity arise
in  warped compactifications from      the 
non-dynamical   constraints  imposed by local symmetries 
and the choice of classical fields ansatz that decouple  moduli   from
Kaluza-Klein
modes~\cite{gidd05,dougba07,dougtorr08,shiu08,frey08,douglas09,dealvisflux03,freyroberts13}. 

Our study   relies heavily on the  analysis  in~\cite{kiroman84}
of  type  $II\ b$ supergravity for  Freund-Rubin type  flux
compactification   on $ AdS _5 \times S^5$.
This is part of a  wider family of supersymmetric   models~\cite{romans85}  obtained by replacing
the transversal  base   manifold $S^5$ by  a suitable  5-d    coset space. 
The 4-d  compactification  on the conifold~\cite{candelas90},      
on which we focus in this work,  asymptotes       
$ AdS _5  \times T^{1,1} $ spacetime. 
The stability  properties  of  general non-supersymmetric
compactifications on  $ AdS _p \times M_q$   spacetimes  was also
examined  in~\cite{dewolfemitra01}  by  means of methods extending
those of~\cite{kiroman84}.   A  large list  of  4-d   flux 
compactifications  utilizing 
complex threefolds with  toric singularities asymptotic  to 
 $ AdS _5 \times X_5 $     spacetimes has  also appeared in the
meantime~\cite{martelli04}.

One   common feature  shared  by the
conic  threefolds is the preservation  of  continuous isometries   along    
the spatial sections  transversal to the  radial direction. 
Although   only discrete  symmetry subgroups 
survive  once   these  manifolds are  embedded 
in a     Calabi-Yau  compactification, the 
approximate  continuous  symmetry is still useful in classifying 
the  normal modes    for  fields in $ AdS _5$ associated to 
harmonics   of the  compact base manifold.   This   is  also 
especially  tailored    in applying  the   AdS/CFT duality
correspondence to   deformations  of the   background  solution. 

Our  goal   is to  apply the   Kaluza-Klein theory 
to 4-d flux compactifications of type $II\ b $
supergravity on $ M_{10} = M_4 \times X_6 $ spacetimes  with
an attached strongly warped region  modeled 
by  the warped deformed conifold.  The resulting
spacetime  asymptotic to $AdS_5 \times T^{1,1} $  admits
Minkowski spacetime $M_4$    as the time   boundary  of
a  conformal   compactification   of $AdS_5 $
at radial coordinate distances $ r = 1/z \to
\infty$   of   the  Poincar\'e  patch. The dimensional reduction is
carried out by  decomposing the 
supergravity fields on harmonic functions of the  conifold
compact base~\cite{ceresole99,ceresoII99,ceresoIII99}. The 
field theory  on $AdS_5 $ consists  of
massless modes separated  
from towers of scalar, vector and 2-tensor massive
modes  by finite   mass gaps of  size set by the inverse curvature
radius $1/ \calr $.   The   subsequent decomposition on plane waves in $M_4 =
\dh (AdS_5) $    yields  additional towers of radially excited
normal modes.  The reduced  4-d $\caln =1 $ supersymmetric 
field theory is approximately symmetric under the superconformal
supergroup $SU(2,2 |1) $ and the conifold isometry group $ SU(2)
\times SU(2) \times U(1)_R  \times  U(1)_b$.

We  shall develop  concrete calculations in  the   hard wall
approximation  of  $AdS_5 \times  T^{1,1} $ spacetime   for a  slice   of $AdS_5
$    ending at   flat   spacetime boundaries,   similarly  to  the   earlier works~\cite{kofman05,berndsen07,kofman08,freydacline09}.  
The wave functions and mass spectra of harmonic modes in  $ T^{1,1}$ with   radial excitations in $ M_4$  are  defined in terms of     Sturm-Liouville  type eigenvalue problems  given by second order differential equations along the radial direction of $AdS_5$ with appropriate boundary conditions.   All  
wave equations   are solved  in terms of   Bessel functions.  
Along with the massive graviton modes in $M_4$, we consider  scalar,
vector and 2-form massive modes and calculate the mass spectra and
interactions for a set of singlet and charged modes under the isometry
group of $ T^{1,1}$.  The background deformations by compactification
effects are  described within a perturbative approach invoking the
AdS/CFT correspondence.  
 
The information   gathered in the present work  
motivates us to reexamine the  impact  of  warped modes   on the early universe
cosmology.
The fate of massive metastable particles   produced  during 
compactification  confronts us  with the  universe overclosure
paradigm~\cite{bailin87}.   Two   examples 
are  the pyrgon~\cite{kolbky84}  and  
crypton~\cite{benakli98} candidates.  We   start with a    
review   of the  energy transfer
mechanisms    after    $D3-\bar D3$-brane inflation. 
We  next  use  standard   methods to discuss  the thermal evolution of  the
produced gas of weakly interacting  metastable   warped modes  and  its 
ability       to accomplish  the   post-inflation reheating of the  universe~\cite{barnaby04,kofman05,chialva05,firouz06}.  Barring the
threat that stringy effects  come into play~\cite{frey05}, it is
safe to  utilize  a  field theory   framework 
in  discussing the decay and pair
annihilation reactions occurring within    the inflationary
throat~\cite{chen06,kofman08}   and  the  tunneling processes
transferring warped modes to distant throats  hosting the Standard Model~\cite{chialva05,langfelder06,harlinghn07,harling08,halterRE09,chialva12}.

The graviton spin 2   modes  on which we   mostly  focus  
benefit  from universal  features.   However, as    shown in      previous studies~\cite{berndsen07},
other components of the  supergravity  multiplet give rise to 
lighter   warped modes   of spin $0, \ 1 $, charged under
the isometries.  The   lightest charged  Kaluza-Klein
particles (LCKPs)  forbidden    to  decay   to massless modes   
constitute    natural   candidates    for  a left over
population of     cold dark matter relics~\cite{kofman05,chen06,kofman08}.    
If the      throat isometry breaking      from  compactification effects 
is small enough  to   guarantee  metastability,  one could test  the   metastable modes late  decays   through   existing astrophysical  constraints~\cite{berndsen07,kofman08,freydacline09}. 

The contents   of the present  work are organized into 6 sections.  In
Section~\ref{sec2},  we   develop  Kaluza-Klein theory for 4-d flux compactifications 
of type $II \ b $ string theory on a warped target  spacetime $ M_4 \times
X_6$    preserving  $\caln =1$ supersymmetry. 
The dimensional reduction is carried out   for a  conifold throat   
embedded in a    conic Calabi-Yau  orientifold $X_6$.  We   consider   in
Subsection~\ref{subsec2.1}       the
field equations for  bosonic   field  components  of the  supergravity
multiplet,  in Subsection~\ref{subsec2.2} 
the linearized equations  for  field fluctuations  around
the  GKP   background  solution,    in Subsection~\ref{subsec2.3} 
the decompositions  on harmonic    modes 
of the conifold base
$T^{1,1}$  in   the large radial distance regime     where  
spacetime asymptotes $ AdS_5 \times T^{1,1}$,   and in   
Subsection~\ref{subsec2.4} deformations  of  the background
solution  by     compactification effects      within   
a perturbative approach   based on   gauge-string duality~\cite{gandhi11}.

In Section~\ref{sec3},  we  study   the    wave equations for linear field
fluctations  within the hard wall approximation.
Subsection~\ref{sub3sec1}  examines  the
graviton  and scalar  fields,  Subsection~\ref{sub3sec2} the
vector and 2-form  fields and   Subsection~\ref{sub3sec3}   reviews 
main  properties   of   wave functions.  
In Section~\ref{sec4},  we  study  the  warped  modes    interactions.
Subsection~\ref{subsec4.1}  examines local couplings   between
bulk graviton modes,  Subsection~\ref{subsec4.3}    a sample of 
trilinear  couplings between   bulk modes and
Subsection~\ref{subsec4.2}  boundary couplings of gravitons  
with $D3$-branes. In Section~\ref{sec5},  we present  results for  mass spectra 
and local couplings of singlet and charged warped modes 
as a function of  the Calabi-Yau   effective radius $ L $, the 
throat curvature radius  $\calr $ and
the warp factor  parameter  $w$,
with  Planck mass $ M_\star $ setting the energy scale. 
In Section \ref{sec6},  we  study   in the context of  
$D3 -\bar D3 $ brane inflation  the impact
of a gas of massive gravitons in   reheating the universe and
leaving a cold dark matter relic.
Subsection~\ref{subsec6.1} examines  the thermal evolution
and  Subsection~\ref{subsec6.2}    the constraints  from thermal relaxation
and  abundance  in the warped modes  phase. The main conclusions  are 
summarized    in Section~\ref{sec7}.

The  present  paper also includes two appendices dealing with  formal properties
of type $II\ b $ supergravity  flux compactifications   on warped
spacetimes.   In Appendix~\ref{app1}, we review aspects of
differential geometry  and list useful formulas
for the    field equations and  wave equations and the $Dp$-brane action. In Appendix~\ref{app2}, we review algebraic
and harmonic analysis  properties of   the warped deformed conifold.
 
\section{Flux compactifications of type $II \ b $ string theory}
\label{sec2}

We    consider  in this section  
flux compactifications of type $II\ b $ string theory 
on  a  warped spacetime $M_4 \times X_6 $ with an attached 
throat asymptotic  to $AdS_5 \times    X_5 $
spacetime.  Our interest is on  strings localized  deep inside  the  
throat  whose   mass spectra   get redshifted       
below the string  mass scale $ m_s$  by warping.  This   justifies a description  
of the lowest-lying   states     using     supergravity.     
We  focus on classical   backgrounds with   4-d $\caln =1 $ supersymmetry   
consisting  of a  conic Calabi-Yau orientifold $X_6 $   whose
moduli space contains   conifold    singularities at which one can embed  
large stacks of $D3$-branes and/or  3-fluxes sourcing strongly  curved   regions.  
After  discussing  the  vacuum  solution   and the   Kaluza-Klein  
dimensional reduction  of   supergravity   fields,  first on
harmonic modes of $ X_5 $  in  $ AdS_5$  and next on normal modes in $ M_4$, 
we   examine    how  compactification   affects the properties of
warped modes. 

\subsection{Supergravity  theory in warped spacetimes} 
\label{subsec2.1}

Our  discussions  will  mostly  focus  on the  massless supergravity 
multiplet of type $II\ b $ supergravity    in 10-d.  
(Supergravity  theories in  various spacetime 
dimensions are reviewed in~\cite{deWit02}.)    
The      string theory   NSNS and   RR  bosonic sectors include  the  fields  
$ (g _{MN}, \ B_{MN} ,\ \phi ) $ and  $ ( C _0 ,\ C_{MN} ,\ C_{MNPQ}
)$. We  
restrict  consideration to the  classical   action of  
lowest quadratic order in  fields  derivatives in  
Einstein frame 
\bea && S_{10}  =  {1\over 2\kappa ^2 } \int 
[\star R - \ud ( d \phi \wedge \star d \phi + e ^{2\phi } F_1  \wedge
\star  F_1 )  - {g_s \over 2} ( e ^{-\phi } H_3 \wedge \star
H_3 + e ^{\phi } \tilde F_3 \wedge \star \tilde F_3 )  - {g_s ^2 \over
4} ( \tilde F_5 \wedge \star \tilde F_5 +2 C_4 \wedge H_3 \wedge F_3 )
 ] ,  \cr &&  [F_1= d C_0,\  H_3 = d B_2  ,\  F_3 = d C_2,\   F_5 = d
  C_4   ,\ \tilde F_3 = F_3  - C_0   \wedge   H_3 ,\  
\tilde F_5 = F_5 - \ud C_2 \wedge H_3 + \ud B_2 \wedge F_3
  = \star _{10}  \tilde F_5 ,  \cr &&   
 2 \kappa ^2 \equiv  2 \kappa _{10}^2 g_s ^2 = 
(2\pi )^7 \a ^{'4} g_s ^2  = {\hat l _s ^8  g_s ^2
    \over 2\pi }  , \  \hat l _s = 2\pi \sqrt {\a '} = 2\pi  l _s = 
{2\pi  \over m_s} ,\  <e^{\phi } >  = g_s ]    \label{sub2eq0}   \eea 
where $\kappa _{10}^2 $ is the physical gravitational   scale,   independent
of  the string  coupling constant $ g_s$. 
The   special   Einstein frame   that we use is 
obtained  from   the string frame   via  the substitution  $ g_{MN}
\to e^{\phi   /2} g_s ^{-1/2}  g_{MN} $    aimed at  removing  the dilaton   field 
prefactor   in the scalar curvature  term. This choice is   same 
as the string frame, if  the  dilaton field  is constant      
$e ^{\phi}  = g_s$~\cite{polchler00,herzogov01},     and  
differs from  the  standard  Einstein frame,
defined via $ g_{MN} \to e^{\phi /2
}g_{MN}$~\cite{polchb,gkp01},     by  explicit factors of  $ g_s$. 
In   the system of units $\hbar = c =1$  that we  use, all  the basic 
fields are   dimensionless. 
A brief review  of Riemann manifolds, including a summary of  our
notational  conventions,    is  presented in Appendix~\ref{subapp0} 
while  warped  Riemann manifolds   are briefly discussed  in
Appendix~\ref{subapp2}. The  equations of motion and Bianchi
identities   inferred  from  the variational principle and    
gauge   symmetry constraints of  type $II\ b$  supergravity   
are  presented in Appendix~\ref{subapp1}. 

\subsubsection{Background  solution  with  fluxes and branes}

The  GKP vacua~\cite{gkp01} of type $ II\ b $ supergravity 
consist    of a warped product of a 
maximally symmetric  (de Sitter, Minkowski  or  anti-de Sitter)  
spacetime $ M_4$,  of constant positive, zero or negative scalar
curvature $\tilde R ^{(4)} $ (in our  metric signature convention  with
mostly  positive eigenvalues)  times a conic Calabi-Yau threefold
$X_6$.  The 4-d $\caln =1 $  supersymmetry is preserved upon   
specializing to  a Calabi-Yau orientifold. The  symmetry    under the product
of  the orientation twist  by a  $ Z_2$  isometry,   $\O \s
(-1)^{F_s^L} $,   introduces $ O3/O7$-planes. The massless modes  are arranged
initially      into $\caln =2 $ (gravity,  vector, hyper,
tensor) multiplets  
$ G= (g _{\mu \nu }, \  V^0)  \oplus  h ^{2,1}  ( V^\a , S^\a
)\oplus  h ^{1,1}  ( V^A, b^A, C^A, \rho ^A ) \oplus (B_2, C_2, \tau ).
$      The orientifold projection reduces the massless spectrum~\cite{grana03,grimmlouis05}   to the $\caln = 1$  gravity  and  axio-dilaton chiral
supermultiplets from $ g _{\mu \nu } $ and  $\tau $, 
$h^{2,1} _+ (V_\a )  $ vector supermultiplets
from $ C_4 $ and $h^{2,1} _- (S_\a  ) \oplus 2 h ^{1,1} _- (b _a , c_a )
\oplus 2 h^{1,1} _+ (v_a , \rho _a ) $  chiral supermultiplets from $
g_{i j} \oplus (B_2 , C_2) \oplus (g _{i\bar j} , C_4 ) $, where 
 $ h_\pm ^{p,q}$  are Hodge  numbers  for the cohomology    $(p, q ) $-forms even/odd under $\s $.  The
two allowed ($SO$ and $USp$ type)  projections produce 
fixed (non-dynamical) orientifold planes $ O ^{\mp } p , [p=3,7]$ of opposite
sign tensions and RR charges $ \tau ({O^{\mp } p } ) = \mp 2 ^{p-5}
\tau _{p} $ and $\mu ({O^{\mp } p}) = \mp 2 ^{p-5} \mu _{p} $, 
counting branes and their $ Z _2 $ images separately. (The 
anti-orientifold planes $ \bar O ^{\mp } p$ have same    tensions  but
opposite  sign  charges. Another $ Z_2$ charge also arises from the NSNS sector.) When the $ N\ Dp$-brane stack is on top of an
$ O^\mp p$-plane, the gauge symmetry is enhanced to $ SO(2N) $ or $
USp (2N) $.  

The warped throat  is sourced by RR and NSNS 3-fluxes,  $\int _{A
} F_3 = \hat l _s ^2 M , \ \int _{B} H_3 = - \hat l _s ^2 K $,  across a
pair of dual cycles $A,\ B$ embedded near the conic singularity of
$X_6$.  The internal space manifold consists then of a bulk  to
which is glued a non-compact conic manifold of 5-d base $ X_5$. 
(In the dual brane picture, 3-fluxes are replaced by
regular $N \ D3 $-brane and fractional $M \ D5$-brane stacks with $
K=N/M$.)  The  classical vacuum  is  described
by the unwarped metric  $ \tilde g _ {mn} $ of the underlying    Ricci flat
manifold $X_6$,  equipped with an antisymmetric tensor $\e _6$ and a  volume
form $ vol (X_6 ) $,  the constant axio-dilaton profile $\tau (y) = i/g_s$
and  the  warp  and 5-form field strength profiles , $ h (y)
\equiv e ^{- 4 A (y)} $ and $\a (y) $.    The corresponding   background  fields are  of similar  form   to that of    $D3$-branes~\cite{polchb},  
\bea && \bullet
\ ds ^2 _{10} = g_{MN} d X^M d X^N = e ^{2 A(y) } d\tilde s ^2 _4 + e
^{-2 A(y) } d\tilde s ^2 _6 ,\ [ d\tilde s ^2_4 =\tilde g_{\mu \nu }
  dx^\mu dx ^\nu ,\ d\tilde s ^2_6 =\tilde g_{mn} dy^m dy ^n,\ h^{-1/2
  } (y) = e ^{2 A(y) } ] \cr && \bullet \ g_s \tilde F_5 = (1 + \star
_{10} ) \calf _5 = d \a (y) \wedge \ \e _4  + e ^{-8 A} i_{ d \a }
\e _6 ,  \ g_s \tilde F_{\mu \nu \rho \s m} = \tilde \e _{\mu
    \nu \rho \s } \dh _m \a (y) ,\ g_s \tilde F_{mnpqr} = e ^{-8 A}
  \tilde \e _{smnpqr} \dh ^{\tilde s} \a (y) ,\cr && 
[\e _4  \equiv   vol (M_4)  =
\sqrt {\tilde g _4} dx ^0 \wedge \cdots \wedge dx ^3 ,\ \e _6  \equiv  vol (X_6)
= \sqrt {\tilde g _6} dy^1 \wedge \cdots \wedge dy^6 ,  \ 
(i_{ w } V ^{(p)} )_{m_1 \cdots m_{m_{p -1 } } } = w^m V^{(p)} _{m m_1 \cdots m_{p
    -1} } ] \label{sub2eq1} \eea 
where $i_{ w }$   denotes  the  interior product   with the vector $
w^m$. Aspects of  the  differential calculus  for
general relativity  on Riemann manifolds  are detailed    in Appendix~\ref{subapp0}  and 
on warped Riemann manifolds  in Appendix~\ref{subapp2}. 
The field equations for   the 
bosonic components of type $II\ b $ supergravity,    
adapted to the background in Eq.~(\ref{sub2eq1}),  are    listed in
Eqs.~(\ref{appeq21}). For convenience,  we also  list
in  Eqs.~(\ref{appeq24}) the equivalent set of   field equations   
for  the    warp profile fields  $  \Phi _\pm  $   and 
imaginary self dual and  anti-self dual (ISD/IASD)  combinations  $
G_\pm  $  of the  complex 3-form  field  strength, 
\bea && \Phi _\pm= e^{4 A } \pm \a ,\   \L = \Phi _+ G_- + \Phi _- G_+
,\ [G_\pm = (\tilde \star _6 \pm i) G_3 ,\ G_3= F_3-\tau H_3 ,\ \tau =
  C_0 +i e ^{-\phi } ] .   \label{sub2eqx3}   \eea 
The background spacetime   may also imbed localized probe branes (with
negligible back-reaction on the geometry).  The 
$Dp$-brane of world volume $ M_{p+1}$,  extended   along $(p+1)$
longitudinal directions $ x^{\hat \mu } $ and  localized in $(9-p) $
transversal directions $y^{\hat m} $, contributes  source terms   in
the field equations   involving the stress energy-momentum
tensor and  RR charge scalar density
\bea && T^{loc} _{\hat \mu
  \hat \nu } ( X) = - T_p g _{\hat \mu \hat \nu } (X) \d  ^{(9-p)}(y)
,\   \rho ^{loc} _p (X)= \d  ^{(9-p)}(y)  ,\cr && 
[\d  ^{(9-p)}(y) =   {\d  ^{(9-p)} (y  ^{\hat m } -
y  ^{\hat m } _0 )  \over \sqrt {g _{9-p } } } ,\ T_p
  = {\tau _p } e ^{(p-3) \phi  /4 } ,\ \tau _p = {\mu _p \over
    g_s^{(p+1)/4 } } ,\ \mu _p = {2\pi \over \hat l_s ^{p+1} } ] \eea
where $\tau _p  $ and $\mu _p $ are  brane tension and RR charge
parameters  and the covariant Dirac  delta-function $ \d  ^{(9-p)}(y)
$  fixes  the world brane  location along the  transversal directions. 
For a  $Dp$-brane   whose  $(p+1)$  world  brane  directions
extend along $ M_4 $ and wrap  around the cycle $\S _{p-3} \subset X_6 $, 
of  coordinates $ (x^\mu \in M_4 ,\ y^m  \in  \S _{p-3} )$, 
the energy-momentum stress tensor  components are
expressed in terms of the covariant   delta-function $\d  ^{(9-p)} $
and the  metric tensors $ g_{\mu \nu } $ and $\Pi ^\S  _{mn}  $  induced
on $ M_4 $ and $\S = \S _{p-3} $,
\bea && T^{loc} _{\mu \nu } = - T_p g _{\mu \nu
} \d ^{(9-p)} (y)  ,\ T^{loc}_{mn } = - T_p \Pi  ^\S _{mn} \d ^{(9-p)}(y).    \eea     
 In the  trace reversed Einstein  field
equation for the Ricci curvature scalar $  R^{(4)} = g^{\mu \nu }
R_{\mu \nu } =  \kappa ^2  \hat T^{loc} $  along $ M_4$, the source term from  $Dp$-branes enter through the  linear combination  of  stress tensor  components 
\bea && \hat T
^{loc} (y) \equiv - \ud g^{\mu \nu } ( T_{\mu \nu } - {1\over 8 }
g_{\mu \nu } T ^M_M )^{loc} = {1\over 4 } (T^m _m - T _\mu ^\mu
)^{loc} = { 7-p \over 4} T_p \d ^{(9-p)}(y).  \eea
For $Dp$-branes wrapped on  homology $(p-3)$-cycles   that 
support   gauge or gravitational (flux or instanton) bundles, the
string and sigma-model radiative corrections to the Dirac-Born-infeld
(DBI) and Chern-Simons (CS)  parts of the    action  produce, upon integrating 
 over the wrapped cycles  volume,  contributions  to  
the effective $D3$-brane tension and charge 
 \bea &&   \d S (Dp) = \int _{M_4\times \S _{p-3}}  (-T_p k _p (R,F) + 
  \mu _p f _p(R,F) ) \simeq \int _{M_4} (-T _3   k_3 ^{Dp}  +\mu _3   
f_3 ^{Dp} ) ,\cr &&    [\int _{\S _{p-3}} k_p (R,F) \simeq  (\hat l_s
  g_s ^{1/4} ) ^{p-3}
  k_3 ^{Dp} ,\  \int _{\S _{p-3}} f _p (R,F) \simeq \hat l_s ^{p-3}
  f_3 ^{Dp} ] \eea 
where $ k _p( F, R)  $ and $ g _p(F, R) $ are  some functionals of the
spacetime and gauge field strengths $R$ and  $F$ whose volume 
integrals    are of order  $ Vol (\S ) \to \hat l_s ^{p-3 } .$
Upon taking  the volume  integral over  $X_6$ of 
the effective  $D3$-brane tension  and charge  density, 
$ \hat T ^{loc} = T _3 k_3 ^{Dp} \d ^{(6)} (y)  ,\  
\rho _3^{loc} = \mu _3 f_3 ^{Dp} \d ^{(6)}  (y),\ [\d ^{(6)}  (y)= {\d ^{(6)}
    (y - y_0) \over \sqrt {g _6} } ]$
one can define  the effective  $D3$-brane mass and    charge  as,  
$ \int _{X_6} \hat T ^{loc} = T_3  M_3 , \ 
\int _{X_6}  \rho _3 ^{loc} = Q_3 , \ [ \int _{X_6} =   \int  d ^6 y
  \sqrt {g_6} ] . $  For illustration,  
the $D 7$-brane of world volume   $ M_4 \times \S _4 $    
acquires   from  the  radiative corrections  to the  string (tree level)  disc
amplitudes~\cite{bachasgreen99} and    one-loop anomalous
amplitudes~\cite{harvey99,morales98}
 contributions  to the  effective action in string frame   of  $ O(l_s ^4)$  
\bea && \d S (D7) = - {\mu _7
  \hat l_s ^4 } \int _{M_4} [e ^{-\phi } vol (M_4) \wedge ({ \int
    _{\S _4} tr( R \wedge \star R ) \over 4 ! \ 32 \pi ^2}) - C_4 \wedge ({
    \int _{\S _4}  tr( R \wedge R ) \over 4 ! \ 32 \pi ^2 }) ] , \eea
which  induce   finite $D3$-brane tension and charge terms if  $ \S _4 $ 
supports  a gravitational instanton.  For  instance, if   
the 4-cycle $ \S _4 \sim K3 $ embeds   a single instanton  
that sets    the values of the volume integrals in the
factors inside parentheses    to $ \pm 1 $,  the  $D7$-brane  acquires
the effective $ D3 $- or $\bar D3 $-brane source terms, 
$\d S(D7 /\bar D7)= - \mu _3 \int _{M_4} (e ^{-\phi } \e _4 \mp C_4 )$.

In the   construction of vacuum solutions   including localized  sources, 
an  important r\^ole  is played  by probe  $ Dp$-branes of 
effective $D3$-brane tension and charge density  that satisfy    the BPS type   inequality,  
\bea && g_s \hat T ^{loc}(y) \geq {\mu _3} \rho
_3 ^{loc} (y) \ \Longrightarrow \ 
\int d^6 y \sqrt {\tilde g_6} (g _s  \hat T ^{loc}(y)  -\mu _3 \rho _3 ^{loc}
  (y) )    = \mu _3 (M_3 - Q_3) \geq 0 .\label{sub2eq3} \eea  
The  above relation between  the total   mass $M_3 $    and  charge    $
Q _3 $   is  analogous    to   the Bogomolny-Prasad-Sommerfeld
(BPS)   upper    bound    on      mass spectra in topological 
sectors  of  finite central  charges   preserving a  fraction of
the  bulk supersymmetry.    As  discussed in~\cite{gkp01}, 
the BPS inequality in Eq.~(\ref{sub2eq3}) is   saturated   by $D3$-branes or $O3$-planes (where both sides are equal to $\mu
_3 $ or $-\mu _3 /4$) and by $D7$-branes wrapped on a $ K3$
sub-manifold of $ X_6$, once one takes the radiative and anomalous
corrections discussed above  into account.  The inequality is 
satisfied by $\bar D3$-branes (in the form $ \mu _3 \geq -\mu _3 $)
and  by  $D5$-branes wrapped on collapsed 2-cycles of $X_6$ (in the
form $\mu _3 / 2 \geq 0 $), but it is violated  by $\bar O3 $-planes 
(since $M_3 = -\mu _3 /4 ,\ Q_3 = + \mu _3 /4 $).

Two  important properties of GKP vacua~\cite{gkp01}  are 
revealed  by  the   classical equations  for the field profiles $\Phi _\pm $, 
\bea && \tilde \nabla ^2 _6 \Phi _\mp ={e ^{8 A} \over 24
  \tau _2 } G_\mp \tilde \cdot \bar G_\mp + \tilde R ^{(4)}+ e ^{-4 A}
\nabla _m \Phi _\mp \nabla ^{\tilde m} \Phi _\mp + {2 \kappa ^2 \over
  g_s } e^{2A} (\hat T ^{loc} \mp \tau _3 \rho _3^{loc} ), \label{sub2eq3p} \eea  
obtained in Eq.~(\ref{appeq24})     by  combining 
Einstein field  equation for $ R_{  \mu \nu }$ and Bianchi field
equation for $F_5$.  Consider  first the equation for the warp
profile,   given by the linear combination $ \tilde \nabla ^2 e ^{4 A
  (y)} \equiv \tilde \nabla ^2 (\Phi _+ + \Phi _-) /2 $.  
If the uncompactified spacetime $M_4$ is Minkowski or de Sitter
($ \tilde R ^{(4)} \geq 0 $),   the    flux and curvature  bulk  terms   on the right hand side,  including   the last term for localized source branes,
are all  non-negative.
(This is still valid  for anti-de Sitter  spacetimes  with 
absolute value of the curvature negligible compared to the  3-fluxes  contribution $\vert \tilde R ^{(4)}  \vert << \vert G \vert ^2 $.)      
Since the volume integral of the Laplacian  on
the left hand side  vanishes for a compact $X_6$,  one   concludes that the
warp  profile  must    vanish, $ e ^{A(y)} =1 $, 
unless one introduces in the background unusual source terms of
negative tension, $ 
2 \kappa ^2 _{10} \mu _3 \hat T ^{loc} = \hat l_s ^4 \hat T ^{loc} < 0
$.   The  means to circumvent   this  no-go condition on
warped  (or on unwarped de Sitter)  vacua    is    available in string theory 
compactifications  on orientifolds  with $SO$ type
projection producing   $Op $-planes of negative tension $ \tau
_{Op}= - 2 ^{p-5} \tau _p $    and    negative charge  $\mu _{Op}=
- 2 ^{p-5} \mu _p $.  

The  second  property is derived by taking the volume integral  of   Eq.~(\ref{sub2eq3p}) for   $\Phi _-  $ or $\Phi _+  $ over a
compact manifold $ X_6$     for a Minkowski or anti-de Sitter
spacetime $M_4$ ($ \tilde R ^{(4)} \leq 0 $).   In either cases, matching
the net non-negative  contributions from  the (first and third) bulk terms 
on the right hand sides  to  the vanishing integral on
the left hand sides    requires localized sources (fourth term) 
that satisfy     the     BPS inequality 
\bea && g_s \hat T ^{loc} (y) \mp \mu
_3 \rho _3 ^{loc} (y) \geq 0 \ \Longrightarrow \ M _3 
\mp Q_3 \geq 0 .  \eea    The  upper sign case refers to the  backgrounds  with
positive 3-fluxes  of the type      considered so far, 
while the lower sign case refers to   the conjugate type backgrounds with
negative 3-fluxes $ M <0 \ \Longrightarrow \ N= MK <0$. (The    
corresponding  dual   gauge theories    are  supported by   $ N\ D3
$-brane  and  $ \vert N\vert \ \bar D3 $-brane stacks  
of  charge densities $ \pm \vert \rho _3\vert  $.)  
These   two  cases  are related  by  trading the profiles 
$\Phi _\mp \to  \Phi _\pm $,  the   complex  3-form fields $  G_\mp
\to  G_\pm  $   and exchanging probe $D3$-branes   for $\bar
D3$-brane $ \rho _3   ^{loc} \leftrightarrow  - \rho _3  ^{loc}
$. (This is  consistent   with the  mnemotechnic  
rule that $  D3$-branes source 
$\Phi _+$ and $\bar D3$-branes source $\Phi _-$, while $ D3$-branes feel the
potential $\Phi _-$ and $\bar D3$-branes feel $\Phi _+$.)    
The constraints  imposed   in    Eqs.~(\ref{sub2eq3p})  for    $\Phi
_\mp $, respectively,   are solved by the   families of  GKP backgrounds
preserving  4-d $\caln =2 $ supersymmetry (reduced to $\caln =1$ by
the orientifold projection) characterized   by 
imaginary self dual or anti-self dual (ISD/IASD) 3-fluxes  and   
probe $D3$- or $\bar D3$-branes saturating the BPS inequalities   
\bea && \Phi _\mp =0,
\ G_\mp =0,\ g_s \hat T ^{loc} \mp \mu _3 \rho ^{loc} _3 =0 ,\ [\hat T
  ^{loc} (y) = {1\over 4} (T^m_m- T^\mu_\mu ) ^{loc}
]. \label{sub2eq4} \eea

We  now  discuss the   constraint on  3-fluxes and
$D$-branes   derived from the condition that the $ C_4$ potential  does not develop a   tadpole
VEV   when the   embedding  manifold $ X_6$  is  compact.
Applying Stokes theorem to the
volume integral of the 5-form Bianchi identity  in  Eq.~(\ref{appeq2})
over $ X_6$    
\bea && 0= \int _{X_6} d \star \tilde F_5 \equiv \int
d^6 y \sqrt {\tilde g_6} {1\over 5!} \tilde \e ^{m_1 \cdots m_6} \dh
_{m_1} \tilde F _{m_2 \cdots m_6} = - \int _{X_6} F_3 \wedge H_3 +
{2\kappa ^2 \tau _3 \over g_s} \int d^6 y \sqrt {\tilde g_6} \rho
^{loc} _3 = {2\kappa _{10} ^2 \mu _3}(N_{flux}  + Q_3 ) \cr && 
\Longrightarrow 0= N_{flux} + N_{D3} -   N_{\bar D3} -{ N_{O3} \over 4} +Q
^{D7} _3 ,\ \  [N_{flux} = {1\over   \hat l_s ^4} \int _{X_6} H_3
  \wedge F_3 = MK ]  \eea 
shows  that the net  total  RR $D3$-charge from 3-fluxes and from $ N_{O3}\ O3^-$-planes  and
$ (N_{D3} D3 + N_{\bar D3} \bar D3 + N_7 \ D7 ) $-branes must
vanish.  For supersymmetric vacua with 3-fluxes  of positive charge, 
$ M> 0 $   and  $N_{\bar D3} =0 $,  the compensating   $D3$
charge     of negative    sign can arise     from   $O3 ^-$-planes or from the 
effective   charge   $Q ^{D7} _3$ of  $D7$-branes wrapped  on homology
cycles.  

 F-theory  compactifications  on   elliptically
fibred Calabi-Yau fourfolds~\cite{vafa96,beasley08}   provide a   wide
 family of  solutions     generalizing   the
 orientifold   vacua.    The warped flux compactifications  
in~\cite{becker96,verlinde99,dasgupta99,greene00,gubserF00} 
are constructed  from    M-theory compactified on elliptically fibred
Calabi-Yau fourfolds $X_8$,   involving   $T^2$-tori bundles over Calabi-Yau
threefold  base manifolds $X_6$~\cite{denef08}.   The starting point is
 11-d supergravity on $T^2 \times M_9,\ [T^2 = S^{1} (r_{11}
  ) \times S^{1} (r_{9}) ]$ 
for a $ T^2$-torus of  complex  coordinate $ z = x +\tau y  $, complex
structure   modulus $ \tau = \tau_1 +i \tau _2$ and  volume $v
= (2\pi )^2r_{11}r_{9} $. This is   connected to   type $II \ a $
theory   on $  S^{1} (r_{9})\times    M_9 $, 
through  the  familiar correspondence   between the  respective   metrics
\bea && ds ^2 _M \equiv {v \over \tau _2 } ( (dx + \tau _1 d y ) ^2 + \tau
_2 d y^2 ) + ds ^2 _9 = \call  ^2 e ^{4\phi /3} (dx + C_1) ^2 + e ^{-2 \phi
  /3} d s ^2 _{II a} \cr && \Longrightarrow \ C_1 = \tau _1 d y ,\ e
^{4\phi /3} = {v \over \call  ^2 \tau _2 } ,\ d s ^2 _{II a} = {\sqrt v
  \over \call  \sqrt {\tau _2} } (v \tau _2 d y ^2 + ds ^2 _9 )
. \label{eqMIIa} \eea 
Applying  $\calT $-duality   along $ S^1 (r_{9}) :\ g _s \to g _s l_s / r_9
,\ r_9 \to l_s ^2 / r_9 $,     transforms this to  type  $II \ b$
theory  with Einstein frame  metric
\bea && d s ^{ E 2} _{II b} = {1\over  g _s ^{1/2} }  d s ^2 _{II b} = {
  \sqrt v \over \call } ( {\call ^2 (l ^M_s )^4 \over v^2 } d y ^2 + ds ^2 _9
) ,\ [ l_s ^M = l_s g_s^{-1/3} ] .  \label{eqMIIb} \eea
We  have  denoted   the string theory scale in M-theory metric  by $ l_s ^M $ 
and retained    the  standard notation $l_s $  for  the   string    scale in 
type  $II \ b $  theory   metric. This  coincides    with    the 
convention in~\cite{dealvis96,morrison04}, identifying   the fundamental
length scales of M- and type $II $-theories,   $ l_s = l_M
,\ [ ( 2 \kappa ^2 _{11} )^{-1} = 2\pi / \hat l_M^{9}]$  but  differs from the   convention    of~\cite{denef08} and
other studies   where one identifies  the M-theory
scale  parameter    $ l_M$   with $ l_s ^M $ instead. 
If the arbitrary scale parameter $\call $  in  the metric 
Eq.~(\ref{eqMIIa}) is now set at $ \call  = \sqrt v $ and the
limit $ v \to 0 $  of a shrinking torus  is taken at fixed $ l_s ^M $, the
direction $y$ decompactifies  and the  resulting 
metric  reduces to that of  the  flat background spacetime $M_9 \times
S^1(r_9) \to M_{10} $.    

The dynamical  implications are  inferred by   considering
the    11-d bosonic  supergravity action,  including the $ O(l_s ^8) $  
gravitational couplings~\cite{vafawitten95,duffliumina95} and  the
contribution from probe $ M2$-branes sourcing the potential $C_3$,
\bea && S_{11} = 
{2\pi \over \hat l_M^9} \int _{M_{11}} [ \star _{11} R - \ud (G_4 \wedge
  \star _{11}  G_4 + {1\over 3} C_3 \wedge G_4 \wedge G_4 ) + \hat l_M^6 C_3
  \wedge X_8 (R) - \hat l_M^6 \sum _{i=1} ^ {N_{ M2 } } \d ^{(8)} (y
  -y_i^{(0)} ) C_3 ] , \cr && [X_8 (R)= {1\over (2\pi )^4 192} (tr R^4
  - {1\over 4} (tr R^2 )^2 ) ,\ G_4 \equiv d C_3 ,\ \hat l_s^M = 2  
\pi  l_s^M  ,\   \hat l_M = 2 \pi  l_M  ] . \label{eqM11}  \eea
Matching    onto   the reduced supergravity actions of type $ II \ a $
and $b $ theories is accomplished  through      the following  correspondence
between  fields  and parameters
\bea && \bullet \ II\ a:\ d s^2 _M = g_s ^{4/3} d x_{11}^2  + g_s ^{-2/3} d
s^2 _{II a}  ,\  C_3  \to R_{11}  B_2 ,\  {2\pi  \over \hat l_M^3 } 
\int C_3 \wedge  X_8 \to  {2\pi  \over \hat l_s^2}  \int 
B_2  \wedge  X_8 , \cr &&
[{R_{11} \over 2 \kappa ^2_{11} } \equiv { 2\pi r_{11} \over
  \hat l_s ^9 }= {2 \pi  g_s ^{2/3} \over \hat l_s ^{8} } ={2 \pi \over
  (\hat l_s ^{M })^8 g_s ^2} , \ r_{11}={l_s ^3 \over (l_s^M)^2 } =
  l_s g_s ^{2/3} = l_s^M g_s,\  \hat  l_s^M = 2\pi   l_s^M ]. \cr &&
\bullet \  II\ b:\ C_1 \to C_0 =\tau _1
,\ \tau _2 = {r_9\over r_{11} } \to e ^{-\phi } ,\ C _{mnx} \to
B_{mn},\ C_{mny} \to C_{mn},\cr && C_{mxy} \to g_{my},\ C_{\mu \nu \l
} \to C _{\mu \nu \l \rho } ,\ G_4 \to {i \over 2 \tau _2} (G_3\wedge
d \bar z - \bar G_3 \wedge d z) = (F _3 \wedge dy+ H _3 \wedge dx )
. \eea 
The  M-theory $M2$-branes of tension  $ T_2 ^M = 2\pi / \hat l_M^3  $,  wrapping the circles $ (S^1 _{11} , \ S^1 _{9} ) $,  give rise to  $  (F1,\ D3 ) $-branes of type $ II\ b$  theory   while the
$ M_5  $-branes of tension  $T_5 ^M = 2\pi / \hat l_M^6 ,\ [2 \kappa ^2 _{11} =
  \hat l_M^9 / (2\pi ) =  2\pi  /( T_2 ^M T_5 ^M) ]$,  wrapping the circles $ (S^1 _{11} , \ S^1 _{9} ) $    give rise to   $( D4 ,\ D6) $-branes of type $ II\  a $ theory. (The     resulting branes tensions refer to the scale $ l^M_s $  of  the M-theory metric.)  
The lifting  from string theory to F-theory  compactification  associates the 
axio-dilaton field $\tau (y)$ to the complex structure modulus of the
$T^2$ fibre and the complex 3-form field $ G_3$ to the real 4-form
field $G_4$.  The singular (degeneration) loci in $X_6$,  where the
$T^2$ fibre collapses to 1-cycles, are (complex codimension unity)
4-cycles $ \S _4 $ of $X_6$  that support $7$-brane solitons of
world volume  $ \S _4 \times R^{3,1}$ (in the limit $v \to
0$). The $(H_3,\ F_3 )$ fluxes, threading  with
integer  winding numbers $ (p,q)$
the  inequivalent 1-cycles of $T^2$    wrapped  by $ 7$-branes,
induce NSNS and RR charges    on  the corresponding   $ (p,q)\ 7$-branes. 
The distinct     vacua   for the axio-dilaton and
3-form fields     are then  arranged  into 
equivalence classes   under   the  action of the modular 
group   $ SL(2,Z)$ that realizes     the  unbroken   
duality symmetry of type $II\ b$ theory, 
  \bea && \tau \to {a \tau + b \over c \tau + d } ,\ {H_3 \choose F_3}
  \to \pmatrix{d & c  
  \cr b & a} {H_3\choose F_3} , \ [G _3 \to {G _3 \over c \tau + d}
  ,\ a,b, c, d \in Z ,\ ad -bc =1] . \eea 
The   backgrounds   are  thus     described by 
multiple valued axio-dilaton and 3-form field strengths
 that  jump by monodromies $ M_{p,q} \in SL(2,Z) $ upon
circling  a degeneration point in $X_6$ where a $ (p, q)
\ 7$-brane is located.  The 4-fluxes are  also   interpreted as
domain walls from $ M5$-branes wrapped on $\S _4 \in X_6$.   In
addition to the 3-fluxes of trivial monodromy supported by 3-cycles
$\S _3 \in X_6$,   one can also  consider  3-fluxes of non-trivial
monodromy $ G_3 \sim F_2 \wedge d\tau /\tau _2 $ supported by magnetic
2-fluxes $ F_2$  on $D7$-branes wrapped on 4-cycles of $X_8$.  
The  cancellation  of tadpoles of the $C_3$ potential   
is ensured by  invoking Stokes theorem  for  the volume integral
of  the  field equation in Eq.~(\ref{eqM11})  
over the compact fourfold $ X_8$ 
\bea && 0= \int _{X_8} d \star _{11} G_4 =
{1\over 2 \hat l_M^6 } \int _{X_8} G_4 \wedge G_4 + N_{ M2 } - {\chi (X_8)
  \over 24} \cr && \Longrightarrow \ 0 = N _{flux}  + N _{D3} - N _{\bar
  D3} - {\chi (X_8) \over 24} ,\ [\int _{X_8} X_8 (R)= + {\chi
    (X_8)\over 24} = {1\over 24}\sum _{p=0} ^4 (-1)^{p} b_p ] \eea
where $\chi (X_8) $ denotes the Euler character  
and $ X_8 (R) $  the  Euler  characteristic class of $X_8$. We  have
utilized      the  relation  $ N _{flux}  = MK$  between 4-   and
3-fluxes   and     $ N _{D3} - N _{\bar D3}\sim N_{ M2 } $
between $M2$-branes  and $D3$-branes wrapped on $S^1 (r_{11}) $.
If a single $D7$-brane is introduced per  4-cycle  homology class   of $X_8$,  
the  resulting  effective  $D3$-charge $ Q_{3} ^{D7}= - \chi (X_8) / 24 $
would  then play   the same  r\^ole  as  the negative charge of $ O3^-$-planes in type $ II \ b$ orientifolds   in compensating the   large
$ C_3$-charge  $ N _{flux}  $  from  4-fluxes  sourcing warped    throats.
The   large  (positive)  effective  charge  from $7$-branes 
wrapping   homology 4-cycles of $X_8$  require    complex fourfolds of large     (positive) Euler  character.
Two    examples of complex  fourfolds   toric  varieties 
with Euler characters $16 848  $ and $13 248 $ are   constructed    in~\cite{deneflorea04}.    
For instance, in  the    case of complex fourfolds
built  from  pairs of $ K3$  manifolds,  the character     evaluates
to~\cite{greene00}, $\chi ( K3 \times K3 ) = 24 ^2 ,\  
[\int _{K3} X_8 (R) =  (R\wedge R) /(16\pi ^2)  ,\
\int _{K3} tr (R\wedge  R)/(16\pi ^2) = 24 ] .$ 

\subsubsection{Effective scalar potential}

 The scalar potential  is  that  part of the quantum 
effective action  of  a field theory   
related to its  functional path  integral  by  a Legendre transformation
with respect   to    sources  terms  added to  the action.  
The potential $V( \phi _i ) $  represents  the  energy density
functional in the  vacuum expectation values (VEVs) 
of   (elementary or composite) bosonic fields  $ \phi _i  $. 
Beyond   the  dominant classsical part,  it 
acquires   quantum   corrections   that  can be evaluated in
perturbation theory.   The   vacuum solutions  
arise as local minimas  in static backgrounds, but
one can also use   the potential    to   track  down  the flow
of classical  field  solutions   in  time-dependent backgrounds.  
For string theories compactified on  target spacetimes $ M_D =M_d
\times X_k ,\ [D= d+k] $   of effective action $ S_D$,  the   classical gravitational potential
 is a functional of  the (metric, branes, fluxes)
pseudo-moduli and massive modes    fields   defined   by the  
integrand of  the  effective action,
$\G  ( \phi _i )  = - \int d^4 x \sqrt {-\tilde g_4}  V  (\phi _i )$.
  In flux compactifications, the  contributions from gravitational interactions     and     localized sources yield typically   discretum (landscape) of  vacuum solutions parameterized by
quantized fluxes across cycles of
the internal  manifold that span  multi-dimensional discrete vector
spaces.    The  potential  in warped compactifications   
on  a spacetime with  external and internal  parts $M_d ,\ X_k$  
involving    warp and conformal profiles $A $ and $B$, defined as in Eq.~(\ref{appeq5}), 
\bea &&  d s ^2 _{D} = e ^{2 A(y)} \tilde g_{\mu \nu} d x^\mu d x ^\nu
+ e ^{2 B(y)} \tilde g_{ mn} d y ^m d y ^n  , \eea  
was examined       in  several
works~\cite{buchel03,dealvisflux03,gidd05,douglas09}   for  
maximally   symmetric (Einstein)   non-compact    spacetimes   $ M_d$
of constant scalar curvature $ \tilde R^{(d)} $ with respect to  the unwarped metric $\tilde g_{\mu \nu} .$
We      shall  mostly  review here   the approach
of~\cite{douglas09}  which   focuses    on  the  warp  factor $ A(y)$
  for an arbitrary conformal factor $ B(y)$    absorbed
inside the   warped  metric  $g_{mn} = e ^{2 B(y)} \tilde g_{mn}
$. (The supersymmetry   preserving    vacua of type $ II\ b $
supergravity  require $ A= - B$.) 
The gravitational  theory for the  metric and  moduli fields $ u ^I (x) $
in $ M_d$  is  assumed to be described by an effective action $ S _d $
including    curvature, moduli    kinetic energy  
and   scalar potential terms, 
\bea && S_d (g _{\mu \nu } , u ^I ) = \int d ^{d} x \sqrt {- \tilde
  g_d } ( {M_ \star ^2 \over 2}\tilde R ^{(d)}  -\ud G_{IJ} \dh
  _\mu u ^I \dh ^\mu u ^J -   2 V  ).  \label{MD0p}  \eea    
 The potential $ 2  V  $  (where the   numerical factor   $2 $  here  is conventional) subsumes the  effects of integrating out  massive  pseudo-moduli and Kaluza-Klein  modes of the underlying $D$-dimensional theory  whose action
\bea && S_D \equiv  \int _{M_d} d ^d x \sqrt {- \tilde g_d }S_k(g_{\mu \nu }
,g_{mn} )  =  {1\over 2 \kappa ^2 } \int _{M_d} d ^d x \sqrt {- \tilde g_d }  \int _{X_k} d^ky \sqrt { g_k } (R^{(D)} -\ud \sum _p \vert F^{(p) } \vert ^2 +  2
\kappa ^2  L_m ) ,\label{Action1}   \eea  
includes   curvature  and  $p$-form flux
terms $ F^{(p)} $ along  with   matter  and  localized source
terms in $ L_m$.     It is convenient to  consider  magnetic type fluxes    
 only, with       $ F_{0 m_1\cdots  m_{p-1} } =0 ,\  F_{m_1\cdots  m_{p} }  \ne 0 $, 
and   to restrict to  a non-dynamical  Lagrangian   $ L_m$  
satisfying   the condition, $ \d L_m /\d g^{\mu \nu } =0 \ \Longrightarrow \  
T_{\mu \nu } = g_{\mu \nu } L_m \ \Longrightarrow \  
L_m = T ^\mu _\mu / d \equiv T ^{(d)} / d $.  
Both  $T^{(d)} $ and $ F_{mn \cdots  }  $  are then independent of  $ A(y)$.
 Since the effective action   $   S _d  $  is   required  to reproduce the  
Einstein equations   derived  from  the  action $S_D $, it is natural 
to  identify the classical potential $V$  to the   action    $ S_k$ in Eq.~(\ref{Action1})   involving   curvature,    flux and  localized sources  contributions integrated     over the  internal   space  volume.    
The dependence of $ R^{(D)} = R ^{(d)}   + R ^{(k)} $
   on $A(y)$   can be    made explicit  
by  means of the   identities  in  Eq.~(\ref{appeq11}) (for $ B\to 0$),  
\bea && R^{(d)} = \tilde R^{(d)} e ^{-2 A} - d (\nabla _k ^2 A + d
(\nabla A )^2 ) ,\  \  \ R^{(k)} = \hat R^{(k)} - d(\nabla _k ^2 A
+(\nabla A )^2 ) ,   \label{eqscalpot} \eea
where $ \hat R ^{(k)} (\tilde  g_{\mu \nu } , g_{mn} )  $
is the scalar curvature with respect to the unwarped 
metric $ \tilde g _{\mu \nu }$  and the warped      metric $
g_{mn}$   including the factor $ B(y)$. (The notation
$ \tilde R ^{(k)} (\tilde g _{\mu \nu } ,\ \tilde g _{mn }  ) $    
is reserved for the curvature  in the unwarped metric of the
underlying  manifold   $ X_k$.)    The resulting  formula for the
internal space action 
\bea &&  S_k(g_{\mu \nu } ,g_{mn} )  = {1\over 2 \kappa ^2 }  
\int _{X_k} d^ky \sqrt {g_k} e ^{d A}
[e ^{-2 A} \tilde R^{(d)} + \hat R^{(k)} - 2 d \nabla _k^2 A - d
  (d+1) (\nabla A )^2  + {2\kappa ^2 \over d}  \hat T^{(d)}] , \cr && 
[\hat T^{(d)} =    T^{(d)}- {d\over 4 \kappa ^2  } \sum _p
\vert F^{(p) } \vert ^2 ,\ T ^{(d)} = g^{\mu \nu } T _{\mu \nu }
= T^\mu _\mu ,\  \nabla _k^2 = (\nabla ^2  _D )_k  ,\
T_{\mu \nu  } = -{2\over \sqrt {- g } } {\dh L_{m} \over \dh g^{\mu \nu }} ]   \eea
includes curvature and warping terms  along  with 
magnetic field  and matter   terms     entering   through the linear
combination  of the  stress energy  tensor $ T_{\mu \nu }$ and
$p$-form fluxes.
Let us now  consider the  path integral   for the
generating functional of  Green's functions
$ Z [J, \Phi ] = e ^{W [J, \Phi ]} $  with the additional   
source term $ \int _x J \phi  = \int _x \tilde R ^{(d)} {M_\star ^2  (A)  } / 2  ,\ [\int _x = \int d ^{d} x \sqrt {- \tilde
     g_d } ] $   involving a Planck mass  parameter depending on  the warp
profile,    with       $\Phi $  standing  for  
all fields   in the action  $S_D (g_{\mu \nu} (x),\Phi  )$ other than $g_{\mu \nu} (x)  $.  One   can then  define the effective action $\G  (A, \Phi )$,    generating functional of  one-particle irreducible  (1PI) amplitudes,   through
the  Legendre transformation,   trading 
 $ \tilde R ^{(d)} $ for  $ M_\star ^2  (A)  $    and  the  
 stationarity conditions  on  $ W$   for those on $\G $, 
 \bea && W  [\tilde  R ^{(d)} , \Phi ] = \int _x {M_\star  ^2  (A) \over 2 }
 \tilde R ^{(d)}    - \G  (A, \Phi ) ,\ [{\d  W \over \d \Phi } = 0,\
 {\d  W \over \d      \tilde R ^{(d)} } =  \ud  {M_\star ^2  (A )} \ \Longleftrightarrow  \  {\d  \G \over   \d \Phi } =0 ,\  {\d  \G \over  \d A (y)  } =0   ] . \eea
The equation ${\d  \G   / \d  g^{mn}  } =0  $  reproduces Einstein
equation   ${\d W  / \d g^{mn}  } =0  $. 
The  tree level (classical)   potential  is evaluated by 
substituting   the identities in Eqs~(\ref{eqscalpot})
for $  R ^{(d)}  ,\ R ^{(k)}  $    in  the defining equation,   
 \bea && 2\int _x V (A, \Phi )= \G ^{(0)}  (A, \Phi )  =
 \int _x    {M_\star ^2  (A )  \over 2} \tilde R ^{(d)}
 - W ^{(0)}  [\tilde  R ^{(d)} , \Phi ] 
 \simeq \int _x  ( {M_\star ^2  (A )  \over 2} \tilde R ^{(d)}  - S_k  ) ,\
[\int _x = \int _{M_d} d ^d x \sqrt {- \tilde g_d }  ]  .    \eea  
The terms  depending    on $\tilde R ^{(d)} $ 
cancel out  from  $ V  $    once     one  invokes the relationship
  between   the    gravitational and    supergravity  mass  scales 
$\kappa _d^2 = 1/ M_\star ^2 $ and $\kappa ^2   $ and   the
internal space warped  volume  $V_W $, 
\bea && V_W / \kappa ^2   = 1/\kappa ^2_d  ,\  [V_W
\equiv \int _{X_k} d^ky \sqrt {g_k} e ^{(d-2) A}  ] .\eea  
To  complete the definition of $  V  $,   it is then  necessary to 
impose the   requisite  condition through    the Lagrange multiplier term 
$ {C \over 2} (\kappa ^2 M_\star ^2 - V_W )$ with  a constant  $C$. The resulting    classical   potential   from   the   constrained    Legendre 
transformation   is 
\bea && 2\kappa ^2 V  = -\ud \int  d^ky 
\sqrt {g _k} e ^{dA}(\hat R ^{(k)} + d(d-1) (\nabla A )^2+ {2 \kappa ^2 \over d}
\hat T ^{(d)} )  + {C\over 2}   ( V_W - \int  d^ky \sqrt {\tilde g _k}  
e ^{(d-2) A(y) }  )  
\cr && = -\ud \int   d^ky \sqrt {g _k} 
[ u^2 ( \hat R ^{(k)} + {2 \kappa ^2  \over d} \hat T ^{(d)}  ) + 4 {d-1 \over
d} (\nabla u )^2 ] +  {C\over 2}   ( V_W - \int  d^ky \sqrt {g _k}  
u ^{2 -4/d} )   . \label{MD1p}  \eea 
In going to the second line  entry in Eq.(\ref{MD1p}), we  have  traded 
the warp profile  for the  convenient   auxiliary function  $u(y) = e
^{d A (y) /2 } $  and   have performed   an
integration by parts  in    the   volume integral 
dropping   the boundary terms   which vanish if $ X_k$ is   compact. 
The stationarity   of the   effective   action with respect to   
variations of   $ u (y)$ yields the
constraint equation for the  warp  profile   $u(y)$,
\bea && 0= {\d  V  \over \d u }\ \Longrightarrow \ 
{d-2\over 2} C u ^{1-4/d} - 2 (d-1) \nabla ^2 (X_k)   u + ({d \over 2} \hat R
^{(k)} + \kappa ^2 \hat T ^{(d)} )  u(y)  = 0,  \eea 
showing that  the constant  $C $  sets   the normalization of the
solution   for  $u(y)$.  The integration by parts in 
$ \d  _u \int   \sqrt {g _k}  (\nabla u )^2 \to   - 2\int  
\sqrt {g _k}  \nabla ^2 (X_k) u  $ has introduced   the $X_k$ manifold
Laplacian $\nabla ^2 (X_k) $,    independent   of $ u (y)$,  which  
must not  be confused    with   $ \nabla _k^2 =
(\nabla _D )_k^2 $.  Specifically,  
$\nabla  ^2 (X_k)  =  \nabla _k^2  -e ^{-2B} d (\nabla _m A) \nabla
^{\tilde m}  =   e ^{-2B} ( \tilde \nabla _k ^2 + (k-2) \nabla
_m B \nabla ^{\tilde m} ) .$      Upon   combining    the  above variational  equation with Eq.~(\ref{MD1p}),   one  obtains the  simple  formula for the  on-shell   potential (extremal value),   
$  V  _{min} = (d-2)M_\star ^2   C  / (4 d)   $. 
Returning  now    to the  truncated  effective action Eq.~(\ref{MD0p}) with the potential  $ V\to  V _{min}$  interpreted as a cosmological term, 
$  T^{(d)} = - {(d-2 )  M_\star ^2  C /  2} ,$  one deduces the  relation $\tilde R ^{(d)}   = C $ from the  reverse trace part of   Einstein equation
\bea &&  {\d S_d\over  \d g _{\mu \nu } } =0,\ [S_d = \int d ^{d} x
  \sqrt {- \tilde 
    g_d } ( {M_ \star ^2 \over 2}\tilde R ^{(d)} -   2 V _{min}   ) ]
\ \Longrightarrow \  \tilde R ^{(d)} = -{2 T^{(d)}  \over  M_\star ^2
  (d-2)} = C. \eea  

Since the  potential   is intrinsically an   off-shell quantity,  it
was important to define it 
without  making use of  Einstein equations along $ M_d$  and $ X_k$.
Indeed, the naive   construction utilizing  the 10-d
supergravity  field equations   to build  the
potential for  the universal volume modulus is known to  
give results  inconsistent with    the  4-d  description~\cite{buchel03}.   The  problem  is traced  to the fact  that the   dependence of fluxes and     $Dp$-branes on   the     fields is  valid only   for  the    on-shell potential.  We see from Eq.~(\ref{MD1p}) 
that internal manifolds of negative curvature $ \hat R^{(k)} $   and orientifold  sources of  negative  energy $\hat T^{(d)} $
give  positive contributions to $V$,
while the warping factor, brane sources and fluxes give  negative
contributions. This property   confirms the familiar fact that the      curvature action of gravitational theories is  exposed   to  destabilizing effects.   For instance, the  path integral of Euclidean
quantum gravity  acquires divergent contributions  from rapidly varying conformal   rescalings of the metric~\cite{gibbhawk91,gibbhawkperry78}.
The    analogous  instability     arises here  through  contributions  unbounded from below to the classical potential   from   conformal
transformations  of the  $ X_k$ metric.  This was   the reason   that
motivated~\cite{douglas09} to  examine whether
the  vacuum  instability from      rescalings $ g_{mn } \to e ^{2 B(y)}  \tilde  g_{mn}$  belonging to  a fixed conformal  equivalence   class
could be  avoided in warped compactifications.
Expressing  $ \hat R ^{(k)} $  in Eq.~(\ref{MD1p})   in terms  of   the    curvature and Laplacian  $ \tilde R^{(k)}   (\tilde g_{\mu \nu } ,\tilde g_{mn } )$  and $  \tilde \nabla  ^2 _k$,  referred  to  the     unwarped metric  tensors,
by means of  the identity   
\bea && \hat R ^{(k)} =  e ^{-2B} 
(\tilde R ^{(k)}  - 2  (k-1) \tilde \nabla _k ^2   B - 
(k-1)(k-2) (\tilde \nabla B)^2 )    , \eea 
and   integrating by   parts to  simplify the dependence on   the warp
and conformal  profiles as,    $- 2 d (k-1) ( \tilde \nabla A)
(\tilde \nabla B) - (k-1)(k-2) (\tilde \nabla B)^2 - d (d-1)
(\tilde \nabla A)^2    =  + 2 (k-1) \tilde \nabla  _k^2 B + (k-1) (k-2) (\tilde
    \nabla B )^2  $, one obtains the  formula  for the potential
\bea  && 2 \kappa ^2  V  =  \ud \int _{X_k} d^k y \sqrt {\tilde g_{6}
} e ^{d A + (k-2) B} 
  [-\tilde R^{(k)} + 2 (k-1) \tilde \nabla  _k^2 B + (k-1) (k-2) (\tilde
    \nabla B )^2\cr && - d (d-1) (\tilde \nabla A )^2 
- {2\kappa ^2 \over d} \hat T^{(d)} ]   + {C \over 2} 
(V_W - \int _{X_k} d^k y \sqrt {\tilde g_{6} } e ^{2 A
    + k B} ) .  \label{MD1} \eea 
We see that internal manifolds of negative curvature ($ \tilde R
 ^{(k)} <0$)   favour de Sitter spacetimes  ($ V  > 0$), in
contrast to  the contributions to  the    potential    from  the warp and
conformal  profiles      which involve  an explicitly  non-positive definite
 quadratic form  of  $ \tilde \nabla A ,\ \tilde \nabla B$  with a  quadratic form  of  negative   determinant   $ -d (k-1) (D-1)$.   Despite the presence 
of  stable/unstable  directions associated  to  the two eigenvalues
of positive/negative, it   was argued in~\cite{douglas09} 
that the  net  contribution  could  turn  out positive after imposing the
constraint  equations on $ A $ and $  B $.

So far, the     derivation  of $ V$
was     independent  of   Einstein equation     along  the internal
space    directions.    On  the other hand, it  was  found
 in~\cite{dougkallosh10}   that the  combined  equations  for 
the  Ricci tensor  traces in $ M_d $ and $ X_k$  yield useful
relations   between  the scalar curvatures $ R^{(d)}  ,\  R^{(k)} $
and   the   energy-momentum stress tensor traces     $ T^{(d)}
,\  T^{(k)} $  showing that   the no-go  conditions on de Sitter spacetime    
  can be circumvented  by    considering   internal manifolds of negative  
curvature       including warping or higher order curvature  terms.   
This   motivated~\cite{douglas09}  to  pursue a detailed    analysis  of the    warp  and
conformal factors  dynamics  in terms of    the   effective  potential  
$V (u, v),\ [u = e ^{d A    /2 } ,\ v= e ^{(k-2) B/2 }  ] $ 
for the  constrained   Legendre transform  with respect to  both $ g_{\mu  \nu } $ and  $ g_{mn } $. The  resulting classical potential  
\bea &&  2 V ( \tilde R^{(d)} , u, v) = 
{M_\star ^2 \tilde R^{(d)} \over 2}  - S_k + 
 {C}  ( V_W - \int d ^k y \sqrt {\tilde g _k}  u ^{2 -4/d} v^{2 + 4 /(k-2)} )
+{D }  ( V _{k}   - \int  d ^k y \sqrt {\tilde g _k}  v^{2 + 4 /(k-2)} )
,  \eea  
includes  the  additional   Lagrange multiplier $D$   designed to 
fix      the  volume $V_{k}  = \int d ^k y \sqrt {g_k}  $
independently  of   $ V_W.$  We shall not  pursue this discussion
except to note that the constraints from  the traces  of
Einstein equations   along  $ M_d $ and $X_k$, inferred  from  
the stationarity  equations,  $ \d V/\d u =0,\ \d V/\d v =0,$ 
yield      coupled   equations  for $u,\ v$ which  were derived 
in~\cite{douglas09}.   

For the    supersymmetric GKP vacua, $ d=4,\ k=6$   with $B= -
A$, the  scalar potential  is  given by   the simpler formula  
\bea && 2  V  = {1\over 2 } {1\over 2 \kappa ^2 }\int _{X_6}
  d^6y \sqrt {\tilde g_{6} } ( -\tilde R^{(6)} - {\kappa ^2 \over 2} 
\hat T ^{(4)}  - 10 \tilde \nabla _k^2 A + 8 (\tilde \nabla A )^2 )  + {\tilde
    R^{(4)}}   (V_W - \int _{X_6}  d^6y \sqrt {\tilde g_{6} } e ^{-4
    A} )   ,  \label{MD2} \eea 
where $\hat T ^{(4)}  = T _\mu ^ {\mu } +  \vert F^{(p) } \vert ^2/
\kappa ^2      $  and   $\tilde R^{(6)}  =0$, if one restricts to  (Ricci  flat)
Calabi-Yau manifolds.   We 
see that   warping  contributes  a  net positive potential energy,  
since the  Laplacian term drops out for a compact manifold $ X_6$. 
Fluxes and brane  (unlike  
orientifold)   sources give negative contributions to $V$. 
The  warp   profile   satisfies  the  Schr\"odinger type equation  
\bea && 0= \nabla  ^2   (X_6) u - {2 \calu (y) \over 3} u (y) - { \tilde R ^{(4)} 
\over 6} ,\ [\calu  =    {\hat R^{(6)} \over 2} + 
{\kappa ^2 \over 4} \hat T ^{(4)}  
,\ \hat T ^{(4)}   =  (T _\mu ^\mu )   - {1 \over
   \kappa ^2    } \sum   _p \vert F ^{(p)} \vert ^2 ]   \eea
where     $ \calu   $    coincides with 
the    gravitational potential which was  identified  in~\cite{gidd05} with 
the time component of Einstein  tensor,  $ G_{t} ^t= - \kappa _4^2
\calu   \sim  T_{00}$.   
The  ground state  solution  for $ u(y)$   
corresponds   to a wave  function  whose  support is 
concentrated     at wells of the  potential  $\calu  $. 
The approximate solution $ u \simeq  -\tilde R^{(4)} / (4 \calu  )$,  in regions  where $u$  varies slowly,   shows that the potential 
must  have    opposite sign to the curvature  in order  that 
$ u(y) \geq 0$~\cite{douglas09}.

 Finally, we consider    the    contributions 
from 3-fluxes. Since these arise in the 10-d  action     from    both 
kinetic  energy and Chern-Simons terms,  one cannot  treat  
electric  fields    as    Hodge   duals   of magnetic  fields. 
In   GKP vacua, the  kinetic and
topological   terms are  found to   combine into the
single  wedge product of the complex 3-form field 
strength,     $ (g_s /\tau _2 ) ( e^{4A} G _3 \wedge \star _6 \bar G_3 + i\a
G _3 \wedge  \bar G_3  ) =  (g_s e^{4A} /\tau _2 )  ( G_3  \wedge  \bar G_- )
,\ [ \bar G_- = (G_- )^\star = (\star _6  +i ) \bar G_3  ] $    where the 
Hodge star duality refers  to the unwarped metric $ \tilde g_{mn
}$.  The resulting 3-flux potential  splits  into   tension
and (topological) charge parts
\bea &&  V  _{flux} (x) = {g _s \over 4 \kappa ^2 } \int _{X_6} { e ^{4A} \over \tau _2} G_3 \wedge \bar G_- = \tau _3 (T
_{flux} - N _{flux}),  \cr && [T _{flux}= {1\over 4 \kappa ^2_{10} \mu _3 }
  \int _{X_6} {e ^{4A} \over \tau _2} G_3 \wedge \star _6 \bar G_3
  ,\ N _{flux}= - {i\over 4 \kappa ^2 _{10} \mu _3 } \int _{X_6} {e ^{4A}
    \over \tau _2} G_3\wedge \bar G_3 ]  \label{sub2eq17}\eea
whose  net  contribution   is positive     provided the  effective  $D3 $- or
$\bar D3 $-brane tension $ T_{flux} $ and charge  $N_{flux}$ 
satisfy  the BPS like inequality, $V  _{flux} (x) = \tau _3  (  
T _{flux} - \vert N _{flux} \vert  ) \geq 0$.

\subsection{Type  $ II\ b$  supergravity compactification  
on  warped   $ M_4   \times X_6$  spacetime}
\label{subsec2.2}

The  top most  priority  in    Kaluza-Klein  theory  is in analyzing 
the   classical action  variation   under    linear   perturbations 
of the various fields. As is
known, the background solution and the field
equations are derived by cancelling the linear order variations,
the vacuum stability  is established   from the  wave equations  of
field fluctuations, which are  derived   by   cancelling the second
order variations,   while the  local couplings between    modes
are derived  from the higher  order variations. For a fixed
background, the  linearized wave equations  can also be 
evaluated  from     the field functional derivatives of  the classical
equations. 

 The wave equations for the bosonic supergravity fields   are quoted
for    a general spacetime in Eq.~(\ref{appeq2})  and 
for the warped spacetime $ M_4 \times X_6$ in Eq.~(\ref{appeq31}).  
The     normal    modes   in $ M_4$  are   defined by decomposing the 
wave  functions in $ X_6$  on  orthonormal bases solutions   subject
to appropriate boundary conditions, while the   vacuum   
solution in the presence of background deformations  are obtained by
solving the wave  equations in Eq.~(\ref{appeq35})
for $x$-independent (constant in $ M_4$)  fields fluctuations. 

The  classical action reduction    
can  be   carried out    at various levels of
sophistication. For the metric tensor  fluctuations 
along $ M_4$ and $ X_6$,  it is tempting in  a first
approximation  to disregard  mixings    between    
the spin $2, 1, 0 $ fields, $ h_{\mu   \nu } ,\ h_{\mu m
} ,\ h _{mn}$. However, mixing terms    between   massive  modes   and       
massless moduli  are expected on general  grounds to  arise  
as a manifestation of Higgs mechanism from  the  spontaneous  breaking of  
 10-d  diffeomorphisms   induced  by   finite background VEVs  for the
 metric tensor components, $ g _{\mu \nu },\  g _{mn } .$ 
For  instance, in a   flat  (toroidal)  spacetime of dimension $ D$ 
with   de Donder gauge, the     metric
tensor   $ g _{MN}$  reduction~\cite{han98}    
gives  rise to a  massless  sector with graviton,
$U(1)$ graviphoton and scalar modes  and massive towers
involving    a single graviton, $ (D-5) $  vector 
and $ (D-4) (D-5) /2$  scalar  modes.   

The invariance of gravitational  field theories under general
coordinate  and $ U(1)$ gauge transformations  is    
conveniently taken into account by representing the 
non-dynamical  constraint equations
through    compensator fields. (A
brief general   presentation of the compensator field method is  provided
in~\cite{gates8301}.)  Making   use   of an  extended classical ansatz  
 with    gauge    compensator fields allows  separating the  constraint
equations on   classical profile    fields  from the  dynamical equations  while    
eliminating   kinetic mixings   between moduli    and   massive modes. 
For   warped compactifications of     supergravity
theory,    the initial   construction  in~\cite{gidd05}  was systematized
in  the   context  of the Hamiltonian formalism
in~\cite{frey08,dougtorr08,freyroberts13}  and  the
Lagrangian     formalism    for the  axial gauge in~\cite{shiu08}. 
We   shall   start   the discussion    with   
the  simplified   treatment   for  the 
spin $2$   metric tensor field   $\d g_{\mu \nu } (X)  $  
in the   harmonic gauge~\cite{firouz06},  
ignoring  mixing with     other   fields and with moduli. 
The     refined  treatment    for   the metric tensor field  
components $\d  g_{MN } (X)  $ and  the  other supergravity  multiplet
component fields  is next  developed  following~\cite{shiu08}.  

\subsubsection{Simplified formalism in  harmonic   gauge} 

We  examine the    dimensional  reduction     of the metric tensor 
components along $ M_4 $      by  focusing on 
the    Einstein-Hilbert curvature action $ S_g$  combined
with the non-dynamical matter action  $S_m$ for  the  bulk fields and brane sources~\cite{firouz06},
\bea && S (g) \equiv S _{g} (g)+ S _{m} (g) =  \int d^{10 }
X \sqrt {- g _{10} } [  {1\over 2 \kappa ^2 } 
g^{MN} R _{MN} (g) + L_{m} (g) ] . \label{sub2eqm10} \eea 
The  expansion   of the  perturbed action $  S (\tilde g + h
) $     in powers of $\d \tilde g _{\mu   \nu } = h _{\mu \nu } $
contains  a first order   variation,  encoding    the 
field equations,  and  higher order variations, encoding the wave
equations and   self couplings of  fluctuations,
\bea && S (g + \d g ) = \ {m_D ^8\over 2} \int d^4 x \sqrt { -(\tilde
  g _4 + h ) } \int d^6 y \sqrt {\tilde g _6} e ^{-4 A (y) } ( \tilde g ^{\mu \nu }     R  _{\mu \nu }   ( \tilde g _4 + h) + 2 \kappa ^2 L_{m}( \tilde g _4 + h) ) \cr && = {m_D ^8
  \over 2} [ V_W \int d^4 x \sqrt { -\tilde g_4 } \tilde R ^{(4)}
  (\tilde g ) - \int d^4 x \sqrt { -\tilde g_4 } \int d^6 y \sqrt
  {\tilde g_6}  e ^{-4 A     (y) }   ( h^ {\mu \nu } 
 (G_{\mu \nu } - \kappa ^2 T _{\mu \nu } )   \cr &&  + {1\over
    2! }  h^ {\mu \nu } \d (G_{\mu \nu } - \kappa ^2 T _{\mu \nu } ) 
+ {1\over 3! }  h^ {\mu \nu  } \d ^{2} (G_{\mu \nu } - \kappa ^2 T _{\mu \nu } ) 
 + \cdots ) ] , \cr && [V_W
  = \int d^6 y \sqrt {\tilde g_6} e ^{-4 A(y) }
 ,\ G_{\mu \nu } = R _ {\mu \nu } - \ud g_ {\mu \nu } R ,\ T_{\mu \nu
  } = -{2\over \sqrt {- g } } {\d S_{m} \over \d g^{\mu \nu }} =
  g_{\mu \nu } L_{m} - 2 {\d L_{m} \over \d g^{\mu \nu } },\ m_D ^8 =
  { 1\over \kappa ^2} ] .  \label{sub2eq10} \eea
The  classical field equations and the  wave equations   involve 
the    linear    combinations of   Einstein tensor 
$G_{\mu \nu } $ and matter stress energy-momentum tensor $T_{\mu \nu }
$, 
\bea && {1\over \sqrt {-g} }  {\d \over  \d g^{\mu \nu }  }
  (\sqrt {- g} ( R + 2\kappa ^2 L_m ) ) 
=  -(G_{\mu \nu } - \kappa ^2 T _{\mu \nu } ) =0 ,\ \  \d (G_{\mu \nu
} -\kappa ^2 T _{\mu \nu } ) =0  .\eea   
Matching  the   zeroth order  classical  term 
in Eq.(\ref{sub2eq10})   to the   familiar   4-d curvature  action 
relates  the  gravitational     mass parameter $ m_D$,  given by a known  
function of   the string theory mass scale   and coupling
constant parameters   $ m_s,\ g_s$,    to  the (reduced) Planck mass 
$ M_\star $  and  the  internal  manifold  warped volume $ V_W $,  
\bea && m_D ^8 \equiv { 2 m_s ^8 \over (2\pi )^7 g_s ^2 } = { M_\star ^2 \over
  V_W } \ \Longrightarrow \ m_s= ( {\pi g_s ^2 M_\star ^2 \over L^ 6
}) ^{1/8} , \ [M_\star = {M_P \over \sqrt {8\pi } } = 2.43\ 10^{18}
  GeV,\ V_W \equiv (2\pi L )^6]   \eea 
where  $ L$   denotes the effective  radius of $X_6$. 
The Ricci   curvature  tensor  components  $ R_{\mu \nu }  (M_{10} ) $
consist  of  a part  arising  from the 4-d spacetime   curvature tensor
$R ^{(4)} _{\mu   \nu }  (M_4) = \tilde R ^{(4)} _{\mu   \nu }    = R _{\mu \l \nu } ^{\ \ \ \ \l } $, independent of the warp profiles,  and another part $R _{\mu p  \nu } ^{\ \ \ \  p} $      involving the internal  space  covariant  derivatives      and  the warp profiles.  This  decomposition is  explicit  in   the identity,   quoted  from  Eq.(\ref{appeq7}),   
\bea && R^{(4)} _{\mu \nu } (M_{10} )    = \tilde  R ^{(4)} _{\mu \nu }(M_4 )  - \ud 
[\nabla _{6} ^2 g_{\mu \nu }   + e ^{-4A} (\nabla _m e ^{2A}) (\nabla ^m
  e ^{2A})  g_{\mu \nu } ] \cr &&   
= \tilde R ^{(4)} _{\mu \nu } (M_4 ) - \ud [ \nabla _{6} ^2 
( e ^{2A}   \tilde g_{\mu \nu } )   + e ^{-2A} (\nabla _m e ^{2A}) (\nabla ^m
  e ^{2A}) \tilde   g_{\mu \nu }  ] . \label{eqRicci4} \eea  
The  tilde  symbol refers as usual  to the unwarped  metric  tensors.
 We also need the  Ricci  tensor  variation  in order to evaluate
the    second order    variation  of the  perturbed  reduced  action. 
Taking the  differential    of Eq.~(\ref{eqRicci4}) with respect to 
the unwarped metric  $\d /\d \tilde g _{\mu \nu } $, 
at fixed $ A(y)$ and $ \tilde R ^{(4)} _{\mu \nu } $, gives  
\bea &&  \d R
_{\mu \nu }  \equiv  E ^R_{\mu \nu } = \d  \tilde  R ^ {(4)} _{\mu \nu } -\ud
     [\nabla _{6} ^2 (e    ^{2 A}  
\d  \tilde g_{\mu \nu } )  + e^{-2A} (\nabla _m e ^{2 A} )  (\nabla ^m e ^{2 A} 
  ) \d \tilde  g_{\mu \nu } ] , \cr &&  
[\d  R ^{(4)}_{\mu \nu   }  = {e^{2A} \over 2}  \d \tilde R ^{(4)}_{\mu \nu
  } ,\ \d \tilde R ^{(4)}_{\mu \nu} =  {1 \over 2} 
( - \tilde \nabla _4^2 \d \tilde
  g_{\mu \nu } - \nabla   _\mu \nabla _\nu \d \tilde g + 
2 \nabla ^\l \nabla _{( \mu } \d \tilde g _{\nu ) \l } )     
,\ \d \tilde g = \tilde g ^{\mu \nu } \d \tilde g _{\mu     \nu }  ,  \cr &&    
\nabla ^2 _4 = {1\over \sqrt {g_{10}} } \dh _\mu \sqrt {g_{10}} g
  ^{\mu \nu } \dh _\nu     ,\ \nabla ^2 _6 = {1\over \sqrt {g_{10}} }
  \dh _m\sqrt {g_{10}} g   ^{mn}\dh _n ] .    \label{sub2eq11} \eea
 The same result can also  be obtained    by  invoking  
 identities in  Eq.(\ref{eqsubvarR}).  
  For the sake of  brievity, we    shall always adhere    
to the    condensed   notation for the 4-d and 6-d 
components of the 10-d Laplacian   $\nabla ^2 _{4,6}  \equiv  (\nabla ^2 _{10} )_{(4,6)}  ,\  [\nabla ^2 _{10} = \nabla ^2 _4 + \nabla ^2 _6 ] $. The corresponding  operators 
with  respect to the unwarped metrics are  denoted   $\tilde \nabla ^2 _{4,6}  $.  It is   very important not to confuse $ \nabla ^2 _{4,6}
$      with   the  scalar  Laplacians $
\nabla ^2 (M_4) ,\ \nabla ^2 (X_6)$   of the
manifolds     $  M_4$ and $ X_6$,  as  we  explain in more detail
near  Eq.~(\ref{appeq11}). 
In the harmonic (or transverse-traceless) gauge for the unwarped
metric tensor,  
$h_\mu \equiv  \d  \tilde g  _\mu \equiv \tilde \nabla ^\nu \d \tilde g
_{\mu \nu } =0,\ h \equiv  \d  
\tilde g  \equiv \tilde  g^{\mu \nu } \d \tilde g  _{\mu \nu } =0 $,
to which we  specialize  in  the  remainder of this subsection,
$ \d R ^{(4)}  _{\mu \nu }= - \ud e ^{2A}\nabla ^2 _4 h
_{\mu \nu } $, 
the  Ricci tensor variation  simplifies to
\bea && \d R _{\mu \nu } = - \ud (e ^{2A} \nabla ^2 _4 h
  _{\mu \nu } + \nabla _{6} ^2 ( e ^{2A}h _{\mu \nu } ) + e ^{-2 A}
h_{\mu \nu } (\nabla _m e ^{2A}) (\nabla ^m e ^{2A} )) . \label{sub2eq13p} \eea

The wave  equation 
$ \d R _{\mu \nu } =\kappa ^2 \d (T _{\mu \nu  } - g_{\mu \nu } T_M^M / 8 ) $  
can be put in  a    convenient form   if one restricts      to   classical
backgrounds that are independent of the 4-d spacetime metric. 
The   ensuing condition  on the  matter
Lagrangian  density,    $ \dh  L_m  /\dh g^{\mu \nu } =0 $,
shows that  the  stress tensor along $ M_4$
is proportional   to the  metric tensor 
$ T _{\mu \nu } =   g_{\mu \nu } L_m - 2 (\dh  L_m  /\dh g_{\mu \nu
} ) \to  g_{\mu \nu } L_m $~\cite{firouz06}.    Using 
this    condition  in  the trace-reversed  10-d Einstein equation,   
$ R_{MN} = \kappa ^2 ( T_{MN} - {1\over 8 } g_{MN} T ) $, 
entails   the proportionality between  the Ricci   and  metric tensors,  
\bea && R_{\mu \nu }  
=  - {\kappa ^2 \over 8 } (2  L_m  +\calT ) 
g_{\mu \nu } ,  \ [\calT = \calT ^m_{m} = g^{mn} \calT
  _{mn},\ \calT _{mn }= -2 {\dh L_m \over \dh g^{mn} } ] 
\label{sub2eqp13} \eea
where we note  that  $\calT _{mn} $ is only  a part of the  
stress energy-momentum tensor  along $ X_6$.  
Applying  the above  proportionality relation
to the case of  perturbed     metric and  Ricci  tensors,   $
g_{\mu \nu  } \to  g_{\mu \nu } +  \d g_{\mu \nu
},\ R_{\mu \nu }  \to   R _{\mu \nu } + \d R _{\mu \nu }$,  about  the
background   with $\tilde R ^{(4)}
_{\mu \nu } =  \tilde  R ^{(4)} \tilde  g_{\mu \nu } / 4$,   
yields the  pair of equations
\bea && -{\kappa ^2\over 8 }  e ^{2A} (2 L_m  +\calT )
= [{1\over 4}  \tilde R ^{(4)}  -\ud (\nabla _{6} ^2 (e ^{2A}) +
  e ^{-2A} (\nabla _m e ^{2A}) (\nabla ^m e ^{2A}) ) ] , \cr &&
 \d  R_{\mu \nu } =   
-{\kappa ^2\over 8 }  e ^{2A} (2 L_m  +\calT ) \d \tilde  g  _{\mu \nu
}  = [{1\over 4}\tilde   R ^{(4)}  -\ud (\nabla _{6} ^2 (e ^{2A}) +
  e ^{-2A} (\nabla _m e ^{2A}) (\nabla ^m e ^{2A}) ) ]  \d \tilde g  _{\mu \nu }
, \label{sub2eq13px}   \eea
where the first  equation is interpreted as a constraint  on  the
warp  profile 
$A(y)$  in the absence of  background  perturbations  and the second
exhibits  the  ensuing proportionality  between  variations of the   Ricci and metric  tensors.   Combining the latter equation 
with  Eq.~(\ref{sub2eq13p}),  one  can use the following  steps to  
put the wave equation for $ h  _{\mu \nu }  $ in  a convenient
form~\cite{firouz06},  independent  of the stress tensor, 
\bea && - \ud ( e ^{2A} \nabla _4 ^2 h_{\mu
  \nu } +\nabla _{6} ^2 (e ^{2A} h_{\mu \nu } ) + e ^{-2A}(h_{\mu \nu
} \nabla _m e ^{2A}) (\nabla ^m e ^{2A}) h _{\mu \nu } ) = h_{\mu \nu }
[{1\over 4} \tilde  R ^{(4)} - \ud (\nabla _{6} ^2  (e
  ^{2A} )  + e ^{-2A} (\nabla _m e ^{2A}) (\nabla ^m e ^{2A})
  )  ]   , \cr && \Longrightarrow \ 0=- \ud e ^{-2A}
\tilde  R ^{(4)}   h _{\mu \nu }  + \nabla ^2 _4 h _{\mu \nu }+ e ^{-2A}
\nabla _{6} ^2 (e ^{2A} h _{\mu \nu } ) - e ^{-2A} h _{\mu \nu }
\nabla ^2 _6 e ^{2A}
= (\nabla ^2 _{10}  -\ud e ^{-2A} \tilde  R ^{(4)} ) 
h _{\mu \nu } (X) ,  \label{sub2eq14} \eea
where $\nabla ^2 _{10} = \nabla ^2 _4 + 
\nabla _{6} ^2  =  (e ^{-2A} \tilde \nabla ^2 _4 +e ^{2A}  \tilde
\nabla _{6} ^2  ) .$ 
The    corrresponding    wave equation  in an arbitrary gauge  reads, $
e ^{2A} \nabla ^2 _{10} h _{\mu \nu } +  \nabla _\mu \nabla _\nu h - 2 \nabla
^\l \nabla _ {(\mu } h _{\nu ) \l }  -\ud \tilde  R ^{(4)} h _{\mu \nu }  =0.$
We learn from  Eq.~(\ref{sub2eq14})      that  the graviton modes,  for  sources akin to
an effective 4-d  cosmological constant term   and  for $  \tilde  R
^{(4)} =0$,   arise as zero modes  of the 10-d
scalar Laplace wave operator, just as   for  scalar fields.  This
property was discovered in 
previous studies of the gravitational interactions aimed at
extensions of Randall-Sundrum model  with thick branes, bulk scalar fields, smeared or 
intersecting distributions of branes  and  higher dimensional
backgrounds~\cite{brandsfet99,csakishirman00}.

\subsubsection{Lagrangian  formalim in axial gauge}  

The   dynamical description of warped compactifications 
greatly simplifies if one  introduces     compensator  fields  to   separate
out the kinematical  constraints   imposed  by    general
coordinate  invariance.     In the  axial  gauge   
formalism that we shall  adopt,  the     metric  tensor
mixed  components $g_{\mu m }$   are introduced   as  compensator  fields  
that  are  set to  zero   by   choosing the gauge  $g_{\mu m } =0$,  after
imposing  the   action   stationarity under variation of $g_{\mu m }$.    
We   here   briefly review   the   Lagrangian formalism
of~\cite{shiu08},  with   technical details      relegated to
Appendix~\ref{subapp4}. 
The   variation of  the   curvature action
$S (g) \equiv S _{g} (g)+ S _{m} (g) $  in Eq.~(\ref{sub2eqm10})   
perturbed    by  small fluctuations of the warped metric  tensor $ \d g _{MN } $,  including $\d A(y)$ and $\d \tilde g _{MN } = h_{MN }$,  
is expressed in terms of   the Einstein tensor $G_{MN} $    and the stress 
energy-momentum tensor $T_{MN} $  for  bulk matter fields and
localized sources   as 
\bea && S (g + \d g ) - S (g) \equiv \d  S +
\d ^{2} S + \cdots \cr && = - {1\over 2 \kappa ^2 } \int d^{10 } X
\sqrt {-\tilde g _{D}} e ^{-4 A} [ \d g ^{MN} ( G_{MN} - \kappa ^2
  T_{MN} ) + \ud \d g ^{MN} \d ( G_{MN} - \kappa ^2 T_{MN} ) + \cdots
] , \cr && [G_{MN} = R_{MN} - \ud g_{MN} R ,\ T_{MN} = -{2\over \sqrt
    {- g } } {\d S_{m} \over \d g^{MN} } = g_{MN} L_{m} - 2 {\dh L_{m}
    \over \dh g^{MN} } ]. 
\label{sub2eq8}  \eea  
The linearized wave equations, for a fixed    vacuum
background solution,    are obtained by varying Einstein
equation,   $ 0= \d  (G_{MN} - \kappa ^2  T_{MN}) .$
We have used the identities for the first and second order variations of the curvature scalar~\cite{wald84}, 
\bea &&  \bullet \  \d   (\sqrt {-g_D} )  = -\ud \sqrt {-g_D}  g ^{MN}
\d  g _{MN} = \ud \sqrt {-g_D}   g _{MN} \d  g ^{MN} ,\  
\d  (\sqrt {-g _D } R) = - \sqrt {-g _D }  \d g ^{MN }  G_{MN } ,\  \cr &&  \bullet \  \ud  \d ^{2 } (\sqrt {-g _D   } R) = -\ud \sqrt {-g _D } \d g ^{MN}
(\d G_{MN}  + \ud G_{MN} \d g ),  \eea  
dropping  boundary contributions
from total derivative terms $ \dh _M (\cdots ) $  from $\d R_{MN}$  
and  using the definitions $\d ^{2} = \d \cdot  \d ,\  
\d g = g_{MN}\d  g ^{MN} $.
The overall minus signs  in Eq.~(\ref{sub2eq8}) for the action variation    arise    from the notational convention  $  \d (g) ^{MN} = -  \d g ^{MN} $   defining the inverse of 
the metric matrix  as $ (g) ^{MN} = (g ^{-1} ) _{MN} $,  hence implying 
$\d (g ) ^{MN} = \d (g ^{-1} ) _{MN} = - g ^{MM'} g ^{NN'} \d g_{M'N'}
= -  (\d g )^{-1}  _{MN} = - \d g ^{MN} $.  In particular, 
under local rescalings of the
background metric, $ g_{MN} = e ^{a} \tilde g_{MN}$, one has  $\d
g_{MN} = e ^{a}  (\d a \tilde g _{MN} +   \d \tilde g^{MN} ) 
 ,\ \  \d g^{MN}= -\d ( e ^{-a}  (\tilde g ^{-1} )_{MN} ) = 
 e ^{-a} (\d a \tilde g ^{MN} +   \d \tilde g^{MN} ).$  

 The vacua of compactified  theories     typically involve
 free  constant  moduli parameters  $ u^I $  that arise  
as  zero modes of the equations $\d _I G_{mn}  =0 ,\ [\d _I  g_{mn} =
  \d u ^I \dh  g_{mn} /\dh u^I ]$.  In  the   reduced
effective action   in $ M_4$, the moduli are 
promoted to quantum fields  $ u ^I (x) $. For instance, the  axio-dilaton  field  is  split into   moduli  and    Kaluza-Klein modes  as $ \tau = \tau _0 (x)   + \d _K  \tau $.   It is  important  to  choose the     classical ansatz for  $\d g_{MN} $  by   carefully  distinguishing      the  massive  Kaluza-Klein   modes from  the    massless geometric  moduli
while    allowing   terms  depending on  $ u^I(x) $ and their    gradients
$\dh _\mu u^I(x) $.  As is known, the  4-d dynamics  of moduli  fields  is described  by a  non-linear sigma model action    with 
the moduli space metric  set by  kinetic energy
terms.   The  compactifications on  Calabi-Yau manifolds is discussed in~\cite{grimmlouis05}.

In warped compactifications, the   use of a generalized classical   ansatz  involving moduli  dependent warp
profiles $  \d A(y, u ^I (x) ),\  \d \a (y, u ^I (x) ) $,   along with 
compensator   fields  depending on spacetime  derivatives  of the
moduli  fields, is  necessary  to ensure  the consistency of   
field  equations~\cite{gidd05}.     A  suitable     ansatz for
linear perturbations of the 10-d metric tensor is \bea && \d ( d s
_{10} ^2 ) = e ^{ 2 A (y) + 2 \d _u A (y, u) } \ (\tilde g _{\mu \nu }
(x) + \d \tilde g _{\mu \nu } (X) ) dx ^\mu d x ^\nu + 2 e ^{ 2 A
  (y) } \d \tilde g _{\mu m } (X, u) \ d x^\mu d y ^m \cr && + e ^{-
  (2 A (y) + 2 \d _u A (y, u) ) } ( \tilde g _{m n } (y) + \d \tilde g
_{m n } (X, u) ) \ d y^m d y ^n , \cr && [\d _u A = u^I {\dh A(y,
    u) \over \dh u^I } ,\ \d \tilde g _{\mu m } (X, u ) = \dh _\mu u
  ^I (x) B_{I m } (y) ,\ \d \tilde g _{\mu \nu } (X, u ) = \d _K
  \tilde g _{\mu \nu } (X) + 2 \dh _\mu \dh _\nu u^I (x) K_I (y)
] \label{sub2eq5} \eea where the differential  $ \d _K $  generates
massive  modes only,   $\d _u $ generates  
moduli,  as   $\d _u A (y , u ) = u^I \dh A / \dh u^I,$  for instance,
and   $\d \tilde g _{m n } (x,y)= \d _{K} \tilde g
_{m n } (x,y)+ \d _u \tilde g _{mn } (x,y).$  
The   $x$-dependent derivative   variation  terms,  
$ \d g _{\mu \nu }  = 2 K _I (y)   \dh _\mu  \dh _\nu  u^I (x)  ,\   \d g _{\mu
  m } = B_{I   m}(y) \dh _\mu u ^I(x) ,$  
are   gauge     compensators  fields in the sense that they 
 can   be   eliminated by  diffeomorphism transformations, 
$\d g _ {MN }   = \nabla _{(M } \xi _{N)} $.  
(The  corresponding  variations  of the 2- and
4-form potentials involve the  $ U(1)$ gauge  transformations, 
$\d C_2  =  d  t^I (x) \wedge T^{(1)}
_I   (y),\  \d C_4  =  d  v^I (x) \wedge S^{(3)}_I  (y)$.) 
In the axial gauge $g _ {m \mu } = 0$,
however,  $ B_{I m}$  may   be set to zero only after implementing the
stationarity constraint $ \d G_{m \mu } =0$ induced by the variation $
\d g _ {m \mu } $.  The  analysis of  gravitational  dynamic
makes an important use      of the  identities   for 
the first and second order variations of  Ricci curvature tensor~\cite{wald84}
\bea && \bullet \ \d R_{MN} \equiv E ^R_{MN}  
= \ud (- \nabla ^2 \d g_{MN} - \nabla _M \nabla _N \d g
+ \nabla ^P \nabla _M \d g_{NP} + \nabla ^P \nabla _N \d g_{MP} )
\cr && =\ud (- \nabla ^2  \d
g_{MN} -\nabla _M \nabla _N \d g + 2 \nabla _{ ( M } \nabla ^P \d g_{N
  ) P}  + 2 R _{P (MN) Q } \d g ^{PQ}   + 2 R _{ ( M } ^{\ \ \ P} \d g
_{N ) P}   ) , \cr && 
\equiv \ud (\D _L \d g_{MN} -\nabla _M \nabla _N \d g + 2 \nabla _{ ( M
} \nabla ^P \d g_{N ) P} ) ,\ 
[\D _L \d g_{MN} = - \nabla ^2 \d g_{MN} + 2 R _{P(MN)Q}
  \d g ^{PQ} + 2 R _{ (M} ^{\ \ P} \d g _{N) P} ]   
\cr && \bullet\ \d G_{MN } \equiv E _{M N } = \ud ( - \nabla ^2
\d g_{MN } - \nabla _M \nabla _N \d g + 2 \nabla ^P \nabla _{( M } \d
g_{N ) P } -R ^{(D)} \d g _{M N } + g_{M N } (\nabla ^2 \d g - 
\nabla ^P \nabla ^Q \d g _{PQ} ) ) , \cr && [\nabla ^2 = \nabla ^P \nabla _P = g^{PQ} \nabla _P
  \nabla _Q ,\ \d g _M = \nabla ^N \d g_{MN} ,\ \d g = g^{MN} \d g
  _{MN}  ]  \label{sub2eq9} \eea 
where $ E^R  _{M N } $ denotes the inverse graviton propagator,
$E _{M N } $ is a  related wave operator and  the Lichnerowicz
operator  $\D _L $  
was introduced  by using the identity,  $\nabla ^P \nabla _{( M} \d g _{N) P}
- \nabla _{( M} \nabla ^P  \d g _{N) P} =  R ^{P \ \ \  \ \ Q}_{ \ \ (MN ) }  
 \d g _{PQ} + R _{(M } ^{\ \ \ Q} \d g _{N) Q}  $. 
The dependence on     warp profiles can be   
extracted    out  by  invoking  the standard  identities for  conformal
rescalings   of the metric tensors~\cite{wald84}. To this effect, 
we  have compiled in
Appendix~\ref{subapp2}  several  useful formulas adapted
from~\cite{douglas09,shiu08}.  
For spacetimes $M_D = M_d \times X_k $ of dimension $ D=d+k
$ and metric tensor $ ds ^2 _D= e ^{2 A (y)} d\tilde s^2_d +e ^{2 B
  (y)}d\tilde s^2_k $, the dependence of Laplace operator and Riemann
and Ricci curvature tensors on the     profile
functions $A (y), \ B(y)$ in the external and internal 
space metric tensors was discussed    in~\cite{douglas09} and
that for second order variations of the curvature tensors 
in~\cite{shiu08}.    We mostly    specialize  in the  text  to the case $ B=-A$.     Let us first consider the   variation 
of Einstein  gravitational equation along $M_4$  for $ (MN) = (\mu \nu
)$.  Setting  the  
purely longitudinal metric tensor  field  gradients to zero   gives  
the   algebraic  constraint equation  relating  fluctuations of
the warp profile  to   the trace of the  internal space metric tensor, quoted from the first entry in Eq.~(\ref{appeq31}), 
\bea && 0= \d (G _{\mu \nu }
-\kappa ^2 T _{\mu \nu } )  \supset \tilde \nabla _{\mu }
\tilde \nabla _{\nu } ( 4 \d A - \ud \d \tilde g ) \ \Longrightarrow
\ \d A = {1\over 8} \d \tilde g \equiv {1\over 8} \tilde g ^{mn} \d
\tilde g _{mn} . \label{sub2AXIAL1} \eea 
This     condition  is  also  obtained  by requiring  that the
warped internal manifold volume  remain constant, $0=\d V_W = \int
\sqrt {\tilde g_6 } e ^{-4A} (- 4\d A + \ud \d g ) $.  
The projection along the mixed directions $ (MN) =(\mu m ) $  yields 
the    second  constraint equation, quoted from the last entry in Eq.~(\ref{appeq31}),     
\bea && 0= \d G _{ \mu m } = \ud \ \tilde \nabla _{\mu } [ \nabla ^ {\tilde
    n} \d \tilde g _{mn} - \ud \nabla _m \d \tilde g - 4 \d \tilde g
  _{mn} \nabla ^ {\tilde n} A ] =0 \ \Longrightarrow \ \nabla ^
{\tilde n} ( \d \tilde g _{mn} - \ud \tilde g _{mn} \d \tilde g ) = 4
\nabla ^ {\tilde n} A \d \tilde g _{mn} , \label{sub2AXIAL2} \eea
which  is natural   interpreted as    a warped version of  de
Donder gauge.   Imposing the above two conditions yields the simpler 
formula   for the 4-d  graviton field wave equation
   \bea && 0= \d (G _{\mu \nu } -\kappa ^2 T _{\mu \nu } ) = \d  _K \tilde
   G ^{(4)} _{\mu   \nu } - \ud e ^{4 A} \tilde \nabla ^2 _6 ( \d _K
   \tilde g_{\mu \nu } 
- \tilde g_{\mu \nu } \d _K \tilde g _4) - \ud e ^{4 A} \tilde g_{\mu
  \nu } \d \tilde R^{(6)} - \kappa ^2  (e ^{2A}  \tilde g_{\mu \nu }
\mu _3 \d \rho _3 ^{loc} + \d T ^{loc} _{\mu \nu } ) ,\cr && [\d _K 
  \tilde G ^{(4)} _{\mu \nu } =
\tilde E_{\mu \nu } \equiv \ud ( - \tilde \nabla ^2 h_{\mu \nu } -
\tilde \nabla _\mu \tilde \nabla _\nu h +  2 \tilde \nabla ^\l  \tilde
\nabla _ {( \mu }  h_ {\nu ) \l } +
\tilde \nabla _\nu h_\mu - h_{\mu \nu } \tilde R ^{(4)} + \tilde g _{\mu \nu
} (\tilde \nabla _4 ^2 h -\tilde \nabla ^\l h _\l ) ) ] \label{sub2AXIAL3} \eea 
where we have separated out the  contribution from $ D3$-branes. Note that 
the    wave operator   $\tilde E_{\mu  \nu } $      propagates   also
longitudinal and trace  components of the metric, 
$h _\mu = \tilde \nabla ^\nu  h _{\mu \nu } ,\ 
h = \tilde  g^{\mu \nu } h _{\mu \nu }  $   and  that the variations of the effective $D3$-brane charge
density and stress tensor  arise through the   metric  dependent  terms 
$\d \rho _3 \propto \d (1 /\sqrt {g _6 } )  ,\   \d T _{\mu \nu } \propto \d
(g_{\mu \nu } /\sqrt {g _6 } )  $. 
 If one    imposed at this stage the transverse-traceless gauge condition on the  unwarped  metric,  defined   near Eq.~(\ref{sub2eq13p}), assuming
 Ricci flatnesss   $ \d \tilde R^{(6)} _{mn} =0 $,  
 the   wave  equation   in Eq.~(\ref{sub2AXIAL3})  
 would       exactly coincide with the  previously derived   
Eq.~(\ref{sub2eq14}).   This conclusion  would be invalid, however, because the
transverse-traceless and axial gauge  choices are not mutually
compatible, as    follows   from the impossibility to simultaneously solve 
the equations for the variation vector $ \xi _M $, 
\bea && \ 0= \d g_{\mu m } = 2 \nabla _{(\mu } \xi
_{m )} = \nabla _{\mu } \xi_{m } +  \nabla _{m } \xi_{\mu } 
,\ 0= \d \nabla ^{\mu } g_{\mu \nu} = 2 \nabla ^{\mu } \nabla
_{(\mu } \xi _{\nu )} ,\  0= \d g^\mu _\mu = g^{\mu \nu } \d g_{\mu \nu }=
2 \nabla ^{\mu } \xi _\mu . \eea

We next consider   fluctuations  of  the  internal metric tensor   
in  the  gravity  Lagrangian,  
$L^{(2)} _{grav} = -  {1\over 2\kappa ^2 } \sqrt {g _{10} }  \d g
^{MN} \d (G_{MN} -  \kappa ^2T_{MN} ),$  along with  
those of the axio-dilaton.   The  study can be  developed 
systematically by  utilizing    the quadratic   order  terms of  the 
perturbed     action, given  in Eq.(\ref{appeq34}). 
Upon imposing  the     constraint equations 
$ \d A = \d \tilde g /8 $ and $\d G_{\mu m}=0$  from
Eqs.(\ref{sub2AXIAL1}) and (\ref{sub2AXIAL2}),   the resulting  part 
of the  reduced  kinetic   action  simplifies  to 
\bea && S ^{(2)} = {1\over 4 \kappa ^2 } \int d ^ {
  10 } X \sqrt {\tilde g _{ 10 } } [- e ^{-4 A(y) } \d \tilde g ^{ \mu
    \nu } \tilde E_{\mu \nu } + \ud \d \tilde g ^{\mu \nu } \tilde
  \nabla ^2 _6 ( \d \tilde g_{\mu \nu } - \tilde g_{\mu \nu } \d
  \tilde g _4) + \ud \d \tilde g _4 \d \tilde R ^{(6)}
- {3 \over 8} \d \tilde g _4  \nabla ^2 _6 \d \tilde  g _6 
  \cr && + \ud e
  ^{-4A} \d \tilde g^{mn} \tilde \nabla ^2 _4 \d \tilde g_{mn} - \d
  \tilde g ^{mn} \d \tilde G ^{(6)} _{mn} + {1\over 2 \tau _2 ^2 } e
  ^{-2 A } \d _K \bar \tau ( e ^{-2A } \tilde \nabla ^2 _4 + e ^{2A }
  \tilde \nabla ^2 _6 ) \d _K \tau - {1\over 12 } e ^{-2 A } \d _K
  \bar \tau \dh _\tau \dh _{\bar \tau } ({ G\cdot \bar G \over \tau _2
  })\d _K \tau \cr && + e ^{-2 A } \d \tilde g _m ^n \hat \d [{ 1
      \over 4 \tau _2 } ( G_{npq} \bar G ^{mpq} - {1\over 6} \d _n ^m
    \vert G \vert ^2 ) ] + \kappa ^2 e ^{-2 A} (\mu _3 \d \tilde g _4
  \d \rho _3^{loc} + \d \tilde g _{\mu \nu } \d (T ^{loc} )^ { \tilde
    \mu \tilde \nu } + \d \tilde g _m ^{\tilde n} \d (T ^{loc })
  ^{\tilde m}_n - {1\over 4} \d \tilde g _6\d T ^{loc } ) ]
, \label{sub2eq16} \eea 
where  the functional
differential  stands for  $\hat \d \equiv \d _{\tau , g, G} =\d \tau (\dh / \dh
\tau ) + \d \tilde g_{mn}(\dh / \dh \tilde g_{mn} ) + \d G_{mnp} (\dh
/ \dh G _{mnp}) $.  Ignoring momentarily      the mixing  terms  to    $\d
\tilde g _4,\ \d G _3 ,\ \d \tilde g_{mn} $,  one  obtains
the diagonal wave equations for the internal metric
and axio-dilaton fields   $h_{mn}  =  \d \tilde g_{mn} ,\  \d _K\bar
\tau $, 
\bea && \bullet \ 0= (\tilde \D _L - e ^{-4 A} \tilde \nabla _4 ^2 ) h_{mn} ,\ 
[\d  \tilde G_{mn} ^{(6)}  = \tilde E _{mn}= \ud \tilde \D _L h_{mn} 
,\  \tilde \D _L h_{mn} = 
  -\tilde \nabla _6 ^2 h_{mn} + 2 \tilde R ^{\tilde k \ \ \ \ \tilde
    l} _{\ (mn)} h_{kl} + 2 \tilde R ^{\ \ \ \tilde k} _{(m } h_{n) k}
] \cr &&  \bullet \ 0= (2 \kappa ^2)
{\d S ^{(2)} \over \d (\d _K\bar \tau ) } = {e ^{-2 A }} ( {1\over 4
  \tau _2 ^2 } \nabla ^2 _{10} - \dh _\tau \dh _{\bar \tau } ( {
  G\cdot \bar G \over 24 \tau _2 } ) ) \d _K \tau = {e ^{-2 A }\over 4
  \tau _2 ^2 } ( e ^{-2A } \tilde \nabla ^2 _4 + e ^{2A } \tilde
\nabla ^2 _6 - {e^{6A} \over 12 \tau _2 } G \tilde \cdot \bar G ) \d
_K \tau  .\label{sub2eq16p} \eea 

The linearized  equations for the 2- and 4-form potentials  are
obtained by taking the functional derivatives  of the  field
equations, given   in  Eq.~(\ref{appeq33}). The wave equations  and
Bianchi  identities (enclosed   inside  brackets)   read 
\bea && \bullet \ d \d (\star
_{10} G - i C_4 \wedge G ) =0 ,\ [d \d F_3  =0,\ d  \d  H_3 =0 ]
\ \Longrightarrow \  \d
( d \L + i {d \tau \over 2 \tau _2 } \wedge (\L + \bar \L ) ) =0
,\ [\L = \Phi _+ G_- +\Phi _-  G_+ ]  
\cr && \bullet  \ \d (\star \tilde F_5)  = \d \tilde   F_5, \ 
[d \d  ( \tilde  F_5)   = \d ( H_3 \wedge F_3 ) + {2 \kappa ^2 \tau_3
    \over g_s}     \d \rho ^{loc} ]  \label{sub2eq20} \eea 
where we   used   the   relations  $ \d ((\star _6 \pm i ) G_\mp ) =0,\
\d ((\star _6 \pm i ) G_\pm ) = \pm  2i \d   G_\pm ,\
d (\d G_+ - \d G_-) = -2i d  (\d \tau )  \wedge  H_3 .$  
In GKP  vacua, $ e ^{4 A} =\a ,\  G_-=0 ,\ \nabla _{6}  \tau  =0 $,  
the   equation for the 3-form    becomes 
$ 0=  \d d \L = d (  \Phi _+ \d G_- + \d \Phi _-  G_+  )  =
d \d ( e^{4A} (\star _6  G - i  G ) )  .$    
To  simplify the structure of   wave  equations, one usually   decomposes
the  fields    on bases of (closed and non-closed)  forms   in $X_6 $, 
\bea &&   \d (C_2 -\tau B_2)= x^I (x) P^{(2)} _I (y) +  
y^I _{(1)}(x) \wedge  Q^{(1)} _I(y) +   r^I_{(2)} (x)  R_I (y),\cr &&  
\d C_4 =  v^I (x) V^{(4)} _I (y) +  
w^I _{(1)}(x) \wedge  W^{(3)} _I(y) +   p^I_{(2)} (x)\wedge   X _I ^{(2)}(y)
+   q^I_{(3)} (x)\wedge  Y _I ^{(1)}(y)+   r^I_{(4)} (x)  Z _I(y)
, \label{sub2eqp20} \eea 
retaining   the  coefficient fields in $ M_4$      that couple
through   the wave equations. 
Massless  scalar and vector field modes  can arise   from 
$P^{(2)} _I (y),\  W^{(3)} _I(y) $ in $ X_6$, 
but  massless vector  modes   cannot arise  from       
 $ \d B_{\mu   m },\ \d C_{\mu m } $   since Calabi-Yau manifolds  support
 only    closed   2- and 3-forms.  The  $ U(1)$ gauge compensators  
correspond  to   the      derivative terms 
$ \d (C_2 -\tau B_2) = d v^I (x) \wedge Q_I ^{(1)} (y) 
,\ \d C_4 = d t^I (x) \wedge W_I ^ {(3)} (y)   $.  
The   dynamical and  constraint   equations for the coefficient fields
in $ M_4$    are analyzed  within the
Lagrangian  formalism  in~\cite{freyroberts13}    by substituting
Eqs.~(\ref{sub2eqp20}) in Eqs.~(\ref{sub2eq20}). One can  also draw  useful  inspiration   from  applications    to   the   deformed  conifold  developed       in  several studies  that can be traced  from~\cite{benna07}. 
 We shall  not  elaborate   further on this  point except to  record 
the    classical ansatz proposed for the universal axion $a _0  (x) $   
$ \d C '_4 =\ud a_0 \tilde J^2 + a_2 \tilde J -
d a_0 \wedge K_3 - d a _2\wedge K_1 ,\ \d (C_2 -\tau B_2) = - d a_0
\wedge \L _1  , $  
which involves   the gauge  compensators $ K_{1, 3} $ and $\L _1 $ in
the  2-form  and  4-form potentials.     

The  Kaluza-Klein reduction   is   now systematized by   
decomposing  the  fields fluctuations on bases of harmonic
 functions of $X_6 $~\cite{shiu08}.  Although several 
choices   might be envisaged,  it  is  important  
to  select   bases  that simplify calculations.
Let us  start, for illustration, with the 
decomposition   of    scalar field fluctuations  given by 
 zero modes of the 10-d scalar Laplacian,
$\nabla _{10} ^2  = e ^{-2 A} ( e ^{4 A}
 \tilde \nabla _6 ^2 + \tilde \nabla _4 ^2 ) $.
The orthonormal basis  provided  by 
the warped  type   harmonic functions $Y^I(y) $  for the
Sturm-Liouville  eigenvalue problem,    
 \bea && \tilde \nabla ^2_6 Y^{I_1} (y) =
 - e ^{-4 A (y)} E_{I_1}^2 Y^{I_1} (y) ,\ [ \d \phi ( X ) =
   \sum _{I_1} \phi ^{(I_1)} (x) Y^{I_1 } (y) ] 
 \label{sub2eq15p} \eea  is especially  convenient 
because the kinetic energy action    for the  coefficient fields 
$ \phi ^{(I_1)} ( x )$   takes  then the   diagonal structure 
\bea && S^{(2)} = {1\over 2 \kappa ^2 }
 \int d^{10} X \sqrt {-g_{10} } \d \phi (X) \nabla ^2_{10} \d
 \phi (X) = {1\over 2 \kappa ^2 _4 } \int d^4 x \sqrt
 {-\tilde g_4 } \sum _{I_1} \calm ^K_{I_1} \bar \phi
 ^{(I_1)} (\tilde \nabla _4 ^2 - E_{I_1} ^2 ) \phi ^{(I_1)} , 
\cr && [\calm ^K_{I_1} \d _{I_1J_1} = {1
 \over 4 V_W} \int d^6 y \sqrt { \tilde g_6} e ^{-4
A (y)} \bar Y^ {I_1} (y) Y^{J_1} (y) ] \eea 
where  the
eigenvalues $ E^ 2_{ I_1} $ are seen  to  identify to the 
4-d mass parameters.  (For
a system of coupled fields fluctuations $ \d \phi _i ( X ) $, the
analogous reduction 
procedure would result in a matrix matrix $ ( E_{I_1}^2) _{ij} $ whose
diagonalization would define  a basis of   decoupled massive
eigenvector fields.)   Similar  
definitions  apply to  the metric tensor components
along $ M_4 $ and  the axio-dilaton modulus fields.  The metric
tensor components along $X_6$ in the harmonic gauge  are  decomposed on the
warped basis of eigenfunctions of Lichnerowicz operator, 
given   for (Ricci flat) Calabi-Yau manifolds   by~\cite{shiu08} 
\bea &&  \tilde \D _L Y^{I_2} _{mn} (y) \equiv 
(- \tilde \nabla ^2_6  +   2  \tilde  R _{p (mn) q} \d g^{pq}  ) Y^{I_2}
_{mn} (y) = e ^{-4 A (y)} E_{I_2}^2 Y^{I_2} _{mn} (y) , \eea  
where  the index $ I_2$  labels the basis of  harmonic symmetric 2-tensors. 
Substituting   the harmonic decompositions  for the metric  and axio-dilaton  fields  
\bea && \d \tilde g
_{\mu \nu } (X) = \sum _{I_1} h _{\mu \nu } ^{I_1} (x) Y^{I_1 } (y)
,\ \d \tau = \sum _{I_1} t^{I_1} (x) Y ^{I_1} (y) ,\
\d \tilde g _{mn } = \sum _{I_2} u ^{I_2} (x) Y^{I_2 } _{mn} (y) ,
\eea    
in Eq.~(\ref{sub2eq16}),   yields  the    reduced kinetic action of
canonical  form 
\bea && S _{EFF} = {1\over
  2 \kappa ^2 _4 } \int d^4 x \sqrt {-\tilde g_4 } [\sum _{I_1} \calm
  ^K _{I_1} ( {1\over 2 \tau _2 ^{2} } \bar t^{I_1} ( \tilde \nabla _4
  ^2 - E_{I_1}^2 ) t ^{I_1}
\cr && + \ud \bar h^{I_1} _{\mu \nu } ( - 2 \tilde E^{ \mu \nu
} - E_{I_1}^2 ( h ^{I_1 \tilde \mu \tilde \nu } - \tilde g ^{ \mu \nu
} h^{I_1} ) ) + \sum _{I_2} G^{u} _{I_2} \bar u^{I_2} (\tilde \nabla
_4 ^2 - E_{I_2}^2 ) u ^{I_2} \cr && + \sum _{I_k, J_k} [- \g _{I_1
J_2} \bar u^{J_2} h ^{I_1} + (\a _{I_2 J_2} - \b _{I_2 J_2} ) \bar
  u^{I_2} u^{J_2} + B _{I_2 J_1} \bar u^{I_2} t^{J_1} + A _{I_1 J_1} \bar t^{I_1} t^{J_1} ] , \cr &&
[G^{u} _{I_2} \d _{I_2J_2} = {1 \over 4 V_W} \int d^6 y \sqrt {
\tilde g_6} e ^{-4 A (y)} \bar Y^ {I_2} _{mn} (y) Y^{J_2
\tilde m \tilde n} (y) ,\cr && \tilde E_{\mu \nu } = \ud (
-\tilde \nabla ^2_4 h _{\mu \nu } -\tilde \nabla _\mu \tilde
\nabla _\nu h + 2 \tilde \nabla ^\l \tilde \nabla _{( \mu } h
_{\nu )\l } + \tilde g_{\mu \nu } (\tilde \nabla ^2 h - \tilde
\nabla ^\l h _\l ) )] \label{sub2eqm20} \eea 
where  $\a ,\ \b ,\ \g
,\ A,\  B$ are mass terms induced by the   geometry  and  3-fluxes. These
are given in~\cite{shiu08}  by volume integrals over the basis functions where 
$\a _{I_2 J_2} \sim \int Y^{I_2} _{mn} 
 \bar Y^{J_2 m r }  G^{n pq} \bar G_{r pq},\   \b _{I_2 J_2} \sim \int Y^{I_2  r}
_r   \bar Y^{J_2m n }  G_{mpq} \bar G_{n} ^{\ \  pq} $ refer to the  internal  metric, $  \g _{I_1I_2} \sim \int  \bar Y^{I_1}\tilde \nabla ^m \tilde \nabla ^n  
 Y^{I_2}_{mn}   $ refers    to   the  external-internal metrics  mixing,  $ A _{I_1J_1} \sim \int e ^{-2A} Y^{I_1} \bar Y^{J_1} G\bar G   $ refers
 to  the axio-dilaton and  $ B_{I_2J_1} \sim \int  e ^{-2A}
Y^{I_2m n} \bar Y^{J_1 } G _{mpq} \bar G _n ^{\ \ pq} 
$ refers      to  the axio-dilaton   internal metric mixing. 
The harmonic analysis for antisymmetric
tensor fields is developed in a similar way in terms of eigenfunctions
of Laplace-Beltrami operators.   We restrict,  for   illustration purposes, to the   decompositions of $ \d C_{mn} ,\ \d B_{mn} $ and $ \d
 C_{mnp\mu } $ on bases of 2-forms $ \o _a $ and 3-forms
 $\chi _I $,  \bea && \d ( C_2 -\tau B_2) = \sum _a ( c^a (x)
 -\tau b ^a (x)) \o _a (y) ,\ \d C_4 = \sum _I V^I (x) \wedge
 \chi _I (y)  .  \eea      
Substituting   into   the 10-d
 kinetic terms  $ ( G \wedge \star \bar G ) $ and $ (F_5
 \wedge \star F_5 ) $,    yields   the kinetic and mass
 terms for the 4-d scalar  fields $ c^a (x) ,\ b^a
 (x) $ and vector fields  $V^I_\mu (x)$~\cite{shiu08} 
\bea && S ^{(2)} (c^a,
 b^a ) = - {2 g_s ^2 \over 2 \kappa _4 ^2} \int d^4 x \sqrt
 {- \tilde g_4 } \sum _{a, b} G_{ab} ( d c^a \wedge \star _4
 d c^b + g_s ^{-2} d b^a \wedge \star _4 d b^b  + M_{ab}^{2}
 (c^a c^b + g_s ^{-2} b^a b^b )  ) ,\cr && S ^{(2)} (V^I) = -
 {2 \over 2 \kappa _4 ^2} \int d^4 x \sqrt {- \tilde g_4 }
 \sum _{I, J}  ( G_{IJ} ^V d V ^I\wedge \star _4 d V ^J +
 M_{IJ} ^{2} V^I\wedge \star _4 V^J ) , \cr && [{G_{ab}
     \choose M_{ab}^{2} } = {1\over 4 V_W} \int _{X_6} {\o _a
     \wedge \star _6\o _b \choose e ^{4A} d \o _a \wedge
     \star _6 d \o _b } ,\ {G_{IJ} \choose M_{IJ} ^{2} } =
   {1\over 4 V_W} \int _{X_6} { \chi _I \wedge \star _6\chi
     _J \choose e ^{4A} d \chi _I \wedge \star _6 d \chi _J
 }].  \label{sub2eqx20} \eea 
We   finally   state briefly  the main  conclusions drawn   from the above discussion~\cite{shiu08}. The   choice  of warped type bases of   eigenfunctions,   $ \tilde \nabla _6^2  \to -  E_m ^2 e ^{-4A} $, 
is seen  to  remove  automatically  kinetic mixings  in
Eqs.~(\ref{sub2eqm20})  between graviton  modes
along $ M_4$ and geometric moduli from $\d g _{mn}$.
Mixing  terms between the scalar modes from $ a_0 \in C_{mnpq} $
and $(c^a - \tau b_a ) \in C_2 -\tau B_2 $ and the complex structure
moduli $S$  also  cancel out.   Nevertheless,  the geometrical mass mixing term
$ \d \tilde g _4 \d \tilde R ^{(6)} $ and       mixing    terms 
induced  by   3-fluxes still occur also  in Eq.(\ref{sub2eq16}).

Due  to the non-linear dependence of the scalar curvature on
the metric,  the expansion of the   
perturbed action $ \sqrt {(\tilde g + h   )} R ^{(4)} 
( \tilde g + h ) $     in powers of  the metric  tensor fluctuations 
$h_{\mu \nu } $   contains    cubic and higher
order local couplings between  graviton modes $ h^p \dh h \dh
h,\ [p\geq 2]$.     We  provide in Eq.~(\ref{appeq37})  of
Appendix~\ref{subapp5}   a closed formula  in an arbitrary gauge 
for the perturbative  interactions of gravitons,   exhibiting
the      various saturations of  spacetime indices. 

\subsubsection{Effective action   for  moduli fields} 

The moduli  fields  of the  internal  manifold, 
defined  as solutions of the  linearized  equations,  $ \d \tilde  G
^{(6)} _{mn} = \tilde E_{mn} =0 $, are  expected to be    among the
lightest  excitations of the  compactified  theory.  We  examine
first  the case where   3-fluxes  effects   can be treated  as
perturbations  that   lift     the moduli space of 
vacua without    affecting  the   background geometry.     
Let us consider a    Calabi-Yau manifold $X_6$, equipped with a fixed
Ricci flat metric of given    complex structure, specified by complex
holomorphic and    anti-holomorphic coordinates $z^i 
,\ z^{\bar i}, \ [i=1,2,3] $. The      moduli space  of  vacua is  
parameterized by complex and Kahler structure  cohomology  classes.
The   complex structure    and Kahler    moduli  fields $ S_\a ,\ T _a $ are  dimensionless  fields defined   from   the decomposition of  the (unwarped)  metric tensor  fluctuations      on the bases of (exact, modulo closed)  
complex 3-forms  and real 2-forms,  
\bea && \d \tilde g _{kl} = \sum _{\a =1} ^{h^{(1,2)}
} \d _\a \tilde g _{kl} (y) S_\a (x) + H.\ c.  ,\ \d \tilde g _{k \bar
  l} = \sum _{a=1}^{h^{(1,1)}} \d _a \tilde g _{k \bar l} (y) T _a (x),
\cr && [\d _\a \tilde g _{kl} (y) = {1\over \vert \vert \O
    \vert \vert ^2 } \bar \O _{k} ^{\ \bar i \bar j } \chi ^\a _{ l,
    \bar i \bar j } (y) ,\ \d _a \tilde g _{k \bar l} = \o ^a _{k\bar
    l} (y) ]  \label{sub2eq6} \eea 
where  $\O (y) $ and $\bar \O (y) $ are complex conjugate $(3,0) $ and
$(0,3) $-forms and $ (\chi ^\a (y),\ \chi ^{\bar \a } (y) ) ,\ \o ^a
_{r\bar s } (y) $ are bases in the vector spaces of harmonic $(1,2)
\oplus (2,1)$-forms and $(1,1)$-forms on $X_6$,  of dimensions given by
the Hodge numbers $ h ^{p,q}$.
The $ S_\a $ join with  fermion  fields into chiral  supermultiplets  and
the     $T_a$  pair up with the
axion moduli  $ a_a  \in C_{mnpq}$   into  bosonic components
of chiral supermultiplet fields.   When  
promoted to (massless)  superfields,  
$S_\a (x)  $ and $ T_a  (x) $   are used    to  construct the
effective action   in terms of supersymmetric
sigma  models on  Kahler manifolds~\cite{candelas91}.  
The kinetic energy  terms    are   described  by  the 
metric tensors $ G ^S _{\a \bar \b } $ and $G ^T _{a \bar b }$ 
generated  by   the  Kahler potentials  $ K (S_\a ,\ T_a) $  of  the moduli
spaces~\cite{dewolfe02},  
\bea && S^{(2)}=- 
{1\over 2 \kappa _4 ^2} \int d^4x \sqrt {- \tilde g_4} \tilde g ^{\mu
  \nu } (G ^S_{\a
  \bar \b }  \dh _\mu S^\a \dh _\nu S^{\bar \b } +G ^T_{a \bar b } \dh
_\mu T^a \dh ^\mu T^{\bar b } ) ,\cr && [ G ^S_{\a \bar \b } (S) =
  {1\over 4 V_W} \int d^6y \sqrt {\tilde g_6} e ^{-4 A (y)} \d _\a
  \tilde g ^{kl } \d _{\bar \b } \tilde g _{kl} ,\ G ^T_{a \bar b } =
         {1\over 4 V_W} \int d^6y \sqrt {\tilde g_6} e ^{-4 A (y)} \d
         _a \tilde g ^{k\bar l } \d _{\bar b } \tilde g _{k\bar l} ]
\cr && \ \Longrightarrow \ \kappa _4 ^2 K (S_\a ,\ T_a) = - \ln (-i
\int _{X_6} e ^{-4 A} \O (S_\a ) \wedge \bar \O (\bar S_\a ) ) - \ln
(-i \int _{X_6} \sqrt {(\tilde g_6 + \d \tilde g _6 (T_a) ) } e ^{-4
  A} ) . \label{sub2eq15} \eea  
The     above formulas  are   of standard   form  except for the 
warp   profile factors under the  volume integrals  over $X_6$.
Note that the compatibility  with the constraint equation 
$\d G_{\mu m }=0$ in Eq.~(\ref{sub2AXIAL2})  requires~\cite{shiu08}   adding  to the  decompositions  in Eq.~(\ref{sub2eq6})  extra  terms $ \d ' \tilde g _{kl }
 ,\ \d ' \tilde g _{k   \bar l } $  to the  wave  functions of moduli.  

We   next   consider the universal  dilaton  and  volume moduli 
associated to   Weyl rescalings    of the 10-d   spacetime metric 
 and    6-d internal   space  metric~\cite{dewolfe02,gidd05,frey08}.   
The  reduced  dilaton   field $ \phi (x) $  joins  
with the axion field  $C_0(x) $  in  the  scalar     component of   the 
4-d axio-dilaton  superfield,    $ \tau (x) = i \tau _2 + \tau _1 = i e
^{-\phi (x) } + C_0(x) $.   The    volume   field      $ u(x)$,    
introduced  through   the       internal and external metrics
rescalings   $ g_{mn } \to e ^{2  u(x) } g_{mn } , \ g_{\mu \nu } \to e ^{-6 u(x) }
g_{\mu \nu }  $,    designed to   restore   the  4-d  Einstein frame,
partners with   the   axion     descending from  the  4-form potential,    
\bea && \d C_4 = \ud a_0 \tilde k \wedge \tilde k
+ a_2 \tilde  k , \ [\tilde k _{mn} =\tilde J_m ^p \tilde g _{pn},\ 
\tilde J ^2 = -1,\ 
\ V_{CY} = \int _{X_6}\sqrt {\tilde g_6
} ={1 \over 6} \int \tilde k \wedge \tilde k\wedge \tilde k  = {1\over
    3} \int \tilde k \wedge \star _6 \tilde k]    \eea  
into   the scalar component  of the   volume chiral  superfield
$ \rho (x) = i \rho_2 + \rho_1 =
 i e^{4 u(x)} + a_0(x) ,\  [d a _2 = e ^{-8 u (x)  } \star
  _4 d a_0 ] $.   The $0$- and $2$-form fields $ a_{0, 2}$  are   
related   by  the 4-duality mapping implied  by the  5-form
self-duality,   $\star  _{10} \tilde F_5  = \tilde F_5$,  
and the   Kahler 2-form  $\tilde k = \ud \tilde  k_{mn} d y^m  \wedge d
 y ^n$,     built  from the complex structure  matrix $ \tilde J$
 and  metric of $X_6$, is used to  construct the  unwarped volume $V_{CY} $.
The   4-d kinetic energy  actions  and Kahler potentials
of the  dilaton and volume  moduli are   then given by  
\bea && S^{(2)} = {1 \over 4
  \kappa ^2_4} \int d^4x \sqrt {- \tilde g_4} [{1\over 2\tau _2^2 }
  \bar \tau \nabla _{4}^2 \tau +{3 \over 2\rho _2^2 } \bar \rho \nabla
  _{4}^2 \rho ] \  \Longrightarrow \ \kappa _4^2 K (\tau , \rho )
= - \ln (-i (\tau - \bar \tau ) ) - 3 \ln (-i (\rho - \bar \rho ) ).
\eea 
In warped compactifications,   however,  the    above
rescaling  definition  of the universal volume modulus is 
incompatible with the classical  field equations.  A  consistent
definition  in   GKP vacua  (where $ \a = e ^{4A}$),  is provided   by
the additive shifts  in the metric tensor and 4-form field   classical  profiles   along $ M_4 $ and $X_6$~\cite{gidd05},
\bea && e ^{ \pm 2 A (y) } \to e ^{
  \pm 2 A (y, c(x) ) } = (e ^{- 4 A (y) } + c(x)) ^{\mp 1 / 2 } ,\cr
&& \tilde F_{5} \to d (e^{-4A} + c(x))^{-1} vol(M_4) + (e^{-4A} +
c)^{2} d (e^{-4A} + c(x))^{-1} vol(X_6).   \label{volmod1} \eea
This   follows from the  equation for the inverse warp  profile,  Eq.~(\ref{appeq25}), 
$ \tilde \nabla ^2 _6 (e^{-4 A} ) = - e^{-8A} \tilde R^{(4)}
- {1\over 12 \tau _2} G  \tilde \cdot \bar G + {2 \kappa ^2 \over g_s}
e ^{-6 A} \hat T^{loc}  ,\ [\hat T^{loc} \propto 1/ \sqrt {g_6} = e
  ^{6A} / \sqrt { \tilde  g_6}  ]  $  which is  found to  admit 
(for  $  \tilde R^{(4)} =0 $ or   very small   relative to other terms)     
a  one-parameter family of solutions   
generated  by    $y$-independent shifts of the   warp and 5-flux   
profiles,     $ e ^{-4 A (y)}
\to e ^{ - 4 \hat A (y, c)} = e ^{-4 A (y)} + c(x) ,\ \a ^{-1} (y) \to
\a ^{-1} (y) + c(x) $. The comparison  to  the metric  
$ d s^2 _6 \simeq  c^{1/2}  d  \tilde s^2 _6 $  
in the dilute flux case, $ c(x) >> e ^{-4 A (y)} $, 
relates  the  universal  volume modulus $ c(x)$  to 
 the multiplicative ansatz $ u (x) $ in the large volume  limit
     as $ c(x) \simeq e  ^{4u(x) } \sim \calv ^{2/3} $. 
The   effective   action for $ c(x)$   was derived   
in  the Hamiltonian  formalism~\cite{frey08} 
by  introducing    a modulus-dependent warp profile $ e ^{ - 4
  \hat A (y, c)} $  and  the gauge compensator   field    $ \d \tilde g
_{\mu m} = \dh _m \hat B (y)\dh  _\mu c (x) $. The  constraint
equation,   $ \tilde \nabla _6^2 \hat B (y)= e 
^{-4 A(y)} + V_W /V_{CY} $,   is  then 
found to  admit  an  algebraic  solution for $
\hat B(y)$,    resulting    in   the modified Kahler potential 
\bea &&\kappa ^2 _4 K (\rho )  = -3 \ln ( -i   (\rho - \bar
\rho ) +2 {V_{W} \over V_{CY} } ) \equiv -2 \ln \calv ,\ [\rho = i
  \rho_2 + \rho_1 = i c(x) + a_0(x) ].\label{volmod2} \eea  

We now briefly   consider    the  effects of  3-fluxes for GKP vacua
in F-theory flux  compactifications  on 
$M_4 \times X_8$. The  $G_4$-fluxes  here  are classified in  cohomology
classes  dual   to   domain walls in $ M_4$  associated to 
 wrapped 5-branes  of world volume   in
 $ R^{2,1}   \times \S _3  \subset M_4  \times X_6   $. The  domain
 walls  tensions are   interpreted  as    F-terms 
from the   flux  superpotential~\cite{gvw99}  
$ W _{GVW}  \propto  M_\star ^8 \int  _{X_6} G_3 \wedge \O  $.
Holomorphy  requires  that the superpotential   is insensitive   to warping.
Since the field equation      $ d \L  = 2 d  (e ^{4 A} G_-)=0  $,
from Eq.(\ref{sub2eq20}),  implies  
that $e ^{4 A} G_-$ is a  (closed) harmonic    3-form,   
decomposing      $ G _3$ in  $ W _{GVW}  $ on       the   basis  of
 cohomology       $ (\O _{3,0} ,\   \chi   ^\a _{1,2} )  $-forms  
produces from the supergravity  action the  scale  free 
scalar potential,   $ V_{flux} = \sum _{A= S_\a , \tau }  \vert D_A  W
_{GVW} \vert ^2 $,   where the summation  over $A$  runs only 
over the  complex structure and dilaton moduli~\cite{dewolfe02}. 
 The resulting  flux potential is  formally analogous to the  formula 
 in Eq.~(\ref{sub2eq17}).    

\subsection{Dimensional  reduction  of supergravity    theory 
on warped conifold}
\label{subsec2.3}

For  obvious  practical  reasons, it is  desirable  to  modelize  the deformed
region glued to the bulk    by a solvable  conic manifold.
One of the best understood   candidate  is the conifold $ \calc
_{6}$~\cite{candelas90}. This is a non-compact    Kahler 
manifold,   part      of the  family of
Stenzel spaces~\cite{pufu10} $ \calc _{2d -2 } $,  
defined as loci of
the complex quadric equations $ \sum _{a=1} ^{d} w _a ^2 =\e ^2
,\ [w_a \in C,\ \e ^2 \in R ^+] $.  Away from
the conic singularity  at the apex  $ w_a =0$,  the deformation
parameter  $\e $ is   negligible    and  $ \calc _{6}$
asymptotes to the undeformed (singular) conifold,  equivalent    to a
real cone over the compact base manifold $ T^{1,1}$. 
We present a brief review of algebraic properties of the conifold in Appendix~\ref{app2}.

\subsubsection{Flux compactification on   $ AdS_5 \times T^{1,1}$ spacetime} 
\label{subsubsec23.1}

The   modelization by the conifold $ \calc _{6}$ is
motivated by the solvability of 10-d supergravity in the target
spacetime $M_4 \times \calc _{6}$  embedding  space-filling $N\ D3$-brane
stacks.   The  Klebanov-Strassler solution~\cite{klebstrass00} describes the
background for  the warped deformed conifold with 
a quantized $F_3 $-flux $M$ and a  quantized 2-form potential 
$B_2 \propto K \equiv N/M$   embedded near the  apex.  (In the dual gauge
theory picture, the $F_3 $-flux corresponds to a fractional $
M\ D5$-brane stack wrapped on the collapsed $S^2$-cycle near the tip.)
At large radial distances away from the   apex, $ r >>
\e ^{2/3} $, the    geometry asymptotes  $ AdS
_5\times T^{1,1}$ spacetime,  where the radial coordinate    joins with the uncompactified directions   along  $
M_4$ to build the $ AdS _5 $ spacetime transversal 
to the conifold compact base   $T^{1,1} $. 
We  shall restrict  consideration to the large  radial distance  limit of  the
Klebanov-Strassler   solution     which corresponds   to   the   
Klebanov-Tseytlin  solution   for the undeformed
conifold~\cite{klebse00}. (This is also asymptotic  to the  solution
in the resolved conifold background~\cite{zayas00}.)  
We refer for more details to the
review~\cite{herzog01}.    The   classical 
metric tensor and 5- and 3-form field
strengths   are given in terms of a radial warp
profile $ h (r) $ and the    harmonic 2- and 3-forms  $\o _{2, 3} $ on $ T^{1,1}$  
as  \bea && \bullet \ ds ^2 _{10} = h^{-1/2 } (r)
d\tilde s ^2 _4 + h^{1/2 } (r) (dr ^2 + r ^2 ds ^2 ( T ^{1,1} ) )
,\ \ [h (r) \equiv e^{-4 A (r) } = (\calr / r )^{4} (c_1 + c_2 
\ln {r \over   \calr } )   , 
\cr && c_2 ={3 g_s M \over 2\pi K },\ c_1 = 1 + {3 g_s M
    \over 8\pi K } ,\ \calr _+ ^4 \equiv \calr ^4 = {27 \pi \over 4}
  g_s N \a ^{'2} ,\ {\calr _- \over \calr _+ } = ({3 g_s M \over 8 \pi
    K }) ^{1/4} ] .  \cr && \bullet \ F_3 = dC_2 = {M \a ' \over 2} \o
_3 , \ H_3= dB_2 = {3 g_s M \a ' \over 2 r} dr \wedge \o _2, 
\ [\o _2= \ud ( g^{(1)} \wedge g^{(2)} + g^{(3)} \wedge g^{(4)} )
  ,\ \o _3= \o _2 \wedge g^{(5)}   , \cr &&    
B_2 = {3 g_s M \a ' \over 2}\ln {r \over   \calr } \o _2 
= {\hat l_s ^2 \over V
    (S^2)} K (r) \o _2  ,\ K (r) = {3 g_s M \over 2 \pi } \log (r/\calr )
 , \ \int _{S^2} \o _2 = V(S^2) = 4\pi, \  \int _{S^3} \o _3 = V(S^3) =
8 \pi ^2 ]. \cr &&
\bullet \ g_s \tilde F_5 = d h^{-1} (r)  \wedge \ vol (M_4) - r^5 d h (y) \wedge
\ vol (T^{1,1} ) ,\ g_s \tilde F _{\mu \nu \l \s m} = \tilde \e _{\mu
  \nu \l \s } \dh _m h ^{-1} ,\ g_s \tilde F _{mnpqr} = -\tilde \e
_{smnpqr} \dh ^s h ,\cr && [ \int _{T^{1,1}} \tilde F_5 = \hat l_s ^4
  N_{eff} (r) ,\ N_{eff} (r) = N (1 + c_2 \ln (r / \calr ) ) ,\ vol
  (T^{1,1}) = {1\over 108} g ^{(1)} \wedge \cdots g ^{(5)} = {1\over
    54} \o _2 \wedge \o _3 ] .  \label{sub2eq22} \eea 
The   basis of  left-invariant  1-forms  
$ g ^{(a)} $   of $ T^{1,1}$ is defined in Eq.~(\ref{app2eq9}),
the ultraviolet  scale in the  logarithmic terms    is identified  to 
the   curvature radius  parameter $ r _{uv} = \calr $,  
the differential $p$-form   fields and their
tensorial  components  are assigned  the scaling     dimensions,    $
[F_{M_1 \cdots  M_p} ]   = E ^{1} , \   [F _{(p)} ] = E ^{1 -p } ,\ [vol
  (AdS _5) ] = E^{-5},\  [vol (M_4)] = E ^{-4} ,\   [vol (T^{1,1} ) ]
=E^0 $  and  the  non-covariant   5-flux (unlike
$\tilde F_5$)   is integer  quantized, $ \int  _{T^{1,1}} F_5 =  \int  _{T^{1,1}} 
(\tilde F_5 - B_2 \wedge F_3  )=  \hat l_s ^4 N$. 
The warp profile  can  be  equivalently  expressed 
by   trading  the maximal   (ultraviolet)  radial  scale $
r_{uv} $    for the  minimal (infrared) radial  scale  $ r _{ir}$  where   
       warping   is   minimal, by writing   
\bea &&  e ^{-4 A}= (\calr / r )^4  
[1 + c_2 (1/4  + \ln  (r/r_{uv} ) ) ]   = (\calr / r )^4   c_2 \ln  (r/r_{ir} )  ,\    [ r_{ir} / r_{uv}= e ^{-2\pi K / (3 g_s M) -1/4}  \sim w_s]
  .  \label{eq.warpA1} \eea  
The  inequalities $ \sqrt {\a'} < \calr _- < \calr _+ $   between      the 
curvature radii at the conifold apex and mouth $\calr _\mp $  and the
string  length scale are needed  to   
justify the classical   supergravity description. 
The logarithmic radial  profiles proportional to the RR  3-flux  
$M$  are associated to       the breaking of    conformal   symmetry 
breaking.  At $ M=0$, the  geometry   identifies
to the   $ AdS _5  \times T^{1,1} $   spacetime,    direct product of
constant negative and positive  curvature 
manifolds of  curvature radii  $\calr $. We usually set
the   curvature radius parameter to unity   since the
dependence on   $\calr $  is easily  reinstated  via  naive
dimensional analysis.
The classical solutions  for the warp  profile and     5-flux   
 near  $r \to 0 $   are  given (ignoring  logarithms)    by the limiting  formulas, 
\bea && h ^{-1}  (r) \simeq  ({r \over  \calr })^4 ,\  g_s \tilde F_5
 \simeq {4 \over \calr } ( vol (AdS _5) + \calr ^5 vol ( T^{1,1} ) ) ,\ [vol
  (AdS _5) =({r  \over \calr })^3 dr \wedge vol (M_4) ].\eea
We also  quote, for the sake of comparison, 
the   classical  solution  in the near horizon 
background  spacetime $ AdS_5 \times S^5 $  for the 
$ N\ D3$-brane stack  supported  by a constant $5$-flux 
\bea && \bullet \ ds ^2_{10} =
e^{2A(r)} \tilde g_{\mu \nu } dx^\mu dx ^\nu + e^{- 2A(r)} ( d r^2 +
r^2 \tilde g_{ab } d\T ^a d \T ^b ) = ds^2 (AdS _5 \times  S^5 ) ,
\cr && [ds^2 (AdS _5 \times S^5 )= { \calr ^2\over z^2 } (d \tilde s
  ^2_4 + dz ^2) 
  + \calr ^2 d  \tilde s^2 (S^5 ) ,\ d\tilde s^2 (S^5 ) = \tilde g _{ab }(S^5
  ) d\T ^a d \T ^b, \cr && e^{2 A(r) } = \a ^{1/2 } (r) = ({ r\over
    \calr } )^2,\ \calr = (4 \pi g_s N \a ^{'2} )^{1/4} ,\ {\calr
    \over z} = {r \over \calr } ] \cr && \bullet \ g_s F_5 = N (1
+\star ) d \a (y) \wedge vol (M_4)= {4 N \over \calr } ( vol (AdS _5)
+ \calr ^5 vol ( S^5 ) ) ,  \  [\int _{S^5} F_5 = \hat l_s ^4
  N] .   \label{sub2eq23} \eea     
This consists of   a pair  of  sub-manifolds  of   equal, opposite
 (negative and  positive) sign    curvature   radii  related to those   of the
 conifold as $\calr  
(T^{1,1} ) / \calr (S^5 ) = V (S^5 ) / V (T^{1,1} ) = 27/16$.  The
exponential warping in $AdS _5$ becomes manifest in the
parametrization of the metric utilizing the  flat proper distance
radial variable $Y$, \bea && ds^2 (AdS _5) = e ^{-2Y/\calr } d\tilde s ^2_4 + dY^2
+ \calr ^2 d \tilde s^2 (S^5 ) ,\ [\calr /z = r/\calr = e^{-Y /\calr }
]. \eea 
The redshift effect  from   warping, setting     the scale hierarchy
relative to the string theory  scale,  is usually   defined 
  by   the minimum   value of the warp profile,   
$w_s \equiv m_{eff } / m_s  = min \ ( e ^{ A ( r ) } )  \simeq e ^{ A
  ( r_0 ) } \simeq r_0 /\calr  $. 
With this definition    of the  warp factor $w_s$, analogous to 
that in Eq.~(\ref{eq.warpA1}), 
the proper  energies and energy densities in $ M_4$   near 
the  apex    get rescaled  by  powers  of the warp factor    $ w_s$ and $ w
_s^4$.   

The   Klebanov-Strassler  solution  describes the classical string
theory background   corresponding  (in the sense of AdS/CFT duality)
to the $\caln = 1$   supersymmetric gauge theory 
on $N D3 + M \ D5 $-brane stacks   embedded   near the  conifold 
tip. The  associated  quiver gauge theory  was  first identified
in~\cite{klewit98} and tested in~\cite{gubser98,ceresole99}.   
The  local   symmetry group is    
$\calg  = SU(N+M) \times SU(N) $ and   the   matter   content
consist  of two pairs of  chiral superfields $ A_i,\ B_i ,\ [i=1,2] $ in
(conjugate) bi-fundamental representations  of  $\calg $ 
and    in  doublet representations of the   anomaly  free global 
(flavour, R and  baryon  number) 
symmetry group $G = SU(2)\times SU(2)\times Z^R _{2M} \times U(1)_b
$,   interacting  through the quartic order  superpotential $ W $,     
\bea && W= h Tr _\calg (A_1B_1 A_2 B_2 - A_1 B_2 A_2 B_1) 
,\ [\calm = \{A_{i} \in (N+M , \bar N)_{2,1, 1/M, 1} \oplus  B_{i}  \in
(\overline{N+M} , N)_{1,2, 1/M , -1} \}  ]   \label{sub2eq25} \eea 
where the lower  suffix labels refer  to  representations of 
the global  symmetry group $G$.  The   coupling constants  of 
the  gauge theory     on   $  N D3$-branes  and 
$M D5$-branes wrapped  on $ S^2$  near the  conifold tip 
are related to the  supergravity theory  modulus from $ B_2$ as~\cite{lawrence98,klebnekrasov99}, 
\bea && [g_1 ^{-2}  \pm   g_2 ^{-2}]= { e ^{-\phi} \over 4\pi  g_s}
[1, (   {2 \over \hat l_s ^2 } \int _{S^2} B_2  -1)  \ \text{mod}
  (2\pi )] . \eea  The  energy-distance    relationship $Q = r /  (\a '
\sqrt { g_s N_{eff} } ) $  connects  the radial distance $r$  of the
supergravity theory   to  the running momentum  $Q$ of the gauge theory  
on the  $ N\ D3$-brane stack that decouples  from gravity
in the   limit of large 't Hooft   running coupling    constant
$g^2 _{eff} (Q)  = g^2 _{YM} N_{eff} ,\ [N_{eff} \simeq   N]$.
The moduli  space of vacua   includes  mesonic 
and baryonic branches~\cite{strasslercascade05,dymarberg05}  along
which  the   symmetries are  broken   typically   as       
$ \calg \to SU(M),\    Z^R _{2M}  \subset U(1)_R \to Z_2 $, possibly  with 
residual chiral matter. The renormalization  group running of  
the gauge  theory couplings at finite $M$ 
is reproduced  by the 3-flux and   5-flux   logarithmic 
terms in classical radial profiles   
which   jump by   integers  periods as   the radial  distance  
decreases  by  appropriate  steps~\cite{klebnekrasov99}, 
\bea &&   K (r) \to  K (r) -k  ,\ N_{eff} (r) \to N_{eff}  (r)  - k M ,\ 
[\d  _k (\ln r ) = -{2\pi  k  \over 3 g_s M}  ,\  
\int _{S^2}  B_2 =\hat l_s ^{2}  K(r)  
\equiv  {3 \hat l_s ^{2} g_s M \over 2 \pi }
\ln (r /\calr ) ]  . \eea    
The  5-flux   vanishes, $ N _{eff}  (r_0)= N - KM =0$,
when one reaches   the conifold apex  after  $k= K$  steps~\cite{gkp01},   
$ r_0 / \calr = e ^{-2\pi K   / (3 g_s M ) }  \simeq w_s $.   
At each    step, the gauge  theory factor     
with  the  larger  numbers  of colour   and
flavour    undergoes   self-similar     Seiberg dualities  replacing, say,  
the        theory  with $ N_c= N+M ,\  N_f = 2N $    by  
the    dual   theory  with $N_c'  = N_f -N_c = N-M ,\  N_f '= 2N $.      
For   $M = NK$, the cascade of Seiberg dualities 
$  SU(N+M )\times SU(N) \to SU(N) \times SU(N-M) \cdots \to
SU(2M)\times SU(M)   $ ends  with the  $ SU(M)$  theory.  
A   continuous family  of  supersymmetric  solutions for type $
II\ b$     supergravity   on  the resolved warped   deformed  conifold,  
interpolating     between   Klebanov-Strassler and
Maldacena-Nunez~\cite{malnez00}  solutions,   
was   meanwhile constructed  in~\cite{butti04}  and  further tested in~\cite{maldatelli09}.

\subsubsection{Harmonic  analysis  of fields on  conifold}
\label{subsubsec23.2}

The  calculations are  greatly   simplified if  the  deformed conifold is  
replaced      by a truncated  undeformed  conifold
bounded  by  hard walls~\cite{polstrassler02}. 
The  $ M_4 \times \calc _6$  spacetime in the  interval    between  
the conifold  apex and   mouth   consists     of 
an  $AdS_5 \times  T^{1,1}$  slice  along  the 
 $r$-direction of $AdS_5$ bounded by  the horizon   and 
boundary  at  $r _{ir} \simeq r_0 $ and   $ r _{uv} \simeq \calr $.
The end    point  radial distances of the supergravity theory 
are related to the  dynamical   (confinement and     $ U(1)_R$  chiral
symmetry breaking)    mass  scales $\L \simeq   S ^{1/3}  $,   and the
ultraviolet  cutoff mass  scale $\L _0 $  of the $N D3$-brane   gauge   theory 
by $ r_{ir} =\a ' \L \simeq  \a ' S^{1/3} ,\   r_{uv} =\a ' \L _0 .$ 
 In the holographic interpretation~\cite{verlinde99}  that we
adopt, the  infrared  and  ultraviolet   boundaries   
delimit the safe   interval beyond which     the  $ N \ D3$-brane
decoupling  from  closed  strings  is invalid. 
Additional  sectors  could be  introduced by  hosting 
 probe $D3 $- or $D7$-branes  in appropriate   configurations 
near $ Z_2$-orbifold points, say.     
In the holographic gauge   mediation  model of~\cite{benini09},
for instance,   GUT scale  Standard Model  branes are    
introduced near the boundary and 
hidden  $D3$-branes    near the horizon    with messenger type
$D7$-branes  in between. This interpretation allows  one to   dispense  from  
a dynamical  stabilization mechanism    of   the 
boundaries~\cite{goldbergerwise99},   hence  avoiding    the
existence  of a  light radion  moduli  that   could   
back-react on  the   background~\cite{brummer05}.   

The dimensional reduction proceeds in
two  main stages.  One first  decomposes    the 10-d field  fluctuations  on
bases of scalar (and solid) harmonics $ Y^{\nu , M} _q ( \T ) $ of the conifold 
compact base manifold $ T^{1,1} $,   satisfying the
orthonormalization and completeness conditions in
Eq.~(\ref{app2eq11}).  The coefficient fields in $AdS_5$ are in turn 
decomposed in Fourier series  on plane wave modes in $ M_4 = \dh
(AdS_5) $.  For  given 5-d geometric mass $ M_5$, 
the    normal  modes  of   4-d squared masses $ k_m^2 = - E_m ^2$ 
are assigned wave functions  in the radial and   angle    directions
of $ AdS_5$  and   $ T^{1,1} $, labelled by quantum numbers $\nu $ and $m$ 
associated to representations of the conifold isometry group $G $ and
the radial excitations in $ M_4$. 

We  here briefly discuss the harmonic analysis on the  base $T^{1,1}$
of the conifold, sending  the reader for    
more details    to Appendix~\ref{suc2app2}.      The isometry
group of the (undeformed) conifold is $ G= SU(2) \times SU(2)
\times U(1) _R $, where the $ U(1) _R $ factor identifies to the
R-symmetry group  of  the supersymmetric  dual gauge theory. The  fixed radial    sections of
the  conifold  are copies
of the coset space $ T^{1,1} \sim G/H= SU(2)_j \times SU(2) _l / U(1)
_H $.  The states of conserved angular momenta $ (j, \ l ) $
for  the  Lie algebra generators $
T_a ,\ \tilde T _a , \ [a=1,2,3] $ of the 
groups $ SU(2)_{j,l}$  are labelled by the magnetic
quantum numbers $M\equiv  (m,n) = (j_3 , \ l _3 ) $, 
\bea && U(1)_H:\ T_H = T_3 + \tilde T_3 = j_3 + l_3 \equiv  {m+n } =q
,\ U(1)_R :\ T_5 = T_3 - \tilde T_3= j_3 - l_3 \equiv  {m-n } = r ,\eea 
related to the charges $ (q ,\ r)  $   as 
$m \equiv j_3 = { (q +r)/  2}  ,\ n \equiv l_3 = {(q - r)/  2} .$ 
The  scalar fields on $T^{1,1}$  are decomposed on   
 harmonic  functions $ Y^{j l, M} (\T ) $, eigenfunctions of 
the scalar Laplacian  $-\tilde \nabla ^2 _{T^{1,1} }  $   of eigenvalues,   
$ H_0 (jlr)=  6 (j(j+1) +    l(l+1) -{r^2 / 8} ) $,   
symmetric under     $r \to -r$.  The  tensor
(spinor) fields in irreducible representations of the  tangent
space   group $SO(5)$  of $T^{1,1}$   are decomposed on 
 solid harmonics $ Y^{j l , M } _a,\ Y^{j l, M} _{( ab) } ,\ Y^{j l ,
  M} _{[ab]} , \ \cdots $   that are   equipped with 
sets of $SO(5)$ group indices
$a,\ b ,\cdots $ subject to   the appropriate symmetries under indices
 permutations  and  traceleness  under pair  summations  
of  indices.  For general fields  in  $AdS_5 \times
T^{1,1}$   in  given     representations   of  the  symmetry   group
$SO(5)$ of the tangent   space of $T^{1,1}$,       labelled by  the
charge  $q$,       the  harmonic   decomposition  reads   
\bea &&  \Phi   _{q}(\hat x, \T ) = \sum _{ (j,l), (mn)} \Phi
^{jlr}_{q, mn}   (\hat x) (Y ^{(jl) } (\T )) ^{mn}_{q}   ,\  
[(m, n) = (j_3, l_3)]    \eea   
where the  correspondence   between  the magnetic  quantum numbers 
and the  charges  $ (q, r )$  can be used to  transform  the summation
over $ (m, n) = (j_3, l_3)$   into a    summation over $r$. 
  The possibility  to trade  $ (j_3, l_3 ) $ with the
(generally non-conserved) charges  $ (q, \ r )$     justifies adopting
  the     non-redundant notation for the basic scalar  harmonics, 
$ Y ^{jl r } _ q (\T ) ,\ [r= 2 j_3- q] $.  
The charge   $r $  is  used to label    distinct  modes  of
same $j, l , q $  quantum numbers  states  but  it is 
not  necessarily  a good (conserved) quantum number.    
The allowed values of $q, \ r$  satisfy the
bounds $ \vert q +r \vert \leq 2 j ,\ \vert q -r \vert \leq 2 l$.
 For scalar  fields, the  $ G/H$ quotient  projection  selects the  $
 U(1)_H$ singlet mode $ q=0$,   with  
$r \in (-\tilde j ,\cdots , \tilde j ),\ [\tilde j = \min (j,l) ].$ 
 The   solid harmonics  can  also  be    represented
by the    vector  spaces    (closed under the action of
 the    Lie algebra   of $G$) generated  by bases of 
column vectors  of  scalar harmonic  components  
$ [Y^{j l r_i } _{q_i} (\T ) ] $ carrying     discrete sets of charges  
$q_i \in [0, \pm 1 ,\cdots   ] ,\ r _i(r)\in [ r, r \pm 1,\cdots   ]
$.   For instance, the singlet, spinor and vector harmonics  
  of   quantum numbers   $(jlr)$   in   
representations $[1,\ 4,   \ 5] $     of the  group $SO(5)$ 
are assigned the    following sets of   $q_i$-  and $r_i$-charges
\bea && {q_i \choose  r_i }  = 
{[(0), \ (0, 0, -1, +1 ), \ (+1 , -
1, + 1, -1 , 0 ) ] \choose  [(r),\ (r-1,r+1, r,r) ,\ (r+1, r-1,
  r-1,r+1, r) ] }   . \eea   Since the matrix embeddings of $T_H$
in $ G/H $ have only zero or integer eigenvalues, it follows that the
angular momenta  $ j, \ l $ are both integer or half-integer  with   $ r$ 
running   simultaneously   over odd or even integers.  

The analysis    of harmonic   modes in 
$ AdS _5\times T^{1,1} $ spacetime is   formally  similar  
to  that  developed~\cite{kiroman84} in the $ AdS _5\times S^5$  
spacetime of    isometry group $SO(6)$. 
The     wave equations   are  nearly of same form, except
for  the   radial  profiles  of  3- and 5-fluxes and 
a few other   instances  documented in~\cite{ceresole99,ceresoII99}.
We shall adapt several results derived
in~\cite{kiroman84}  to our case.  The  coordinates  of $AdS_5$ spacetime  
are denoted    $ \hat x_{ \hat \mu } = (x _\mu ,r) ,\ [\hat \mu = (\mu
  , r ) ] $.  The formalism  simplifies  if the metric tensor  $
h_{\hat \mu \hat \nu } $ is replaced by the metric tensor shifted  by a Weyl transformation~\cite{kiroman84} \bea && h'_{\hat \mu \hat \nu } (X) =
h_{\hat \mu \hat \nu } + {1\over 3 } \tilde g_{\hat \mu \hat \nu } h
_a ^a ,\ [ \d \tilde g_{\hat \mu \hat \nu } = h_{\hat \mu \hat \nu
  },\   h
  _a ^a = \tilde g^{ab} h_{ab} ,\ h ^{\hat \mu } _{\hat \mu } \equiv
  \tilde g ^{\mu \nu } h _{\mu \nu } = h ^{' \hat \mu } _{\hat \mu } -
{5\over 3} h ^a _a] .  \label{sub2eq26} \eea 
Imposing  the Lorentz type gauge choice on the metric
tensor and de Donder type gauge choices on the 4- and 2-form potentials
\bea && \nabla ^{\hat \mu } ( h' _{\hat
  \mu \hat \nu } -{1\over 3} \tilde g_{\hat \mu \hat \nu } h ^{' \hat
  \l } _{\hat \l } ) =0 ,\ D^a C_{\hat \mu \hat \nu \hat \rho a
}=0,\ D^a C_{\hat \mu \hat \nu ab } =0,\ D^a C_{\hat \mu abc} =0,\ D^a
C_{\hat \mu a} =0 ,  \eea 
removes  spurious modes associated to  the terms involving covariant
derivatives of harmonic functions. We display  in     
Table~\ref{tabx1} the  correspondence  between  bosonic   fields 
of the  10-d  supergravity multiplet and  descendant  $ AdS _5$  fields in  
harmonics modes of $ T^{1,1}$,  based on~\cite{ceresole99}  and, 
for the 2-form field $ B_{MN } $,   based   on~\cite{jatkar99}. For
illustration, we quote   below the 
 decompositions  for the metric tensors,  axio-dilaton and  4-form fields 
\bea && h' _{\mu \nu } (X) \equiv \d \tilde g '_{\hat \mu \hat \nu } (X) =
\sum_{\hat \nu , M} H _{\hat \mu \hat \nu } ^{(\nu )} (\hat x ) Y^{\nu ,
  M} ( \T ) ,\ \d \tilde g_{a}^{a} (X) = \sum_{\nu , M} \pi ^{(\nu )}
(\hat x ) Y^{\nu , M} ( \T ) , \ \d \tilde g _{a\hat \mu } (X) =
\sum_{\nu , M} B ^{(\nu )} _{\hat \mu } (\hat x ) Y^{\nu , M } _{a } (
\T ) ,\cr && \d \tilde g _{ab} (X) =\sum_{\hat \nu , M} \phi ^{(\nu )}
(\hat x ) Y^{\nu , M} _{(ab)} ( \T ) ,\ \d \tau (X) = \sum_{\nu , M} t
^{(\nu )} (\hat x ) Y^{\nu , M} ( \T ) ,\ C _{a\hat \mu } (X) =
\sum_{\nu , M} a ^{(\nu )}_{\hat \mu } (\hat x ) Y^{\nu , M } _{a } ( \T ) ,
\cr && C _{ abc \hat \mu } (X) = \sum _{\nu , M} \e _{abc} ^{
  \ \ \ de} (\phi ^{(\nu )}_ {\hat \mu } \cald _d Y^{\nu , M} _e  ( \T )+ \tilde
\phi ^{(\nu )}_{\hat \mu } Y^\nu _{[de]} ( \T ) ) , \ C _{abcd}(X ) = \sum _{\nu
  , M} b ^{(\nu )} (\hat x) \e_{abcde} \cald ^e Y^{\nu , M} ( \T ) , \cr &&
     [\nu = (j, l, r) ,\ M =\{ m = j_3 , n = l_3 \} ,\ j_3 \in (-j_3
       ,\ -j_3 +1,\cdots , j_3),\ l_3 \in (-l_3 ,\ -l_3 +1, \cdots ,
       l_3) ] .    \label{sub2eq27} \eea 

  The AdS/CFT duality~\cite{malda99}  establishes a  1-to-1  
correspondence   between  massive fields in  $AdS_{d+1} $  and 
operators in $ M_d$   in       same unitary
representations of the   superconformal  group~\cite{gunaymarcus85}. 
The mass-dimension relationship  between  the bulk  $p$-form 
fields   of squared masses $M^2_{d+1,\nu } $ (in the supergravity convention)  
and the boundary  $p$-form  operators of  
scaling  dimensions $\D  _\nu $   is~\cite{wittenholo98}  
\bea && M^2_{d+1,\nu }  = (\D _\nu  -p ) (\D _\nu  + p-d)
.\label{sub2eqscal} \eea  

\begin{table} \begin{center}
\caption{\it \label{tabx1} Harmonic decompositions of  the  
supergravity multiplet  components of type $II\ b$ theory 
on $AdS_5\times T^{1,1} $ spacetime.   We display in the first
three arrays   the correspondence   between 
the 10-d  bosonic fields,  the scalar    and solid harmonic 
functions of $ T^{1,1}$ and   the    5-d   fields  in $AdS_5 $, where  the antisymmetric,   symmetric and symmetric-traceless combinations 
with respect to    the   base manifold   tangent
space group  $SO(5)$  indices $a, \ b ,\cdots $   are denoted  by $
[ab] ,\ (ab),\  \{ab\} $.
The    mass parameters  $ M_5 ^2 $ of  the  various  bosonic fields  $ M_5 ^2 $
($ M_5$ for   2-form fields)  
in  the supergravity theory convention are  displayed in the fourth
array, using  the   notations    of~\cite{ceresoII99},  
$ H_0 = 6 (j(j+1) +
  l(l+1) -{r^2 / 8} ) ,\ H_1 = H_0 +16\pm 8 \sqrt {H_0+4} , \ H_2 =
  H_0 +8\pm 4 \sqrt {H_0+4} ,\ H_3 = H_0 +4 \pm 2 \sqrt {H_0+4},\ H'_3
  = H_0 +12 \pm 6 \sqrt {H_0+4},\ H^{''} _3 = H_0^{\pm \pm } +7 \pm 4
  \sqrt {H_0+4} ,\ H^{'''}_3 = 3 + H_0 ^{\pm \pm } , \ H_4 = H_0 +4
  \pm 4 \sqrt {H_0+4},\ H'_4 = H_0^{\pm \pm } +1 \pm 2 \sqrt {H^{\pm
      \pm } _0+4},\ H_5 = H_0 +4,\ H'_5 = H^{\pm \pm }_0 +5 \pm 2
  \sqrt {H^{\pm \pm }_0+4},\ [H^{\pm } _0 = H_0 (j,l,r\pm 1) ,\ H^{\pm
      \pm } _0 = H_0(j,l,r\pm 2)] . $ Only a representative   subset of  
the mass eigenvalues  are  specified  for systems   of  coupled fields, as
for instance    $ (\pi ^{(\nu )} , b^{(\nu )}) $  or $  (B_{\hat \mu }
^{(\nu )},\phi _{\hat   \mu }^{(\nu )} )$.}
\vskip0.5 cm 
\begin{tabular}{|c|ccccc   c||ccc  c||ccc   c|} \hline & && Scalars
  &&&& Vectors && &&   Tensors & & & \\ \hline 10-d & $ g^a _a$&$
  C_{abcd}$&$\tau $&$C_{ab}$&$g_{(ab)}$ &$B_{ab} $ &$ g '_{a\hat \mu }
  $&$C_{\hat \mu abc } $&$C_{\hat \mu a } $& $B_{\hat \mu a } $
  &$g_{\hat \l \hat \mu }$ &$C_{\hat \l \hat \mu }$ &$C_{
    \hat \l  \hat \mu ab } $ &$ B_{\hat \l \hat \mu }$ \\ $ T^{1,1}$ & $ Y^\nu
  $ & $ Y^\nu $ &$ Y^\nu $ &$ Y^\nu _{[ab]} $ & $Y^\nu _{(ab)} $ &$
  Y^\nu _{[ab]} $&$ Y^\nu _a $&$Y^\nu _{[ab]} ,\ Y^\nu _{a} $&$ Y^\nu
  _{a} $ & $Y^\nu _{a} $ &$ Y^\nu $ & $ Y^\nu $&$ Y^\nu _{[ab]}$ &$
  Y^\nu $ \\ 
$AdS _5$ & $ \pi ^{(\nu )} $&$ b ^{(\nu )}$ &$ t ^{(\nu )} $ &$a ^ { \pm
    \nu } $&$ \phi ^{(\nu )} $ & $ \hat a ^{(\nu )} $ &$B ^{(\nu )} _{\hat \mu }
  $&$\phi^{(\nu )} _{\hat \mu } $&$a^{(\nu )}_{\hat \mu } $& $\hat a^{(\nu )} _{\hat
    \mu } $& $ H^{(\nu )} _{\hat \l \hat \mu }$&$ a ^{(\nu )} _{\hat \l \hat \mu
  }$&$ b^ {\pm \nu } _{ \hat \l \hat \mu }$& $ \hat a^{(\nu )} _{\hat \l
    \hat \mu }$ \\ 
$M_5^2 \ (M_5) $ & $H _1$&$H_1$&$H_0^{--} $&$(H_4 ,
  H' _4) $&$ (H_3 , H^{',''} _3) $& - & $ (H_3 , \ H ^{', ''}) $& $
  (H_3, H ^{'''} _3) $& $H_2 $& - & $ H_0$ &$H_2$ & $(H_5, H '_5 ) $ &
  - \\ \hline
\end{tabular} \end{center}  \end{table} \vskip 0.1 cm
The wave equations  of the harmonic modes    are  encoded  within the  
perturbed   reduced action  of quadratic order   in the  fields
fluctuations. Along with the kinetic     energy terms, this includes   
mass  terms arising from the   Laplace-Beltrami operators of $
T^{1,1}$~\cite{kiroman84}.  The graviton $ h ' _{\hat \mu \hat \nu
} $ (modulo a suitable field redefinition), the axio-dilaton
and the 2-form $a_{\hat \mu \hat \nu } \in \d C_{\hat \mu \hat \nu } $
modes all satisfy  diagonal type wave equations.    The scalar fields
from   partial  traces  of the metric tensors  and from  
the 4-form potential, $\pi
(X)\in \d g ^a _a ,\ H _{\hat \mu \hat \mu } (X) = \tilde g ^{\hat \mu
  \hat \nu } H _{\hat \mu \hat \nu } ,\ b (\hat x) \sim \d C_{abcd} $
are coupled  through non-diagonal wave equations.  While the charged (non-singlet $\nu \ne
0$) modes are related algebraically  as  
$H _{\hat   \mu \hat \nu } ^{\nu } ( \hat x)  = a  \pi ^{\nu } ( \hat
x)  $,     one can    impose the same  proportionality  relation also
  on the singlet  mode as a gauge choice.   
  (In  $ AdS_5 \times S^5  $ spacetime, $a=  {16 /15} .$)
The  resulting  wave equation  for  the 
coupled field  system $( \pi ^{(\nu )}  (\hat x) ,\ b ^{(\nu )} (\hat x) ) $ 
of  charge  $\nu $, adapted  from~\cite{kiroman84}, is given in matrix     
form  by  
\bea && (\nabla ^2 _{AdS} - M^2 _5 ) {\pi ^{(\nu )}
 (\hat x) \choose b  ^{(\nu )}  (\hat x) } =0 ,\ [M_5^2 = \pmatrix{+ H_0 +32 &
    +80 H_0 \cr +4/5 & + H_0} ] . \label{sub2eq28p} \eea The
 $2\times 2$   squared mass matrix  diagonalization
yields the pair of eigenvectors and eigenvalues
\bea && S _{\mp } ^{ \nu } (\hat x) = 10 ( (2 \mp \sqrt {H_0 ^{ \nu }  +4} ) \pi
^{(\nu )} (\hat x) + b ^{(\nu ) } (\hat x) ) ,\ M^{\nu 2} _{5 \mp } = H_0 ^{ \nu } 
+16 \mp 8 \sqrt {H_0 ^{ \nu }  +4} , \  [\D ^\mp _\nu = {-2 \choose 6} +  \sqrt {H_0  ^{ \nu } +4}  ]   \label{sub2eq28} \eea 
where the dimensions of the corresponding  dual field
theory operators, $ \D ^-_\nu = \D ^+_\nu -8 = -2 + \sqrt {H_0 ^{ \nu
  } +4} $,   are evaluated  by solving  the equation $ M_5 ^2 =\D (\D
-4)$~\cite{wittenholo98,klebwitten99}. 

The  2- and 4-form potentials give rise to vector and
tensor modes in $ AdS_5 $ (omitting   $\nu $ labels),
  \bea && [ b _{ \hat \mu \hat \nu} ,\ a_{
    \hat \mu \hat \nu} ,\ \phi ^a _{ \hat \mu }, \ c_{ \hat \mu \hat
    \nu} ,\ c_{\hat \mu \hat \nu \hat \rho } , \ B^a_{\hat \mu } ] \in
[\d B_{\hat \mu \hat \nu} ,\ \d C_{\hat \mu \hat \nu} , \ \d C_{\hat
    \mu abc} , \ \d C_{\hat \mu \hat \nu ab } ,\ \d C_{\hat \mu \hat
    \nu \hat \rho a } ,\d h_{\hat \mu a} ] \eea  
where the 2- and 3-form potentials in $
AdS_5$   are Hodge duals  to vector and scalar potentials.   
No additional propagating modes arise from $ \d C_{ \hat \mu \hat \nu \hat
  \rho a } ,\ \d C_{\hat \mu \hat \nu \hat \rho \hat \s } $, due to  
the 5-form field self-duality.  The
reduced action for the pairs of vector fields $ V ^a _{\hat \mu } =
(B^a_{\hat \mu } , \phi ^a_{\hat \mu } ) $,  
satisfying the transversality condition $ \tilde \nabla ^{\hat \mu } V _{\hat
  \mu } =0$,    involves  non-diagonal mass terms determined by the
action of Hodge-de Rahm operators $ \D _1 $ on the $SO(5)$ vector
harmonics $ Y^{jlr, M } _{ a } ,\ [a=1,\cdots , 5] $.  The resulting
wave equation,    adapted from~\cite{kiroman84}, 
   is given   in  a  $ 2 \times 2 $ matrix notation   by  \bea &&
( \calm ^2 _{AdS} - M^{\nu 2} _V ) 
 {B^{(\nu )}_{\hat \mu } \choose \phi ^{(\nu )}  _{\hat \mu } }
 =0,\ M^{\nu 2} _V= \pmatrix{- \D _1 + 8 & -16 \D _1 \cr 1 & -
 \D _1 } = \pmatrix{\l _1 + 8 & 16\l _1 \cr 1 & \l _1
 }, \label{sub2eq29} \cr && [\calm ^2 _{AdS} V _{\hat \nu } =
 \tilde \nabla ^{\hat \mu } F _{\hat \mu \hat\nu } = \tilde
 \nabla ^{\hat \mu } (\tilde \nabla _{\hat \mu } V_{\hat \nu }
  - \tilde \nabla _{\hat \nu } V_{\hat \mu }) = \tilde \nabla
  ^2 V_{\hat \mu } - [\tilde \nabla ^{\hat \mu } ,\tilde \nabla
   _{\hat \nu } ] V_ {\hat \mu } ,\cr && \D _1 Y ^\nu _a = -
   (\nabla _ { T^{1,1} } ^2 Y^\nu ) _a + R^b _a Y^\nu _b = - (\l
   _1 ^\nu Y^\nu ) _a,\ R^b _a = 4 \d _{a b}]
\label{sub2eq30} \eea  
where $ \calm _{AdS} $ is Maxwell operator for the $ U(1)$ gauge
theory in $AdS _5$ and $\l _1 $   denotes the $ 5\times 5 $ matrix 
in the basis of vector harmonic column vectors~\cite{ceresoII99}
\bea &&  
\{ Y^{ (j l r_i ) } _{q_i} \} ,\ [a= [(r _i , q _i)] \in [(r+1,
  1), (r-1, -1), (r-1, 1),(r+1,-1), (r,0)] ,\ i\in 1,\cdots , 5] .\eea
The matrix  $\l _1 $  admits   five eigenvectors $ \caly ^{ j l r } 
_a = \sum _{i=1} ^5 c_i^a Y^{ j l r_i } _{q_i} , \ [a = 1,\cdots , 5]
$ of which  a single  eigenvector of
eigenvalue $ H_0 $, given by  the longitudinal solid harmonic 
$\caly ^{ j l r } _{L, a}  = D_a Y^{ j l r }$,  vanishes  in the transverse
gauge.  The remaining four eigenvalues for  fixed $(jlr)$
consist of the two pairs \bea
&& \{ \l _{1 \pm } ' = 3 + H_0 ^{\pm \pm } ,\ \l _{1 \pm } '' = H_0 +
4 \pm 2 \sqrt {H_0 +4 } \},\ [H_0 ^{\pm \pm } = H_0 (j, l, r\pm 2 )]
. \eea The $ 2 \times 2 $ matrix $ M_V^{\nu 2
}$   associated  to  each of these  eigenvalues  
admits      the pair  of eigenvectors and
eigenvalues \bea && V ^{(\nu )} _{\hat \mu , \mp } = (4 (1 \mp \sqrt {1 + \l _1 } )
B^{(\nu )}_{\hat \mu } + \phi ^{(\nu )} _{\hat \mu } ) ,\ M _{V \mp } ^{\nu
  2} = \l _1 + 4 \mp 4 \sqrt {\l _1 +1} . \eea 
With the four  $ \l _{1 \pm } {' },\ \l _{1 \pm } {''} $ generating two
eigenvalues each,  the  mass spectrum for the coupled system of vector fields
$ (B_{\hat \mu } ^a, \ \phi _{\hat \mu } ^a) $   in $AdS_5$ and $ T^{1,1} $ spaces
consists  of the 8    mass
eigenvalues~\cite{ceresoII99} \bea && M ^{\nu \ 2} _{5 \pm } = { H_0
  ^{++} + 7\pm 4 \sqrt { H_0 +4} ,\ H_0 ^{--} + 7\pm 4 \sqrt { H_0 +4}
  \choose H_0 +12 \pm 6 \sqrt {H_0 +4} ,\ H_0 +4 \pm 2 \sqrt {H_0 +4}
} .  \label{sub2eq32} \eea    
The  dimensions of the dual conformal operators are
determined by solving the mass-dimension  relationship 
$ M_5 ^2= (\D -1 ) (\D -3 )$.  

The  second  stage    of the dimensional reduction 
involves  the  decomposition of   5-d fields   
on plane wave modes in $ M_4 $  of four-momenta $ k_m $ 
and  squared mass  $ E_m ^2= -k_m ^2$,  
\bea && \phi ^{(\nu )} (\hat x ) = \sum _{k_m} \phi ^{(\nu )} (k_m) 
e ^{i k_{m} \cdot x } R _{m} (r) ,\ [k^2_{m} = - E^2_{m},\ 
m = [ (\nu ,  k_m ) ,\   \hat x = ( x, r) ,\ x\in M_4  ]  
\label{sub2eq33}  \eea 
where we have lumped together the labels $\nu ,\ k_m $ for the angular
and radial excitation modes in $ T^{1,1}$ and $ M_4$ into a single
label $m$. 
As will be discussed  in Subsection~\ref{sec3}, the radial wave
equations involve  Sturm-Liouville eigenvalue  problems  on  the 
radial interval for $ r \in [r_0, r _{uv}] $, 
equipped with a Hermitean scalar product and
boundary conditions  that select  a discrete infinity of quantized
eigenvalues $ E^2_{m} $.  
The   resulting  solutions   belong   to Hilbert  spaces   generated by orthonormal  bases of radial wave functions  $R _{m} (r) .$    

\subsubsection{Warping effects on   compactification moduli}
\label{subsec24.3}

The throat embedding   in a compact  manifold 
is expected to  affect  the properties     of  bulk moduli. 
To illustrate this point, we  first  examine the  effects
on the geometric moduli   within the warped  conifold model. 
The correction  to the   kinetic term of the complex  structure   
$S$-modulus, hinted at after  Eq.~(\ref{sub2eq15}), was examined  in  the
Hamiltonian and Lagrangian formalisms in~\cite{dougba07,dougtorr08}.   
Evaluating the contribution to the metric tensor 
in the  same  setup and same notations
as~\cite{dougtorr08}  for the classical warp  profile,
Eq.~(\ref{eq.warpA1}), 
$ e ^{-4 A}\simeq  c +  {81   / (8r^4)} {(g_s M \a ' )^2} \ln
(r/r_{ir} )  $   and  the cohomology  3-form,    $ \chi ^S  =(\o _3
    - i (dr /r )  \wedge \o _2  ) / (8\pi ^2)  ,$     gives  
\bea && G_{S\bar S} = - i e ^{K_0}  \int _{X_6} e ^{-4 A}
\chi _S \wedge  \bar \chi _S = {1\over c \int \O  \wedge  \bar \O }
 \int _{\calc _6}  [c  + ({\calr \over r} )^4 c_2  \ln {r \over r_{ir} }
 ] \chi _S \wedge  \bar \chi _S   \cr && \Longrightarrow \   
  G_{S\bar S}  =   - { i\over \pi V_W \vert  \vert 
\O \vert  \vert ^2} [c \ln \call ^3 + {3\over 4} {81 \over 32} 
    {(g_s M  / \a' )^2  \over  \vert S\vert  ^{4/3} } (1- {1\over \call ^{4} }
( 1 +4 \ln \call ) ) ]  ,\ [\call = {\L _0 \over  \vert S\vert  ^{1/3
  } }] . \label{eqGcs} \eea   
The first  term  $c (x) \ln (S) $  exhibits  the familiar  logarithm 
dependence of   special  Kahler geometry   
with   the  coefficient   identified to  the universal  volume,   $ <
c (x) > \simeq  \calv   ^{2/3}  $,    as    explained  near  
Eq.(\ref{volmod1}).    The warping effect   from      3-fluxes 
contributes   the second power term that blows up at $ S\to 0$ as $
(\a  ^{'3/2} \vert S \vert )^{-4/3}$.  

The   conifold  embedding   in  the compact  manifold can also   
affect  the    wave functions of  the 
universal  volume  modulus.   The  resulting profiles
  for   the metric tensor     components  $ \d _c \tilde g _{MN } $ 
  along $ M_4 $  and $ \calc _6$    near the   conifold apex~\cite{frey08},
  $(\d _c   g_{\mu \nu } , \d _c g_{rr } ) \simeq  ( r^2 /
\sqrt N   , \sqrt N   / r^2 )  \ c(x) , $ 
 are  both  amplified by the multiplicative   factor $ (N/ r ^4) $ 
relative to the naive  estimates obtained in the treatment ignoring the 
compensator  fields,  $ (\d _c   g_{\mu \nu } , \d _c g_{rr } )^{(0)}
\simeq  ( (r/  \sqrt N)^6    ,  (r / \sqrt N  )^2)  \ c(x) .$

We now   focus  on the conifold  base breathing  modulus,    represented  by 
the    field  $\G (X)$ in   the ansatz  for the deformed metric
of the (undeformed)   conifold
\bea && d \tilde
s ^2 _{6} \equiv \tilde g_{mn} d y ^m d y ^n = d r ^2 + r^2 e ^{2 \G
  (X ) } d \hat s ^2 _5 = d r ^2 + e ^{2 a (X ) } d \hat s ^2_5 ,\cr
&& [d s ^2 _{6} \equiv e ^{-2 A} d \tilde s ^2 _{6} ,\ d \hat s ^2 _5
     = \tilde g_{ab} d \T ^a d \T ^b ,\ a (X) = \ln r + \G (X) ,\  
\sqrt {- \tilde g _{10} } =  \sqrt {- \tilde g _{4} }  
\sqrt {\tilde g _{5} }  e ^{5a}   ]  
\label{sub2eq43} \eea 
where   we have set   $ \calr =1$,  for convenience.      
The    linear  variations   $ \d \G (X) \equiv
\g (X) $ identify       to those of the metric tensor  trace, 
$ \g (X) ={1\over 10} \d \tilde g^a _a ( X) = {1\over 10} \pi
(X) $.  We wish  to examine   the  angular and radial 
excitation  modes  of the breathing    modulus  in isolation from
the 4-form potential $ \d C_{abcd} (X) $ and  3-form field strength
$ \d  G_{mnp} $  to which they  might  couple. 
The   second order variation  of the reduced action for $ \G (X)$ in the  spacetime $AdS_5 \times T^{1,1}$
can be evaluated    from   the terms in  Eq.~(\ref{sub2eq16}) involving the
gravitational and 3-form  fields, which we rewrite for  convenience,  
\bea && S^{(2)}   = {1 \over 8 \kappa ^2 } \int d^{10
} X \sqrt {- \tilde g _{10} } [e ^{-4 A (r)} \d \tilde g ^{mn} \tilde
  \nabla ^2 _4 \d \tilde g _{mn} - \ud e ^{-4A} \tilde g _{mn}  \tilde
  \nabla _4 ^2 \d \tilde g _6 -2 \d \tilde g ^{mn} \tilde \d \tilde
  G^{(6)} _{mn} + \d \tilde g_4 \d \tilde R ^{(6)} \cr && + {3 \over 2
    \tau _2 } e ^{4 A} (\d \tilde g_{mr} \d \tilde g^{ms} G ^{ \tilde
    r \tilde p \tilde q} \bar G _{rpq} - {1\over 6 } \d \tilde g_6 \d
  \tilde g^{mn} G _{mpq} \bar G _{ n} ^{\ \ \tilde p \tilde q} ) ] . \label{sub2eqCURFLUX} \eea
We shall ignore the transversal and trace  terms in $\d \tilde G ^{(6)}_{mn}= - \ud  (\tilde \nabla ^2 _6\tilde g _{mn} 
+ \nabla _{(m} \nabla _{n)} \d \tilde g_6  - 2   \nabla ^{\tilde p} \nabla
_{(m}  \d \tilde g  _{n )p } )  $  
and also  the curvature terms  in $\D _L$  by setting 
$\d \tilde G ^{(6)}_{mn} \to \ud \D _L \d \tilde g _{mn} \to 
- \ud  \tilde \nabla ^2 _6 \d \tilde g _{mn} .$  
With the substitutions  $ \d \tilde g_{mn} = \d ( e
^{2 \G } ) r^2 \tilde g_{ab} \to 2r^2 \g_{ab} \g (X) ,\ \d \tilde g^{mn}
\to + 2r^{-2} \tilde g^{ab}  \g (X), $   one obtains for the
kinetic, curvature and 3-flux terms   of quadratic order   in  $ \g $
\bea &&  S^{(2)} = {1 \over 8 \kappa ^2 } \int d^ 4 x \sqrt {- \tilde g
  _{4} } \int dr d ^5 \T \sqrt {\tilde g_5 } [ 20\ e ^{-4 A } r^5 \g \tilde
  \nabla ^2 _4 \g - 2 r ^3 h ( ( \tilde R ^{(5)} + 4 \tilde \nabla _5
  ^2 ) \g + 30 r \g ' +5 r^2 \g '' ) \cr && + {2 g_s } r ^{-1} e ^{4
    A} (r^2 \vert G _{abr} \vert ^2 + \vert G _{\t \phi \psi } \vert
  ^2 ) \g ^2 (r,\T ) ] ,\ [\g ' = {\dh \g \over \dh r } ,\ h =h ^\mu
  _\mu ] . \label{sub2eq45}  \eea  
At  this stage,  we consider a more detailed all order  calculation  of  the
non-kinetic  part   of the reduced  action  for $\G  (X)$
going  beyond the linearized description.   
The curvature and flux contributions
to  all orders    of  $ \d \G  (X)$   can be evaluated 
by  invoking   known identities for
conformal transformations of the metric tensor. For this purpose,   we move 
one step backwards to the    terms  in the reduced  action  of interest 
\bea && S _{10} (\G )   = S_{g}  (\G )  + S_{flux} (\G )   = 
{1\over 2\kappa ^2 } \int d^{10} X \sqrt {-\tilde
  g_{10}} [e ^{-2 A} R ^{(10)} - {g_s \over 2 \tau _2 } e ^{4 A}
  {1\over 3 !} G_{mnp} (\bar G - i \star _6 \bar G ) ^{\tilde m\tilde
    n\tilde p} ] ,  \label{sub2eqREDACT} \eea 
expressing  the dependence
on the warp profile by means of the formulas \bea && R = R ^{(10)}
\equiv g^{MN} R_{MN} = R ^{(4)} + R ^{(6)} ,\ [R ^{(4)} = g^{\mu \nu }
  ( R _{\mu \l \nu }^ {\ \ \ \ \l } + R _{\mu n \nu }^ {\ \ \ \ n } )
  ,\ R ^{(6)} = g^{mn} (R _{m\mu n } ^ {\ \ \ \ \ \mu } + R _{m p n }
  ^ {\ \ \ \ \ p} )] \cr && \Longrightarrow \ R ^{(4)} = e ^{-2A }
\tilde R ^{(4)} - 4 e ^{2A }\tilde \nabla ^2 _6 A ,\ R ^{(6)} = e ^{2A
} ( \tilde R ^{(6)} + 6 \tilde \nabla ^2 _6 A - 8 (\tilde \nabla _6
A)^2 ) , \cr && [ \tilde \nabla ^2 A (r) =E ^{-5a } \dh _r e ^{5a }
  \dh _r A (r) = 5 (\dh _r a ) (\dh _r A) + \dh _r ^2 A ]
.  \label{sub2eq46} \eea 
The covariant derivatives of the warp profile   add extra 
 contributions  shown in the last entry above.  To get the  contributions from $x$-independent  rescalings  of the  breathing field  $\G (r, \T ) $, it is only necessary to
consider $\tilde R ^{(6)} $.   The  dependence  on $r$ and
$\T $  arising  from the independent conformal transformations of
the 6-d and 5-d parts of the internal metric $ d \tilde s ^2 _6 = d r
^2 + e ^{2 a (r,\T ) } d \tilde s^2 _5 ,$ are  then evaluated by means of
the identity 
\bea && \tilde R ^{(6)} = \tilde R ^{(5)} e^{-2 a } -2 k
\tilde \nabla _r ^2 a - k (k+1) ( \tilde \nabla _r a ) ^2 - e^{-2 a }
(8 \tilde \nabla _5 ^2 a + 12 ( \tilde \nabla _5 a ) ^2 ) \cr && = r
^{-2} [\tilde R_{5} ( e ^{-2 \G } -1) - ( 30 (\dh _Y \G )^2 + 10 \dh
  _Y ^2 \G + 50 \dh _Y \G ) - e ^{-2 \G } (8 \tilde \nabla _5 ^2 a +
  12 ( \tilde \nabla _5 a )^2 ) ] , \label{sub2eq47} \eea
where   for $T^{1,1}:\   k=5 ,\ \tilde R ^{(5)} = 4k =20 $.
(We have exhibited in the second line entry above the alternative
formula as a function of $Y = -\ln r $,  in order to call attention to
mismatches by  factors $-1/2 $ and  $ -1/4$ in 
the terms $ \dh _Y ^2 \G $ and $\tilde \nabla _5 ^2 \G  $ 
needed to  match  our results  to  those  of~\cite{freydacline09}.)
Evaluating the 3-flux term  by means of  the  formulas 
\bea && (G \wedge \bar G _-) \equiv G
\wedge (\star \bar G + i \bar G ) = G \wedge \star (\bar G -i \star
\bar G ) = (G\tilde \cdot \bar G) ,\ [(G\tilde \cdot \bar G) \equiv
  {1\over 3! } G\tilde \cdot \bar G = {1\over 3! }  G_{mnp} ( \bar G
  -i \star \bar G ) ^{\tilde m \tilde n \tilde p} ,  \cr && (\bar G - i
\star _6 \bar G) ^{r\t \phi } = e ^{-4a } (1 - e ^{-\G }) \bar G _{r\t
  \phi } ,\ (\bar G - i \star _6 \bar G) ^{\t \phi \psi } = e ^{-6a }
(1 - e ^{\G }) \bar G _{\t \phi \psi } ]  \label{sub2eq49} \eea 
one obtains the  expanded formula for the reduced action in
Eq.(\ref{sub2eqREDACT}),  
\bea && S ^{(2)} (\G )= {1\over 2\kappa ^2 } \int d^{10} X \sqrt {-\tilde
g_{10}} [ (e ^{-4 A} + c(x) ) \tilde R ^{(4)} + \tilde R ^{(6)} + 2
\tilde \nabla _6 ^2 A - 8 (\tilde \nabla _6 A )^2 \cr && - {g_s
\over 2 \tau _2 } e ^{4A (r) } (1 + c (x) e ^{4A} )^{-1} {1\over 3
!} G_{mnp} (\bar G - i \star _6 \bar G) ^{\tilde m \tilde n\tilde
p } ] \cr && = {1\over 2\kappa ^2 } \int d^4 x \sqrt {-\tilde g_4}
\int d r \int d^5 \T \sqrt { \g_5} [ e ^{3a} (\tilde R ^{(5)} - 8
\tilde \nabla ^2 _5 a - 12 (\tilde \nabla _5 a ) ^2 ) \cr && + e
^{5a} ( e ^{-4 A} \tilde R ^{(4)} - 10 \dh _r ^2 a -30 ( \dh _r a
)^2 + 2 ( 5 (\dh _r a) (\dh _r A) + \dh _r ^2 A - 4( \dh _r A )^2 )
) \cr && - {g_s ^2 \over 2 } e ^{4A (r) } (1 + c (x) e ^{4A} )^{-1}
 [ e ^{-4 a} (1 - e ^{- \G } ) \vert G_{r ab } \vert ^2 + e
   ^{-6 a} (1 - e ^{\G } ) \vert G_{abc } \vert ^2 ] ,\label{sub2eq48}
 \eea   where the volume modulus $ c(x)$  was introduced
 through the replacement,  $ e ^{-4 A} \to e ^{-4 A} + c(x) $.     
(As an aside, we note that   for de  Sitter spacetime $ M_4$ in the static, non-comoving
 metric $ ds ^2 _4 = -f (r) d t ^2 + f^{-1} (r) d r ^2 + r ^2
 d \O _2 ^2 ,\ [f (r) = 1- r^2/ \calr ^2 ,\ H= 1/\calr ] $ 
or for the vacuum energy dominated time-dependent, flat space
  Robertson-Walker spacetime with $ a (t) \propto e ^{H t}
 ,\ H = (\L / 3 )^{1/2} ,\ [T_{\mu \nu }= - \rho _V g _{\mu
     \nu } ,\ \rho _V = \L M_\star ^2 ] $ 
the  scalar    curvature  $ R  ^{(4)} = 12 H^2 $     during the  early universe
 inflation  takes  values   $ R ^{(4)} \sim (10 ^{-1} - 10^{-6}) M_\star ^2 $
 that  are considerably larger than   the  present day  values  attributed  to the cosmological constant.)  It is convenient to  evaluate the contribution from
 3-fluxes in  terms of  the  components of $ G_3$ in Eq.~(\ref{sub2eq49})
 referred to   the diagonal basis of 1-forms $ (dr , g^{(a)} ) ,\ [a=1,\cdots , 5] $.      This  can be   parameterized by  the two
 dimensionless   functions  $ f _r ,\ f_a $ as 
\bea && (G \wedge \bar G _- ) =  {\a ^{'2} \over   \calr ^6 } 
[{f_r ^2 \over r^2  } e ^{-4a } (1- e ^{-\G
   }) +{f_a ^2  } e ^{-6a } (1 - e ^{+\G }) ] = {\a ^{'2} \over
  \calr ^6 } e ^{-5a } r ^{-1} (e ^{+\G } -1 ) ( f_r ^2 -
 f_a ^2 e ^{- \G }) , \cr && [{f_r ^2 \over r^2  } = {6^2
     \over 2!}  \vert G_{r a b }  / \a ' \vert ^2 ,\ {f_a ^2 } = {6^2
     \cdot 9 \over 3!}  \vert G_{a bc } / \a ' \vert ^2 ]
 ,  \label{sub2eq50}  \eea 
where we  have  reinstated  the  dependence on  $\calr $. 
The classical solutions  for  3-fluxes in  Eq.~(\ref{sub2eq22})    
contribute    the constant values  $ f_r ^2 = f_a ^2 = {81\over 2} M^2 .$  
Collecting the  kinetic  term of  quadratic order in $\g $, 
 Eq.~(\ref{sub2eq45}),   with the potential term  of all orders in $\g
 $   from gravitational and flux   contributions in Eq.~(\ref{sub2eq48}), and 
introducing   the  dependence on the  volume modulus  through
the warp  profile factors $ e^{\mp 4 A} \to   (e^{-4 A} + c(x))^{\pm 1}
$,  yields  the final formula for the breathing mode  reduced action 
\bea &&   S  ^{(2)} _{kin} (\g ) + S_g  (\G ) + S_{flux} (\G ) 
= {1\over 2\kappa ^2 } \int d^4
 x \sqrt {-\tilde g_4} \int d r \int d^5 \T \sqrt { \tilde \g_5 } [ e
   ^{-4 A} ( 1 + c (x) e ^{4 A} ) ( 5 r^5 \g \tilde \nabla _4
   ^2 \g + e ^{5a(r,\T )} \tilde R ^{(4)} ) \cr && + e^{3 a }
   (R ^{(5)} - 8 \tilde \nabla _5 ^2 \G - 12 ( 3 \G (\tilde
   \nabla _5 \G )^2 - \G \tilde \nabla _5 ^2 \G ) ) + e ^{5a}
   (-10 \tilde \nabla _r ^2 a - 30 ( \tilde \nabla _r a ) ^2
 + 10 ( \tilde \nabla _r a) (\tilde \nabla _r A) + 2
   \tilde \nabla _r ^2 A - 8 (\tilde \nabla _r A)^2 )  \cr && 
- { (g_s\a ' )^2   \over 2  \calr ^6 } e ^{4A}   r ^{-1} ( 1- c (x) e ^{4A} ) (e
   ^{\G } -1)( f_r ^ 2 - f_a^2 e ^{-\G }) ] . \label{sub2eq51}  \eea

\subsection{Warped throat   deformation   by    compactification effects}
\label{subsec2.4} 

The   deformations  of the  background spacetime  $AdS_5\times T^{1,1} $  produced  by    embedding  the  conifold in  the
compact  manifold  are expected to  affect  the properties of warped modes.  
The  tools  to   describe    these effects by means  of the   AdS/CFT
correspondence   closely     mirror  those   used for  the 
back-reaction     of  embedded branes.  The gravity  fields 
in  $ AdS_5 $  bulk   source   operators  of the  superconformal
gauge theory  on   the $ N \ D3$-brane   stack  of world     volume 
along   the  time boundary   $ M_4 = \dh ( AdS_5)$.  For   large $N$,  at  
fixed 't Hooft   coupling constant   $\l = g_s N $, 
the  gravity theory  is in   regimes of  weak or strong  couplings 
(large or small curvatures),    when the  gauge theory  is in  regimes  of
 strong or weak  couplings.   We       start   the discussion   with 
a  general  introduction of     the perturbative  approach  
initially developed~\cite{baumann10}  for  $D3 /\bar D3 $-brane
inflation      and  later systematized~\cite{gandhi11}     for 
deformations of  GKP  vacua of type $II \ b$ supergravity  theory.    
We next consider   general deformations   of  the 
conifold  background     that  preserve or  break
supersymmetry.   Finally, we  examine    how 
the  compactification effects affect  the  wave  functions    and
interactions of  warped modes.  

\subsubsection{Introduction to  AdS/CFT   correspondence}
\label{subsec24.1}
The  string-gauge  duality  links   gravity  theories    in      
$ AdS_5   \times X_5 $ spacetime to 
gauge theories  on   the      time    boundary of
$ AdS_5 $.  The isometries  of  $ AdS_5  $ and $ X_5 $ 
map  to the  conformal  and flavour symmetry groups $  SO(4,2) $ and  $
G $ of the  gauge theory.  For bulk and boundary  theories  preserving 
$ \caln =1 $ and  $  2 \caln =2 $ supersymmetry, respectively, 
the conformal and  flavour groups  are  embedded  in  the      
superconformal supergroup $  SU(2,2 |  \caln )$    and  
the  R-symmetry   group. The  field  content of the
supergravity   theory in  $ AdS_5  $  consists of 
supermultiplets of (massless and massive)  fields 
of spins  $S \leq 2$   for the  harmonics of $X_5$   labelled by 
  quantum numbers $\nu $      of the isometry group $G$,    
which are in 1-to-1 correspondence with  conformal operator
supermultiplets  of the  dual gauge theory of flavour group $G$.
Each   unitary   representation  $\cald (E_0, j_1, j_2 ; r) $ 
of  $ SU(2,2| 1)$    splits into  finite direct  sums of  unitary
representations    $\cald (E_0, j_1, j_2)$  of   $SO(4,2)$, 
expressed  in terms of the  respective  
highest weight representations~\cite{ferrarazaff98}, $ \cald (E_0,
j_1, j_2 ; r) \to  \oplus _s \cald (E  ^{(s)}_0 , j ^{(s)}  _1, j ^{(s)} _2)
$.   

Consider the    decomposition of  linear  perturbations  $  \varphi
(\hat x ,\T )  $    of  the 10-d  supergravity fields    on       
harmonic functions of $ X_5 = T^{1,1} $ with     coefficient  fields 
$\varphi  _{\nu } (\hat x ) = \varphi ^{j_1  j_2 r } _{\nu } (x, r)  $   in $AdS _5$ 
of squared mass $M_{5  } ^{\nu 2}  $. To each  component is    
associated   a  composite operator $ O _{j_1 j_2 r
} ^ {\nu } $    of   scaling dimension  $ \D _{\nu } $ 
related  to the mode  mass  by a simple    algebraic    formula  $ \D
_{ \nu  }  ( M^2 _{5    \nu  }) $ 
depending on   the  modes  spin and  $r$-charge,    $ j_1,\ j_2 ,\ r$.
In   coordinate patchs  where    the   radial and 4-d spacetime
coordinates  $\hat x  =  (r , x)$  separate,    the   
fields  decompose    into  bases of $ M_4$  fields  times 
radial   wave   functions.  The mass-dimension  relationship   $ r  =
\mu  \a ' $ links   the      radial coordinate $r$ to the gauge
theory renormalization group momentum scale $  \mu  $,  with 
the  radial   positions of    end points    identified  to      the dynamical
and ultraviolet cutoff mass    scales   $ r_{ir} = r_0 \sim \L \a ' $
and $r_{uv} = \calr \sim \L _{uv} \a ' $.  The deformed background is
described    by  $x$-independent   
zero mode  solutions of the Laplace-Beltrami   operators  of $
X_6$ which  are      decomposed   on    eigenfunctions of the
Laplace-Beltrami   wave operators  of $X_5$  with 
radial  coefficient  functions   $\varphi _\nu (r) $. 
The    free  10-d  scalar  fields  decompose into    
two    independent towers     of   
non-normalizable (NN) and normalizable (N)  radial solutions  in
$AdS_5$,    labelled   by  the representations
$\nu $  of the  compact base   $X_5$  isometry group $G$. 
The   resulting harmonic decomposition   is  given  near the boundary 
(at large $  r$)  by 
\bea && \varphi  (y ) = \sum _{\nu ,M} \varphi _\nu (r) Y ^{\nu, M}
(\T ) = \sum _{\nu ,M} ( c^{NN} _\nu (\varphi ) ({r \over \calr } ) ^{\D _\nu
    -4} + c^N _\nu (\varphi ) ( {r\over \calr } ) ^{-\D _\nu })  Y ^{\nu, M}
 (\T )  ,  \label{sub2eq34} \eea  
where the    coefficients $ c^{NN} _\nu $ and $c^{N}
_\nu $  dominate in the  ultraviolet and infrared at large and small $r$ 
(in the generic  cases  with $\D _\nu > 2$). The non-normalizable
field     components    source
perturbations  of the    gauge theory    by    composite operators
$ O^\nu $ that    break  the conformal  symmetry      in the ultraviolet 
\bea && \varphi  (r, \T ) =  \sum _\nu \hat c_\nu ^{NN} (\varphi )
(r/ r_{uv} ) ^{\D  ({O_\nu})  - 4}  Y ^\nu (\T )  \ \Longrightarrow \   
\d L  (x, \mu  ) = \sum _\nu \hat  c ^{NN} _\nu (\varphi ,  \L _{uv})  
(\mu / \L _{uv} ) ^{\Delta  ({O_\nu})  - 4} O
_\nu (x) ,   \label{eqadscftdual} \eea 
while   normalizable field components source    the 
 VEVs       of  the perturbed theory operators  in the infrared    
$c _\nu ^{N}  (\varphi ,  \L _{uv})   \propto <O _\nu >$~\cite{wittenholo98,klebwitten99}.    
Therefore, non-normalizable solutions   encode  the effects  from  
gluing    to the   bulk,   while  normalizable  solutions  encode 
the   back-reaction  from branes near the  throat tip.   
(Both   solutions    must be included    for  branes  
localized   inside the throat.)        
The scaling dimensions   of operators    dual  to  
scalar  fields in $ AdS_{d+1}$  of  mass  $M_{d+1,\nu } $
are   given  by the  two  real  solutions 
of Eq~.(\ref{sub2eqscal}), 
\bea && \D _{\nu  \pm  }  = {d \over 2}   \pm  ( ( {d \over 2 } )^2 + 
M_{d+1,\nu }  ^2  )^{1/2}   , \ [\D _{\nu   + } >   {d \over 2 } ,\ \D _{\nu  -   } <
  {d \over 2 } ]   .  \eea 
The   lower  bound   on the solution  $\D _+ >   {d  /2 } $   is less  
stringent than the   lower  bound $\D > ({d\over 2 } -1) $ imposed on scaling dimensions     in  unitarity representations of the conformal   group $ SO(d,2)$.       The lower     bound  imposed on   masses    
$ M^2_{d+1 } > - ({d\over 2})^2 + 1$,     in the 
standard   $ AdS _{d+1} $ invariant quantization~\cite{breitfreed82},    
would then   select    the unique solution   $\D = \D _ +$,  
since  the   other   solution  ($\D _-  <  {d\over    2 } -1 $)  lies
outside the    unitarity bound.    However,     in   gravity
theories   involving   modes  of  squared mass  in the interval,   
$ - ({d\over 2})^2 <  M^2_{d+1 } < - ({d \over 2})^2
+ 1$, also the  lower solution    satisfies the unitarity 
bound   $\D _- > {d\over   2 } -1 $, so  one    would   have  the two 
admissible  solutions   with  $\D _+$    and  $\D _- = d -\D _+ $.  
There exist  in these cases    two   distinct    dual   field theories 
where   the operators  associated to  the normalizable and 
non-normalizable  radial   scalings,       $  r^{-\D _\pm } $ and 
$r^{\D _\pm -d } $,    exchange  r\^oles~\cite{klebwitten99}. 

The  deformed vacuum  background 
is  defined  as    the   solution of  the classical equations   
including small perturbations  of  the various  bosonic fields given
by the   VEVs,     $\varphi _A  (y) =[\d   \Phi _\pm (y)   ,\ 
\d \tau (y) ,\ \d G_\pm (y) ,\ \d g _ {mn} (y) ] $ constant in $  M_4$.   
The linearized  coupled   equations   are  
represented       schematically by separating the free 
wave  operators on the left  hand sides    from  source    terms   on 
the  right hand sides, 
\bea && \tilde \nabla _6 ^2 \d \Phi _\pm = S _{\Phi }  (\d
{\Phi _\pm },\d G _\pm ) ,\ \tilde \nabla _6 ^2 \d \tau = S _{\tau } ( \d {\tau },
\d G _\pm ) ,\ d (\d G_\pm ) = S _{G}  (\d {G _\pm } ) ,\cr && \tilde \nabla _6 ^2
\d \tilde g _{mn} + \tilde \nabla _m \tilde \nabla _n \d \tilde g - 2
\tilde \nabla ^p \tilde \nabla _{(m } \d \tilde g _{ n ) p} = S_{g}  ({
  \tilde \d g _{mn} })  .  \label{sub2eqx34} \eea  
The   complete  formulas     are provided  in 
Eqs.~(\ref{appeq35}).  In the approach of~\cite{gandhi11}   that we
follow, the fields   variations are  split into homogeneous
and inhomogeneous parts    $ \varphi - \varphi ^{(0)} = \varphi _{H}
+\varphi _{IH} $,   where the former   are  
zero mode  eigenfunctions of the    Laplace-Beltrami or Lichnerowicz
wave operators  of $\tilde \calc _6$.  
The homogeneous part  of a  canonically normalized  scalar zero
mode  field
$\varphi (r,\T )$ in   $ M_4 \times \calc _6 \to AdS _5 \times
T^{1,1}$,  decomposes     over  towers   of    harmonic  functions, 
labelled by the conserved  quantum numbers   $\nu = (j,l,r ) $ of the isometry  group, 
\bea && 0 = \tilde \nabla ^2
_6 \varphi (r,\T ) = (r^{-5} \dh _r r^5 \dh _r + r^{-2} \tilde \nabla
^2 _5 ) \varphi (r , \T ) , \cr && \Longrightarrow \ \varphi (r,\T ) =
\sum _{\nu , M} \bigg ( c^{NN} _\nu (\varphi )({r\over r_{\star } } )
^{\D _\nu -4} + c^N _\nu (\varphi ) ( {r\over r_{\star } } ) ^{-\D
  _\nu } \bigg ) \ Y^{\nu , M} (\T ) ,\ [\D _\nu = 2 + \sqrt {H_0 ^\nu
    + 4}]    . \label{sub2eq36} \eea  
To each harmonic mode   is associated    a pair of  
non-normalizable  and normalizable radial
solutions   determined  by   a  single   scaling 
dimension parameter      $\D _\nu $    depending    algebraically 
  on  the eigenvalue  of  the    base manifold  Laplace operator  
 $-\tilde \nabla ^2 _5 =   -\tilde \nabla ^2 _{ T^{1,1}} 
=  H_0 ^\nu = H_0 (j,l,r )$.   The constant  
coefficients   $  c^{NN, N } _\nu  $   parameterize  the  field
 perturbations at  the floating  radius   $ r_\star $. 
Consistently    with  the holographic
(renormalization group)    correspondence,  the  coefficients  
obey the   scaling  laws   $ c^{NN } _\nu (\varphi ) \equiv c 
_{\nu } ^{NN} (\varphi , r_\star )\sim r_\star ^{\D _\nu - 4 }  $ and $c^{N} _\nu  \sim   r_\star ^{-\D _\nu } $. 
We  shall   mostly  focus  on  the non-normalizable
perturbations induced  by    compactification effects and   hence will
usually omit hereafter  the  upper suffix $NN$.  
It is  very convenient then  to characterize the coefficients  
by their  values at the ultraviolet  matching scale, distinguished  by
a hat,    $\hat c ^{NN}_{\nu } (\varphi )\equiv c ^{NN}_{\nu } (\varphi , r_{uv}
) $. The  values of  the  ultraviolet scale  coefficients
are  of   natural order  unity,  $\hat c ^{NN}_{\nu } (\varphi ) =
O(1)$, unless    the background   satisfies    unbroken  approximate 
(super)symmetries   that force  them to take vanishing  or    suppressed  values. 

The linearized  description for 
the   deformed background   is valid    inside  the 
range of  radial distances,   $ r_{ir} < r < r_{uv} $,   
far   from    the   boundary  and horizon
regions   where  perturbations grow   rapidly. 
Warped  compactifications  let us dispose of  two small 
parameters, given by   the  matching radius ratio 
$ r _\star / r _{uv}$  and   the  warp factor ratio 
$ w = e^{ A } \vert _{min} = r_0 / r _{uv} \simeq r_0 /\calr $.  
The   inhomogeneous parts      are then   evaluated as double    
power expansions  in  these   parameters   $\varphi _{IH} =  
\sum _{n  \geq   1}  \varphi ^{(n)} $  that  allow 
solving the   resulting coupled field equations
iteratively.  The  differential equations   at  a given order $n$, given schematically   by  
\bea && r^{-5} \dh _r r^{5}   \dh _r \varphi _A ^{(n)}= N_{A}^ B (\varphi _C ^{(m)}
) \varphi _B^{(n)} + S _A^{(n)} , [m< n,\ A=1, \cdots , 6] \eea  
with     $N_{A}^ B $  depending  on  source terms 
evaluated at previous orders $ m< n $, 
are solved  by    means of the radial wave   operators Green's
functions.    The triangular structure of the matrix $N$ in
the vector space of fields $\varphi ^{(n)} _A$ allows,   for  a  suitable    
ordering     of the equations,  an algebraic resolution   at each  order
$n$. 

The  dominant contributions  from   operators sourced in the  ultraviolet
by compactification  effects are those of lowest  scaling dimensions
$\D _\nu $  with a radial dependence $r  ^{\D -4} $   
growing fastest   in  the  infrared.  The  important
operators  are  fortunately    few in  
numbers.   To avoid   destabilizing the  background    
by   perturbations  $ r ^{\D _\nu - 4 }$   growing     in the
infrared,  one    could    be  tempted to  discard  the relevant
operators of dimensions  $\D _\nu < 4$.    However, this   objection
is    really justified  for   semi-infinite   (non-compact) throats.     
For    throats   embedded  in compact manifolds,  
only the operators whose growth in the infrared has a
disruptive effect on the background  must be forbidden.  
With  an inclusive criterion of
this kind, one  could   also   include    the relevant operators  
whose  coefficients  are sufficiently suppressed   by    warping.  
Any  relevant     operators with  unsuppressed coefficients  must  
then   be  ruled out  in some way, for instance,  
by  invoking a  remnant discrete symmetry or    an
unbroken supersymmetry of the   warped throat.   

\subsubsection{Compactification effects  on   background solution} 

The  AdS/CFT   duality    in  the conifold  case 
relates the   $T^{1,1}$  harmonic fields   $\varphi _\nu  (x,r) $
in $ AdS_5 $       to      the gauge singlet composite  operators $O _\nu (x) $ 
of  Klebanov-Witten    conformal  gauge  theory in $M_4 = \dh
(AdS_5) $.    The  $\caln =1 $  supermultiplets in irreducible
representations $\nu  = (jlr)$  of  the  isometry/flavour  group $ G  $
combine into  a finite number of 
unitary representations   of the  conformal   group $ SO(4,2 ) $, 
filling      unitary   representations   of the
superconformal   group $ SU(2,2 | 1)$.
The    various modes   fit  into
nine superconformal multiplets $\cald (E_0,  j_1, j_2 ; r)
$~\cite{ceresole99}   consisting  of   a single
graviton    multiplet  $ g,$   four
gravitino  multiplets $ G.I \  - \ G.IV $  and 
 four vector multiplets  $V.I \ - \  V.IV  $. 
Each  of these       multiplets    includes
 a    finite collection   of      unitary irreducible 
representations   of the conformal  group $SO(4,2 ) \supset    U(1)
_{E_0} \times SU(2)_{j_1}  \times SU(2)_{j_2}$,  labelled  by 
conformal primary   representations  $\cald (E _0^{(s)}   , j_1^{(s)}
, j_2 ^{(s)} ) $  and flavour  group  $G$   representations $\nu  $. 
The   descendant conformal  operators
are  constructed  by   repeated  action of  the translation operator $
P_{\mu } \sim i\dh _\mu $ on the   primary   representations.  

At  leading   $ O(1/ N)$ of the dual  gauge theory, the  representations
consist of    composite operators     constructed  from 
single traces over colour indices  of  the fundamental   gauge  and
matter  fields products,  symmetrized over flavour group indices.
The   dimensions and    $r$-charges  at the conformal fixed point 
are $  \D   [W_i, A_i , B_i]  =  [3/2,\ 3/4,\ 3/4 ], \  r  [W_i, A_i , B_i]
= [1,\ 1/2 ,\ 1/2 ] ,    \ [i=1,2].$  
The supersymmetry   multiplets   in $AdS_5$ are 
either  short    representations   that  map to protected chiral and
semi-long (conserved current)  superfield operators in $M_4$ 
of   rational dimensions  stable under   renormalization, or 
non-short  representations  that 
map to unprotected non-chiral (real or vector)
operator superfields    in $M_4$,     acquiring    anomalous    dimensions 
 from quantum effects.   To the    latter  class  belong  the     
operators    of   irrational     dimensions  that lie    
outside of   the  weakly  coupled  perturbative  regime. 
 
We     start by discussing   the radial  scalings for   
harmonic modes of   
the various supergravity   fields.    The    warp  profile   fields  $\Phi
_\pm (X) = e ^{4A   (X) } \pm  \a (X)  $, defined   
by the longitudinal   components  of the deformed  metric
and 4-form field components along $ M_4$,  as  in   the classical ansatz  in
Eqs.~(\ref{sub2eqx3}) and~(\ref{sub2eq22}),    are 
non-dynamical  fields  by construction.  
(The warp profile  could  be gauged away  via a  redefinition    of
 coordinate  variables.)   The   variations  of these fields   
are   formally    related   to  the  transversal components 
of the   metric and 4-form field  components along $X_5$,   
$\d \tilde   g _a^a= \tilde g ^{ab} \d \tilde g _{ab}
 \equiv  \pi (X),\  \d C_{abcd} \sim   b (X) $  by the  functional 
    relations $\d \Phi _\pm (X) \simeq   e ^{4 A (r)} (\pi   (X) ,\  b
    (X)  ) \simeq  (r/\calr ) ^4 \sum _\nu   (\pi _\nu , b_\nu )
    Y^{\nu } (\T ) $. These      formulas are deduced  from the    
conditions that the   warped volume  of  $ X_5$   and    the 5-form field
self-duality      remain  preserved under  deformations 
\bea && \ 0= \d g_{5}(X)\equiv  g ^{ab} \d g _{ab} = \d \tilde
g_{5} -10 \d A  (X) = \pi  (X) - {5\over 2} e ^{-4A } \d ( e ^{4A (X)}), 
\cr &&  \d C_{0123}(X) =  e ^{-4A} \d \a (X) \propto   e ^{8A} \d C_{abcd}
(X) \ \Longrightarrow \d \a (X) \propto  e ^{4A} b (X) , \eea 
where the    proportionality   constants   
are not needed    since the    two  fields   are    normalized independently. 
The     correspondence  in Eq.~(\ref{sub2eq28p})  
between    the transverse   fields of fixed
masses  $ S ^{\pm } _\nu ( M^{ \nu 2} _{5,
  \pm } )$   and  the operators eigenmodes  of fixed dimensions   
$ O ^{\pm } _\nu (\D ^\pm _\nu ) $,    also   holds   for the  
rescaled    longitudinal     fields    $ r^{-4} \d _\nu \Phi _\pm  $.    
However, the   radial rescalings for $\d \Phi _\pm (y) $   
implies  that these fields  are not canonical
conformal fields in  $ AdS _5$   in the sense 
of Eq.~(\ref{sub2eq36}),   unlike  the  canonical  fields 
$(\pi _\nu  , b _\nu )$  whose  eigenvectors  
satisfy  the standard   radial scalings with  $\D  ^{\mp } _\nu =\D  _\nu \mp
4,\  [\D _\nu = 2  +   \sqrt    {H_0 +    4} ]$.   
For     uniformity of   notations, it is     convenient     to
replace  $\d  \Phi  _+  $   by  the  inverse   field 
$\d  \Phi  _+ ^{-1}   $,    introducing  instead  
the   canonical   field  
$\d  \hat  \Phi  _+ ^{-1}  = r^4  \d  \Phi  _+ ^{-1}$ of 
dimension   $ \D _\nu  (\Phi _{+ } ^{-1}  )$,  as in~\cite{gandhi11}.   
(Note  that $\d \Phi ^{-1} _+ = - {1 \over 4} e ^{-8 A} \d\Phi _+ \sim r
^{-8} \d\Phi _+.$)   The   radial    scaling laws for     
the operators dual    to eigenvectors of the  harmonic modes    
of $\d \Phi _{\pm }  / r^4  $   or    to   eigenvectors of  
the system     $(\pi _\nu  , b _\nu )$, 
\bea &&  \d \Phi _- = \sum _\nu  (c^{NN} _\nu r ^{\D _\nu  -4  } +   
c^{N} _\nu r ^{-\D  _\nu  } ) Y^\nu (\T )  ,  \ \ \    \d \Phi _+ 
= \sum _\nu  (c^{NN} _\nu r ^{\D _\nu   +4 } +   
c^{N} _\nu r ^{-\D  _\nu    + 8 }) Y^\nu (\T )  ,\eea 
can be used  to  obtain the  scaling laws   of  the canonical fields     
$ \d  \hat  \Phi _- \equiv  r ^{-4} \d \Phi _-  ,\   \d \hat  \Phi ^{-1} _+
\equiv   r^4 \d   \Phi ^{-1} _+ $.  
Matching the   resulting   harmonic decompositions  
to  the standard      form of the   radial   scalings  for 
canonical  fields, 
\bea &&  \d \hat \Phi _- =\sum _\nu  (c^{NN} _\nu r ^{\D _\nu (\Phi _-)    -4  } +   
c^{N} _\nu r ^{-\D  _\nu  (\Phi _-)  } ) Y^\nu (\T )  ,  \    
\d  \hat  \Phi ^{-1} _+ = \sum _\nu  (c^{NN} _\nu r ^{ \D _\nu  (\Phi
  _{+ } ^{-1}  )  - 4 } + c^{N} _\nu r ^{-   \D _\nu  (\Phi _{+ }
  ^{-1}  ) })  Y^\nu (\T ) ,   \eea 
leads to  assign    the    scaling dimensions          
$ \D _\nu (\Phi _\mp ^{\pm
  1}  )=\D _\nu  \mp  4 \equiv \D _\nu  ^\mp$  to  NN solutions   and 
$ \D _\nu (\Phi _\pm  ^ {\mp 1} )=\D _\nu \pm  4 \equiv \D _\nu  ^\pm
$    to   N   solutions.     
We rewrite   these   radial scalings  in the contracted   form 
\bea &&  \d  \hat  \Phi _\pm  ^{\mp 1 }   = \sum _\nu   (c^{NN} _\nu 
r ^{\D _\nu ^\pm   -4 }+   c^{N} _\nu   r ^{-\D  ^\mp _\nu } ) Y^\nu
(\T )  ,\ [\D _\nu  ^\pm  =\D _\nu  \pm  4 ,\  \D _\nu = 2  +   \sqrt
  {H_0 +    4} ]    \eea 
to emphasize   that       $\D _\nu ^{\pm
} $  are    exchanged  upon going from   NN and N   solutions 
for either fields or, equivalently,  are    exchanged  upon going from
$\d  \hat \Phi _- $  to $\d \hat  \Phi  _+ ^{-1}  $ fields  for 
either     NN and N   solutions. The operators  dual to N
perturbations of  $\Phi _{\mp }  $  are    
sourced by NN   perturbations of  $\Phi _{\pm }  $. 
This reflects the fact that the operators  whose VEVs are  dual to 
N modes of  $\d \Phi _-$   are   sourced by   NN modes of  $\d  \Phi
_+^{-1}$  and,  conversely,   that the   operators sourced by
NN modes  of  $\d \Phi _-$  have VEVs   dual  to   N modes of $\d \Phi _+^{-1}$. 

An analogous radial rescaling factor $ r^{-2}$ is needed for the
base  manifold   metric  since  the    canonical conformal  field here  
identifies to   the  warped  metric tensor $\d  {\hat
  g}_{ab} \equiv   \d  g_{ab}  = e  ^{-2A} \d \tilde   g_{ab} . $
The dimensions of   operators   dual to the  
 harmonic   scalar   fields $\phi _m ^{(\nu )}  $,  eigenmodes    of the  
5-d mass matrix   for  $\d \hat g _{ab}$,   
 obey the  mass-dimension   relationship
for  scalar   fields.   No rescaling   is needed  for the axio-dilaton
field  $\d \tau  $    or  for the $\d C_{ab}$ fields. 
The  dimensions of the operators   dual to   the 
harmonic  scalar   field     $a ^{(\nu )} \in \d C_{ab}$, eigenmodes of the 5-d mass
matrix,    are   determined through the  mass-dimension   relationship. 
 The  eigenvalues       $ \l ^{I_s} \equiv   H_0 ,\  \l ^{I_t} ,\  \d ^{I_2} $ 
of the   (second order)  scalar    Laplace and   2-tensor
Lichnerowicz operators and  the (first order)    2-form
Laplace-Beltrami  operator in $ T^{1,1}$
 \bea && (\tilde \nabla  ^2  +  \l ^{I_s} ) Y  ^{I_s} =0,\
       (\D _K +  \l ^{I_t} )Y  ^{I_t} _{    \{ab\} }   =0,\ 
( \star _5 d  -i \d ^{I_2} ) Y  ^{I_2} _{ [ab]   }    =  0 , \eea 
are  related to the  scaling dimensions  $  \D _\nu ^{I_s},\   
\D _\nu ^{I_t}  ,\ \D _\nu ^{I_2} $ of the scalar, 2-tensor and 2-form
 operators    by  the formulas
\bea && \D _\nu \equiv  \D _\nu ^{I_s}  = 2 +\sqrt {
  \l ^{I_s} +4} ,\ \D   _\nu ^t   =  2 + \sqrt { \l ^{I_t} -4 } 
,\ \D _\nu ^{I_2} =  \D _\nu (\d ^{I_2} )= \max ( \d  ^{I_2}  , 4- \d ^{I_2} ) , 
\cr &&   
(\pm )  \d ^{I_2} _{I, II, III} = \pmatrix{-1 + \sqrt { H_0^{++}
    +4}  \cr \sqrt {  H_0 +4}  \cr   1 +   \sqrt { H_0^{--} +4}}  = 
\pmatrix{- 3 + \D _\nu (j,l,r+2)\cr   \D _\nu (j,l,r), \cr  - 1 + \D _\nu
(j,l,r-2)  }    , \ [\D _\nu (j,l,r) = 2 + \sqrt {4 + H_0 (j,l,r) }]   \eea     
where  the  three series $ I,\ II,\ III$ of 2-form  operators  in the above 
list were classified  in~\cite{baumann10}     with  the   two signs   
$ ( \pm ) \d ^{I_2} $   referring to    the pairs of  Hermitian
conjugate    operators.   We  summarize    in   the table  below 
the  radial scalings   for NN and   N harmonic  perturbations of   the 
various  canonical    fields, based on~\cite{gandhi11}. 
  \begin{center}   
\begin{tabular}{c|c|c}      
&  \  Non-normalizable (NN)  & Normalizable (N)     \\ \hline  
 $\d  \hat \Phi _-  $ &  $ c ^{NN} _\nu  r ^{\D _\nu ^- -  4} ,\ [\D _\nu ^ - =
    \D _\nu -4 ]   $   &  
$c ^{N} _\nu  r ^{-\D _\nu ^+  }  $   \\ 
  $ \d  \hat \Phi _+^{-1}  $ &   $ c ^{NN} _\nu r ^{\D _\nu ^+ -4 } ,\ 
[\D _\nu ^ + = \D _\nu +4 ]  $  & $  c ^{N}
 _\nu  r ^{- \D _\nu  ^-}  $ \\  
$\d \tau  $  &     $ c ^{NN} _\nu  r ^{\D _\nu  (\tau )  - 4}  ,\ [\D _\nu  (\tau )   =  \D _\nu ] $  & $c ^{N} _\nu  r ^{-\D _\nu  (\tau )   }  $  \\ 
$\d   G_-  $  &     $ c ^{NN} _\nu  r ^{\D _\nu  (G_-)  - 4} ,\ 
[\D _\nu  (G_- )  =  \D (\d _\nu  ^ {I_2} \geq 2  ) ] $  & 
$c ^{N} _\nu  r ^{-\D _\nu  (G_-)} ,\ [\D _\nu  (G_- ) =   \D (\d ^ {I_2}
  \leq   2  )   ] $  \\     
$\d   G_+  $  &     $ c ^{NN} _\nu  r ^{\D _\nu  (G_+)  - 4} ,\ 
[\D _\nu  (G_+ )  = \D ( \vert \d  _\nu ^ {I_2} \vert  \geq 2  ) ] $  & 
$c ^{N} _\nu  r ^{-\D _\nu  (G_+)} ,\ [\D _\nu  (G_+ )  = 
 \D ( \vert \ \d ^ {I_2}\vert \ \geq  2  )   ]  $ \\    
$\d \hat g _{ab} $  &     $ c ^{NN} _\nu  r ^{\D _\nu  (g )  - 4}  $  & 
$c ^{N} _\nu  r ^{-\D _\nu  (g)} ,\ [\D _\nu  (g)  = \D
  _\nu ^t ]  $  \\ \hline     
\end{tabular} \end{center} 
The  dominant operators     in the
infrared correspond to  the      harmonic modes  of lowest  radial scaling
dimensions  sourced  in the   ultraviolet, as seen on
Eq.~(\ref{eqadscftdual}). The  effect is     clearly more pronounced 
for        larger warping hierarchy,  $r_{ir} / r_{uv}  = w  << 1
$.     In  the dual  theory    preserving $\caln =1 $   supersymmetry, 
the  operators    occur  as   superfields        in
  chiral and semi-long (semi-conserved)    representations  
or  non-chiral representations.  The  scaling dimensions   for the  former    
supermultiplets   are  protected      by 
non-renormalization theorems  while those  of  vector 
supermultiplets      can be affected  by    quantum
corrections. The dimensions of   anomalous operators are  small 
in the perturbative regime of  small  't Hooft coupling $\l $,    but
grow  significantly at  strong coupling. The  contributions from
string mass    excitations    
$\D \sim M_5 \calr  \sim m_s  \calr \sim \l ^{1/4} $, for instance, 
diverge at  $\l \to \infty $.    The   leading operators at large 
$ N $    are    built from single   traces   over   
colour   of   products  of the basic  (singleton representation) fields 
$ W_i,\ A_i,\ B_i ,\ [i=1,2]$     in the 
 symmetric-traceless  combinations  of flavour  indices  
associated to     irreducible representations of $G$.   A  complete
classification    is  provided  in~\cite{ceresole99}.  
As  for the  chiral  operators, these     arise    in 
three infinite   series of superfields   $ S^k,\ T^k, \ \Phi ^k
$ listed in the   following table.     
\begin{center}  \begin{tabular}{c | ccc}   
& $  S^k    $ &    $  T^k _\a  $ &  $ \Phi ^{ (\pm )  k}  $ \\ \hline 
$ O _k $ & $ Tr   (  (AB)^k  )$ & $ Tr   (W  _\a   (AB)^k  ) $  & $ Tr
( ( W  _1 ^2 \pm  W  _2 ^2 )    (AB)^k  ) $ \\  
$(j_k,l_k,r_k) $ & $({k\over 2},{k\over 2}, k)   $ & $ ({k\over 2},{k\over 2},
k+1) $&  $  ({k\over 2},{k\over 2}, k+2) $ \\    
$\D _k $ & $ {3\over 2} k  $&$ { 3\over 2} (k +1)  $& $ {3\over 2} (k
+2) $ \\ \hline  \end{tabular} \end {center}
The dimensions and $r$-charges  satisfy  the   proportionality
relation   $ \D _k = 3 r_k/2 ,\ [k\geq  0] $ and  appear    
sequentially as $ (\D , r )= [ (0,0),\ ({3\over 2} , 1),\  (3, 2)
  ,\  ({9\over 2} ,   3),\cdots ] $.   For instance, the  values  
$(\D , r ) = (3, 2)$ is common to   the  three operators      
$S ^{2}_ {(112)}   , \   T^1 _{(110)} , \ \Phi ^{(\pm  ) 0 }_{(000)} $.

The semi-conserved  superfield  operators  for   the    towers  
$  J ^k _A = Tr (A e^V \bar A e ^{-V} (AB)^k ) $  (and 
$ J^k _B $   of same form)    of dimensions $ 2 + 3k/2$,  built 
on the    flavour singlet  (Konishi and Betti)
operators $ K _{(000)} = J _A + J_B  ,\   \calu _{(000)} = J _A - J_B
$,  are expected  to    acquire anomalous dimensions.  
For instance, the  approach to  criticality within  the conformal  window
 of  the Konishi   axial current  in QCD like theories~\cite{anselmisenI96},
$ <J _\mu (x)J _\nu (0)> \sim {g _{\mu \nu }  / \vert
  x  \vert  ^{6 + 2  \b ' (g^2_\star )  } }$,   involves  an  anomalous  dimension      proportional to the beta function   slope. 
 Two  other   important examples  of non-chiral    superfield operators
 arise  via   the    tower of rational dimensions 
$    O _{(k/2,k/2, k)}   =  Tr (W^2 _1 \bar W^2 _2
(AB)^k )      \in V.II :\    (\D , r )    = (6 + 3k/2, k) $ and
 the operator (absent in the weak coupling  phase)
 of  irrational dimension $ O _{(110)} \vert _b
\in V.I :\  (\D , r )      = (\sqrt {28} -2 , 0)  $.   

The operators   are  usually classified  by  
their transformation  properties  under   symmetries.  
Lorentz symmetry     in  $ M_4$  is preserved by scalar operators
only. Unlike  the  singlet operators  of   the  flavour symmetry  group $ G$, 
the  non-singlet  operators   induce      an explicit  breaking 
$ G \to  G'$.   Since   continuous isometries 
are not allowed  in   Calabi-Yau manifolds  and   would produce
 non-stringy  gauge symmetries in  general  manifolds,  
only   discrete   subgroups of $ G  $     should survive  compactification.    
The    supersymmetry  preserving      operators       must 
belong to       highest  (F- or D-term)   components of  (chiral
or vector)  supermultiplets.   In   the  superspace representation  
of $\caln = 1 $ superfields,      using     
fermionic variables  $\t ,\ \bar \t  $ of (energy) dimension and $r$-charge 
 $ \  \D (\t ) =  -1/2,\  \D (d \t ) = 1/2,\  r (\t ) =
1  ,\   r (d \t ) =- 1  $,   the  dimensions and   $r$-charges 
of the   supersymmetric  components of chiral or vector  operators  
$ O^{C} ,\ O^{V} $,   
\bea &&  \D (O^C\vert _{\t ^2} ) =\D (\int d ^2 \t O ^C ) =  \D (O ^C
) +1,\ r (O^C\vert _{ (\t ^2,\ \bar \t ^2)  } )  = r (O^C\vert  _b)  \mp 2 ,\cr && 
\D (O^V\vert _{\t ^2\bar \t ^2}  ) =\D (\int d ^4 \t O ^V  ) = \D (O
^V ) +2 ,\ r(O^V\vert _{\t ^2\bar \t ^2}  ) =  r (O^V\vert _b ) ,\eea
get   shifted  by 1 or 2 units   and   $\pm 1$
or    $0$  units     relative to those of the bottom
components, signalled  by the lower  subscript $b$.  One learns whether   some      operator preserves or  breaks
supersymmetry   by  examining  how its dimension and $r$-charge $ E_0 ^{(s)} ,
\ r^{(s)} $  are related  to those  of  the highest weight
representation $ E_0  , \ r  $    of the  superconformal multiplet.
The (chiral, non-chiral)  superfield operators   obey  then the
characteristic relations     
$  E_0^{(s)} - E_0 = (1 , \ 2 )  $ and $r^{(s)} - r =  (\pm 2  ,\ 0 )
$. The  combined Lorentz symmetry and supersymmetry    conditions
thus select       only  operators  in the   $V.I - V.IV $ multiplets.  

The relevant     information is   extracted   from Tables 7-9
of~\cite{ceresole99}  with     helpful assistance  from  Tables 1-4
of~\cite{aharony05}.    For instance, since  $\d  \Phi _- \sim  S^- (\pi ,
b )   \in V.I  $  and  $\d  \Phi _+  ^{-1} \sim   S^- (\pi , b )
\in    V. II $   belong to bottom   and top components
of   non-chiral supermultiplets, 
these are   associated   to  supersymmetry breaking and  conserving  
operators, respectively.   
The  perturbations   $\d \tau  \in  V. IV  $  arise from  
$\t ^2 $  components of  chiral  supermultiplets,      hence   are
 supersymmetry preserving. The   modes    $\phi ^{(\nu )}  $   of  $\d g
_{ab}$      occur in  all   four $V. I \ -  \ V. IV $  multiplets   
as   bottom  $ (b),\  \t ^2 $ or $\t ^2\bar \t ^2 $  
components   of  chiral   and  vector 
supermultiplets.  The modes   $ a ^{(\nu )} $  modes  of $\d C_{ab}$ 
occur in   several  $V$  multiplets  as $ b, \ \t^2$   components of 
 chiral    operators  and  were   classified in~\cite{baumann10} 
in  three series of chiral and  non-chiral 
operators $ O_I,\ O_{II}^\a ,\ O_{III}$   dual to 3-fluxes $\d G_\mp
$.       (The   latter   work  also signals    that    
the   states  $ (E_0 +1,\ r-2) \in V.I  $ and $(E_0
+1,\ r\mp 2) \in V. II $ which were    listed   as    
$\phi ^{(\nu )} \in g_{ab}$ modes  in~\cite{ceresole99} 
should instead be  counted as $a  ^{(\nu )} \in C_{ab}$ modes.) 

The  fact  that    regular  $ N\ D3$-brane stacks have 
vanishing potential  energy,  hence move
 freely   in  Klebanov-Strassler geometry~\cite{klebstrass00},
requires   discarding  the      operators
that    lift  the    moduli space of vacua of Klebanov-Witten gauge
theory.   The   dominant   contributions from
conformal   symmetry breaking  in the infrared 
clearly arise   from the  relevant and  marginal  operators 
of dimensions $\D  ^{top}  _\nu \leq 4$. 
 If one  restricts  to  supersymmetry preserving
perturbations,  then  it  would be  necessary to rule out  
all the    relevant and 
marginal   operators  in top components of superfields.   
The  requisite  conditions are 
automatically  satisfied   for    compactifications  
  involving   ISD   3-fluxes  of   $ G_{2,1}$  
type  only  or     preserving  a suitable     
discrete  global symmetry    subgroup of $G$~\cite{baumann08}.  

In addition to the       homogeneous parts 
proper to    each  field  $\varphi  ^{(1)} $,    
contributions   of same  order  are  induced  by source
terms  arising at   successive  levels of the   system of coupled
equations~\cite{gandhi11}. 
For instance, the 3-form  and  axio-dilaton   fields  acquire   the induced 
corrections   $\d G _- ^{(1)} \sim r ^{-4}  \d \Phi  _- ^{(1)} , \ 
\d \tau ^{(1)}  \sim  r^{-4} \d  \Phi _- ^{(1)}  + \d G _- ^{(1)}  ,$
while    the $\d (\Phi _+ ^{-1} )^{(1)}$ field  is    corrected  
by  all other  fields.   
It was  shown in~\cite{gandhi11} that the  radial scalings
for  higher order  perturbations     have a multiplicative
structure  given by  the  schematic  relation~\cite{gandhi11} 
$     \varphi ^{(n)}= \prod _{i=1}^n 
[c ^{\nu _i }(\varphi ) (r/r_\star ) ^{\D _{\nu _i } (\varphi ) -4 } Y ^{\nu _i } (
\T ) ] .$       The       corrections   to $ \d \Phi _-  $  sourced  by  
3-fluxes $ \d G_- ^{(1)}  $ are  obtained   by solving  the 
following equation in  terms  of the Green's function $  G(y, y') $ 
of  the scalar Laplace operator~\cite{baumann10},  
\bea && \tilde \nabla ^2 \d \Phi _-^{(2)}
= {g_s \over 96} \vert \d \L ^{(1)} \vert ^2 ,\ \ \ [ \d \L ^{(1)} (y)
  = 2 e ^{4 A} \d G_-^{(1)} \simeq  \sum _i c ^{\nu _i} r ^{\d _{\nu _i} } Y
  ^{\nu _i }(\T )] \cr && \Longrightarrow \ \d \Phi _-^{(2)} ={g_s
  \over 96 } \int d ^6 y' G(y, y') \vert  \d \L ^{(1)} (y') \vert ^2
\simeq \sum _{\nu , i, j} h ^\nu _{ij} r ^{ \D _{ij} -4 } Y ^{\nu } (\T
), \cr && [\D _{ij} = \d _{\nu _i} + \d _{\nu _j} ,\ \ h ^\nu _{ij}
  \simeq - {g_s \over 192 }{c _{\nu _i}   \bar c _{\nu _j}  \over
    \sqrt {H_0^\nu +4} }     \int d ^5 {\T '} Y ^{\nu \star } Y ^{\nu
    _i }Y ^{\nu _j \star } ]   .   \label{eqdefphi2}  \eea
The resulting second  order corrections
$\d  _\nu \Phi _{-} ^{(2)} $     involve 
a  double  sum over complex conjugate pairs of  operators of conformal
dimensions $\d _{\nu _i} ,\  \d _{\nu _j}$   whose quantum
numbers $ \nu _i ,\ \nu _j $   are  restricted by the selection rule
$\bar \nu \otimes \bar \nu _j \otimes \nu _i \sim (000) $.

\subsubsection{Backgrounds with   broken  supersymmetry} 

To  make  contact   with phenomenology, it is  highly 
desirable  to   consider  backgrounds with   broken supersymmetry. 
Several constructions  have been proposed.  
One could consider  quotients  of the     deformed orbifold
by discrete  symmetries~\cite{kachtein98,lawrence98}  that yield     
stable   solutions  with   explicit  supersymmetry    breaking 
at energy scales    of   $ O(1/\calr )$.     For instance, dividing by 
the  discrete  orbifold  group $ \G  =  Z_2 ^R $  part of  the deformed  conifold  flavour group     $ G=  SU(2) \times  SU(2)
\times Z_2 ^R,\ [Z_2 ^R \subset    Z_{2M}^R \subset U(1)_R ]$    
is found to leave   a discrete   flavour
subgroup $ G' \subset  G $   that  effectively  disallows
all  the  global  singlet relevant   operators~\cite{kachsimtri09}.
On  the other hand, stable and regular   solutions    with
spontaneously  broken supersymmetry    were obtained   
for the  deformed  conifold  in    several   studies  that 
may   be traced   from~\cite{brsusyco03,dewolfe08,benini09}.
Pursuing     along either  of these lines, however,   
would   involve the heavy task of
identifying   the dual  operators   selected  by   the  deformed  backgrounds.  

For this reason, we     shall   consider   the  simpler  approach   
where   the spontaneous supersymmetry breaking  in   an external    (hidden)
sector   is transmitted     by gravitational  interactions to the
(visible) dual gauge theory sector  via  a 
spurion   chiral superfield $X$.   
Restricting to   the    F-term   breaking case,  then   the
superspace   decomposition,    
$  X = x + F_X \t ^2 \equiv X {\vert _b } + X {\vert _{\t ^2} } \t ^2,
$ associates   $F_X  $   to  the supersymmetry breaking  scale  
while   $x $     in the  gravity mediation (in  contrast to the gauge mediation)   case  vanishes or   is  assumed  to
be  hierarchically   smaller  than    the  warped    energy
scale $ w m_s$.  Warping   allows one  to   accommodate a  
wide hierarchy      relative to  the fundamental  scale, 
$F_X /  m_s^2  \sim w ^2  << 1$.  
The  soft supersymmetry breaking  effects  from 
 integrating out   bulk   massive   modes 
are  represented  in  the  dual   theory  by products  of the   
superfield  operators   with   the  spurion  superfield. 
The  lowest order couplings    for chiral and non-chiral
supermutiplet operators  $O^C,\ O^V$ consist of  
   (superpotential  and kinetic)  F-terms  and    D-terms  with 
associated  coupling constants  $ f_C $ and   $ d_C,\ d_V$,
\bea && \d L =  (f _C [O ^C X ]_F
+ d _C[ O  ^C X X^\dagger ] _D + H.\ c. )  +  d_V [O ^V  X X^\dagger  ] _D
\cr &&  =  (f _C( O ^C {\vert _b } F_X + O ^C \vert
_{\t ^2} x ) + d  _C ( O ^C \vert _b F_X \bar F_X + O ^C \vert _{\t ^2} x \bar
F_X ) + H.\ c. ) +  (d_V   O ^V \vert _{\t ^2\bar \t ^2 } x x^\star
+ O^V \vert _b F_X F_X^\star ) .\eea
The   holomorphic couplings  (with $ f_C  \ne 0$)    of the   
chiral  supermultiplet  operators $ O^C$     cause suppressions by 
one power of $ F_X$ less than the  non-holomorphic  
couplings  (with $ d_C  \ne 0$).   
In line with  the    discussions in~\cite{baumann10,gandhi11},  
we    shall assume  
that  the chiral supermultiplet operators  for $\d \tau $  have 
holomorphic couplings    ($ f_C  \ne 0$) only,  while  the chiral
supermultiplet  operators  for   $\d \Phi _\pm  ,\  \d g _{ab}$   and $ \d G _\pm
$ have    non-holomorphic couplings   ($ f_C  = 0,\ d_{C, V}  \ne
0$)   only.  The $b ,\ \t ^2   $  components 
then pick up   factors    $F_X^1,\ F_X^0$   for $\d \tau $  and 
factors    $F_X^2,\ F_X^1$   for all other fields.
The $b, \ \t ^2  ,\ \bar \t ^2 $  components of  non-chiral  supermultiplet
operators  pick up  factors    $F_X^2,\ F_X^1,  \ F_X^0$.   
In summary, the  supersymmetry  breaking  insertions for 
  operators  sourced by  
$\d \Phi _-  ^{(1)} $  and  $\d \Phi _-  ^{(2)} $  are  $ F_X^2$,   
those  sourced  by $ \d G_- ^{(1)}$   are  $ F_X$ (except the mode $
C_b (002)$  with $ F_X^2$),  those  sourced  by  $\d g_{ab}^{(1)}   $
are  $ F_X$  (except the mode $V_D (112)$  with $ F_X^0$) and those
sourced  by   $\d \tau  ^{(1)} $ are $ F_X^0$.
 
One also needs to   associate     suppression   factors  
to the various  perturbations   by  assigning mode dependent  
exponent  parameters $ Q _\nu (\varphi )  \geq  0 $ 
(vanishing  for supersymmetric operators) 
for   the  warp factor dependence  of  the various operators, 
  namely,    setting  the  constant coefficients  as 
 $\hat c _{\nu } (\varphi ) \propto  
w ^ {Q _\nu (\varphi )} ,\ [w  = e^{ A } \vert _{min} = r_0 /\calr
]$~\cite{gandhi11}.    
For instance, the   NN perturbations    of homogeneous  parts  of 
the  warp profiles   are   defined as
\bea && \d \hat \Phi _\mp ^{\pm 1 } \equiv r ^{\mp 4}  \d \Phi _\mp
 ^{\pm 1} (r) \simeq \sum 
_\nu c _\nu ^\mp ({r \over r_{\star } })^{\D ^\mp _\nu -4 } Y ^\nu
(\T ) \ \Longrightarrow \ \d \Phi _- =  \sum _\nu  c _\nu ^-
({r \over r_{\star } }) ^{\D _\nu ^- } Y ^\nu (\T )   ,\ \d \Phi _+
 ^{-1}  =   \sum _\nu  c_\nu  ^+ ({r \over r_{\star } }) ^{\D _\nu ^+
   -8 }  Y ^\nu (\T )  ,  
\cr && [\D _\nu ( \Phi _- ) \equiv \D _\nu ^- =\D _\nu - 4 ,\ \D
  _\nu ( \Phi _+ ^{-1} ) \equiv \D _\nu ^+ = \D_\nu + 4 , \ \D _\nu =
  2 + \sqrt { 4 + H_0} ,    \cr &&   
c _{\nu } ({\Phi _- }
) = \hat c _{\nu } ^-  w ^{ Q_\nu (\Phi _-) }  ({r_\star
  \over r_{uv} } )^{\D _\nu   (\Phi _- ) }   =  \hat c _{\nu } ^- 
w^{Q _\nu ^-  + P \D _\nu ^- }  
, \  c _{\nu } ({\Phi _+ ^{-1} }) = \hat c_{\nu }^+  w ^{ Q_\nu (\Phi _+ ^{-1} ) } 
({r_\star \over r _{uv} } )^{\D _\nu (\Phi _+ ^{-1} ) - 8 } = \hat  c_\nu ^+  
  w ^{Q _\nu  ^+  + P ( \D _\nu  ^+  -8 ) }  ]  , \label{sub2eq352} \eea
where  the constant auxiliary  parameter $ P = O(1) $     defines the
ratio    of  the matching   radius distance relative to the
ultraviolet  cutoff  radius,  $ {r_\star \over r
  _{uv} } = w ^ P, \ {r_\star \over r _{0 } } = w ^ {P-1} ,\ [w ={r_0
    \over r _{uv} } ] $. To    obtain the above result,  we
have  expanded     the fields perturbations in powers of $r / r_\star $,
setting  $ r_{uv} = \calr $, replacing the unperturbed profiles as $ e ^{\pm
  4 A (r) } = (r/\calr ) ^{\pm 4} $, and exhibiting  the dependence on
$w$ as    prefactors to  the  ultraviolet scale coefficients  
$\hat c ^\nu ( \Phi _\mp ^{\pm 1} )  $  of  natural $ O(1) $. 
The     radial scaling  laws  
$c_\nu  ^\varphi = \hat c_\nu  ^\varphi  w ^{ Q^\varphi _\nu  +
\D _\nu ^\varphi -4}$  show  that one   eliminates the threat   from 
perturbations  growth  in the  infrared  by  imposing the bounds
$ Q ^{\varphi }_\nu  \geq - \D ^{\varphi }_\nu  + 4 $.
The multiplicative     structure of   radial scalings
for  higher order perturbations~\cite{gandhi11}, 
$  \varphi ^{(n)} \approx   
\prod _{i=1}^n  [ \hat c ^{\nu _i } (\varphi )  w  ^{Q_{\nu _i } (\varphi
  ) +  \D _{\nu _i }(\varphi )   -4 } ]  , $ 
implies  that   all  higher   order   corrections  get   
protected  once the lowest  order conditions are respected. 

To get some practice  with  the the present  formalism, 
we  briefly illustrate    the implications from the above
discussion for  three     applications involving a $\bar D3 (r_0) $
near the tip and a mobile $ D3(r)$-brane  embedded in the conifold.   
First,    we recall  that the  interaction potential for the  static 
$ (D3 (r) -\bar D3 (r_0)) $-brane system~\cite{baumann208} 
can be   evaluated    either  from 
a  NN perturbation sourced  by $D3$   and felt  by $\bar D3 $ 
or    from a    N perturbation sourced  by    $\bar D3  $     
and felt  by    $ D3$.   The    interaction 
potentials evaluated in the two descriptions 
(omitting  for simplicity the harmonic function  factors  $ Y^{\nu }
(\T)  $),   give the same scaling law  
\bea &&\bullet \  \d V (\bar D3 (r_0) / D3(r)) /\tau _3  
= \d  \Phi _+ (r_0) =  ({r_0\over \calr })^4 \d  \hat \Phi _+ (r_0)  
 \simeq   c^{NN} _\nu ({r_0\over \calr })^4  (r_0/r) ^{\D _\nu (\Phi _+ ^{-1} )   - 4}  
=  w ^4 c^{NN} _\nu (\Phi _+) ({r_0 \over r}) ^{\D _\nu } . 
\cr && \bullet \   \d V    (D3 (r) /  \bar  D3(r_0) ) /\tau _3   
=  \d  \Phi _-  (r) = ({r \over \calr })^4 \d  \hat \Phi _- (r)  
\simeq  c^{N} _\nu ({r \over \calr })^4 (r/r_0 )^{-\D _\nu (\Phi _-)}
=     w^4 c^{N} _\nu (\Phi _-)    (r/r_0 )^{ - \D  _\nu}  . \eea  
The leading tower  of   operators $ Tr (F^4 (AB)^k )  \in  V.II  $, 
dual  to $NN / N$ modes of $\d   \Phi _\mp ^{\mp 1}  $, of 
dimensions $\D _\nu ^+= \D _\nu +4= 8 + 3k/2 ,\ [k\geq 0 ]$ 
contribute at  $k=0$  the  standard        Coulomb 
potential,    $ \d V \propto   w^4 \tau _3   (r _0/r)  ^{4 }$. 

The second application     concerns    the transmission of
supersymmetry breaking   in the same  setup   from  a hidden $\bar D3
(r_0)$-brane  sector   to   a visible  $D3(r)$-brane 
sector.  The   cross coupling  between the  
hidden sector spurion superfield $X = \t ^2 F_X $  and  the visible sector  
quark superfields $ Q_i$, from  integrating out     massive bulk modes,
is  obtained     by  replacing  in the dual  gauge theory  the spurion 
D-term by     the vector superfield operator $X^\dagger X \to  O^V_\nu $~\cite{kachrum07},        
\bea && \d L   (D3 (r)) = {M_\star ^ {-2}} 
\int d^4 \t   X^\dagger X    Q^{\dagger }  _i Q_j \simeq m_{3/2
} ^2  \tilde Q^{\dagger } _i \tilde Q_j ,\  [m_{3/2} ^2 = {F_X^2 /M_\star ^ 2}  ]  
\cr &&  \Longrightarrow \    
\d L   (D3 (r))  =    \tilde m^2 _{ij}\tilde Q^{\dagger } _i \tilde
Q_j ,\ 
,\   [\tilde m^2 _{ij}  =   m ^2 _{3/2}  c _{ij}  <\ O^V _\nu  
\vert _{\t ^2 \bar \t ^2} >  ({r \over r_0 })^{ - \D _\nu  (\Phi _- ) }
\simeq   m ^2 _{3/2}  c_\nu ^N  c _{ij}  ({r \over r_0 })^{\D _\nu  +4
}  ]   \eea 
where  the selected   operators $ O^V_\nu $    sourced  by $\bar D3
(r_0) $  are   those dual    to     normalizable    
perturbations of $\d \Phi _- $.
The top  component  of the 
singlet   superfield operator $  O^V_{ (000)} = Tr ( W_1 ^2 \bar W_2^2 )  $,
of dimension  $\D  _\nu +4   =8$,  gives the  leading   contribution (at
$ r = r_{uv} $)  to the       soft   squark   mass       $ \tilde m^2
_{ij}  \approx m ^2 _{3/2}  w ^ 8  $.    The   resulting strong   suppression       relative to the dimensional  scaling    estimate  
($ \tilde m^2 _{ij} \sim w ^ 4$)   illustrates  
the    effectiveness  of warping   in  
sequestering   supersymmetry  breaking. The  operator $ O_{(110)} $,    whose       top component
has the  lower     dimension    $\D ^{top} _\nu = {\sqrt {28} } $, is forbidden
because   its   VEV is  dual to a N  perturbation of     $ \d \Phi
_+$ (sourced   by  a NN perturbation of     $ \d \Phi  _- \in  V.I$)   
that cannot be induced  by $ \bar D3$~\cite{berg10}.

Finally, the third      application deals   with  
the supersymmetry preserving contributions from compactification effects  
to  the     Lagrangian  of a $\bar D3 $-brane  
near the conifold tip  due to    perturbations of the 
 warp   profile,
   $ \d L (\bar D3) = -\tau _3 \d \Phi _+ = -2 \tau _3 w ^4 \d A (y) .$ 
The  metric   perturbations  $ \d g _{ab} $  in the 
3-flux   source term     for $\d A  (y) $      
induce   a mass term for the open string moduli
fields~\cite{aharony05}  
\bea && \nabla ^2 \d A \simeq \d (G\cdot \bar G ) 
\sim {f^2 \over r^2  (S^3)  }   w ^{\D _\nu (g ) -4
} \   \Longrightarrow \ \d A \sim {f^2\over  g_s M }  
w ^{\D _\nu (g ) -6} \tilde g _{ab} X^a (x) X^b (x)  ,\ [r^2  (S^3) =
  g_s M \a '] \cr &&  
\Longrightarrow \  \d L (\bar D3 (r_0)) =  - m^2
_{ab} X^a X^b ,\ [m^2 _{ab} \sim {f^2\over   g_s M \a '}  
w ^{\D _\nu (g ) -2} ]   . \eea   
The  leading irrelevant  operator  from the  top component of  
the superfield $ O_{(110)} \in V.I $,  of   dimension  $\D _\nu (g ) = 5.29 $,  
contributes  the    mass   $ m^2 _{ab} \sim w ^{3.29}$.  
The    contributions   to $   \d L(\bar D3
(r_0)) =- \tau _3 r^8 (r_0 /r )^{ \D _\nu (\Phi
  _+^{-1} ) -8 } Y^\nu (\T ) \sim - \tau _3 (r_0 /r) ^{\D _\nu +4}
Y^\nu (\T )  $ from     operators   dual  to  $\Phi
  _+ \in   V.II$   of     dimensions $\geq 8$ are sub-dominant.  
(For an   orientifolded conifold, a      larger   mass term could
arise   from the Mobius   surface  string amplitude~\cite{aharony05}.) 

\begin{table}   \begin{center} \caption{\it \label{tabdef1}
 Deformations  of the  classical background    from embedding the conifold in     a flux compactification    with    broken supersymmetry.   
 The  perturbation theory orders  are
 indicated  by  the  upper suffices $ (1),\ (2)$   attached to  the  supergravity    conformal fields  $\varphi  = (\d  \hat \Phi _ \pm 
^{\mp 1}  , \d \tau ,\  \d  G _\pm  ,\  \d  g_{ab} ) $ in the  first
line entry. We list the leading operators $O  ^\varphi _{b, F , D}
(\nu  ) ,\ [\nu =(jlr)]  $,  
their associated  dimensions $\D _\nu ^\varphi $ and supersymmetry
breaking suppression parameters $ Q _\nu ^\varphi $       in the
second, third and fourth  line entries.  The   radial scaling  laws of
NN  perturbations  are          $ \varphi _\nu \sim c_\nu
^{NN} r ^{\D _\nu ^\varphi - 4}  ,\  [c_\nu =\hat c_\nu ^\varphi  w ^{Q
  _\nu ^\varphi  +\D _\nu ^\varphi  - 4   } ] .$   
The symbols  $ C _x, \ V _x  $  specify whether the operator  is part
of   a chiral  or  vector   supermultiplet  with the lower  suffix   $
x =  (b , \ F ,\ D ) $  indicating whether  it is   a bottom,   $\t ^2 $ or
$  \t ^2 \bar \t ^2 $ component.  The  operator  denoted  $ S
_b (010) $    refers  to    the bottom component of  
the conserved  current   supermultiplet $ Tr  (A A ^ \dagger ) .$  
The $r$-charge in the modes quantum  numbers   $ (jlr) $  refers to   the multiplet $r$-charge     whereas    the    proper charges  for
$ (b , \ F ,\ D ) $ superspace   components are    given by    
$  r ^{(s)} = r, r \pm  1  , r  $.}  
\tiny 
\begin{tabular}{|c|ccccc||c|} \hline 
$ \varphi $ &$ \d \hat \Phi _- ^{(1)} $ &$ \d \tau ^{(1)} $&$ \d G_\mp
 ^{(1)} $&$ \d \hat  g_{ab}^{(1)} $ & $ \d (\hat \Phi _+ ^{-1} )^{(1)}
    $ & $ \d \hat \Phi _-^{(2)} $ \\ \hline 
$ \D _\nu ^ \varphi      $&$\D _\nu ^-=  \D _\nu -4 $&$ \D _\nu (\tau
  ) $&$ \D  _\nu ( G_\mp )  
    $&$ \D _\nu (g _{ab} )     $&$\D _\nu ^+=   \D _\nu +4 $&$ \D
  _{ij} = \d _{\nu _i} + \d     _{\nu _j} $  \\   \hline 
$\varphi  _\nu  (jlr) $ & $ {C _b (\ud  \ud 1),\   S _b
    (\underline{01}0)    \atop  C _b(112) ,\   V _b (110)}   $  & 
${ C _F(000) ,\   C _F(\ud  \ud 1)  \atop   C  _F(112) }   $ & 
$  {C _F (\ud \ud  -1) ,\  C _b(002)   \atop  C  _F (000) ,\ V  _F(110) } $ & 
$ {  C _F(\ud \ud - 1) , \   C _F(00  -  2 ) \atop  
C_F (\ud \ud 1)   ,\ V _D    (112)}  $  &
${  V_D(000)  , V_D (\ud \ud 1) \atop  V _D   (100), V_D (112)}   
 $ &  $ {C_F (\ud \ud 1) ^2 ,\ C_F (002)^2 \atop C_F (\ud \ud 1)
\times  V_F (000)  ,     V_F(112) \times  C_F(\ud \ud 1) } $  \\   \hline 
$ \D _\nu ^\varphi    $ & $ ( {3\over 2} ,2 ,  3 , 3.29) $&$ (4,
{11\over 2} , 7 )$&$ ({5\over 2}, 3, 4,  4.29 ) $&
$ ( {5\over 2}  , 4 , {11\over 2} ,  5.29)$&$( 8, 9.5, 10, 11) $&$ 
 (5,  6,  {13\over 2},  6.79 )  $ \\ \hline 
$ Q _\nu  ^\varphi   $&$ (4 ,4,4,4)
$&$ (0,0,0)  $&$ (2,4,2,2) $&$ (2, 2, 2,0)  $&$ (0,0,0,0) $ & $
(4,4,4,4) $  \\   \hline 
$X_\nu ^\varphi $  &  $ {1.5 ,2, 3, 3.29}  $  & $  {0, 1.5}   $  &  $ { 0.5, 3, 2, 2.29 }  $  &  $ {0.5, 2, 3.5, 1.29}  $  &  $ {4, 5.5 , 6, 7 }  $  &  ${1, 2, 2.5, 2.79}$ \\  \hline 
 \end{tabular} \end{center} \end{table}    \vskip 0.3 cm

We  now  return to    the  discussion  of    compactification effects 
on the  GKP   background,  focusing     on   metastable supersymmetry breaking     vacua     that are lifted  to $dS$   spacetime   
by  introducing a  $p\ \bar D3$-brane    near the conifold tip~\cite{kachpearson01,kklt03}. 
(One  should be aware  that  the  option  that   dS spacetimes are  generic in the landscape  of    flux  vacua has been challenged  in recent
studies~\cite{danielriet18}.
Although     the  classical  description for (smeared  or localized)   
  $ \bar D3$-branes~\cite{borokhov02,dewolfelinde04,browndW09,benini09}
seem  to develop    unphysical  singularities in the 
infrared~\cite{bena11,dymarsky11,bena1215},
and the back-reaction   from  volume  stabilization   
might     choose   $AdS_4$  instead  of $ dS_4$ vacua~\cite{moritz18}, 
the effective field theory construction    was  found  to be   well behaved perturbatively~\cite{polchsb15}.)   Identifying the small positive interbrane   potential   energy    $  \D V ( D3/ p \bar D3)  \simeq    
p \tau _3  \Phi _- =  2 p \tau _3 w
^4   $  on the gravity theory side    with
the    vacuum energy density 
$   [ X^\dagger X  ] _D   \sim     F_X^2  $      on the gauge
theory side,  yields $ F_X / m_s ^2\sim w^2
$~\cite{baumann208}.      The   resulting   
suppression factors    from supersymmetry breaking  of  the various perturbations       
\bea && \d \Phi _-  ^{(1)}   \sim F_X ^\dagger F_X \sim w^4 ,\  \d
\tau ^{(1)} \sim   w^0 ,\  \d g_{ab} ^{(1)}  \sim w^2    ,\
\d G_\pm ^{(1)}  \sim F_X \sim w^2   ,\   (\d \Phi  _+ ^{-1}  ) ^{(1)}  \sim  w^0 
, \  \d \Phi _-  ^{(2)} \sim  w^4   , \eea
correspond to  the      parameter assignments 
$Q _\nu (\d \Phi _- ,\ \d  \tau , \ \d
G_\pm , \ \d  g_{ab}  , \ \d  \Phi _+^{-1} , \d \Phi _-  ^{(2)} )  =
(4, 0, 2, 2  , 0, 4) $,   where we have used for         
$\d \Phi _-^{(2)} :\ Q= Q_{\nu _i}  + Q_{\nu _j}
=4$. 
We list in Table~\ref{tabdef1} the     relevant  and irrelevant
operators  of  lowest dimensions,         dual to   the 
various   supergravity  fields,   based  on~\cite{ceresole99}
and~\cite{aharony05,baumann08,baumann208,baumann10,gandhi11}.  
With the adopted   conventions  for
$\D _\nu ^\varphi $, the radial  scalings for the various  fields
$ \varphi _\nu  (r) = \hat c _\nu w ^{Q_\nu ^\varphi } (r/r_{uv}
)^{\D _\nu ^\varphi  -4} $   yields the
warp  factor dependence of  fields   near the apex,  
$\d  \varphi (r _0) = \hat c _\nu w ^ {X_\nu ^\varphi } ,\
[X_\nu ^\varphi  = Q_\nu ^\varphi  +  \D _\nu ^\varphi - 4].$ 
The  leading  contributions   arise then from   perturbations of  
lowest $ X _\nu  ^\varphi $. 

The selected  operators dual  to   $\d \hat \Phi _-  ^{(1)}  \in V.I $ 
arise    from   bottom components of two relevant chiral  operators  
and   one irrelevant   non-chiral operator contributing wih
suppression     factors   $  X_\nu = [3/2, 2, 3.29] $. 
The  marginal singlet operator    in  $\d \tau  \in V.II
$    with  $ X  ^\tau = 0$    can be gauged away. 
The contributions from $ \d \hat \Phi _+ ^{(1)}\in V.II $ 
are  negligible since the  leading operator from  the top    component   of the  singlet non-chiral  
 supermultiplet  $  Tr (W_\a ^2 \bar W_{\dot   \a} ^2) $  has 
dimension  $ \D  _{(000)} =  6 + \sqrt {H_0 +4} =8 $. 
The  second order  contributions   in $ \d \hat \Phi _- ^{(2)}$  
in  the last column  arise from       $ (ij)$ 
pairs of operators dual  to  3-fluxes~\cite{baumann10}  of     effective 
dimensions $\D  = \D _{ij}  $ in the notations  of
Eq.~(\ref{eqdefphi2}).   These  are
accompanied   by smaller  prefactors involving products of coefficients. 
The selected  operators dual  to $\d G _\pm ^{(1)}   $  belonging to 
$ V.I,  V.IV,  V.III, V.I$     correspond in
the classification of~\cite{baumann10} to  $ O_X ( jlr
) = [O_I (\ud \ud -1) , O_{III} (002 ) , O_{I} (000) , O_{I} (110)
] $.   The operators dual  to   $\d \tilde g_{ab}^{(1)}  $ belonging to 
$V.I,  V.III,  V.IV, V.I$   correspond to the harmonic 
 modes $\phi   $  of dimensions  $ \D   ( \phi ) = [\sqrt {H_0 +4 } ,  
\   -1  + \sqrt    {H_0+4}]$~\cite{ceresole99,ceresoII99}. 
 (The  dimensions  $  5/2, 4$  of  the   first  two 
   operators  dual to $\phi  _\nu $      differ from 
those    of~\cite{gandhi11}     quoted in~\cite{moritz18}    as $\D
= 2, 3 $.)    

We see that the leading contributions with $ X _\nu  = 1/2  $ 
arise  from 3-flux  and metric   deformations  associated 
to  relevant operators while those  with   $ X _\nu  = 3/2 $ 
arise from $\d  \hat \Phi _- ^{(1)}  $ and $\d \tau .$ 
It thus appaears  that once one   accounts   for the  warp  factor
dependence     none of  the  relevant operators induced  by  
compactification effects    can  destabilize    the  
metastable  supersymmetry breaking vacuum. 

\subsubsection{Warped modes interactions from  warp profile deformations}
\label{subsec24.2}

We  wish  to   examine   how   the background deformations  affect    
the   warped modes   interactions.  
For the graviton   modes  $ h^{ (m) } _{\mu
  \nu } $, these  arise in the   reduced  curvature   action  from   
cubic and higher order   local couplings  of  schematic form    
\bea && S ^{(2)}   (h)  \simeq \sum _{n\geq 3} \int d^4 x \sqrt {-\tilde
  g_4} h ^n \dh h \dh h \int d^6y \sqrt {\tilde g_6} e ^{-4 A (r, \T )
} \prod _{i=1} ^n \Phi _{m_i } (r ,\ \T ) , \cr && 
[e ^{-4 A (r, \T )} = e ^{-4 A (r)} (1 +\ud e ^{-4 A (r)} ( \d \Phi _+ + \d
\Phi _- ) )^{-1} = e ^{-4 A (r)} ( 1 +{e ^{-4 A (r)} \over 2 } (- \Phi
_+ ^{2} \d \Phi _+ ^{-1} + \d \Phi _- ) ) ^{-1} ] \label{sub2eq37}\eea
by evaluating the   coupling  constants   from  
overlap  integrals  of wave  functions weighted  by the 
deformed warp  profile     $ e ^{-4 A  (r, \T ) }.$
The   deviations  from the  classical values   $\Phi _-^0 =
0,\ \Phi ^0_+ = 2 e ^{4   A(r)} $ originate  at  leading
order  from the  non-normalizable  harmonic  modes  in 
the homogeneous parts of $\d \Phi _\pm $, 
\bea && 
\d \Phi _ - = \sum _\nu c _\nu ^- ( {r  \over  r_\star } ) ^{\D _\nu -4} Y ^{\nu
} (\T ) ,\ \d \Phi _+
^{-1} = \sum _\nu c _\nu ^+ ({r  \over   r_\star }) ^{\D _\nu -4}Y
^{\nu } (\T ) ,\cr &&   [c _\nu ^\pm \equiv  c _\nu (\Phi _{\pm }  ^{\mp 1} ) 
= \hat c _\nu ^\pm  w ^{Q_\nu ^\pm } ({r_\star \over \calr })
^{\D _\nu ^\pm - 4 } = \hat c _\nu ^\pm w ^{Q_\nu ^\pm + P(\D _\nu
^\pm - 4 ) } ,\  \D _\nu ^\mp = \D _\nu \mp 4 ]    \eea
where the   scaling  dimensions $\D  ^\pm _\nu $  are same    as 
 those of eigenvectors of the $ (\pi ^{(\nu )} , b^{(\nu )} )$ system  in
 Eq.~(\ref{sub2eq352}). The     harmonic   
decomposition  of  the warp profile   
\bea && e ^{-4 A (r, \T )} =e ^{-4 A (r)} (1 +\sum _\nu h
_\nu (r) Y^\nu ( \T ) ) ,\  [h _\nu (r) = 2 \hat c _\nu ^+ w ^{Q_\nu ^+} ({r\over \calr })^{ \D _\nu ^+  - 4 } -\ud
\hat c _\nu ^- w ^{Q_\nu ^-  } ({r\over \calr
})^{ \D _\nu ^- - 4} ]    \label{eqsubII4spurion}  \eea
defines      $ h _\nu (r) $      as   radial wave  functions  
of  harmonic spurion    fields    of   scaling  dimensions 
$\D  ^\pm _\nu $.  Similar  effects  arise from deformations  of
the base manifold breathing    mode  discussed in~\cite{berndsen07,freydacline09}.  For instance,  the
non-normalizable modes in the  harmonic decomposition  of the   breathing
field  perturbations, 
\bea && \D \G (r, \T ) = \sum _\nu \g _\nu (r) Y^\nu (\T )
 = \sum _\nu \hat c_\nu ^\g w ^{Q _\nu ^\g } ({r\over r _{uv}})^{\D
   _\nu ^- -4 } Y^\nu (\T ) ,\  [\D _\nu ^- = 2  
   + \sqrt {4 + M_{5-}^2 } = -2 + \sqrt {4 + H_0 }]   \label{eqbreath}   \eea
include  lowest  dimensional   operators   with $ Q _\nu ^\g   =4 $
that     satisfy      the  scaling laws 
\bea &&   \g _\nu (r)  =  w ^{4 }    [ r ^{-5/2}
  ,\ r^{-2},\  r^{-1},\ r   ^{-0.71} ]   ,\ 
\nu = (jlr) = [(\ud  \ud  \pm 1) ,  \ (100), \  (112), \ (110) ] .\eea 

We see that the   selection rules  imposed  on  the conserved
 charges   of coupled     warped modes    
 are    modified  by  the spurion deformation   field     in  an easily
 identified  way.  Upon inserting  the  harmonic decompositions in
 Eq.~(\ref{eqsubII4spurion})  or (\ref{eqbreath}) for the deformed profiles 
in the   overlap integrals    of  wave functions,  
the disallowed $n$-point couplings  
$ C_{m_1} \cdots  C_{m_n} $  get replaced   by     allowed deformed  $
(n+1)$-point  couplings    $C_{m_1} \cdots  C_{m_n}  h _\nu  $ or 
$C_{m_1} \cdots  C_{m_n}  \g  _\nu  $.  

The background deformations  can also  affect  the   wave functions 
$\Psi _m (r, \T )  $  themselves. Starting from the  definition of 
graviton modes as zero modes  of the    10-d  scalar Laplace  operator
with the deformed warp profile,  while   dropping 
the   mixed derivative  terms in   the radial and angular   variables,  
one obtains the  simplified  formulas  for the deformed  wave equations 
\bea && 0= (\tilde \nabla _6 ^2 + e^{-4 A(r, \T )} \tilde \nabla _4 ^2 ) \Psi _m
(r, \T ) , \ [ e ^{-4 A(r ,\T )} \simeq e ^{-4 A(r )} ( 1 - \ud e ^{-4
    A(r )} \d \Phi _- + 2 e ^{4 A(r )}  \d \Phi _+ ^{-1} ) ,\ e ^{-4 A(r) } = r^{-4}
] \cr && \Longrightarrow \ [\dh _r r^5 \dh _r  +  r^{3} \tilde \nabla _5
  ^2 + r ^{5} e ^{-4 A(r )} (1 - \ud e ^{-4 A(r )} \d \Phi _- + 2 e ^{4 A(r )} \d
  \Phi _+ ^{-1} ) E_m ^2 ] \Psi _m = 0 .   \label{sub2eq40} \eea
The  eigenvalue problem     can be translated to a stationary  Schr\"odinger   equation  with   a  free Hamiltonian $H_0 (r,\T) $  and  a radial and angular  potential $V(r, \T ; E_m )  $, depending  on the mass eigenvalues $ E_m$. Upon 
decomposing $V$ and $\Psi _m $  on harmonics of  the
base manifold,  one obtains the coupled  system of 1-d  radial
equations   
\bea &&( H_0 + V (r, \T ; E_m ) \Psi _m = E_m ^2 r \Psi_m  (r, \T ) ,\ [H_0 =
  - (\dh _r r^5 \dh _r + r^{3} \tilde \nabla _5 ^2 ),\ \cr && V (E_m,
  r, \T ) = r E_m ^2({1\over 2 r^4} \d \Phi _- - 2   r ^4 \d \Phi _+
  ^{-1} )   = r E_m ^2 \sum _\nu ({1 \over 2} c^-_\nu w ^{Q _\nu ^- }
  r ^{\D _\nu ^- - 4}   - 2 c^+_\nu w ^{Q _\nu ^+ }  r ^{\D
    _\nu ^+ -4} ) Y ^{\nu   } (\T ) ] . \eea 
(The   factor $r$ on the right hand side
can  be eventually  absorbed within  wave functions.)   
The solutions  span a  vector  space of wave functions equipped 
with   the scalar product,    $ \int d r r \int d^5 \T \sqrt {\g _5 } \Psi _m
^\dagger \Psi _n = \d _{mn}$.    The  perturbation theory  solution,
providing $\Psi _m $ and   $ E_m$  as   series expansions of  the 
corresponding unperturbed eigenfunctions and eigenvalues $ \Psi _k
^{(0) } (r)  $ and $ E_m^{(0)}$  of $ H_0$, 
is naturally   derived  within     a variational   
approach   where $ E_m^2$  are    
calculated  by solving the corresponding  secular equations~\cite{lowdin51}. 
If   we  neglect   the   potential readjustment    at   
successive  orders  by   setting $V( E_m ^{2} , r , \T )  \to  
V( E_m ^{(0)2}, r , \T ) $,   then  applying the 
familiar procedure  from quantum mechanics~\cite{LLqm}, the  perturbed 
wave function  and    eigenvalue   for the   state  $m$
are evaluated  from    the    power series expansions  
\bea && \Psi _m  (r, \T ) = \sum _k c_k^{(m)} \Psi _k ^{(0)} = \Psi _m^{(0)} +
\Psi _m ^{(1)} + \cdots ,\ [ H_0 \Psi _m ^{(0)} = r E_m^{(0) 2 } \Psi
  _m ^{(0)} ] \cr && E_m ^2 = E_m^{(0) 2 } + E_m^{(1) 2 } +\cdots
,\ c_k^{(m)} = c_k^{m (0)} + c_k^{m  (1)} +\cdots ,\ [c_m^{m (0)}
  =1,\ c_{k\ne m} ^{m (0)} =0 ] . \eea  
The     first order correction  are  given by  
\bea &&   E_m^{(1) 2} = V_{mm} ,\ \ \Psi _m ^{(1)}  (r, \T ) = \sum _{k\ne m}
{V_{k m} \Psi _k ^{(0)} \over E_m^{(0) 2 } - E_k^{(0) 2 } } ,
\  [\Psi ^{(0)} _k (r, \T ) = R_k (r ) Y ^{\nu _k
} (\T ) ,\ R_k (r ) = {r^{-2}  \over N_k }  J _{\nu _k } ({E ^{(0)} _k
 \over r } ) ,\  \cr &&    V_{km} 
 \simeq  E ^{(0)2} _m  \int _{r_{ir} }^ {r_{uv}} dr r \int d^5 \T \sqrt
{\tilde \g _5} \Psi ^{(0) \star } _k  \sum _\nu ( \ud \hat c _\nu ^- w ^{Q_\nu ^-} 
({r\over  \calr })^{ 1+ \D _\nu ^- -4 } - 2 \hat  c _\nu ^+  
w ^{Q_\nu ^+} ({r\over \calr })^{1+ \D _\nu ^+ -4 } ) Y^{\nu } (\T )  
\Psi ^{(0) \star }_m (r, \T )  ] .  \eea

If we substitute the   formula  
for the  radial wave functions    in the tip  approximation, from
Eq.~(\ref{tiprwfeq1})  below, the   wave function admixtures   from
the  deformation  $\d \Phi _- $  can be evaluated  from the approximate formula 
 \bea && \Psi _m^{(1)} (r, \T )  = \sum _{k \ne m , \nu }  {E_m ^{(0)2} \over E_m
   ^{(0)2} - E_k ^{(0)2} } (\ud \hat c_\nu ^ -)  \b _{k, \nu } ^{(m)}  
\Psi _k ^{(0)} (r, \T ),\cr &&     
[\b _{k, \nu } ^{(m)} =  w ^{Q_\nu ^ - }   A _{k,m } ^{\nu}  \int dr r ^{1 +\D
    _\nu ^ -  -4} R_k (r) R_m (r)  \simeq   {w ^{\a ^-_\nu }  
q_k q_m A_{k,m } ^{\nu} 
\over \vert  2 + \nu ' _k +\nu '_m - \a _\nu \vert  },\cr &&   
q_{m_i} = (x_{m_i} /  2)^{\nu _i}  / (\G (1 + \nu _i  )
J_{\nu _i} (x_{m_i}  )  ( 1 + (4 - \nu _i ^2 ) /  x_{m_i} ^2  )^{1/2}
)   ,\cr &&  \a _\nu ^- = Q_\nu ^ - + \D _\nu ^ - -4  ,\ \nu ' _m = (4 + M_{5}
  ^{\nu _m 2 } )^{1/2} = \D _\nu ^- -2   ,\ 
A_{k,m } ^{\nu}    = \int d ^5 \T \sqrt {\tilde \g _5 } Y^{\nu
  _k  \star } Y^{\nu }  Y^{\nu _m }   (\T )  ]  \eea 
where  $ \b  _{k, \nu } ^{(m)}  $   stand  for the mixing
coefficients due to the warp profile  deformations.    
To make contact with  the  simple  order of magnitude   formula  
for the admixed wave   function proposed  in~\cite{kofman08}, one  could also consider  the    simple   estimate    for the  off-diagonal
matrix  element  of the potential,  $ V_{km} \sim   \hat c_{\nu _k}^- 
w ^{1 +Q_{\nu _k} ^- + \D _{\nu _k}^- -4 }  , \ [Q_{\nu _k} ^- =4  ]$
yielding a  perturbed wave function of  approximate  form   
   \bea && \Psi _m  ^{(1)} (r, \T ) \sim    
{V_{km} \over \sqrt {\l _2 } }\Psi _k  ^{(0)} (r, \T )
 \sim     {2 (2\pi \eta )^3 \over M_\star } 
\hat c _{\nu _k}  ^- w ^{ 1 + \a _{\nu _k}  ^-}  \Psi _k  ^{(0)} (r, \T )
,\  [\a _{\nu _k} ^- = \D _{{\nu _k}  }   ^- 
= -2 + \sqrt {H_0^{\nu _k}   +4} ]  . \eea
The  present description     of   deformation   effects  due to 
wave  function admixtures  is  analogous   to the familiar 
  formalism  used to describe    mixing effects  in field theories. 
Given some initially disallowed   coupling $ C_m C_n C_p $, 
then   inserting   appropriate   wave function admixtures   for  any
one of  the   three   modes, for instance,  $ C_m \to   C_k $, 
replaces the  coupling  by   an allowed one 
$ C_k C_n C_p $  with    finite    angular  overlap
integral.  The  spurion field  formalism 
discussed near   Eq.~(\ref{eqsubII4spurion})  can be  derived 
from the   present  wave  function  mixing  formalism 
by   inserting  inside the   overlap integral
the  completeness     formula  for the   basis
of unperturbed  wave functions. 

\section{Linearized field  equations in $ AdS _5$ spacetime}
\label{sec3}

We  discuss in this section the  Kaluza-Klein  theory for 
 type $ II\ b $ supergravity     on the  warped   spacetime   
$ M_4 \times  \calc _6 $, restricted to   the large radial distance 
region    where   $ \calc _6 $  asymptotes the undeformed conifold.   
The  reduced theory     lives  in   
a slice of $AdS _5 \times  T^{1,1} $ spacetime  
truncated  at the infrared  and ultraviolet  ends along 
the radial direction  of $AdS _5 $  by  hard walls. 
We  consider  the reduced  kinetic energy 
actions  for   10-d scalar,  vector and 2-tensor fields fluctuations
decomposed on   harmonics  of  $ T^{1,1}$.  
The   linearized  wave equations   for  the towers of   descendant     fields 
in $ AdS _5$ spacetime   are   derived as  an initial   step towards
the goal of determining  the  mass  spectra    and   interactions of
normal  modes  propagating in the  warped throat bulk.

\subsection{Wave equations for gravitons  and scalar   modes}
\label{sub3sec1}

The scalar and graviton fields   fluctuations $ \d \phi
(X),\ h _{\mu \nu } (X ) $  in   $ AdS _5 \times X_\d $ 
arise  as zero modes  of the 10-d scalar Laplace 
wave operator $ \nabla ^2 _{10} $. One
could  account  for massive string excitations    by adding   a mass
term   $  \mu ^2$.   The background can be equivalently  described  as
the  warped   spacetime $ M_4 \times R^1_r \times X_\d ,\ [M_4  =
\dh (AdS _5 )]$   parameterized by the  spacetime, radial and angular
coordinates  $ X ^M \to (x ^\mu , \ r , \ \T ^a) $  for  a metric
of  form \bea && ds ^2 _{10}= h ^{-1/2} (r) d \tilde s ^2 _4 + h
^{1/2} (r) (\tilde g_{rr} (r) d r ^2 + \tilde g_{ab} (r) d \T ^a d \T
^b ) ,\ [a =1,\ \cdots , \d ]. \eea 
The    factorized  dependence on the variables  $ x, \ r ,\ \T $  
is  reflected   by the separable structure  of the  scalar Laplace 
wave operator   
\bea && \nabla ^2 _{10} =
h^{1/2} (r) \tilde \nabla ^2_{4} + h^{3 -\d \over 4} (r) \tilde \nabla
^2_{6} ,  \  [\tilde \nabla ^2_{6} \equiv {1 \over \sqrt {
      \tilde g _{10} } } \dh _m \tilde g ^{mn} \sqrt{\tilde g _{10} }
  \dh _n = \tilde g^{r r } h^{\d -5 \over 4} (r) {1\over G(r )} \dh _r
  G ( r ) \dh _r + r^{-2} h^{\d -5 \over 4} (r) \tilde \nabla ^2_{5}
  ,\cr && \tilde \nabla ^2_{5}= {1 \over \sqrt {\tilde \g _5} } \dh _a
  \tilde \g ^{ab} \sqrt {\tilde \g _5} \dh _b ,\ G(r) = \sqrt {g
    _{10} } g ^{rr}=\sqrt {\tilde g _{10} }\tilde g ^{rr} h^{\d -5
    \over 4} (r) = \sqrt {\tilde g _{4} \over \tilde g^{r r } } h^{\d
    -5 \over 4}  r ^\d \sqrt{\tilde \g _{5} },\ 
h^{-1/2} (r) = e ^{2 A (r)} = ({r \over \calr } )^2 ] .\eea 
The Laplacian in the    $ AdS_5 \times X_5 $ spacetime metric,   where
  $ \tilde g_{ab} (r) = r^2 \tilde \g _{ab},\ 
\tilde g_{rr} (r)=1,\ \d =5 ,\ G(r) = r ^5$,    simplifies to 
\bea && \nabla ^2 _{10} = 
h^{1/2} (r) \tilde \nabla ^2_{4} +h^{-1/2} (r) ({ \tilde g^{r
    r }\over G(r )} \dh _r G ( r ) \dh _r + r^{-2} \tilde \nabla
^2_{5} )   =  h^{1/2} (r) \tilde \nabla ^2_{4} + h^{-1/2}
(r) ( r^{5} \dh _r r^{-5} \dh _r + r^{-2} \tilde \nabla ^2_{5} ). \eea 
The decomposition of scalar and    metric  tensor fields
on orthonormal bases  of  harmonic  functions  in  $X_5 = T^{1,1}$     
and plane waves in  $M_4$ is expressed  by  the  
infinite summations  over coefficient fields in $M_4$   times radial
and angular  wave functions  
\bea && \d \phi ( X ) = \sum _m \phi ^{(m)} (x) \Psi _m (r , \T ) ,\ h
_{\mu \nu } (X ) = \sum _m h _{\mu \nu } ^{(m)} (x) \Psi _m (r , \T )
, \cr && [\phi ^{(m)} (x) = \phi ^{(m)} (k _m) e ^{i k_m \cdot x} ,\ h
  _{\mu \nu } ^{(m)} (x) = h _{\mu \nu } ^{(m)} (k_m) e ^{i k_m \cdot
    x} ,\ \Psi _m (r, \T )= R_m (r) \Phi _m (\T )  =R_m (r)  Y^{\nu ,
    M} (\T ) ]   \label{sub3eq1} \eea 
where  the     angular  wave functions  are 
the scalar harmonics $  Y^{\nu , M} (\T )$,    eigenfunctions of the
base  manifold Laplace operator $ -\tilde \nabla ^2 _{X_5} \to M_5 ^{\nu 2}  $.  (For a
system of coupled scalar fields $ \d \phi _i ( X ) $ with a mass
matrix $ M^2_{5} $, one must perform  the   prior diagonalization of
$ M^2_{5} $  on   the  basis   of eigenvector fields.)
The   radial wave equations for modes of 10-, 5- and 4-d  squared mass
parameters,  $\mu ^2 ,\  M_{5}^{\nu 2}  ,\  \tilde \nabla _4 ^2 \to
E_m ^2$,    are solutions of   Sturm-Liouville eigenvalue problems 
equipped with an Hermitean scalar product, 
\bea &&( {1\over \sqrt {g_r} } \dh _r h ^{-1/2}\sqrt {g_r} \dh _r
- M_5 ^{\nu 2}  h ^{-1/2} r ^{-2} + h ^{1/2} E_m ^2 -\mu ^2 ) R_m (r) =0 ,
\ \ [\sqrt {g_r} = h ^{\d -3 \over 4} r ^\d ] \cr && \Longrightarrow
\ (e^{2A} r^{-5} \dh _r r^5 \dh _r + e^{-2A} E_m ^2 - e^{2A} r^{-2}
M_5 ^{\nu 2}  -\mu ^2 ) R_{m} (r) = 0,\cr && [\int _{r_0} ^{r_{uv}} dr e^{-4A} r^{5}
  R_ {n } R_ {n '} = \d _{nn'} ,\ \int d^5 \T \sqrt {\tilde \g _5 }
  \Phi ^\dagger _m(\T ) \Phi _{m'} (\T ) = \d _{m m'}] .  \label{sub3eq2} \eea 
 The boundary conditions at the     points $ r_0,\ r _{uv}$    
define a vector space of solutions with an
orthonormal basis of radial functions $ R_m (r) $ for modes 
of quantized mass eigenvalues $ E_m $.  For a  warp profile  in 
the  near horizon   region, $ h (r) = e ^{-4 A(r)} \to r^{-4}
$, the radial wave functions  are   given by  linear combinations of
Bessel functions \bea && R _m (r) = {r^{-2} \over N_m} (J_\nu ({ E _m
  \over r}) + b_{\nu , m} Y_\nu ({ E _m \over r}) ) Y ^\nu (\T )
,\ [\nu = (4 + M ^2_{5 } + \mu ^2 )^{1/2} ] . \label{sub2eq23p} \eea 
A similar  construction   applies   to  the   other field  components of the
supergravity multiplet.  For the axio-dilaton field   case, which
was discussed previously     in~\cite{frey06}, the relevant terms in
the   10-d   action   of Eq.~(\ref{sub2eq16}), 
\bea && S ^{(2)}  \equiv \ud \d ^{2} S _{10}  (\tau ) = {1\over 2 \kappa ^2 }
\int d ^ { 10 } X \sqrt {\tilde g _{ 10 } } e ^{-2A} [ {e ^{-2A} \over
    4 \tau _2 ^2 } \d _K \bar \tau ( e ^{-2A } \tilde \nabla ^2 _4 + e
  ^{2A } \tilde \nabla ^2 _6 - \mu ^ 2_{flux} ) \d _K \tau - {1\over
    24} e ^{4A} \hat \d  \hat \d ({G \tilde \cdot \bar G \over
    \tau _2 }) \cr && - e ^{4A } {1\over 8} \d \tilde g ^{m n} \hat \d ( { 1 \over \tau _2 } ( G_{npq} \bar G _m ^{\ \ \tilde p
    \tilde q } - {1\over 6} \tilde g _{mn} G \tilde \cdot \bar G ) ) ]
, \cr && [\mu ^ 2_{flux} = e ^{-2 A} {4\tau _2 ^2 \over 24 } \dh _\tau
  \dh _{\bar \tau } ({ G \cdot \bar G \over \tau _2 }) = { g_s e ^{4A}
    \over 12 } G \tilde \cdot \bar G = { g_s e ^{4A} \over 12 } (F
  \tilde \cdot F + {1 \over g_s^2} H \tilde \cdot H ) ] \eea 
yield       the      explicit  formula  for the reduced action 
\bea && S ^{(2)} =
{1\over 2 \kappa ^2 } \int d ^ { 10 } X \sqrt {\tilde g _{ 10 } } [
  {g_s ^2 \over 4} (\d C_0 + {i\over g_s} \d \phi ) ( e ^{-4A} \tilde
  \nabla ^2_4 + {1 \over G( r ) } \dh _ r G ( r ) \dh _ r + e ^{-2A}
  \tilde \nabla _5 ^2 - \mu ^2 _{flux} ) (\d C_0 - {i\over g_s} \d
  \phi ) \cr && + {g_s \over 12} e ^{4A} ( \d C_0 ( H \tilde \cdot \d
  F + F\tilde \cdot \d H) + \d \phi (- F \tilde \cdot \d F + {1 \over
    g_s ^2} H\tilde \cdot \d H) ) - {g_s \over 8 } e ^{4A}\d \tilde g
  ^{mn} (- 2 H _{m}\tilde \cdot F _{n} \d C_0 + (F _{m}\tilde \cdot F
  _{n} - {1\over g_s ^2 } H _{m}\tilde \cdot H _{n} ) \d \phi ) ] ,\cr
&& [H \tilde \cdot F = H _{m pq} F ^{\tilde m \tilde p\tilde q}
     ,\ F_m\tilde \cdot H_n = F _{m \tilde p\tilde q} H _n^{\ \ \tilde
       p\tilde q} ]  \label{sub2eqD16} \eea 
where $\mu ^ 2_{flux} $ denotes the mass term   from  3-fluxes, 
the other  scalar products 
$F\cdot F ,\ H\cdot H , \ F_m\cdot F_n ,\ H_m\cdot H _n $  and $ \hat
\d $ are     defined in a similar  way, and  the  functional
differential  $\hat  \d $  on    field variables   refer to $ \d
_{\tau } + \d_{G}    $ and $ \d _{\bar \tau } + \d _{\bar G}   $.        The  resulting wave equations for the   system of  pseudoscalar and scalar 
fields coupled to  components  of the   3-form
and   metric fields   along the internal manifold   are expressed  in  a  condensed matrix notation  as 
\bea && \cald  { g_s ^2 \d
  C_0 \choose \d \phi } + {g_s \over 6} e ^{4 A} {H \tilde \cdot \d F
  + F \tilde \cdot \d H \choose - F \tilde \cdot \d F + H \tilde \cdot
  \d H / g_s^2 } + {g_s \over 4} e ^{4 A} \d \tilde g ^{mn} { H_m
  \tilde \cdot F_n+ H_n \tilde \cdot F_m \choose - F_m \tilde \cdot
  F_n + {H_m \tilde\cdot H_n / g_s^2}} =0,\cr && [\cald = e ^{-4 A}
  E_m ^2 + r^{-5} \dh _r r^5 \dh _r + e ^{-2 A} \tilde \nabla ^2 _5 -
  {g_s \over 12} e^{4 A} (F \tilde \cdot F + {1\over g_s^2} H \tilde
  \cdot H ) ] \eea 
where the operator   $\cald $   includes  the   diagonal kinetic terms and mass
terms  from geometry and 3-fluxes.    The mixing   with the  3-form field
components $\d F_{r\t \phi } ,\ \d H_{r\t \phi } $    was found
in~\cite{frey06} to have a neglible effect on the axio-dilaton mass spectra. 
We  shall   also ignore  the mixing  with $\d \tilde g _{mn} $ 
and   simplify  the diagonal wave equations for $ \d
C_0 ,\ \d \phi $     by   parameterizing  the contribution   from
3-fluxes in  $\cald $     in terms of   a  constant mass
parameter,   based on the  approximation   discussed near Eq.~(\ref{sub2eq50}),   
\bea && m_f ^2 \equiv r^2 \mu ^2 _{flux} = {g_s  \over 12}
G  \cdot \bar G =  {g_s \over
  12}  e ^{6 A} G _{mnp} \bar G ^{\tilde m\tilde n\tilde p} 
={g_s \over 4} ({\a ' \over \calr ^3 } )^2   (f_r ^2 + f_a ^2 ) ,\ 
\cr &&  [f _r ^2= {6^2 \over 2} r ^2 \vert G_{rab} /\a '
\vert ^2 ,\   f _a ^2= {6^2  9 \over  3 !} \vert G_{abc} /\a ' \vert ^2 
,\    {\calr ^2 \over  \a' } =   ( 27 \pi  g_s N  /4 ) ^{1/2}  ]
  .  \eea
The   classical  profiles for  the 3-forms  in Eq.~(\ref{sub2eq22})
contribute a flux mass  parameter of    approximate  form, 
$ m_f ^2\simeq {81 M^2  g_s  / (  4 (\calr ^2 / \a'  )^3  ) } $,  
behaving parametrically   as  $ m_f ^2 \sim  M^2 /\sqrt { g_s N^3 }   . $
The      harmonic modes   of the  axio-dilaton  satisfy the
simplified wave equation   
 \bea && 0= [E_m^2
  e ^{-4 A} + r ^{-5} \dh _r r^5 \dh _r - M_5 ^2 e ^{-2 A} - { g_s
    \over 12} e ^{4 A} ( F _{m pq} F ^{ \tilde m \tilde p \tilde q} +
  {1 \over g_s ^2 } H _{m pq} H ^{ \tilde m \tilde p \tilde q} ) ] t
_m (r) , \cr && \Longrightarrow \ 0=[\dh _r ^2 + {5\over r} \dh _r - (
  { M_5 ^2 } + m_{f} ^2 ) ({\calr \over r })^2 + E_m ^2 ({\calr \over
    r })^4 ] t _m (r ) ,\ [\d \tau (X) =\sum _{\nu , m} t _m (r )
  Y^{\nu } (\T ) e ^{i k_m\cdot x} ]   \label{sub2eq19} \eea 
where  we have    reinstated  the dependence on $\calr $. 
The orthonormal basis of wave functions is   expressed
in terms of Bessel functions as 
\bea && t_m (r)= { r^{-2} \over N_n} (J_\nu (\rho
) + b _{\nu ,m}  Y_\nu (\rho ) ) ,\ [\nu = (4 + (M_5 ^2 + m_{f} ^2)\calr ^2  )^{1/2}
  ,\ \rho = {E_m \calr ^2\over r} ] . \label{sub2eq19p} \eea 
We see that in the present approximate description, the   constant  flux mass
parameter    $m_f$   simply  adds up  to   
the geometrical mass $M_5$.   However, the   
classical background deformations  are expected to produce a 
flux  mass  of non-trivial radial profile     peaked  
at the   bulk-throat  interface.   The proper  description of these
effects    would require  going beyond  the   hard wall approximation
by keeping  track of the   interplay between the warp profile   and
the volume modulus    in  the  relevant combination $  ( c + e ^{-4 A
} )  $~\cite{frey06}.     

We consider next the  system  of coupled
pairs of  modes $ (\pi ^{(\nu)}  _m (x, r) , \ b^{(\nu)} _m
(x, r) ) $  descending from   the  scalar fields $( \pi , \ b )  \in  
(\d \tilde g ^{a } _{a} ,\ \d C_{abcd} )$.  
The kinetic energy term  $ \vert F_5 \vert ^2 $   contributes 
diagonal and non-diagonal   terms  $\bar b b ,\ \bar \pi b $ to the 
reduced action of   schematic   form  $S ^{(2)} \sim
H_0^\nu (\bar Y^\nu Y^\nu ) [ H_0^\nu \bar b ^{(\nu)} b ^{(\nu)}  -
  {4\over 5}   \bar b ^{(\nu)} \pi ^{(\nu)}  ] $ where we have set  
$ \nabla ^2 _5 = D^e D_e = - H_0 $.
These contributions are    incorporated in  the  lower entries  of 
the mass matrix $ M_5 ^2$  in Eq.~(\ref{sub2eq27}) in which  
the $\bar \pi \pi ,\ \bar \pi b $ terms in the upper entries  of  $
M_5^2$  arise  from  gravitational   terms. 
(In  the more rigorous  analysis    developed in~\cite{kiroman84}, one would  consider   instead        
the   variation of the gravitational equation $ R_{MN} = -{1\over
  6}F_{MPQRS} F_{N} ^{\ \ PQRS}$,     subject to    the 
self-duality condition $F_5 = \star _{10} F _5 $.) 
  The  linearized coupled equations for the 
harmonic   modes  $ (b ^{(\nu)}  ,\ H ^{(\nu)}  _{\hat \mu \hat \mu },\ \pi
^{(\nu)} )$    of $\d C _{abcd} $ and the trace fields $ \d g ^{\hat
  \mu } _{\hat \mu } ,\ \d g ^{a}_{a} $ simplify,  
upon imposing the algebraic conditions 
$ H ^{(\nu)} _{\mu    \mu } \propto  \pi ^{(\nu)}  $, 
to   coupled  pairs of equations for $ ( \pi
^{(\nu)} ,\ b ^{(\nu)}) $ involving the geometrical mass mixing matrix
$ M_5 ^2$ in Eq.~(\ref{sub2eq28p}).  Returning to 
the above  procedure,  we    ignore  momentarily the
mixing terms  and   find the    reduced action for $ b^{(\nu)} _m (x, r)$, 
\bea && S ^{(2)} = { g_s ^2 M_\star ^2 \over 8 V_W } \int d^4 x
\sqrt {\vert \tilde g_4 \vert } \int dr \int d^5 \T \sqrt { \g _5} \ b
^{ (\nu )  \star } _m ( x, r ) [ h ^{-1} r ^ \d Q (r) \tilde \nabla _4 ^2
  \cr && + \dh _r h^{-2} r ^ \d Q (r) \dh _r - h^{-2}r ^ \d P (r) M^2
  _{5,\nu } ] b _m^{(\nu)} (x, r ) Y ^\nu (\cald ^2 )Y ^\nu ,\cr && [M^2
  _{5,\nu } = - \cald ^2 \nabla _5 ^2 = - \cald ^2 {1\over \sqrt
    {\tilde \g _5}} \dh _a \sqrt {\tilde \g _5} \tilde \g ^{ab} \dh _b
  \to H_0 ,\cr && Q(r) = K^{aa} \times \cdots \times K^{dd} = r^{-5}
  ,\ P(r) = K^{aa}\times \cdots \times K^{ee} = r^{-10},\ \d = dim
  (T^{1,1} ) =5] \label{sub3eq3} \eea  
one  finds  a radial wave equation for $ b _m ( r ) $  
of    same   form  as  that of scalar fields,  
 \bea && 0= ( h^{-1} r ^{\d -8} E_m ^2 +\dh _r h^{-2} r
^{\d -8} \dh _r - h^{-2} r ^{\d -10} M_{5} ^2 ) b_m (r) \cr && =
h^{-2} r ^{\d -8} ( h E_m ^2 + h^2 r ^{8-\d } \dh _r h^{-2} r ^{\d -8}
\dh _r - r ^{-2} M_{5} ^2 ) b_m (r) =(r E_m ^2 + \dh _r r ^5 \dh _r -
r ^3 M_{5} ^2 ) b_m (r) . \label{sub3eq3p1} \eea 
For  convenience, we  have absorbed  the  charge and four-momentum 
labels $\nu  ,\ k_m $   within the   single     label $m$.   
If one tentatively   assumes  that the kinetic energy
term  for  $\pi _m^{(\nu)} (\hat x) $  has   same   form  
as  that  of $b_m^{(\nu)} (\hat x) $, then   the geometrical mixing in 
the mass matrix $M_{5} ^{\nu 2}  $ in Eq.~(\ref{sub2eq28p})  is 
taken into account   without   more labour by  going directly 
to    the eigenvector  fields $ S ^{(\nu )} _\pm (\hat x)$ in Eq.~(\ref{sub2eq28}).  The
 orthonormal bases of radial wave functions $ R _{m  \pm
} (r) $ are  again  given  (for $\d =5$) 
by  linear combinations of Bessel 
functions \bea && 0= ( \dh
_r r ^{\d -8} h ^{-2} \dh _r + E_{m \pm } ^2 h^{-1} r ^{\d -8} - M_{5
  \pm } ^2 h ^{-2} r ^{\d -10} ) R _{m \pm } (r) = (\dh _r r ^5 h
^{-2} \dh _r + E_m^2 r - M_{5 \pm } ^2 r^3 ) R _{m \pm } (r) \cr && [R
  _{m \pm } (r) = { r^{-2} \over N_{m \pm } } (J_{\nu _\pm } (\rho
  _\pm ) + b_{m \pm } Y_{\nu _\pm } (\rho _\pm ) ) ,\ \nu _\pm = \sqrt
  {4 + M_{5 \pm } ^2 } , \ \rho _\pm = {E_{m \pm } \over r} ,\ \int dr
  r R _{m \pm } (r) R _{n \pm } (r) =\d _{nm} ]. \label{sub3eq3p2}
\eea 
We now evaluate the effective action for the modes  $\pi ^{(\nu)} \sim
\g ^{(\nu)}  $       by   invoking the reduced action $ S_\G $  in
Eq.~(\ref{sub2eq51})  expanded in  powers of 
$\d \G = \g (x, r, \T )  \propto h^a _a = \pi(x, r, \T ) $,
\bea && S _\G (\g ) =
{1\over 2\kappa ^2 } \int d^4 x \sqrt {-\tilde g_4} \int d r \int d^5
\T \sqrt {\g_5}[5 r^5 (e ^{-4A}  +  c(x))\g \tilde \nabla ^2_4 \g + {1\over 4}
  r^3 h (8 M_5^2 \g - 60 r \g ' - 10 r^2 \g '' ) \cr && + ( r \tilde R
  ^{(4)} + r^3 (-20 + \tilde R ^{(5)})) + (5 r \tilde R ^{(4)} + r^3
  (-100 - 8 \tilde \nabla _5 ^2 + 3 \tilde R ^{(5)} ) - 10 \dh _r r^5
  \dh _r ) \g \cr && - {r \over 2} ( (-25 \tilde R ^{(4)} + r^2 (500 +
  24 \tilde \nabla _5 ^2 - 9 \tilde R ^{(5)} )) \g ^2 + 60 r^4 \g ^{ '
    2 } + 100 r^{-1} \g \dh _r r^5 \dh _r \g ) \cr && - {r \over 6} \g
  ( (-125 \tilde R ^{(4)} + r^2 (2500 + 216 {(\tilde \nabla _5 \g ) ^2
    \over \g ^2 } - 27 \tilde R ^{(5)}) ) \g ^2 + 900 r^4 \g ^{'2} +
  750 \ r^{-1} \g \dh _r r^5 \dh _r \g ) \cr && +{g_s^2 \a ^{'2} 
    \over 2 \calr ^6 } r^3 (-1 + c r ^4 ) [-(f_a^2 - f_r^2) \g + \ud
    (f_a^2 + f_r^2) \g ^2 - {1\over 6} (f_a^2 - f_r^2) \g ^3 ] ],   \label{sub2eq24}  \eea  
where   we omitted boundary terms   arising  via integration by
parts. The first line entry in the formula above 
includes mass mixing terms   for $\g 
\propto h^a _a ,\ h = h ^\mu _\mu $.  There  also occur constant (vacuum)
and linear (tadpole) order terms in $ \g (r) $ that   we shall ignore,
since our interest  is on the  
quadratic   and higher order  terms in $ \g (r) $. Decomposing $\g (r,
\T ) $ on  harmonic modes $\g ^{ (\nu , m ) } (x) $ with 
radial wave functions $\g  _m (r) $  
and substituting $ \tilde \nabla _5 ^2 \to - M_5 ^2 $
and $ \tilde \nabla _4 ^2 \to E_m ^2 $,  yields the 
radial  wave  equations    
\bea && [ 5 r E_m ^2 +
  20 \dh _r r ^5 \dh _r - r^3 ( {9\over 2} \tilde R ^{(5)} + {25\over
    2 r^4 } \tilde R ^{(4)} + 125 (r -3 ) + 12 M_5^2 + m_f ^2 (r) ) ]
\g _{m} (r) =0 , \cr && [m_f ^2 (r) = {g_s^2 \a ^{'2} 
    \over 4  \calr ^6 } (f_r ^2 +f_a ^2) ,\ \d \G (x, r ,\T ) = \sum
  _{\nu ,m} \g ^{ (\nu , m ) } (x) \g  _m (r) Y ^\nu (\T )
] \label{sub2eq21} \eea  
where the warp profile was evaluated   in the  near horizon limit,  $ e ^{A}
\to r $.   While the   kinetic term     has the standard  canonical   form, the
other   terms   in the resulting equation, 
$ (\dh _r r^5 \dh _r + {1\over 4} r E_m ^2 - {3\over 5}  r^3 M_5 ^2
+\cdots ) \g _m (r) =0 $,  
arise with numerical coefficients differing  from 
those expected for scalar fields, 
 $ (\dh _r r^5 \dh _r + r E_m ^2 - r^3 M_5 ^2 ) R_m (r)
=0 $. This  can be explained, however,    by  mixing 
effects  in the $(\pi _m \propto \g _m , b _m ) $  
system that we have not taken  into account. 
One could  introduce  minimally    the   mixing  effect   between
scalar harmonic   modes subsumed in the $2\times 2$ mass matrix     
$ M_{5,\nu } ^2$, 
using the same prescription as in  Eq.~(\ref{sub3eq3p2}) to   
transform   to      decoupled equations   for the eigenmodes $\g ^\pm
 $   of  mass eigenvalues  $ M_{5,\pm } ^2$. Note that the radial
scaling  laws  for normalizable  solutions of the    mixed  modes of
lowest  mass eigenvalue  $ M_{5 - } ^{\nu 2 } $  read 
$  \g ^- _m (r)\simeq  q _m  r ^{- \D  ^ - _\nu } Y
^\nu (\T )  ,\   [\D  ^ - _\nu  = 2 +\nu = \D   _\nu  - 4 ].$

\subsection{Wave equations of  1- and 2-form modes}
\label{sub3sec2}

The   reduced  action  of 10-d supergravity 
in $ AdS _5 $ spacetime  gives rise  to
several towers of massive  1-form fields  $ A_{\hat \mu } $ and
2-form fields $ B_{\hat \mu \hat \nu } $. 
The wave equations  are   of diagonal or non-diagonal types depending on the
structure of geometrical  and  topological  mass terms.  We     shall  restrict  the discussion 
to single 1- and 2-form    fields $ A_{M} $ and $ B_{MN} $  
in the  spacetime  $ AdS _5  \times T^{1,1}$. 
Earlier studies for the spacetime $ AdS_5 \times S^\d $ were presented
in~\cite{davoud02,guirkshiuzurek07,shiuetal07} and~\cite{berndsen07}.
We  start with    the case of  a single vector field $A_{ M } (X) $ 
of squared mass $ \mu ^2 $  and  compute the  
reduced action for the descendent vector  fields $A _{\hat \mu } (\hat  
x) $  through the following steps, 
\bea &&  S (A_M) = - {1 \over 8 \kappa ^2} \int d ^{10} X \sqrt { -g _{10}
} ( {1 \over 2} F_{MN} F^{MN } + {\mu ^2} A _M A ^{M} ) , \cr && =
     {1 \over 4 \kappa ^2}  
\int d ^{10} X \sqrt {- g_{10} } [-{1\over 4} F _{\hat \mu \hat \nu }
  F ^{\hat \mu \hat \nu } -{1\over 2 } F _{\hat \mu a } F ^{\hat \mu a
  } - {1\over 4} F _{ab } F ^{ab } - {\mu ^2\over 2} ( A _ {\hat \mu }
A ^{\hat \mu } + A _ {a } A ^{a } ) ] \cr &&
= {1 \over 4 \kappa ^2}  \int d ^{4} x \sqrt {\tilde g_4 } \int d^5 \T \sqrt {\tilde \g_5 }
dr h ^{\d -3 
  \over 4} r^\d [ - {1 \over 4 } (h F _{\mu \nu } F ^{\tilde \mu
    \tilde \nu } + 2F  
  _{\mu r } F ^{ \tilde \mu r } + 2 r ^{-2} F _{\mu a} F ^{ \tilde \mu a} \cr && + 2
  h^{-1} r ^{-2} F _{r a} F ^{ r a} + h^{-1} r ^{-4} F _{ab} F ^{ a b}
  ) - {\mu ^2\over 2} ( h ^{1/2} A_{\mu } A ^{ \mu } + h ^{-1/2} A_{r
  } A ^{ r} + h^{-1/2} r ^{-2} A_{a} A^{a} ) ] .  \label{sub3eq2p1} \eea 
For a  10-d massless vector field  of $\mu =0 $, one  can specialize
to the unitary gauge,  $ A_r=0 , \ \dh ^\mu A_\mu =0 $.  The
decomposition on harmonic modes in $ T^{1,1}$ and plane wave modes in
$M_4$,  of 5-d and 4-d squared masses $ M_5^2 $ and $E_m^2$,   introduces radial
wave functions   satisfying second-order differential equations in $r$.
The   solutions    for $  A _\mu (x)$ span   a vector space 
generated  by  bases  of orthonormal wave functions,    $ a_m (r) $, 
related   to    Bessel    functions,  
\bea && 0= ( {1\over \sqrt {g_r} } \dh
_r \sqrt {g_r} \dh _r -\mu ^2 h ^{1/2} - M_5 ^2 r ^{-2} + E_m ^2 h (r)
) a_n (r) , \ [A_\mu (X) = \sum _m A_\mu ^{(m) } (x) a_m (r) \Phi _m
  (\T ) ,\ \sqrt {g_r} = h ^{\d -3 \over 4} r^\d ] \cr &&
\Longrightarrow \
0 = (e ^{-6 A} r^5 E_ m^2 + \dh _r r^5 e ^{-2A} \dh _r - M_5 ^2 r^3 e
^{-2 A} - \mu ^2 r^5 e ^{-4 A} ) a_m (r) = (E_ m^2 r ^{-1} + \dh _r
r^3 \dh _r - (M_5 ^2 + \mu ^2 ) r ) a_m (r) , \cr && [a_m (r)= {r^{-1}
    \over N_m} (J_\nu ({E_m \over r} ) + b_{\nu ,m}  Y_\nu ({E_m \over r} ) )
  ,\ \nu = (1 +M _5^2 +\mu ^2 )^{1/2} ,\  \int dr e ^{-6 A} r^5 a_m ^2 (r)
  = \int dr r^{-1} a_m ^2 (r) = 1 ] . \label{sub3eq2p2} \eea 
(We have reinstated $\mu ^2 $ only to indicate that this  amounts to 
an additive  shift     to   $
M_5 ^2$.)   The above result  applies  without difficulty to the coupled 
system $ (B_{\hat \nu } ,\ \phi _{\hat \nu } ) $ in
Eq.~(\ref{sub2eq30})  by including  the  mass mixing terms  within  
a  $2\times 2$ mass matrix $ M_5^2 $.  The   mass matrix  diagonalization leads
to pairs of eigenvector field modes $ V_{\hat \mu }^\pm $ of mass eigenvalues $
M_{5 \pm }^2 $  given in Eq.~(\ref{sub2eq32}).

We   focus  next    on     the    case of 
a   single 2-form field $ B_{MN} (X)$    in $ AdS_5  \times T^{1,1}$.
The reduced    action   expands to  the form
\bea && S ( B_{MN}) = - {1\over  4  \kappa ^2 }  \int d ^{10} X \sqrt { - g_{10}
} ( {1\over \ 3!  } H _{MNR } H ^{MNR } + {\mu ^2\over 2} B _{MN }
B^{MN} ) \cr && = - {1\over  4  \kappa ^2 } 
\int d^4 x \sqrt {-\tilde g_4 } \int dr r ^\d d
^5 \T \sqrt {-\tilde \g _5 } [\ud e ^{-8 A} \dh _{\mu } B _{ \nu \rho
  } \dh ^{\tilde \mu } B ^{\tilde \nu \tilde \rho } + {1\over 2} e
  ^{-4 A} \dh _r B_{\mu \nu } \dh ^r B^{\tilde \mu \tilde \nu } + {1
    \over 2} e ^{-4 A} r^3 \dh _a B_{\mu \nu } \dh ^{a} B^{\tilde \mu
    \tilde \nu } \cr && + e ^{-4 A} \dh _\mu B_{r \nu } \dh ^{\tilde
    \mu } B^{\tilde r \tilde \nu } + {\mu ^2 \over 2} (e ^{-6A} r^5
  B_{ \mu \nu } B^{ \tilde \mu \tilde \nu } + 2 B_{ \mu r } B^{ \tilde
    \mu \tilde r } ) ],  \label{sub3eq5p1} \eea where the tilde symbol
in $\tilde \mu ,\ \tilde r$ is a reminder that upper indices are
raised by the unwarped metric tensor.  For a vanishing 10-d  mass
parameter $\mu =0$,  one can   impose the unitary gauge choice  $
B_{\mu r } =0,\ \dh ^\mu B_{\mu \nu } =0 $.  The  towers    of harmonic modes $ B ^{(\nu , m)}
_{\mu \nu } (x)$ of fixed 5-d and 4-d squared masses, defined by the
decomposition \bea && B_{\mu \nu } (x, r) = \sum _{\nu , m} B^{(n)}
_{\mu \nu } (x) \b _{m} (r) Y ^\nu (\T) ,\ [\tilde \nabla _5 ^2 = -
  M_5 ^2 ,\ \tilde \nabla _4 ^2 = E_m^2 ]  \eea are described by
radial wave functions $ \b _m (r)$   obeying Sturm-Liouville  
eigenvalue problems.  The vector space  of solutions is spanned
by  the    basis of orthonormal wave functions related to  Bessel functions 
\bea &&
0= e ^{-8 A} r^5 (E_m^2 + e ^{8 A} r^{-5} \dh _r e ^{-4A} r^5 \dh _r -
\mu ^2 e ^{2 A} - M_5 ^2 e ^{4A} r^{-2} ) \b _{n} (r) = r^{-3} ( E_m^2
+ r^3 \dh _r r \dh _r - r^2 (M_5 ^2 +\mu ^2) ) \b _{n} (r), \cr && [\b
  _m (r)= {1\over N_m} (J_\nu ( {E_m \over r} ) + b_{\nu , m} Y_\nu (
  {E_m \over r} ) ) ,\ \nu = (M_5 ^2 +\mu ^2) ^{1/2} ,\ \d _{mm'} =
  \int _{r _0} ^\calr dr e ^{-8 A} r^5 \b _m (r) \b _{m'} (r) ]
.  \label{sub3eq5p2} \eea 
The   duality transformation in $ M_4$,  $\dh _\l B ^{(m)}
_{\mu \nu } (x) = \e _{\l \mu \nu \rho } \dh ^\rho \b ^{(m)} (x) $,  may
always be used   to replace   the 2-form fields  by scalar fields
$\b ^{(m)} (x) $.

\subsection{Summary of solutions for  radial wave  functions}
\label{sub3sec3}  

We conclude the present section  by  recording    useful  formulas for
the radial wave  functions    of our selected set of  scalar and
symmetric two-tensor, vector and  2-form massive fields $[ (S, T),\ V
  ,\ 2F]$ of spin  $[ (0, \ 2) ,\ 1,\ 0]$   in $ M_4 $. The
corresponding    massive fields in $ AdS _5$ in  harmonic  modes of $T^{1,1}$,     
labelled  by  the parameter values $
J= [0,\ 1, \ 2] $, obey  the mass-dimension relationship
$M_{5} ^2 = (\D -J) (\D -4 +J) .$ 
The formalism  is  very similar to that  of
the two-brane Randall-Sundrum model~\cite{csak04}. 
While  we    set $\calr = 1 $ throughout   this discussion,  
one can  switch over  to    general   length units  by   
changing $r\to r/\calr ,\ E_m \to E_m \calr.$

The wave functions obey second-order wave equations 
which  are solved by linear combinations of
normalizable and non-normalizable components  related to
Bessel  and Neumann  functions,    \bea && R_m ( r ) \equiv
{\tilde R_m ( r ) \over N_m} = { 1 \over N_m} ( { \rho \over E_m }
)^{-J+2} (J_\nu (\rho ) + b_{\nu , m} Y_\nu (\rho ) ), \ [\rho = {E _m
    \over r} ,\ \nu = ( (2 -J)^2 + M_5 ^2 )^{1/2} , \cr && N_m^2 =
  \int _{r_0} ^{\calr } dr r^{-3} (J_\nu ({E_m \over r} ) + b_{\nu ,
    m} Y_\nu ({E_m \over r} ) )^2 ={E_m ^{-2} } \int _{\rho _V } ^
       {\rho _R } d\rho \rho (J_\nu (\rho ) + b_{\nu , m} Y_\nu (\rho
       ) )^2 ] \label{sub3eq6}\eea where $ N_m $ are normalization
constants. The constant mixing coefficients $\ b_{\nu , m} $ and the
discrete 4-d mass parameters $E_m $ are determined  by the ultraviolet
and infrared boundary conditions for the radial variable interval $
r\in [r_{ir}\equiv r_0   =w\calr  , \ r_{uv} = \calr ]$.  Assigning for
simplicity Neumann boundary conditions at both boundaries, $ R_m '( r
) \vert _{r=\calr }=0 ,\ R_m '( r ) \vert _{r= r_0 } =0 $, then 
the ultraviolet boundary condition determines  $ b_{\nu , m} $ 
as the ratio of  wave functions and the
infrared boundary condition determines  $E_m $ as the zero of a transcendental equation, \bea && R ' _m
( \rho _V ) =0 :\ \ b_{\nu , m} = -{ ( 2- J-\nu ) J_\nu (\rho _V) +
  \rho _V J_{\nu -1} (\rho _V) \over (2- J-\nu ) Y_\nu (\rho _V) +
  \rho _V Y_{\nu -1} (\rho _V) } , \ [\rho _V = E_m = x_m w] \cr && R
' _m (\rho _R) =0 :\ \ (2- J-\nu )J_\nu (\rho _R) +\rho _R J_{\nu -1}
(\rho _R) =0, \ [\rho _R= {E_m \over r _0 } = {E_m \over w} = x_m
]. \label{sub3eq7}\eea 
The solutions associated to Bessel and Neumann functions   
are peaked at the infrared and ultraviolet boundaries respectively.  
Owing to the extreme smallness of $ b_{\nu ,
  m} $ for small $w$, we have omitted in the second line 
entry of Eq.~(\ref{sub3eq7})  the
negligible contribution from the component $ Y_\nu (\rho  _R) $. The
normalization constant contributed   by 
 the Bessel function, $ N_m ^2 = 2 (w x_m )^{-2} \sum _{k=0}^\infty
(\nu + 1 + 2k) J^2 _{\nu + 1 +2k } (x_m) $,    is  approximately
 evaluated by  the fomula $ N_m \simeq J_{\nu } (x_m) f_m / w $.  
In the  two main    radial  intervals  $ I $ and $ II $
inside and   near  the throat horizon,  one  can use the limited
series expansions  
\bea &&  I :\   r <<  x_m w  ,\ \rho  >>   1  ,\ 
[\rho  \geq  \vert  \nu ^2 - {1\over 4} \vert   ] :\ 
\ [J_{\nu } (\rho ) ,\   Y_{\nu } (\rho ) ] = \sqrt {2 \over \pi \rho
} [\cos ( \rho  - {(2\nu +1 ) \pi  \over 4 } ) ,\ \sin 
( \rho  - {(2\nu +1 ) \pi  \over 4 } ) ] .    
\cr && II:\ r  \geq  x_m w  ,\ \rho \leq 1,\ [0 \leq \rho \leq  \sqrt
  {\nu + 1}] :\   J_{\nu } (\rho ) \simeq {1 \over \G (\nu +1)} 
({\rho \over  2 })^\nu   ,\   Y_{\nu } (\rho ) \simeq  - {\G (\nu )\over
  \pi }  ({2\over \rho })^{\nu } .\eea  
In the third  interval  far   from the  horizon, 
$III:\  r >>   x_m w  ,\ \rho <<  1$, the  formulas  are same  as 
those in  $ II$. The asymptotic limits  of the   radial wave functions
in these intervals, 
\bea &&  I : \  r\to
0,\ \rho \to \infty :\ R_m (r) \simeq {r ^J \over J_\nu (x_m ) f _m }
( { 2 w \over \pi x_m r^3 } )^{1/2} (\cos ( {E_m \over r } - { (2 \nu
  + 1) \pi \over 4} ) +b _{\nu , m} \sin ({E_m \over r } - { (2 \nu +
  1) \pi \over 4} ) ) , \cr &&  II , \ III  : \  r\to 0,\ \rho \sim O(1) :
\ R_m (r ) \simeq {(x_m /2)^\nu r ^J w ^{1+\nu } \over J_\nu (x_m ) \G
  (1+\nu ) f _m } ( r^{-2 - \nu } - b_{\nu , m} {\nu \G ^2 (\nu ) 2
  ^{2\nu } \over \pi (x_n w)^{2\nu } } r^{\nu -2} ) ,\ [f_m= (1 +
  {4-\nu ^2 \over x_m ^2 } ) ^{1/2} ]   \label{sub3eq9} \eea 
exhibit    an oscillatory behaviour in   $ I$ (small $r$, large $
\rho $)    followed  by    a power law behaviour in     $ II, \ III$ 
(small $r$,   finite   or large $\rho $). 
Evaluating the  constant  mixing coefficient of  the two  solutions in
the   intervals   $II$  or $ III$ for $\rho _V <<  1$,   one
obtains     the   approximate formula  
\bea && b_{\nu , m} \simeq
{\pi (\rho _V / 2) ^{2\nu } (2 +\nu ) \over \G (\nu +1) \G (\nu -1) (
  (2-\nu ) (\nu -1 ) + \rho _V ^2 / 2 ) } \approx \pi ( {\rho _V ^2
  \over 4 }) ^{\nu -1} = \pi ( {x_m w \over 2})^{2 (\nu -1 )}.   \eea     
Note that  that   the  mixing coefficients 
become sizeable  only    for  modes  of   very large radial
excitations,    $ \rho _V  = x_m w \geq 1 $. 
In the regime of large  $ \rho _V  $, which is not relevant  for our
current applications, we obtain in the case    $\nu = 2 ,\ J=0 $ 
the   following polynomial fit,   
$b _ {\nu , m}  \approx (-7.46 +0.945 \rho _V + 0.0469 \rho _V^2
)\  10^{-3}. $      The radial  wave  functions  near  the  infrared   
and ultraviolet boundaries are  given by the limiting   formulas 
\bea && \bullet \ N_m R_m
(r_{ir})  \simeq  r_{ir}^{-2  +J } ( { (\rho _{ir }  /2 )^\nu
\over \G (\nu +1) }  - {b_{\nu , m} \over \pi } \G (\nu )  (\rho _{ir }
/2 )^{-\nu }  ) \sim  w ^{J-2}  ,\ [\rho _{ir }  = x_m ,\ 
N_m\simeq {J_\nu (x_m)  f_m \over w } ] .  \cr && 
\bullet \  N_m R_m (r_{uv})= r_{uv}^{-2}  
(J_\nu (\rho _V)  +   b_{\nu , m} Y_\nu (\rho _V) ) 
\simeq {(x_m w /2 )^\nu \over \G (\nu +1) } r ^{-2 -\nu +J} 
(1    - \nu \G ^2 (\nu )({2 \over  x_m  w} )^{2} ) .  \eea    
While  $ R_m (r) $ behaves as  
as a     power  $ r ^{J -1}    $  times  oscillating factors in the far infrared   region, outside this region it  involves    normalizable and non-normalizable
components   \bea && R_m (r) \sim 
[ w ^{1+\nu } r ^{-\D _\nu }  ,\ w ^{1 - \nu } r ^{\D _\nu -4 }  ]  ,\ 
[\D _\nu =   2+\nu -J ]   \eea   
with   very small    mixing   coefficients that    scale as    $ b
_{\nu , m} \sim w ^{2 (\nu -1)} $.   
The  Bessel function component $ J_\nu $    
peaks  near the    tip  $r \sim w $ while the  Neumann 
function component  $ Y_\nu $    dominates   
 near the ultraviolet   boundary  only  when 
the  small  value of  $ b_{\nu , m}$  is compensated in
 the   product, $b _{\nu , m}   Y_\nu (\rho _V) / J_\nu  (\rho _V)  
\simeq  b _{\nu , m} - {\nu \G ^2 (\nu ) \over \pi } 
({2 r _V \over x _m w })^{2\nu }  \sim 
\nu \G ^2 (\nu )  ({2  / (x_m w)   }) ^2 $. 
The effective gravitational mass scale is determined by  the 
approximately    mode independent value of the graviton modes radial
wave  functions  at the infrared    boundary,     
$\L _{KK} = M_\star / R_m (r_{ir}) = 
(M_\star  w ) / (w R_m (r_{ir}) ) \sim M_\star w / 1.41  $. 
For illustration, we  record the   numerical   values of the mixing
coefficients and the effective gravitational mass  scale    
for the  first    five    massive  singlet graviton  modes, 
\bea &&   b _ {\nu , m} = [0.46,    1.5 ,   3.2 ,   5.5 ,   8.5 ]
 \   10^{-26} ,\   \L _{KK} / TeV  = {M_\star  \over  R_m (r_{ir} ) } =
     [{28.2},    {-11.3},    {6.47},      {-4.31},    {3.13}]\ 
  {w   \over  2.\ 10^{-14}  }. \eea 
The sign changes    of  $\L _{KK}  $   reflect the oscillating
behaviour  of Bessel functions     at   $ r <  r_{ir} = w$.

The orthogonality   of  the basis of angular
wave functions,  labelled   by quantum numbers $\nu $,  follows
from  invariance under the  isometry group  $G$, while the orthogonality   of  the basis of the  radial
wave functions,  labelled   by the integer   number $ m $, 
follows from the  diagonal  structure    of the  wronskian
$ ( R_{m'} r^5 \dh _r R_m - R_{m} r^5 \dh _r
R_{m'} ) \vert _{r_{ir}  } ^{r _{uv} }  \propto \d _{ m m'} $ which  
holds  for  both Neumann and Dirichlet boundary conditions.  
The    massless  4-d graviton  zero  mode    $ E_0
=0,\ [\nu =2 ]$    descending   from   the 5-d    mode of  $ M_5 =0$, 
singlet under the isometry group,  is  assigned  the
constant  radial wave function $R_0 (r) $. 
The  orthogonality    to   the    radial  wave functions   of   (singlet) 
massive  graviton  modes  can be inferred     by  analyzing the normalization    integral   in  the limit $ E_m \to 0$, which
requires   that the    radial  wave functions of all   finite  mass modes must   yield  a   vanishing integral 
$ I_m \equiv  \int _{r_{ir} }^{r _{uv}} dr r R _m (r)   ,\
[E_m = x_m w /\calr  \ne 0  ]$.
The  numerical   check  of this property 
requires  high  precision calculations,   due to the    strong     cancelations  between  contributions to the scalar   products from 
the    Bessel and   Neumann  radial  components 
of   $ R_m (r) $.   which   are  $ O( w^2)$.  
Nevertheless, we  could  successfully check for $ w= 10 ^{-4} $   
that the normalization integrals for the  first   5 massive 
graviton  radial  modes   take the small values 
$ I_m \simeq [-1.24 ,\  1.66  ,\   -2.00 ,\ 2.28 ,\ -2.54 ] \ 10 ^{-8}
$.    

It is perhaps  the  conditions under  which 4-d (massless) 
zero modes  for scalar and tensor   modes  could occur   in  the present context.
In the untracted $AdS _5$    spacetime, the scalar  fields  of   squared
mass $ M_5^2 =0 $  do not   have     zero modes  
while  massless  spin 1 or  2  fields   do.  
In   the truncated   $AdS _5$  spacetime slice,
zero  modes  do   arise for   massive  fields ($ M_5 \ne 0$)  
provided  one includes    a  boundary action  with a mass  parameter    
carefully   tuned to the   bulk  mass~\cite{csak04}. 
In the  scalar  field case   of action 
\bea && S= -{1\over 2} \int \sqrt {-g_5} (
g ^{\hat \mu \hat \nu } \nabla _{\hat \mu } \phi \nabla _{\hat \nu }
\phi +  M_5 ^2 \phi ^2 + 2 b \ r ( \d (r - \calr ) - \d (r - r_0) )
\phi ^2   ) ,\ [ds ^2 _5= r^2 d \tilde s^2 _4 + {d r ^2 \over r^2 } ] \eea 
the bulk and  boundary   terms   together    enforce     
the  modified  Neumann boundary condition   
$ (r \dh _r +b ) \phi =0$.      A    zero mode 
solution  $\phi   (r) = r^{-b _\mp } = r ^{-2 \mp \nu } $ 
would    exist then     in one of  the   following two cases 
$ b = \ b_\mp   \equiv  2 \mp \nu = 2 \mp (4+  M_5  ^2
)^{1/2} $.   The    corresponding rescaled  wave function   $\tilde
\phi _0 (r) \equiv r \phi  (r) = r^{-b _\mp  +1} = r ^{-1 \pm \nu } $
in   the flat  radial coordinate  $ Y = -\ln  r $     
is  localized  near the  ultraviolet or  infrared  boundaries 
for $ b_\mp  <1 $ or $ b_\mp  >1$ and is   flat     for $ b_\mp
=1 $.   Similar  properties are   found for  the spin 2 and the 
half-integer   fermion   fields. 

 \section{Interactions of warped graviton  modes}  
\label{sec4}

The primary goal   of the present work  was to  
obtain predictions for the mass spectra and interactions of warped
modes.    We     shall mostly     focus   on the  local  self
couplings of  bulk graviton modes  and   
their    boundary couplings  to  embedded $D3$-branes. The 
coupling   of gravitons to  other bosonic components 
of the supergravity multiplet  is  treated   briefly. 

\subsection{Self couplings  of graviton modes} 
\label{subsec4.1}

We wish to   discuss the  mutual interactions of graviton modes obtained by expanding the  
perturbed 10-d action in powers of the metric tensor fluctuations
$h_{\mu \nu } (x, y) $ about the classical vacuum solution. 
The reduced action of quadratic order
contains  the kinetic and mass terms,  discussed in Subsection~\ref{subsec2.2},
\bea && S ^{(2)} _{kin} (h) = {1\over 4 \kappa ^2}
\int d^{10} X \sqrt {-\tilde g_{10} } e ^{-4 A}  h^{\mu \nu } 
(  -   \tilde E _{\mu \nu }  - \ud  \tilde \nabla _6^2 ( h_{\mu \nu } - \tilde
  g_{\mu \nu } h) )  ,\ [\tilde E _{\mu \nu } =\d  \tilde G ^{(4)}_{\mu \nu } ] . \eea 
Due to the non-linear dependence of the scalar curvature on the metric,
there arise  an infinite series of local couplings. We  shall focus here on the  subset  associated to the   4-d part of the perturbed curvature  $ \tilde  R^{(d)} = \tilde g ^{\mu \nu }  \tilde  R^{(d)} _{\mu \nu }  ( \tilde g_4 + h) $  given by  the Riemann tensor  components    $R _{\mu \l \nu } ^{\ \ \ \ \ \ \l  } $, hence ignoring the  part  from  the mixed   components
$  R _{\mu  p \nu } ^{\ \ \ \ \ \ p} $.       The resulting  action 
is  \bea &&  S _{grav} (h)
= {1   \over  2 \kappa ^2 } \int  d ^{10 } X \sqrt {\tilde g_6}
\sqrt { -( \tilde g _4 + h) }
e ^{-4 A}  \tilde R^{(4)} (\tilde g _4+ h)\cr &&  \simeq
{1   \over  2 \kappa ^2 }  \int  d ^{10 } X \sqrt {\tilde g_6}
\sqrt {-\tilde g_4 }  e ^{-4 A} [\tilde R ^{(4)}  + h ^{\mu \nu }
[ \ud (\tilde \nabla _4 ^2 + e ^{4A}\tilde \nabla _6 ^2 ) ( h _{\mu \nu }  -  
  \tilde  g_{\mu \nu } h )   + \tilde \nabla _\mu\tilde \nabla _\nu h ^{(m)}  - 2  \tilde \nabla _{ (\mu }   h ^{(m)} _ {\nu ) }  ]     + \sum _{n > 2} h
  ^{n-2} \dh h \dh h )    ] .  \eea
where  we   have condensed  the higher order  couplings  in a  schematic  form   ignoring  the  contractions of spacetime indices. 
  Substituting the decomposition  of the   10-d     metric 
tensor  field   on  pairs of normalized charge  conjugate
graviton   fields,  
\bea && h_{\mu \nu } (X)= \sum _m (h
^{(m)} _{\mu \nu } (x) \Psi _m (y) + \bar h ^{(m)} _{\mu \nu } (x)
\Psi ^\dagger  _m (y) ) ,\ [\int d^6 y \sqrt {\tilde g_6} e^{-4 A} \Psi
  ^\dagger _m (y) \Psi _{m '} (y) = \d _{m m '} ]  \label{eqeigenf1} \eea  
where the coefficients $\Psi _m (y) $ are eigenfunctions   of the
internal space   scalar Laplacian of    eigenvalues   $ - e^{4 A} \tilde
\nabla _6^2 \to  E_m^2  $  identified   to    the mass parameters, 
yields  the 4-d action for  massive  Fierz-Pauli  spin 2 fields~\cite{derham14} 
\bea && S ^{(2)} _{kin} (h) = {1\over 4 \kappa
  ^2} \int d^{4} x \sqrt {-\tilde g_{4} } \sum _m [
h^{(m) \mu \nu }  ( (\tilde \nabla _4^2 - E_m ^2) h^{(m) }_{\mu \nu }  
 + \tilde \nabla _\mu\tilde \nabla _\nu h ^{(m)}  - 2  \tilde \nabla _{ (\mu }
 h ^{(m)} _ {\nu ) }  ) - h^{(m)} ( \tilde \nabla _4^2 - E_m ^2)
 h^{(m)}  ]  ,  \label{sub4eqY2}  \eea   
where the  orthonormalization  condition on  wave functions   
in  Eq.~(\ref{eqeigenf1}) ensures that  the summation  over modes in
the reduced action   has a diagonal structure.  
We     shall restrict  for simplicity   to  polarization  states of
the classical tensor  field   $ h _{\mu \nu } (X)$ 
that  satisfy  the transverse-traceless
conditions $ h ^{\nu } = \tilde \nabla _\mu h ^{ \mu \nu } =0 ,\  h =
\tilde g _{\mu \nu} h ^{ \mu \nu } = 0$,   hence simplifying   the       
kinetic action    to  $  h^{\mu \nu } \tilde E _{\mu \nu } \to -  \ud
h^{\mu \nu }     \tilde \nabla ^2_4 h_{\mu \nu }  $.   
The interactions  between graviton modes involve  trilinear and
quartic order  couplings \bea && h^ {(m_1)} \dh h^ {(m_2)} \dh h^ {(m_3)}
,\ h ^ {(m_1)} \dh C _{\nu _2}^ {(m_2)} \dh \bar C _{\nu _3}^ {(m_3)}
,\cdots , \ h ^ {(m_1)} h^ {(m_2)} \dh h^ {(m_3)} \dh h^ {(m_4)} ,\ h
^ {(m_1)} h^ {(m_2)} \dh C _{\nu _3} ^ {(m_3)} \dh \bar C_{\nu _4} ^
{(m_4)} , \cdots \eea   in various allowed configurations of the real
(complex) singlet modes $ h^ {(m)} $ in ground  (excited) 
radial excitation  states, labelled by the index 
$m = 0, 1, \cdots $,    and the complex  modes
$ C ^ {(m)} _\nu   $ of charge  under $G$  labelled
by  $\nu = (j,l, r) $ in   ground  and excited  radial  states
labelled by  $m = 0, 1, \cdots $. 
We  provide a closed formula for  a subset of the gravitational
couplings of  the spin 2  modes $ h ^{(m)} _{\mu \nu}$ in a general gauge 
in Appendix~\ref{subapp5}. However, given the unwieldy character of calculations of scattering
amplitudes, we have settled in the present work on ignoring the
spacetime structure,  hence restricting to  orders of magnitude estimates. 
We now    define the effective action of graviton modes
in terms of  the    parameter $\l _2 $ for the kinetic  and mass terms and
coupling  constant parameters  $\l _n $ for the  schematic
form of higher order interactions  
\bea && S (h) = \int d^{4} x \sqrt {-\tilde g_{4} } [
  {\l _2 }  h^{(m) \mu \nu } (\tilde \nabla _4^2 - E_m ^2) h^{(m)
  }_{\mu \nu } + \sum _{n\geq 3} \l _n h ^{(m_1)} \cdots h ^{(m_{n-2}
    )} \dh h ^{(m_{n-1} ) } \dh h ^{(m_{n} ) } ] , \cr && [\l _2 =
  {m_D ^8 \calr ^6 \over 4} ={M_\star ^2 \calr ^6 \over 4 V_W}   
= ({M_\star \over 2 (2\pi \eta )^3} )^2  ,
  \ \l_n =\ud \l _2 J_n = \ud {M^2 _\star \calr ^6 \over 4V_W } J_n
  ,\cr && J_n= \int d^6 y \sqrt {\tilde g_6} e^{-4 A} \prod _{i=1}^n
  \Psi _{m_i} (y) = \int dr r^5 e^{-4 A} \prod _{i=1}^n R _{m_i} (r)
  \int d^5 \T \sqrt {\tilde \g _5 }\prod _{i=1}^n \Phi _{m_i} ( \T ) ] \label{sub4eqY1} \eea 
where the  overall   factor dependence  of $\l _n$ on  $ \calr $ 
was    reinstated  via dimensional analysis, with the understanding
that the wave functions overlap  integrals $J_n$     are evaluated in units of $
\calr $.  Note        the dimensional 
assignments  for  radial wave functions and coupling constants,   
$ [R (r) ] = E ^{-1},\ [\l _n] = E ^{2-n} , \ [n > 2 ]$.   Transforming to
the canonically normalized graviton modes    by means of  the
rescalings,  
$ h _{\mu \nu } \to h_ {\mu \nu } / \sqrt {\l _2 } $   and  $\l _n \to
\l _n / \l _2   ^{n/2},\ [n \geq 3] $   
yields  the  explicit formulas  for the 
 coupling constants   
\bea && \l_n =\ud \bigg ({ m_D ^8 \calr ^6 \over 4 }\bigg ) ^{1- n/2 }
J_n = \ud 
\bigg ({ M_\star ^2 (\calr / L) ^6 \over 4 (2\pi )^6 } \bigg ) ^{1-
  n/2} J_n , \cr && [J_n = { \int dr r \prod _{i=1} ^ n R_{m_i} (r)
    \int d ^5 \T \sqrt {\tilde g _5 } \prod _{m=1} ^ n \Phi _{m_i} (\T
    ) \over \prod _{i=1} ^ n \bigg (\int dr r R_{m_i} (r) R ^\star
    _{m_i} (r) \int d ^5 \T \sqrt {\g _5 } \Phi _{m_i} \Phi ^\star
    _{m_i} (\T ) \bigg ) ^{1/2} } ].  \label{sub4eq3} \eea
Since only the  integrations over polar
angles $\t _1 ,\ \t _2 $ are non-trivial ones, it is   convenient to
absorb   the integrals  over the azimuthal angles
$\phi _{1,2},\ \psi $  inside  the overall volume factor  ${V_5 / 4 }
= {4\pi ^3  / 27} $.    We   define   the  reduced 
dimensionless coupling constants $\hat \l _n  $, of natural order
unity,   by factoring out the  dependence on  the various  physical  parameters  as 
\bea && \l _n = {1\over 2} ({m_D ^8 \calr ^6 \over 4} ({V_5 \over 4} )
) ^{1-n/2} 
\hat \l _n = {1\over 2} ({ 12 \sqrt {12} \pi ^{3/2} (L/\calr )^3 \over
  M_\star w } ) ^{n-2} \hat \l _n ,\ [ \hat \l _n = w ^{n-2}
  J_n].  \label{Yukcc1} \eea  
where  $\hat \l _n  = O(1) $. 
The   dependence  of $\l _n  = O(1) $  on  the   free parameters 
$ M_\star ,\ w  ,  \ \eta = L/\calr  $  is   found  
to be  the same    as  in~\cite{chen06}, as expected.    
The   reduced   coupling constants    in cases 
including   deformation  effects    from the warp profile 
are  given  by the  similar approximate   formula 
\bea && \hat \l _n =  {c_\nu ^- \over 2} w ^{\a _\nu ^- +n-2}  J_n' 
,\  [J_n ' = \int d r r ({r\over w \calr } )^{\D _\nu ^- -4 }
\prod _{i=1}^n  R_{m_i} (r)\int d ^5 \T \sqrt {\tilde \g _5 } 
Y^{\nu }  \prod _{i=1}^n  Y^{\nu _i}(\T )  
 ,\ \a _\nu ^-=Q _\nu ^- +  \D _\nu ^- -4  ] . \label{eqdefcc}  \eea

The selection rules on the $ (j,\ l,\ r)$ quantum numbers in
 $n$-point couplings require that the direct products of
 representations of the participating graviton modes include the
 singlet representation,  \bea && \oplus _i j_i \in 0,\ \oplus _i l_i
 \in 0 ,\ \sum _i m_i = \sum _i (q_i + r_i )/2 \in Z ,\ \sum n_i= \sum
 _i (q_i - r_i )/2 \in Z ,\ [i=1, \cdots , n]
\label{selecteq} . \eea    

\subsection{Graviton modes  couplings  to  bulk      modes and moduli} 
\label{subsec4.3}

The expansion of the reduced supergravity action  in powers of  fields  fluctuations   yields a variety of  tree level  interactions 
among  bulk     modes.   We      here  consider    the   
couplings  of      single gravitons  with pairs of  bulk
bosonic  matter modes   arising from   
the  kinetic terms $ F_5 ^2,\ \nabla \tau ^ 2 ,\ G_3 ^2 $   in the
10-d action. The   symmetry  under  general coordinate   transformations  
yields the  following universal formula for the  lowest order  
reduced action   involving  the   energy-momentum stress tensor  of
matter field modes~\cite{han98},   
\bea &&  \d S _4 (h) =  \sum _{(m np)}  
{ \l _ {m np}  \over M_\star \sqrt 2 }\int  d^4 x 
\sqrt {- \tilde g_4} h^{(m)} _{\mu \nu } (x) T^{\mu \nu } _{(n,p)} (x)
, \cr &&    [\l _{mnp} ={M_\star \sqrt 2 \over (2 \l _2^{1/2} )}  \int d^6 y
  \sqrt {\tilde g_6 }   e ^{-4 A} \Psi  _m  
\Psi  _n \Psi  _p  ,\ {1\over \sqrt {\l _2}}  =  2\sqrt 2 (2\pi
L/\calr  )^3 / M_\star ] \eea 
for  gravitons and      matter  modes   represented  by
canonically normalized fields.  
The cubic  couplings between   massive graviton field  
 $ h^{(m)} _{\mu \nu } $   and       the  scalar, vector    and  
2-form modes $ [ b _m ,\ v _m,\ \b _m]  \in 
[(\d g^a _a ,\ \d C_4 ), \ \phi _{\hat \mu }, \ 
 \d B_2 ]  $, are    described by  the Lagrangian density      
\bea && \d  L _4(h) =  {1 \over M_\star \sqrt 2 } \sum _{m,n,p}   
h ^{ (m) \mu \nu } ( \l _{mnp}  ^b \dh _\mu b
^{(n)} \dh _\nu b ^{(p)}  + \l _{mnp}^v \dh _\mu v _\l  ^{(n)} \dh
_\nu  v  ^{ (p) \l }    
  +  \l _{mnp} ^\b \dh ^\mu \b ^{(n)} \dh ^\nu \b ^{(p)} ) ,
\cr &&  [\l _{mnp}  ^{(b, v, \b )} =  \sqrt{2} (2\pi L/\calr ) ^3    \int dr r^5 ( e ^{ -4 A} R ^h _m R^b _n R ^b _p , e ^{ -6A} R ^h_m
  R^v_n R ^v_p , \ e ^{ -8A} R^h_mR ^\b _n R ^\b _p )  \int d ^5 \T
\sqrt {\tilde \g _5} \Phi _m \Phi _n\Phi _p  ] .  \eea
The  disparity between  the  radial integration measures  $r  ^{[0, -1
    , -3]}  $ in the above three cases  is
compensated by the different normalization conventions of the radial
wave functions,  $  [R ^ h  ,  \   R^b , \   R^v , \   R^ \b ]  
\simeq   r^{[-2, -2, -1, 0 ]}     J _\nu (E_m ^2 / r )     $. The 
resulting    radial     overlap integrals    are of same form 
$ \int d r r ^{-5} JJJ $ in  all cases. 

The    interactions   of gravitons with     geometric moduli 
is obtained by transferring  the factor $ \d \tilde g _{\mu \nu }
(X)  = \sum _m h ^{(m)} _{\mu \nu } (x) \Psi _m (y)  $ 
for the kinetic energies in Eq.~(\ref{sub2eq15})     
inside the integral over  the internal  manifold volume. 
The  resulting   cubic  couplings to pairs of 
complex structure and Kahler   moduli  can be  expressed by 
the effective  Lagrangian  with   coupling  constants 
given by  overlap   integrals involving   the moduli wave  functions,      
 \bea && L (h) = \sum _m h ^{(m) \mu \nu } (x) [\l
  _{m , \a \b } \dh _\mu S^\a \dh _\nu \bar S^{\b } + \l _{m , a b }
  \dh _\mu T ^a \dh _\nu \bar T ^b ] , \cr && [ \l _{m , \a \b } , \l
  _{m , a b } ] = {M_\star ^2 \over 8 V_W \sqrt { \l _2 } } \int d^6y
\sqrt {\tilde g_6} e ^{-4 A (y)} \Psi _{m } (y) [\d _\a \tilde g ^{kl}
  \d _\b \tilde g _{kl} , \ \d _a \tilde g ^{k\bar l } \d _b \tilde
  g _{k\bar l } ] .
\label{sub2eq20p} \eea 
The  interactions   between    the 
massless and  massive  modes  $ \g  (x) ,\     \g _m (x, y)
, \  [m =0, 1,   \cdots ] $   descending  from the 5-d   breathing
 mode  field   $\G (X)$   and the  6-d universal volume
 modulus   $\rho =  i c(x)  +  a _0(x) $  can be deduced  
from the reduced actions  in Eqs.~(\ref{sub2eq51})
and~(\ref{sub2eq24}). 
Beside the      quadratic   diagonal and non-diagonal 
 mass couplings   $ \g _0 ^2, \  \g _0 \g  _m   ,\ \g _m  c ,   $
 there also  occur   cubic and higher order couplings $ c ^3,\ \g _m c
 ^2   
 ,\  \g  a_0^2 ,\  \g _m  a_0^2 ,\cdots $  that were examined in some
detail in~\cite{freydacline09}.

\subsection{Bulk  modes  couplings   to embedded $D3$-branes}
\label{subsec4.2}

The   interactions  of warped   modes       with $D$-branes 
are  conveniently  encoded    through the  construction  of the
action as a    non-linear sigma  model.  The  coupling  constants
refer  to  bosonic fields  for closed strings describing   
the    background spacetime    and the   degrees of freedom   
refer  to   coordinate fields for open   strings  
describing the     brane world volume  embedding in   superspacetime. 
For the supersymmetry preserving (extremal) $Dp$-branes of type $ II $
theories,  the low energy dynamic    involves  
scalar and bi-spinor coordinate fields,  
$X^M (\xi ) \sim 10 ,\ \T (\xi )= ( \t _1 , \t
_2 ) \sim 16 \oplus 16 $,  functions of the brane worldvolume variables
$\xi ^\mu ,\ [\mu = 0, \cdots , p] $    transforming in  vector and
Majorana-Weyl    spinor representations of the Lorentz symmetry
group $SO(9,1) $ of the tangent spacetime.   The bulk fields
projection  on the world brane  is defined  by  the  
 pull-back mappings  from
spacetime using the basis of frame vectors $\dh _\mu X ^M (\xi ) $,
detailed  in  Eq.~(\ref{sub4eq7}).  The brane action is invariant
under  the diffeomorphism     group of the $\xi ^\mu $  and  the
super-Lorentz-Poincar\'e group  of the coordinates $ X^M,\ \T
$ of $ M_{10}$.   The     local  gauge symmetry under
diffeomorphisms is fixed in the static gauge by setting $X^\mu (\xi
^\mu ) = \xi ^\mu $  and the local fermionic $\kappa $-symmetry, 
orthogonal   to  the  conserved supercharges, is fixed by imposing suitable
conditions on the bispinor field $\T (\xi ) = (\t _1 , \ \t _2 ) $.

A brief review of the low energy bosonic and fermionic actions of $Dp$-branes
is provided in Appendix~\ref{subapp6}.  We    shall  focus on
space-filling probe $D3$-branes in $ M_4 \times X_6$ 
of world volumes extended over $M_4$ at points
of $X_6 $.  In this   background the tangent spacetime group 
reduces  to   $ SO(9,1)\to
SO(3,1)\times SO(6) $ and the gauge and matter brane degrees of
freedom consist of   real bosonic fields  and $\caln =1 $
superpartner gaugino and modulino  Weyl spinor fields  of $ SO(3,1)$, descending from $(X^M (\xi )/ \a ') $
and $\T ( \xi )  $,
\bea &&  (A^\mu ( \xi ) , \ \varphi ^m
(\xi ) ) = ({1\over \a '} X^\mu , y^m ) , \ (\psi ^{m} ( \xi ) ,\ \psi ^{g} (
\xi ) ) , \ [m=1, \cdots , 6]   .\eea 
The bosonic action  involves DBI and CS parts given in Einstein frame
by \bea && S _B 
(D3/\bar D3) = - \tau _3 \int d^4 x ( -Det (g + g_s^{1 / 2} e ^{-\phi
  / 2} \calf )) ^{1/2} \pm \mu _3\int C_4 e ^{\calf _2  } ,
\ [\tau _3 = {\mu _3 \over g_s } ,\ \calf = B +\s F  ] \eea 
where the metric $ g $,  4-form potential $ C_4$ and the   NSNS
potential $ B$ induced on the brane via the (omitted) pull-back
transformation  $ \phi ^\star  $,    depend on the open moduli
fields $ y^m (\xi ) $, and  $F $  is the  brane gauge  field strength.      The   determinant is   conveniently  
evaluated by means of the
expansion in Eq.~(\ref{subexpdet}).  The fermionic action 
of quadratic order in $\T $  in Einstein frame, quoted  from
Eq.~(\ref{sub4eq8}),    
\bea && S _F(D3/\bar D3) = i \tau _3 \int d^4 x ( -Det (g + g_s^{1 / 2} e
^{-\phi / 2} \calf ) )^{1/2} \bar \T P_\mp ^{D3}  [(\calm ^{-1})
  ^{\mu  \nu } ( \G _\nu  ( D_\mu+ {1\over 8} \G _\mu  O ) )  - \ud O ] \T ,\eea
involves  the same determinant  factor as in the bosonic action
times  a  quadratic matrix element     in  the  bi-spinors  
involving the   metric  $\calm _{\mu \nu } $, the $\kappa $-symmetry  projector and  the pull-backs of the covariant  operators  $ D_\mu ,\ O$  realizing   the   supersymmetry
transformations of the bulk   gravitinos and dilatino  fields. In the absence of 3-fluxes, the contributions  to the latter operators  in GKP background  are  given   in Eqs.~(\ref{app4EQ7}) while    the  linear combination 
$g^{\mu \nu }  (\G _\nu  \nabla  _\mu  +\G _\nu \G _\mu  O /8  ) - O/2  =
 \G ^\mu \nabla _\mu = \Dslash $  is  seen to  reduce to   the
 Dirac    operator.   The transformation from warped to unwarped metric utilizes $ \g ^\mu
\to e ^{-A} \g ^\mu ,\ \g ^m \to e ^{A} \g ^m $ and that from curved
to tangent spaces utilizes the basis of frame vectors.  In the $\kappa
$-gauge $ \T = (\t , \ 0)$  to which we specialize    hereafter, 
the   brane  modulino and gaugino  fields  arise from  $\t (\xi ) $ 
through   the decompositions in Eq.~(\ref{sub4eq13}). 
The combined bosonic and fermionic actions  are then given  by 
\bea && {S (D3 / \bar D3) } =
 \int d^4 x \sqrt { - (\tilde g _4 + h )} \bigg [{\tau _3 } (e ^{4A} \mp \a
   ) - {\pi \over g_s} \tilde g ^{\mu \nu } \tilde g _{mn} \dh _\mu
   \varphi ^m \dh _\nu \varphi ^n - {g_s \over g _{D3}^2 } e ^{- \phi
   } f_{\mu \nu } f^{\mu \nu } \mp {\t \over (4\pi )^2} f_{\mu \nu }
   \tilde f^{\mu \nu } , \cr && - {i \pi \over g_s}\tilde g ^{\mu \nu
   } [ ( \tilde g _{mn} - i \o _{mn})\psi ^m \bar \s _\mu \dh _\nu
     \psi ^n + {2 e^{-\phi }} \psi ^g \bar \s _\mu \dh _\nu \psi ^g ]\bigg  ]
   ,\cr && [y^m = \hat l_s^2 \varphi ^m, \ \calf _{\mu \nu } = \s
     F_{\mu \nu } = 2\pi \s f_{\mu \nu } =\hat l_s ^2 f_{\mu \nu }
     ,\ g _{D3}^{-2} = \tau _3 (2\pi \s )^2 = 2\pi / g_s , \ \t = 2\pi
     C_0 ] \label{eqredd3} \eea 
where the    gauge theory coupling constant and theta-angle parameter
are   denoted by $ g _{D3}  $ and $\t $ and  the matter and gaugino
Weyl spinor fields $\psi ^{m,g} $ were  redefined via  the
replacements   $\psi ^{m,g} \to \sqrt {i} \hat l_s^2 \psi ^{m,g} $.  
Substituting the perturbed   metric tensor 
$ \tilde g _{\mu \nu } \to \tilde g _{\mu \nu } + h
   _{\mu \nu } ,\ \tilde g ^{\mu \nu } \to \tilde g ^{\mu \nu } - h
^{\mu \nu } $,   yields
the linear and quadratic order  couplings of graviton modes to brane modes   
\bea && S   (D3 / \bar D3) = \int  d^4 x \sqrt { - (\tilde g _4 +h )}
     [ - {\pi \over 
       g_s} h^{\mu \nu } (  \tilde g _{mn} \dh _\mu \varphi ^m \dh _\nu
       \varphi ^n + {i } ( \tilde g _{mn} \mp i \o _{mn}) \bar \psi ^m
       \bar \s _\mu \dh _\nu \psi ^n + {2 i e^{-\phi }}\bar
       \psi ^g \bar \s _\mu \dh _\nu \psi ^g ) \cr &&   - {\pi \over 2 g_s}
     \tilde g _{mn} h ^{\mu \l } h _{\l } ^ { \ \ \nu } \dh _\mu
     \varphi ^m \dh _\nu \varphi ^n ] ,   \label{eqred1d3}  \eea 
 which  agree  formally with  those found in~\cite{han98}   for  flat background   spacetimes.   It is always possible to revert to Dirac spinors and matrices   by  utilizing  the chiral representation in terms of    left and right  Weyl spinors $\psi _R $ and $    \psi _L $, with the  resulting
 representation  for the couplings of   gravitons    equivalent to
Eq.~(\ref{eqred1d3})  given by 
 \bea && -i ( \bar \psi_L \bar \s _\mu \dh _\nu
   \psi_L + \psi _R\s _\mu \dh _\nu \bar \psi _R ) \to i \bar \Psi \g
   _\mu \dh _\nu \Psi ,\ (\psi _R \psi _L + \bar \psi _L \bar \psi _R
   ) \to -i \bar \Psi \Psi ,\ [\Psi = {i\bar \psi _R \choose \psi _L }
     , \ \g ^\mu = \pmatrix{0 & -\bar \s ^\mu \cr \s ^\mu & 0 } ].\eea
Going to canonically normalized fields via the
replacements  $\varphi ^m \to ({2\pi \over g_s } )^{-1/2} \varphi
   ^m,\ \psi ^ {g,m} \to ({2\pi \over g_s })^{-1/2} \psi ^{g,m} ,\ h
   _{\mu \nu } (X) \to h ^{(m)} _{\mu \nu } (x) \Psi _{m }( y_\star )
/\sqrt {\l _2} ,$ one obtains
\bea &&   S ({D3})= \int d^4 x \sqrt { - (\tilde g _4
     +h) } \sum _m {\l ^B _3 \over M_\star \sqrt 2 } h ^{ (m) \mu \nu
   } (x) \cr && \times [ \tilde g_{mn} ( \dh _{\mu } \varphi ^m \dh
     _{\nu } \varphi ^n + {i \pi \over g_s} \bar \Psi ^m \g _\mu \dh
     _\nu \Psi ^n ) + {2 i\pi \over g_s^2} \bar \Psi ^g \g _\mu \dh
     _\nu \Psi ^g ] \ + {\l ^B _4 \over 2 M_\star ^2 } h ^{ (m) \mu \l
   } (x) h ^{ (m) \nu } _{\l } (x) \tilde g_{mn} \dh _\mu \varphi ^m
   \dh _{ \nu } \varphi ^n , \cr && [\l ^B _3 =
{2 \sqrt 2 (2\pi L / \calr ) ^3 } \Psi _m (y_\star ) ,\ \ \l ^B _4
\simeq \vert \l ^{B } _3 \vert ^2 = { 8 (2\pi L / \calr ) ^6 } \Psi
^2_m (y_\star ) ,\ \Psi _m (y_\star ) = R_m (r _\star ) \Phi _m (\T
_\star ) ] .
\label{sub4eq10}   \eea 
The terms in $h ^\mu _\mu $ or $h_{\mu \nu } h^{\mu \nu } $, 
obtained  by expanding  the   determinant  factor  $ \sqrt {-(\tilde g
  _4    +h ) } \simeq \sqrt { -\tilde g _4 } \ ( 1 +\ud h^\mu _\mu -
{1\over   4} h_{\mu \nu } h^{\mu \nu } + {1\over 4} (h^\mu _\mu )^2
+\cdots ) , $ contribute in building up the dependence on the
stress-energy tensor, $ h ^{\mu \nu } \dh _{\mu } \varphi \dh _{\nu } \varphi \to h
^{\mu \nu } T_{\mu \nu } (\varphi ) $.  

We    next  consider   briefly the  interactions of   dilaton   modes $\phi^{
  (m) } (x) $ to $D3$-branes.    Derivative couplings   
to  pairs of  gauge boson or gaugino (singlet) fields appear  
explicitly   in the  Einstein frame  formula of 
Eq.~(\ref{eqredd3}).  
Non-derivative  couplings    with the  $SU(3)$ singlet  and triplet
matter   fermions also  arise     through  the  mass terms   in
Eq.~(\ref{sub4eq9}), which     vanish   in the  GKP
background   but   are  finite  for  deformed   vacua  with $ G_-  \ne
0 $.     It is important to  note that a complete  description of
the  supergravity dilaton   modes   should     
take    mixing  with  the internal metric and 
2-form  fields  into account, as is borne out  from the   dimensional
reduction  analysis  in flat  background spacetimes~\cite{han98}. 
The physical  4-d dilaton field    arises  there   as a mixture of
the  vector and scalar   fields   from  the  metric tensor components,    
$ A_\mu ^{(n)} \in \d g_{\mu m} ,\ \phi ^{(n)} _{mn} \in \d g_{mn} $,
so that     the   modes   of the metric tensor Weyl  rescaling  fields,  
 $\tilde \phi  ^{(n)} =\sum _{m=1}^6   \tilde \phi ^{(m)}_{mm}  $,  
 couple  at linear order,   similarly to  the    
graviton  modes $ h ^{ (m) } _{\mu \nu } $,    through 
terms involving the      trace  of the brane stress-energy tensor. The
resulting   formula  derived in this context is   
 \bea && \d L_4 ( h , \phi ) = 
{\Psi _m (y_\star ) \over   M_\star \sqrt 2 }  ( h ^{(m ) \mu \nu } 
T_{\mu \nu } (x) + \o  \phi  ^{(m )} (x)  \tilde g ^{\mu \nu } T_{\mu \nu
} )  ,\ [\o =\sqrt { 2 \over  3 (D-2 )  } =  
{1\over \sqrt {12} } ] .   \eea   
   
The interactions of bulk modes with  $Dp$-branes  depend  on  
the brane  embedding in the background   spacetime  which is
 described  by    the VEVs  of  open string moduli  fields. 
   The coupling constants  are  given by  overlap 
integrals of   bulk and brane wave functions over the  volume of the 
wrapped $(p-3)$-cycle.    For   $D3$-branes,   no integrations 
are  needed.   On kinematical grounds, introducing a
$D3$-brane at a generic point of the conifold base,  $ T^{1,1} \sim
SO(4) / U(1)_H$,  breaks the isometry group $ G= SO(4) \to SO(3) $, 
hence reducing   the  moduli space of vacua    
to the direct product of the radial direction 
times $SO(4) / SO(3) \sim S^3 $.  (Note that 
the warped deformed conifold  near the conifold apex  shrinks  to 
an $ S^3$ of finite radius    whose  Euler angles     are
non-trivially related to the angles in the
standard parameterization.)  One  could invoke a  variety of
localization mechanisms  for $D3$-branes.  
For instance, the stabilization of
the universal volume modulus $\rho $ and open moduli $y^m$, using the
instanton induced superpotential from $D7$-branes or (Euclidan) $\cale
4$-branes,  is   found to lift the $D3$-brane moduli space to point,
circle or 2-torus ($ S^0,\ S^1, \ T^2 $) loci~\cite{dewolfewood07}.
If the  warped throat is an orientifold, one could localize $D3$-branes by
placing them on top of $ O3^{\pm }$-planes at fixed points.  The
situation for $\bar D3$-branes is different since these feel a
potential that attracts them to the apex.  The localization  is 
then achieved only if the $ \bar D3 $-branes are placed on top of $
O3^+$-planes~\cite{aharony05}.

For a $D3$-brane located  at   the poles   $ \t _{1,2}  = 0 ,\ \pi  $
of    the 3-spheres  $T^{1,1} = S^3 _{1} \times   S^3 _{2} $,  
the only  scalar bulk modes that  couple  
 are those of  vanishing  charge  $ r= m=-n =0$, as 
follows from the property of harmonic functions  $
Y ^{jl r } (\t _i, \phi _i, \psi ) \vert _{\t _{1,2}=0 } \propto \d
_{r 0} $, valid  for arbitrary $ j,\ l$.  
The  coupling  for  modes of  finite $ j,\ l ,\ r\ne 0 $  must 
then occur   through  admixture of harmonic components 
$j' ,\ l',\ r'= 0  $.  For comparison, note  that  in  the  
toy  model of~\cite{kofman08}  where  the compact base 
is  replaced  by an $S^2$ of isometry group $SO(3)$,
the space of vacua of $D3$-branes,   given by the coset   
$ SO(3)/SO(2) \sim S^1$,    allows relocating $D3$-branes at the   
poles without loss of generality.   The   bulk modes of   finite
magnetic quantum number   have  then vanishing couplings,    due to 
 the    familiar property of spherical harmonics 
$ Y_{lm} (0,\phi ) \propto \d _{m, 0}$.

\section{Properties of warped  modes}
\label{sec5}

\subsection{Choice of   parameters}
\label{subsec5.1} 

The parameter space of  our  study of    warped modes properties
consists  of the  string  theory coupling constant and 
mass scale $ g_s , \ m_s =1/l_s $ and the compactification 
data involving   the 3-fluxes    $ M ,\ K ,\ [N= MK]$   the mass
hierarchy  relative  to the string scale~\cite{chialva05} $
w _s \equiv {m_{eff } / m_s } $ and the warped manifold volume $V_W =
(2\pi L)^6 $. There appear a  number of  auxiliary  quantities
that  are explicitly  related to the  basic parameters. Two 
familiar  examples    are the throat curvature radius  $\calr  $ and  the
warp  factor $ w_s  \simeq \min ( e ^{A(r) } ) \simeq e ^{-2 \pi K/(3 g_s M)
} $. One could also  define the mass hierarchy in terms of the warp factor referred  to Planck mass  $ w = {m_{eff } \over M_\star } $.  While there is no  preference in utilizing  $ w $ in place of $ w_s$, it is
important to   settle on    a specific   definition since this 
choice   affects  the parametric dependence  of
various quantities.  We shall present predictions    as a 
function  of   the   warp  factor $w$ since this  
yields a parameterization   closest to that
employed in  phenomenological warped models~\cite{rs9905}. 

Another   useful auxiliary quantity  is the universal volume modulus of the internal manifold volume  $\calv _e $  in string units  which enters
the Kahler potential of the 4-d effective theory in the standard
Einstein frame~\cite{gkp01}, $ K= - 2 \ln \calv _e $.
This is   related  to   the volume $\calv $  in our  special
Einstein frame by $\calv _e =  g_s ^{-3/2} \calv $. 
Although these parameters  are  usually introduced  
for  compactifications  in the regimes  of large volume,   
dilute flux or weak warping, it seems  reasonable  
to relate them   to   the warped volume $ V_W $ 
in our special Einstein frame,  by  assuming     the approximate
relation, $ \calv _e = g_s ^{-3/2} \calv = g_s ^{-3/2} V_W
/ \hat l_s ^6 $.  (Since  $ V_W$ is same as the string frame
volume $ V_W ^s$ for a constant dilaton,  one     could  define
the  latter  by the formal  relationship
$ V_W ^s \approx  g_s ^{-3/2} V_W e^{3 \phi /2}.$) 
All quantities of interest   will  be    expressed in terms 
of the  pair  of  dimensionless parameters 
\bea && z= M_\star \calr = {\l _Ng_s^{1/4} M_\star \over m_s },\ \eta = {L
  \over \calr } = {V_W ^{1/6} \over \hat l_s \l _Ng_s ^{1/4} } =
     {\calv  ^{1/6} g_s ^{-1/4}  \over \l _N} = {\calv  _e^{1/6} \over \l _N} ,\  [\l _N= ({27 \pi N \over 4})^{1/4} ,\ N= MK ] . \eea 
The physical parameters $m_s, \ M_\star ,\ \calr ,\ w ,\  E_m $ 
satisfy the useful   relations  
\bea && \bullet \  M_\star = m_D ^4 \sqrt { V_W} = {2
  \sqrt {\pi V_W} \over g_s \hat l_s ^4 } = {L^3 m_s ^4 \over \sqrt
  {\pi } g_s} = {m_s \sqrt {\calv } \over \sqrt {\pi } g_s} ,
\ \calr = { \l _Ng_s ^{1/4} \over m_s } ,\ w = w_s g_s \sqrt {\pi \over
  \calv }  . \cr &&
\bullet \   m_s = {\pi ^{1/8}g_s^{1/4} M_\star \over (\eta z )^{3/4} }
,\ m_s \calr =   
({\sqrt \pi g_s z \over \eta ^3 })^{1/4} = {\sqrt \pi g_s z \over \sqrt
  {\calv }} ,\ m_K \equiv { E_m } = x_m { w \over \calr }= x_m
{M_\star w \over z}   .\eea
The    following   natural inequalities   among  physical scales,   $1/ M_\star < l_s  < \calr < L $,   entail  $ \eta > 1, \ z > 1 $. 
The typical values assigned to $\eta $ in   warped models  
are of order $\eta = L / \calr \approx 10 $~\cite{chen06}    while  the values   allowed  for $z = 1/c $ by experimental and theoretical constraints    cover  the 
interval~\cite{davoudheriz00,davoudheriz02} 
$z \in  (1 , 100) .$   The empirical value $ m_s L =
L/ l_s \simeq 5 $  selected     in the  brane 
inflation  scenario~\cite{kklt03,kklmmt03} is   consistent with
the value $V_W
^{1/6} = 5 l_s \ \Longrightarrow \  \calv  ^{1/6} = 5 / (2\pi ) $,  
favoured   by  the analysis of moduli stabilization and supersymmetry
breaking~\cite{choi05}.  The large volume scenario  favours larger values~\cite{bala05}, ranging from $ \calv \sim
10^{30}$ in TeV scale models, to $ \calv > 10^{15}$ in partially
sequestered models, to $ \calv > 10^2 $ (or $ \calv > 10^{12} $) in
sequestered models with matter on $D7$-branes (or $D3$-branes) and to
$ \calv \sim 10^{5}$ in studies of flavour constraints using anomaly
mediation of supersymmetry breaking~\cite{spdealwis12}.

It is perhaps  meaningful to  consider the  estimate   for
the warped volume  obtained   by splitting the radial interval into bulk and throat parts.  Setting $ V_W  = V_{W, bulk} + V _{W, thr} = \int d^6 y \sqrt {\tilde g _6} e ^{-4A} , \ [V_{W, bulk}  \simeq c \int d^6 y \sqrt {\tilde g _6} ] $ in correspondence with  the  warp profile
splitting,  $ e ^{-4A} = c + c_2 (\calr / r )^4 \ln {(r/r_{ir})} $,
we  then   evaluate  the conifold throat volume   along same lines as in
the calculation of the complex structure modulus metric in
Eq.~(\ref{eqGcs})     for  a   radial  integral  extended over the segment
 $ r\in [r_{ir}= \a ' \L = \a ' \vert S \vert
  ^{1/3} \simeq w \calr ,\ r_{uv}= \a ' \L _0 \simeq \calr ]$,
\bea && V _{W}^{thr}    \simeq  c_2 V_5 \int _{r_{ir} } ^{r _{uv} } d r r ^5 ({\calr \over r })^4 \ln {r \over r_{ir} }  =  {3\pi ^3\over 2}  (g_s M )^2 \a ^{'      4}  \vert S \vert ^{2/3} [1 + \call ^2 (2 \ln \call -1)
] ,\   [c_2 = {3 g_s M \over 2\pi K} ,\
 \call = {\L _0 \over \vert S\vert ^{1/3} } = { r_{uv} \over r_{ir}  } ].   \eea 
In the   large $\call $ limit  of strong warping, the    throat volume      is dominated  by the term $\call ^2 \ln \call $  inside the square brackets.
Using  the relation  $ c_2 \simeq  1  / \ln \call   $  from Eq.~(\ref{eq.warpA1}), one   can rewrite the  above formula as,
 $V _{W, thr}\simeq ( (2\pi )^3 \calr ^6 /  27 )   (1 + {1\over  2 \ln \call }
 (\call ^{-2} -1  )) .$ Relative   to the total volume $ W_W \approx (2\pi L)^6 $, the    leading contribution   to the   throat volume
is  thus represented by the   small fractional  ratio $V _{W, thr} / V_W \simeq 1 / (27 (2\pi )^3 (L/\calr ) ^6 )$.    This    is strongly suppressed for $\eta \equiv L/\calr = O(10)$   and  exceeds unity  for $\eta  < 4 $. 
Unlike the case of unwarped compactifications, where the dependence
on the complex structure modulus $S$  in the corresponding formula for the 
total volume, $ V_{CY}= V_{bulk} (S) + {8\pi ^3\over 81}
\vert S\vert ^{2} (\call ^6 - 1 )$,    should cancel out  between the
bulk and throat parts~\cite{dougtorr08},
the  possibility of canceling the $S$-dependence in $V_{W, bulk}   (S)  + V_{W,thr}  (S)$  is   less clear. 

\subsection{Results for mass  spectra}
\label{subsec5.2} 

The    particle     spectrum  of the reduced field theory  consists  
of   massless  states    separated   by a wide gap  from the towers  of  massive  states.   The modes     in   $\caln =1 $   supermultiplets 
and irreducible representations  of the throat isometry (flavour
symmetry) group,   $G =
SU(2)\times SU(2) \times U(1)_R$, are  arranged into nine   
 (graviton, gravitino $(G.I,  \ G.II, \ G.III,  \ G.IV)$  and
vector $  (V.I, \ V. II, \ V. III, \ V. IV)$) multiplets  
of the superconformal   group  $ SU(2,2|1)$.   
The  states are  labelled by the conserved 
quantum numbers $ \nu = (j,\ l , \ r  \in (-\tilde j ,\cdots , \tilde
j))  ,   \ [\tilde j = min (j, l) ] $  of the isometry group $G$  and the 5-d  squared mass $ M^2_{5 \nu }  $  with  $ j,\ l, \ r /2 $  being all  either integers  or   half-integers.  

We  shall  consider a set of scalar, symmetric
2-tensor, vector  and 2-form fields $ (S,\ T , \ V,\ 2F ) $  
in low-lying representations   of  $G$. The $T $ field      occurs as a    diagonal mode $ h _{\hat \mu \hat \nu } $ in the  conformal graviton
multiplet  and the $S ,\ V $ fields occur   in the conformal
vector multiplet $V.I $ as lowest mass eigenvectors of the coupled systems 
$ S ^{(\nu )}  (\pi, \ b) $ and $ V ^{ (\nu )} _{\hat \mu }  (B^a_{\hat \mu }
,\ \phi ^a_{\hat \mu } ) $ in  singlet and vector representations of
the tangent  group $SO(5) $ of $ T^{1,1}$.
Amongst the various 2-form  fields in~\cite{ceresole99,ceresoII99}, we have selected   the  field $a_{\hat \mu \hat \nu } $  part of  the $G.I$   
multiplet  of $r$-charge related to that of the highest weight state
by   $ r = r _{hw} -1$.  
The  modes masses and  the  conformal dimensions $\D $ of  the
corresponding  gauge   theory  operators  are given by
\bea &&  S (S_\mp ) :\ M^2 _{5, \mp } =
H_0+16 \mp 8 \sqrt{ H_0 +4},\ \D ^- _\nu = \D ^+_\nu  -8 = -2 + \sqrt{
  H_0 +4} ,\ [\nu _ \mp = \sqrt{ 4 + M^2 _{5, \mp } } ]
. \label{sub5eq1} \cr && 
T (h) :\ M^2 _{5 } = H_0 (j,l,r),\ \D _\nu = 2 + \sqrt{ H_0 +4} , \  [\nu =\sqrt{ 4      + M^2 _{5} } ] .\cr &&
V(V ^a_{\mp} ):\ M^2 _{5 \mp} = H_0 +12 -\mp 6
\sqrt { H_0 + 4} ,\ \D _\nu = -1 +\sqrt { H_0 + 4} ,\ [\nu =\sqrt { 1 + M^2
    _5 } ]. \label{sub5eq2}\cr && 2F (a _{\hat \mu \hat \nu } ):\ M_5
=\vert  - 2 + ( { H_0 (j, l , r _{hw} -1 ) +4 } )^{1/2} \vert
=\vert  - 2 +\sqrt{ H_0   (j,l,r) +4} \vert  ,\ 
\D _\nu = \sqrt{ H_0 +4} ,\ [\nu = \vert M_5 \vert ].
\label{sub5eq3}  \eea  
Each  harmonic   mode     splits  in turn into   sub-towers of  
radially excited normal modes of 4-d mass, 
$m _{K} \equiv E_m = x_m w / \calr ,\ [m= 1,\cdots ] $ where $ x_m$
take discrete values  obtained  from  Eq.~(\ref{sub3eq7})  as roots of
(linear combinations of) Bessel functions.  Since $E_m$ increases monotonically with increasing $ M_5 ^2 $,  the modes 
with  strongest impact  on phenomenology  should arise from $
AdS_5$ fields of lowest (negative,  zero  or positive) squared masses.  
\begin{table}   \begin{center} \caption{\it
      \label{tab1} Mass spectra  for  ten    warped modes 
selected  for the graviton, scalar, vector and 2-form    fields 
$( T, S, V, 2F) $  in   singlet  and charged representations 
of the     conifold  isometry  group $G$.  
The    first column entry lists the   quantum numbers  of the
ten  modes $C_i (j,\ l,\ r) ,\ [i=0,\   \cdots , 9] $    and
the  next   four  groups of column entries for  the $(
T, S, V, 2F) $  fields    split   each into three sub-columns listing  the   modes   5-d mass, the dual operator conformal dimension and the 4-d mass  $M_5, \ \D ,\ x_m = E_m \calr / w $.}
\begin{tabular}{|c| c c  c |  c c  c | c c  c |  
c c c |} \hline 5-d Mode & & Graviton (T) && & Scalar $(S_- )$ & & &
  Vector $ (V_-)$ & & & 2-form (2F) & \\ \hline $ C_i ( j \ \ l \ \ r) $
  & $ M_5 ^2 $ & $ \D _\nu $ & $ x_n $ & $ M_{5 -} ^2 $ & $ \D _\nu ^- $ & $ x_n
  $ & $ M_{5 -} ^2 $ & $ \D $ & $ x_n $ & $ M_{5-} $ & $ \D _\nu ^- $ & $
  x_n $\\ \hline $C_0 \ (0 \ \ 0 \ \ 0) $ & $ 0.  $ & $ 4.  $ & $ 3.83
  $ & $ 0.  $ & $ 0.  $ & $ 3.83 $ & $ 0.  $ & $ 1.  $ & $ 2.40 $ & $
  0.  $ & $ 2.  $ & $ 0.0 $ \\ $C_1 \ (\ud \ \ \ud \ \ 1) $ & $ 8.25 $
  & $ 5.5 $ & $ 5.44 $ & $ -3.75 $ & $ 1.5 $ & $ 2.17 $ & $ -0.75 $ &
  $ 2.5 $ & $ 1.84 $ & $ 1.5 $ & $ 3.5 $ & $ 2.46 $ \\ $C_2 \ (1 \ \ 0
  \ \ 0 ) $ & $ 12.  $ & $ 6.  $ & $ 5.98 $ & $ -4.  $ & $ 2.  $ & $
  1.599 $ & $ 0.  $ & $ 3.  $ & $ 2.40 $ & $ 2.  $ & $ 4.  $ & $ 3.05
  $ \\ $C_3 \ ( 2 \ \ 0 \ \ 0 ) $ & $ 36.  $ & $ 8.32 $ & $ 8.44$ & $
  1.40 $ & $ 4.32 $ & $ 4.18 $ & $ 10.05 $ & $ 5.32 $ & $ 4.96 $ & $
  4.32 $ & $ 6.32 $ & $ 5.67 $ \\ $C_4 \ ( 3 \ \ 0 \ \ 0) $ & $ 72.  $
  & $ 10.71 $ & $ 10.95 $ & $ 18.25 $ & $ 6.71 $ & $ 10.41 $ & $ 31.69
  $ & $ 7.71 $ & $ 7.53 $ & $ 6.71 $ & $ 8.71 $ & $ 8.27 $ \\ $C_5
  \ (2 \ \ 1 \ \ 2) $ & $ 45.  $ & $ 9.  $ & $ 9.15 $ & $ 5.  $ & $ 5.
  $ & $ 4.91 $ & $ 15.  $ & $ 6.  $ & $ 5.69 $ & $ 5.  $ & $ 7.  $ & $
  6.41 $ \\ $C_6 \ (1 \ \ 1 \ \ 2) $ & $ 21.  $ & $ 7.  $ & $ 7.04 $ &
  $ -3.  $ & $ 3.  $ & $ 2.73 $ & $ 3.  $ & $ 4.  $ & $ 3.51 $ & $ 3.
  $ & $ 5.  $ & $ 4.201 $ \\ $C_7 \ ( 1 \ \ 1 \ \ 0) $ & $ 24.  $ & $
  7.29 $ & $ 7.35 $ & $ -2.33 $ & $ 3.29 $ & $ 3.05 $ & $ 4.25 $ & $
  4.29 $ & $ 3.83 $ & $ 3.29 $ & $ 5.29 $ & $ 4.53 $ \\ $C_8
  \ ({3\over 2} \ \ \ud \ \ 1 ) $ & $ 26.2 $ & $ 7.5 $ & $ 7.57 $ & $
  -1.75 $ & $ 3.5 $ & $ 3.28 $ & $ 5.25 $ & $ 4.5 $ & $ 4.06 $ & $ 3.5
  $ & $5.5 $ & $ 4.76 $ \\ $C_9 \ ( 2 \ \ 1 \ \ 0) $ & $ 48.  $ & $
  9.21 $ & $ 9.37 $ & $ 6.31 $ & $ 5.21 $ & $ 5.13 $ & $ 16.73 $ & $
  6.21 $ & $ 5.92 $ & $ 5.21 $ & $ 7.21 $ & $ 6.64 $ \\ \hline
\end{tabular} \end{center}  \end{table}    \vskip 0.3 cm

\begin{table} \begin{center}
    \caption{\it \label{tabmodeszero}
Lowest lying  $AdS_5 $  fields in   harmonic modes of $T^{1,1}$
part of  the graviton, $ V.I $   and   $ G.I,\ G.III  $  superconformal multiplets.         The vector  fields  $\phi _{\hat \mu} ,\
\phi ^a _{\hat \mu } $   occur  as  gauge  bosons of the  Betti gauge
multiplets  and of the isometry   group $ G = SU(2)_j\times SU(2)_l \times U(1)_r   $ in  the graviton  and  $ V.I$     multiplets.  
The   flavour  singlet  fields $a_{\hat \mu  \hat \nu } ,\ a , \  b^-  _{\hat \mu  \hat \nu } $   in  $ G.I,\ G.III  $ multiplets  arise  partly from  Betti
multiplets.  The scalar fields $ S ^-  $ from the lowest eigenvalue of the $ (\pi , b )$ system in $ V.1$ multiplet  include  charged
modes $( \underline {01} 0)
= (100) \oplus (010) $  and   singlet $(000)$   modes.
The  last  line array    lists
the   4-d mass parameter $ x_0 = E_0 \calr / w $     for   
the   massive mode  immediately  above    the     massless    mode.} 
  \begin{tabular}{|c| ccc|| cccccc|| cc|| c|} \hline
& &  Graviton & &     & &  $  V. I $ & & & &    $ G.I $ & &  $ G.III
    $ \\   \hline  
Modes &$ g _{\hat \mu \hat \nu } $&$ \psi _{\hat \mu } ^{L,R}  $&$
\phi _{\hat \mu } $&$  
    \phi _{\hat \mu } ^{a} $&$   \phi _{\hat \mu }  $&$ \phi _{\hat \mu } ^{a} $&$
    \psi _{(L) } ^{L,R}  $&$ S^-(\pi , b) $&$ S^- (\pi , b) $&$
a _{\hat \mu \hat \nu }  $&$ a $&$  b ^-_{\hat  \mu \hat  \nu }$ \\ \hline 
    $(jlr)$ &$ (000)  $&$ (00\pm 1) $&$ (000) $&$ 
    (\ud \ud  1 ) $&$ (000) $&$ (\underline {01} 0) $&$  (\ud \ud  1 )  $& $  (\underline {01} 0) $&$ (000) $&$  
    (000) $&$   (000) $&$  (000) $ \\
 $M_5 ^{2}  \ \  ( M_5)  $  &$ 0 $&$ 0$&$ 0 $&$
- {3\over 4} $&$ 0 $&$ 0 $&$ 0$&$ -4 $&$ 0 $&$
 0 $&$ 0 $&$  0  $  \\  \hline 
$x_0 $&$ 3.83  $&$ -  $&$ 2.40 $&$
 1.84 $&$ 2.40 $&$ 2.40 $&$  - $&$ 1.60 $&$ 3.83 $&$
 0 $&$ 3.83 $&$  0  $  \\ \hline 
\end{tabular} \end{center}   \end{table} \vskip 0.2 cm

\vskip 0.2 cm \begin{figure} [ht] \centering
\begin{center} 
\epsffile{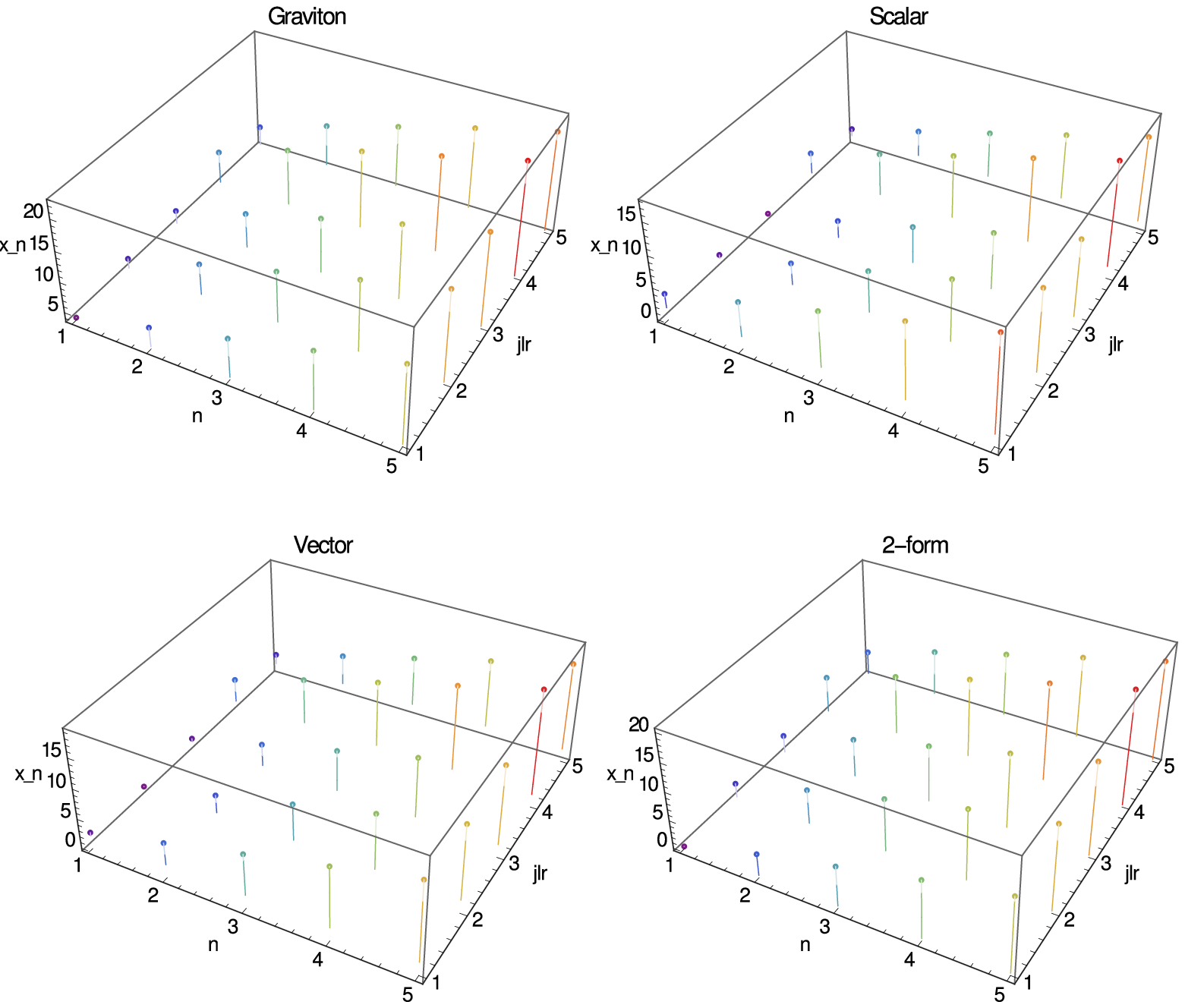}
\end{center}  \vskip 0.2 cm
\caption{\it \label{wf0} The dimensionless mass parameters $ x_n = E_n
  \calr / w $  of the (graviton, scalar,  vector and 2-form)
  $ T, \ S,   \ V , \ 2 F$ fields are plotted in  four  distinct  panels.
Each panel   refers to the 6 modes $ C_0( 0,\ 0,\ 0 )
  , \ C_1 (0.5 ,\ 0.5,\ 1 ),\ C_2 ( 1,\ 0,\ 0 ), \ C_5 ( 2,\ 1,\ 2
  ),\ C_6 ( 1,\ 1,\ 2 ) $  labelled by  $(jlr) =
      [1, 2, 3, 4, 5]$ along the frontal axis  with  the radial
 quantum number $ n = m+1 = [1 ,\ 2, \ 3 ,\ 4,\ 5]$ for the
      ground state and first four radially excited states along the
      sideways axis.} \end{figure}

We  have  computed   the  4-d  mass spectrum  for  10 modes $ C_0
,\cdots ,\ C_9 $  of each  of the $ ( T, S, V, 2F) $ fields
in low-lying $ (j,\ l, \ r ) $ representations of  
$G$. The   ground state mass parameters   $ x_0 =\calr E_0 / w$,
displayed  in Table~\ref{tab1}, 
are seen to  grow  with increasing $j,\ l $  
quantum numbers,    the steepest growth occuring for gravitons. 
It is  instructive to  compare   with the    estimate  for the  
ground state  graviton   mode  mass~\cite{kofman05} 
\bea && m_K  \equiv  { x_0 w \over \calr }   \simeq {w \over
  \calr _- } \ \Longrightarrow \ x_0 \simeq { \calr \over \calr _-} =
c_2 ^ {-1/4} = ({3 g_s M \over 2\pi K}) ^{-1/4} = (\log w_s ^{-1})
^{1/4} \equiv {1 \over \hat \s } . \eea    
For    values   of the   warp factor 
$w_s \in (10^{-5} \to 10^{-14} )  $  or 
$  \hat \s \in (0.55 \to 0.41 ) $,  the  resulting  value
$ x_0 \simeq 2 $    is   about   half the    calculated   one
$ x_0 =3.83$.     From Table~\ref{tab1}, we see that 
 the lightest mode  of mass parameter  
$ x_0 \simeq 1.6$  arises for the scalar field  eigenmode   
 $C^S _2 (\underline{10} 0) =  S^-(\pi , \ b )$   whose squared  mass
 $ M^2_{5 -} = -4 $   saturates   the unitarity bound.
 The $V$ and $ 2 F$ ground state  modes
 $C^V _1 (\ud \ud 1) ,\   C^{2F} _1 (\ud \ud 1)) $  
have also low  masses  $ x_0= 1.8$ and $ x_0= 2.46 $. 
The   corresponding states for graviton   modes     lie
significantly  higher,     $ x_0 (C^T _{ [0,1,2]} ) = [3.83,\ 5.44
  ,\  5.98 ].$  

The mass spectra  of  radial excitation modes of the $ ( T, S, V, 2F)
$    fields  are plotted in Fig.~\ref{wf0}.
The approximate   estimates   for gravitons,     
  $J_\nu (m_{n} / r _{ir} ) =J_\nu (x_{n})= 0 \ 
\Longrightarrow \ x_{n} \simeq \pi (n
+ (2\nu +1) / 4 )   ,\ [w = r_0/\calr = e ^{-Y_{R} /\calr } ]
$,   yield  values $ x_0  \simeq  3.92,\ x_ {m_n} \simeq 3.14  n $  that are numerically close to  the calculated ones, 
$ x_{m}  = (3.83, 7.015, 10.17, 13.32, 16.47) ,\ [m=n-1= 0,1,2,3,4]$.
It is   useful   to determine  the substringy radial excitations  
of warped  modes   of mass   consistent with       the
upper bound  $   m_K \leq w
m_s \Longrightarrow  \ x_m \leq m_s \calr = \l _N g_s ^{1/4} . $ With the   illustrative  parameter
assignments,  $ N= MK \simeq 10^5 , \ g_s \simeq 0.1 $, yielding $ \l
\simeq 10^{3/2}$, the resulting  upper bound  $x_m
\leq 15 $    is seen to  extend  out to the third excited   graviton level, 
$x_4 =13.3 $.  

The  axio-dilaton  mass spectrum   lies  higher   than that of the  graviton due to  the additional  flux   mass term in  Eq.~(\ref{sub2eq19}). The
singlet modulus mode  deserves special  attention if one recalls that 
this the flux  superpotential    in the 4-d description  contributes   a  large  mass insensitive to warping effects.   The    dependence on the  combination $  ( c + e ^{-4 A } ) ,\  [c\simeq \calv ^{2/3}] $   in  the  10-d description  suggests an   interplay between 
warping and  finite  volume  effects.   
Since  fluxes  are localized near the  collapsed   3-cycle 
at  the throat-bulk interface, a  consistent     treatment   
requires    going beyond  the   hard wall approximation. As    noted 
 in~\cite{frey06}, upon  increasing  the  $c $-field VEV  
up  from the  strong warping      regime $  c
<< w ^{-4} $,   the  mass  of the dilaton modulus      decreases   
linearly with $\ln c $      from values 
set by  the  warped string  scale  mass,   $ m_\tau \sim m_f w \sim
w m_s / \sqrt { M g_s } $,  and    drops  sharply to  
the semi-classical value $m_\tau  \sim m_f  / c = n_f m_s /\calv
^{1/2} $     in   the  large  volume regime  $c >>
w^{-4} $.   Meanwhile, the    radial wave function,   initiallly
localized  at the  conifold tip,   gets    non-localized. 
Using a  different normalization for the 4-d metric,  the following 
parametric   predictions for the dilaton mass at small and large volumes, 
$ m_\tau \sim  w m_s /( n_f
 ^{'1/2}  \calv ^{1/3} ) , $ and $  m_\tau \sim n_f m_s /c^{3/2} \sim
n_f m_s / \calv $  were obtained in~\cite{burgessSB06}.
It is also pointed out in this work 
that  the gravitino mass  from  supersymmetry breaking   
in  the  bulk at the breaking scale $ M_S$, 
features  a   similar drop  from  $ m_{3/2}
\sim M_S w /(c n'_f )^{1/2} $ to  $ m_{3/2} \sim M_S / \calv $ upon going  from the throat  to the bulk domination  regimes.  

The pseudo-moduli are mostly supported in the
bulk where couplings are set by the unwarped gravitational mass scale
$M_\star $, while their wave functions in the throat are not well
documented. For this reason, we examine   from a qualitative  standpoint
the stability properties of warped modes.
The open channels for bulk processes are
restricted by selection rules imposed by the throat isometry and the
wave functions orthogonality,
while the decay processes of warped modes
to the open string moduli are sizeable only for branes embedded near
the throat tip.

To learn which decay channels are accessible, one
must have some idea of the quantum numbers of both the lightest
charged Kaluza-Klein particles (LCKPs) and the pseudo-moduli arising
from 5-d fields of vanishing or negative squared mass $ M_5 ^2 $.
Examples of the latter are the massless graviton and Nambu-Goldstone
bosons and (Abelian or non-Abelian) gauge bosons of non-anomalous
symmetries.  A subset of these may acquire finite masses from flux
effects.  For instance, the chiral anomaly in the conifold $U(1)_R$
symmetry produces gauge boson mass proportional to the 3-flux
$M$~\cite{klebow02}.  The scalar and pseudoscalar massless
bosons~\cite{krasnitz00,GHK04} from the spontaneously broken baryon
number symmetry $ U(1)_b$, along the baryonic branch on which the
Klebanov-Strassler background is defined, get lifted by ultraviolet
effects when the conifold is embedded in a compact
model~\cite{benna07}.  
The lowest lying 5-d fields from the analysis~\cite{ceresole99}
are seen  in Table~\ref{tabmodeszero} to belong to singlet, doublet and triplet
representations of the flavour subgroups $ SU(2)_{j, l} $.  
(The  massless singlet scalar mode  $C_1 ^{2F} (000)$  is   disregarded   since it is    likely  to acquire a  mass by mixing with  other scalar modes.)
The most robust candidates for LCKP modes are then $ C_2 ^{S^-} ((100), x_0= 1.6) , \  C_1 ^{V^-} ((\ud \ud 1 ), x_0= 1.8) ,\ C_1 ^{2F }  ((\ud \ud 1 ),   x_0= 2.5)  $.

In anticipation of the  detailed discussion in Section~\ref{sec6},   we
here    try to identify the  allowed two-body decay processes for warped modes, assuming that  all   rates  are of   roughly comparable size.
The large   rest  mass in the decay chains $ H ^{(n)} \to H
^{(n-1)} \to  \cdots ,\ [H= ( h = C_i ^{T}    ,\ C_i ^{S, V, 2F}  ) ]$
suggests that  the radially excited modes fastly  decay, leaving
behind the ground states of the lowest mass charged modes $ H ^{(0)} $
free to decay to massless singlet or charged modes.
While the decay channel  for the first radially excited singlet graviton
mode $h ^{(1)}  ((000),  x_1 =7 ) \to h
^{(0)}  + \bar h ^{(0)}  ( (000), x_0 =3.83) $ is energetically
closed, the decay  channels  of the ground state  graviton
to  pairs of charged scalar or  vector modes $ h ^{(0)}  ( (000) )
\to C^{S^-} _2  + \bar C^{S^-} _2 ,\ h ^{(0)}  ( (000) ) \to
C_1^{V^-} + \bar C_1^{V^-} $ are open.
Other open decay channels occur for $ C_i ^T \to C_i ^{S^-} + C_0
^{S^-} $ for $ i= [2, 6, 7, 8] $ and $ C_i ^T \to C_i ^{V^-} + C_0
^{V^-} $ for $ i= [1, 2] $.
It  thus appear  likely that a population of  LCKPs
might be    produced  and left over as metastable  thermal relics.

\subsection {Results for bulk and brane couplings  of  graviton modes}    
\label{subsec5.3}

\begin{figure} [ht] \centering 
\epsfxsize =6. in
\begin{center}
\epsffile{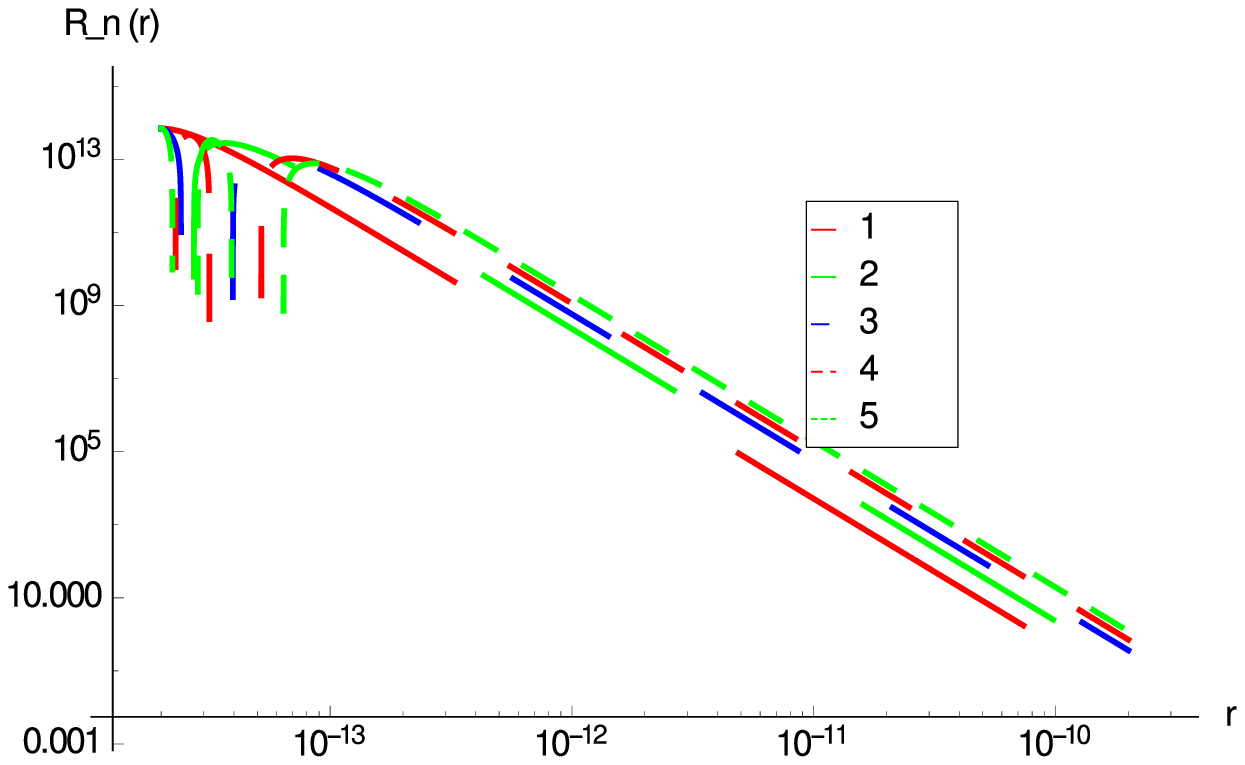} 
\end{center} \vskip 0.2 cm
\caption{\it \label{effpot1}   Double logarithmic plots of  the radial
  wave functions $ R _m (r) $ versus   $r$ for the ground state
and first four radially excited singlet graviton modes.
The  wave functions  in the hard wall model  with Neumann    boundary conditions    at the
infrared and ultraviolet  boundaries  $ r_{ir} =w ,\  r_{uv} =  \calr $
are   shown by    curves   with dashes of decreasing  lengths 
for  the  5 modes  of mass  parameters  $x_m= [3.83 ,
  7.01, 10.17, 13.32, 16.47]$,   numbered   $ 1,\cdots , 5$ in the legend.
We set the  warp factor at the value $ w = 2\ 10^{-14}$
and use   length units  $\calr =1$.
The effective gravitational scales    have  equal    absolute
values   in all  cases,   $\vert \L _{KK}  ^{(m)}  \vert  
= \vert M _\star /    R _m (r_{ir})  \vert    \simeq 28  $ TeV.}  
\end{figure}

\begin{figure} [ht] \centering 
\epsfxsize =6. in
\begin{center}
\epsffile{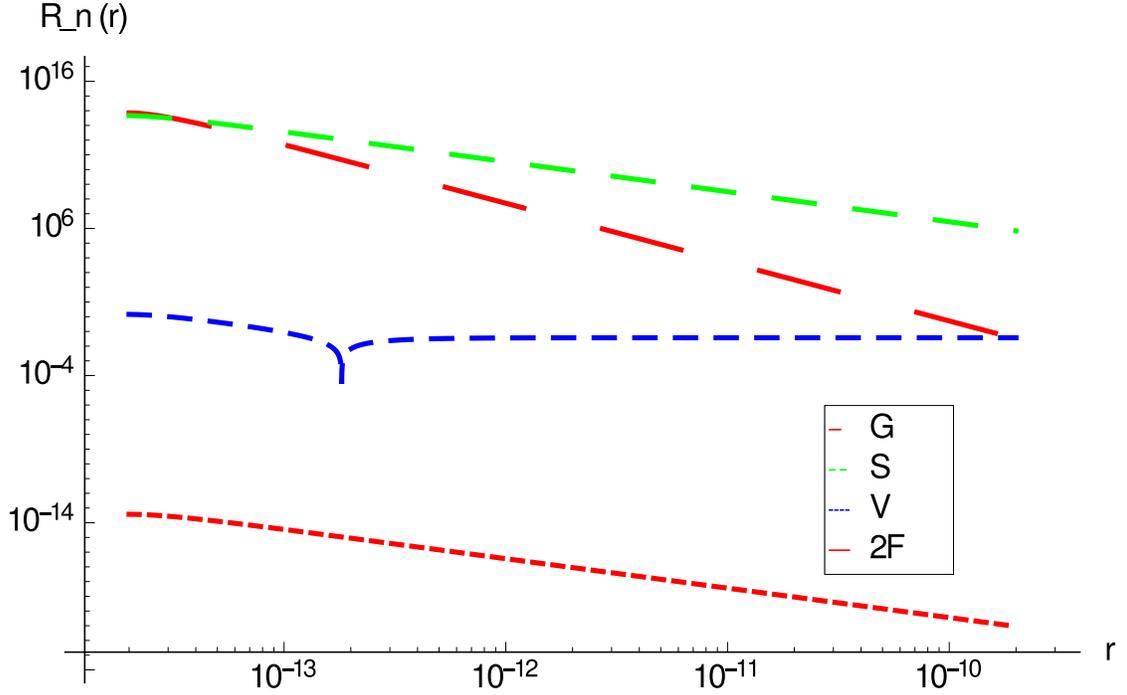}
\end{center} \vskip 0.2 cm
\caption{\it \label{effpot1.2}  Double
  logarithmic  plots  versus $r$ of the radial wave functions $ R _m
  (r) $  near the tip   for  the warp factor value $ w = 2\ 10^{-14}$
  and   length units  $\calr =1$. 
The   4 curves  ordered  from top to botton  refer to the 
$(S, G, V, 2F)$ cases.  
The singlet  graviton mode  $ G:\  (jlr)=(000),\ M_{5}^2
=0 $  and the  charged   scalar,   vector and 2-form    modes 
of quantum numbers $ (jlr)=(100)$ and   parameters
$ [M_{5-}^2  ,\  \nu ,\ x_m ]$   given by $  S^-  (\pi , b):\
[ -4, \ 0,\ 1.59] ,\  V ^- \ (\phi  ^a _{\hat  \mu }  ,\  B ^a_{\hat
  \mu } )  :  \  [ 0, \ 1,\ 2.40 ] ,\ 2F  \ (a_{\hat \mu \hat   \nu }
)  :\ [4  \, 2 ,\  3.05 ] $  are shown by   curves with dashes of
decreasing  lengths.}  
\end{figure}

\begin{figure} [ht]  \centering 
\epsfxsize =6. in 
\epsffile{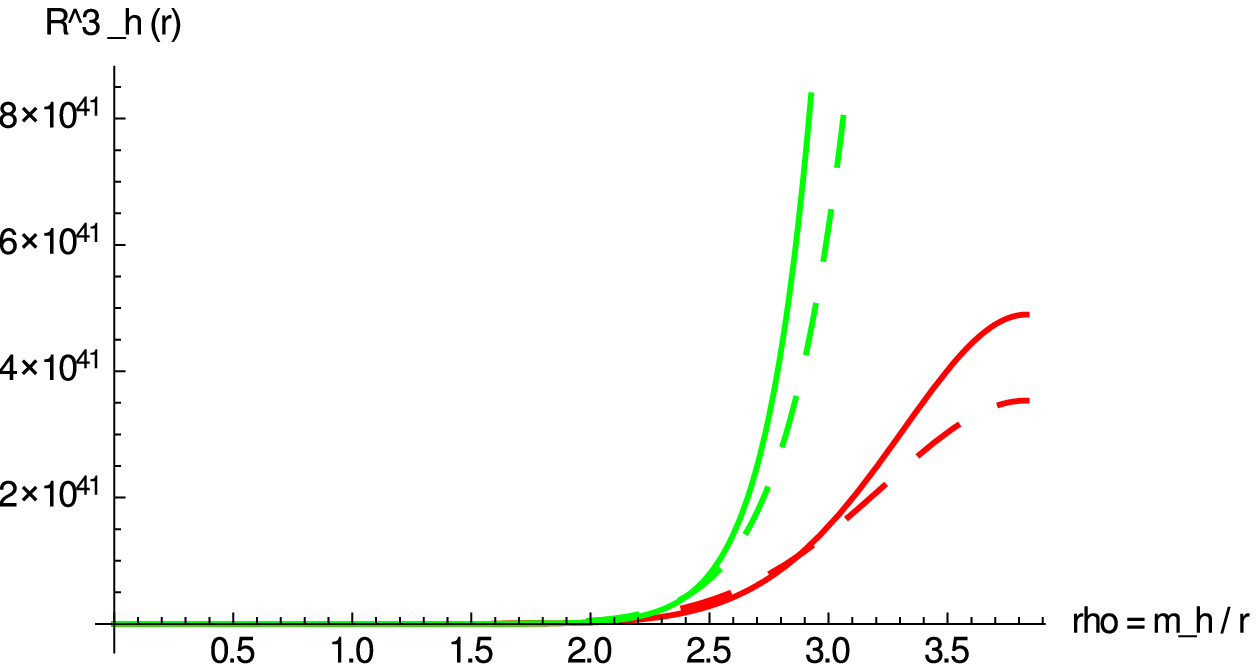}
\caption{\it \label{wf8}
  Logarithmic plots of the radial integrands $ r R_{m_1} (r) R_{m_2}
  (r)R_{m_3} (r) $ (in units $\calr =1$)  versus the radial variable $
  \rho = x _1 w / r $  near the   tip   
for  the warp  factor  value   $  w = 2\ 10^{-14}$ in  two
  configurations of graviton modes. 
The integrand for ground state modes $ h_1h_1h_1$
  is drawn in full lines and that involving two excited radial modes $
  h_1h_2h_2$ is drawn in dashed lines.  The (red) curves on the right
  are obtained with the exact wave functions and the (green) curves on
  the left are obtained in the tip approximation.}
 \end{figure}

One   general feature  of warped        modes 
is  the strong  peaking of their radial wave  functions 
near the  conifold tip. This is   apparent  for  gravitons  from    Fig.~\ref{effpot1}. 
After   rapid oscillations    in the internal  region
of  frequency   increasing   with  the mass $E_m $,  the wave  functions 
reach  maximal values  $\lim  _{r  \sim w} R_m (r) \sim 1/w $,   
followed by   exponentially falling tails  in the  external   region,
of   size  increasing   with   $E_m $,  in agreement with
the  asymptotic formula   
\bea && R_m (r ) \simeq {(x_m w ) ^{1 + \nu } \over
  2 ^\nu J_\nu (x_m) \G (1+\nu ) (x_m ^2 + 4-\nu ^2 ) ^{1/2} } (r
/\calr ) ^{-(\nu +2) +J} ,\  [r  \sim E_m] . \label{tiprwfeq1}  \eea 
The comparison in Fig.~\ref{effpot1.2}  of wave functions   for the  
$ (S,\ T )  ,\   V , \  2F $   fields    reveals large disparities in sizes which are  explained by the   different radial  dependence  $ r ^{J -2-\nu  } $ and normalization  constants  $ N_m = \int dr r ^{[1, -1, -3]} R_m  ^2 (r)$    for $ J=0 ,\  1, \  2$.  
The larger tail for  the scalar field   is due  to  its
smaller mass. The smaller size   of radial wave functions  for  vector and 2-form  fields    are  always  compensated in the  overlap integrals 
for     matrix elements  by   the
warp factors   ($ e ^{-4A} $ or $ e ^{-6A} $),
resulting  in   couplings  of same size   for all  field    types.  
To  assess the  importance  of the tip  region, we show in  Fig.~\ref{wf8}  
plots of the    integrands   for  cubic  couplings of  graviton modes.
The  large   peaks  and   exponential falls beyond  the  infrared
 (for decreasing  $\rho $)   are   clearly  reproduced. 
The   suppression  effect   for   modes  of larger  mass is moderate.
The comparison of the   exact    integrands   with
those evaluated   with  radial  functions  in the tip
approximation, Eq.~(\ref{tiprwfeq1}),  shows  that the   discrepancies
widen  with  increasing  $\rho $. 
\vskip 0.2 cm 

\vskip 0.2 cm 

\begin{figure} [ht]  \centering 
\epsfxsize =6. in
\epsffile{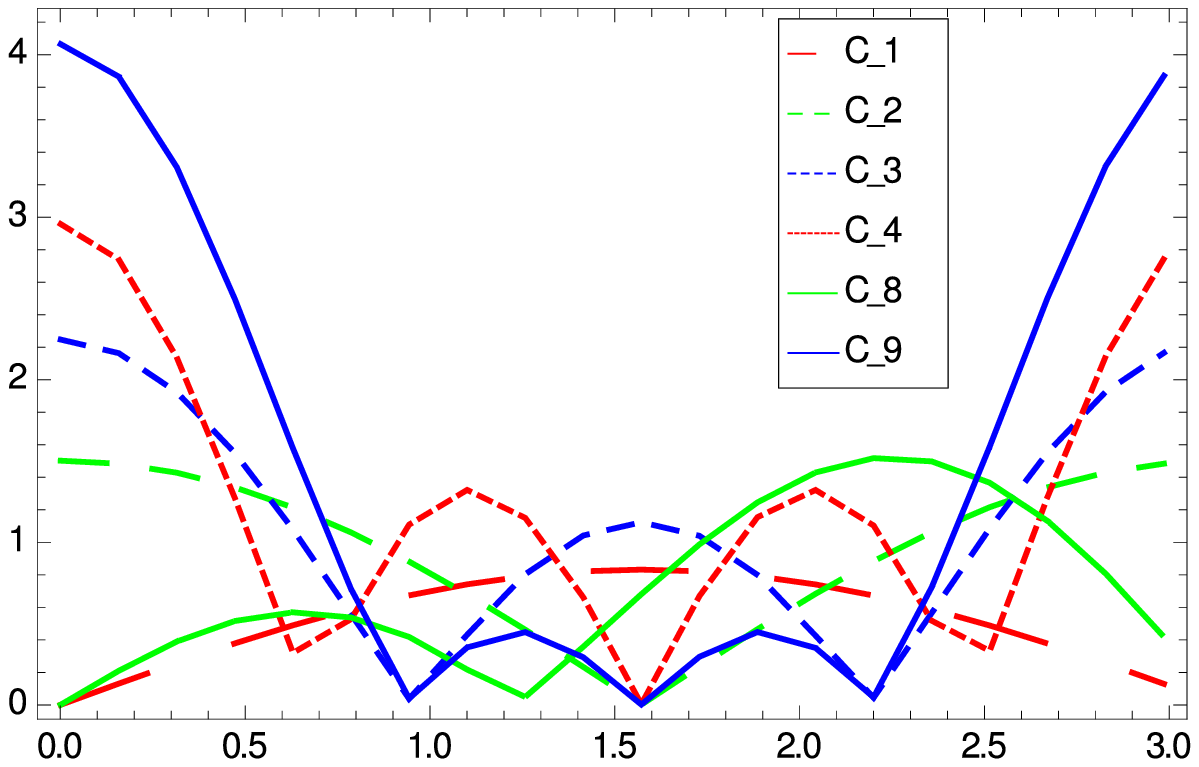}
\caption{\it \label{wf9p}  Plots    of the   coupling constants of
singlet and charged graviton modes  with  $D3$-branes
$w \vert \l _3 ^B \vert   \propto w
  \vert   R_m (r_\star ) Y^\nu (\T _\star )\vert $
  at   the  radial distance  $ r _\star = r_{ir} $
  as a function of the  polar angle  $\t \in (0, \pi ) $   
at points  of the base manifold $ T^{1,1}$ with
equal  polar angles   $\t = \t _1   =\t _2 $  
and   vanishing azimuthal angles  $\phi _1 = \phi _2 = \psi =0 $. 
The six curves
with   dashes of decreasing  lengths 
refer to the   charged   modes 
$ C_1, \ C_2,\ C_3,\ C_4,\ C_8, \ C_9 $    of quantum numbers   
$\nu =  (jlr)= [(\ud , \ud , 1) ,\ (1,0,0) ,\ (2,0,0) ,\
  (3,0,0) ,\ ({3\over 2} , \ud , 1) ,\ (2,1,0)] $. 
Upon  moving  upwards  from bottom to top at $\t =0 $ or $\t =\pi  $,
one crosses successively the  curves for  the modes  $ C_1,\ C_8,\ C_2,\ C_3,\ C_4  ,\  C_9 $.} 
 \end{figure}

It is also instructive to  examine the  sensitivity   of angular   wave functions       to the position of $D$-branes  in the base manifold.   We shall restrict our   study   to   the dependence  on 
polar angles  of  $ T^{1,1}$,   since this is expectedly    stronger  
than that on  azimuthal angles. We  display in Fig.~\ref{wf9p}  plots for  a   set  of charged graviton modes.  The
oscillations    grow    in frequency and  
amplitude    for  increasing angular momenta $ j ,\ l$.
Moderate  peaks  appear   near the poles
$\t =0 $ or $\t =\pi $ except for the modes $ C_1$ and $ C_8$  of 
finite $r$-charge    whose  wave functions vanish  there. 
The rapid angular variations suggest that the mode dependence
of the  overlap integrals  of   wave functions products
should be moderate.

The dimensionless  coupling constants  $\hat \l _n $, 
defined in terms of  the  initial  ones  as,
$\l _n = 2 ^{n/2 -2 } ({164 \ \eta ^3   \over M_\star w })
^{n-2} \hat \l _n ,\ [n=3,\cdots ] $ are given by 
the overlap    integrals   of radial and angular   wave functions
in Eq.~(\ref{sub4eq3}) 
that we      evaluated   numerically   using  `Mathematica' tools.      
The   results for  the  cubic couplings $\hat \l _3  (
h^{(q)} C_l \bar C_l ) ,\ [l=0,\cdots , 9 ,\ q=0,1, 2 ] $  are
displayed   
in  Table~\ref{tab2}   for    a set of  allowed configurations
involving the   list of     modes $h= C_0 , C_1,\cdots , C_9
$.       We see that $\hat \l _3  $         
cluster inside  narrow intervals  that   depend weakly  on the   modes
(isometry group)  charges   and   decrease    by  a factor  $3$  per 
step  $ \D q = 1$  of the radial  excitation    number $q$. 
The change of sign  upon going    from the ground to the first
excited radial    singlet graviton  mode is  explained 
by  the  radial    wave function    oscillation
near the horizon.  Similar    trends  are   found in
Table~\ref{tab3}  for    cubic couplings involving radially excited modes $ h ^{(q)} ,\ C^{(q)} , \ [q= 1, 2 ] $. 

The      quartic and quintic order coupling
constants    $\hat \l _{4 , 5}  $   for ground  and  excited  singlet modes
interacting with  pairs of   charged modes  
$h^{(q) } h^{(q) } hh, \ h^{(q) } h^{(q) } C_i \bar C_i
,\ [i=1, \cdots ,  9 ,\ q=1, 2] $  are  displayed       in
Table~\ref{tab2p}. The dependence on the  modes  charges  is  
weak.  The  empirical rule for quartic 
couplings   $  \hat \l _4 \simeq \vert \l
_3\vert ^{2} $ is   approximately   satisfied  but the   quintic 
couplings  do  not  follow   a   similar power  law.  
The  change  of signs  for $\D q = 1 $,  obtained  previously  
for  $  \hat \l _3$,    is     not  present in cases
where  radially excited states  enter in pairs. 

\begin{table}   \begin{center}  \caption{\it \label{tab2} 
Reduced cubic coupling constants $\hat \l _3$ (in units of
$\calr =1$) for configurations of the singlet and charged modes $ [h=
  C_0 , C_1,\cdots , C_9 ] $ listed in Table~\ref{tab1}. The three
line entries $ q=0, \ 1,\ 2 $ refer to the ground state and the first
and second radial excited states.}
 \begin{tabular}{|c|c|cc ccc ccc c|}
\hline $ q$ & $ h^{(q)} hh $ & $ h^{(q)} C_1 \bar C_1 $ & $ h^{(q)}
C_2 \bar C_2 $ & $ h^{(q)} C_3 \bar C_3 $ & $h^{(q)} C_4 \bar C_4 $ &
$ h^{(q)} C_5 \bar C_5$& $ h^{(q)} C_6 \bar C_6 $ & $h^{(q)} C_7 \bar
C_7 $ & $ h^{(q)} C_8 \bar C_8$ & $ h^{(q)} C_9 \bar C_9 $ \\ \hline
$0$ &$ 0.481 $&$ 0.544 $ &$ 0.558 $&$ 0.603$&$ 0.629$&$ 0.612$&$
0.581$&$ 0.587 $&$ 0.590 $&$ 0.614$ \\ $ 1$ &$ -0.186$&$ -0.291$&$
-0.319$&$ -0.416$&$ -0.479$&$ -0.437$&$ -0.367$&$ -0.379$&$ -0.387$&$
-0.442$ \\ $ 2$ &$ 0.0356$&$ 0.107$&$ 0.131$&$ 0.232$&$ 0.313 $&$
0.258$&$ 0.177$&$ 0.190$&$ 0.199$&$ 0.265$
\\ \hline \end{tabular} \end{center} \end{table} \vskip 0.3 cm

\begin{table}  \begin{center}  \caption{\it \label{tab3}
Reduced trilinear coupling constants $\hat \l _3 $ (in units  $\calr =1$)  of graviton singlet and charged modes $ h
^{(q)} ,\ C_1 ^{(q)} $ in configurations involving the ground state
and first two radially excited states $ q=0,1, 2$.}
\begin{tabular}{|c|cc| cc|} \hline   
$ q$ & $ h^{(q)} h^{(1)}h^{(1)} $ & $h^{(q)} h^{(2)} h^{(2)} $&
  $h^{(q)} C_1 ^{(1)} \bar C_1^{(1)} $ & $ h^{(q)} C_1^{(2)} \bar
  C_1^{(2)} $ \\ \hline $ 0$ & $0.295$&$ 0.268$&$ 0.332$&$ 0.287 $
  \\ $ 1$ &$ -0.0369 $&$ -0.0287 $&$ -0.00340 $&$ -0.0118 $ \\ $ 2$ &$
  0.147 $&$ 0.0536 $&$ 0.0607 $&$ 0.0580 $
  \\ \hline \end{tabular} \end{center} \end{table} \vskip 0.3 cm

\begin{table}   \begin{center}  \caption{\it \label{tab2p}
Reduced quartic  coupling constants $ \hat \l _4 = 2 (M_\star w \eta
^{-3} / 1.64 \ 10^2 ) ^2 \l _4 $ (in units $\calr =1$) of massive gravitons in configurations $ h^{(q)
  2} C_i \bar C_i $   involving  a pair of singlet ground and  radially
excited  modes  with   pairs of charge
conjugate modes  from the list  $ [h= C_0 , C_1,\cdots , C_9 ] $ in
Table~\ref{tab2}.  The bottom line entry refers   to  reduced
quintic coupling constants $\hat  \l _5 $ of  ground state modes.}
 \begin{tabular}{|c|c|cc ccc ccc c|} \hline   
$( l, \ q ) $ & $ h^{(q) l} hh $ & $ h^{(q) l} C_1 \bar C_1 $ & $
   h^{(q) l} C_2 \bar C_2 $ & $ h^{(q) l} C_3 \bar C_3 $ & $h^{(q) l}
   C_4 \bar C_4 $ & $ h^{(q) l} C_5 \bar C_5$& $ h^{(q) l} C_6 \bar
   C_6 $ & $h^{(q) l} C_7 \bar C_7 $ & $ h^{(q) l} C_8 \bar C_8$ & $
   h^{(q) l} C_9 \bar C_9 $ \\ \hline $(2, \ 0)$ & 0.267 & 0.320 &
   0.333 & 0.376 & 0.404 & 0.385 & 0.355 & 0.360 & 0.364 & 0.388
   \\ $(2, \ 1)$ & 0.154 & 0.192 & 0.203 & 0.247& 0.284 & 0.258 &
   0.223 & 0.229 & 0.233 & 0.262 \\ $(2, \ 2)$ & 0.136 & 0.161 & 0.168
   & 0.195 & 0.219 & 0.202 & 0.180 & 0.184 & 0.186 & 0.204 \\ \hline
   \hline $(3, \ 0)$ & 0.159 & 0.196 & 0.206& 0.240 & 0.263& 0.247&
   0.223& 0.227& 0.230& 0.249 \\ \hline
\end{tabular} \end{center}    \end{table}    

\begin{table}  \begin{center}  
\caption{\it \label{tab4} Comparison of cubic local couplings in
  configurations of the singlet and charged graviton modes $h =C_0,\ C_1,
$ listed in  the column entries. The first three columns 
refer to couplings     allowed by the  isometry group 
and  the others to disallowed couplings.   The first
  entry lists the coupling constants $\hat \l _3
  $ values   in the undeformed case and the second lists the  
corresponding   coupling constants   $ \hat  \l _3  / ( -\ud \hat c ^-_\nu  ) $ 
in the  deformed   case with overlap integrals including 
the spurion   harmonic modes  $\d \Phi _- ^ {\nu  _i} ( \bar
  C_i),\ [i=0,\cdots , 9] $ of  $ \d (e ^{-4 A (r,\T ) } )$.   
The   selected  compensating spurion   modes $ \bar
  C_i  $  belong to   the  representations  conjugate  to the products 
of the   participant modes representations,  with the parameters 
set at $Q ^-_{\nu _i}  =4 $ and $\D  ^-_{\nu  _i}= -2 + (H_0 +4)^{1/2} $.}
\begin{tabular}{|c||ccc||ccccc cccc|}
  \hline & $ C_0 C_0 C_0 $ & $ C_1 \bar C_1 C_0  $ & $  C_2 \bar C_2
  C_0 $ & $ C_0C_0 C_1 $ & $ C_0C_0 C_2 $ & $ C_0C_0 C_3 $ & $ C_0C_0 C_4
  $ & $ C_0C_0 C_5 $ & $ C_0C_0 C_6 $ & $ C_0C_0 C_7 $ & $ C_0C_0 C_8
  $ & $ C_0C_0 C_9 $ \\ \hline $\hat \l _3 $ & $ 0.481 $&$ 0.544 $&$
  0.558 $& $ 0 $& $ 0$& $ 0 $& $0$& $ 0$& $ 0 $& $ 0 $& $ 0 $& $
  0$\\ $ - {2\tilde \l _3 \over \hat c ^-_\nu } $ & $ 0.12 w ^0
  $&$ 0.15 w^0$&$ 0.16 w ^0 $&$ 0.17 w ^{3/2}$&$ 0.18 w^2 $&$ 0.27
  w^{4.3}$&$0.39 w^{6.71 } $&$ 0.30 w^{5 } $&$ 0.22 w^{ 3 } $&$ 0.23
  w^{3.29 } $&$ 0.24 w^{3.5 }$&$ 0.31 w^{ 5.21 } $ \\
\hline \end{tabular} \end{center} \end{table}
\vskip 0.3 cm

\begin{table}   \begin{center} 
\caption {\it \label{tab6} Reduced cubic coupling constants $\hat \l _3 ^B
  $ for graviton modes $ h,\ C_i $ with pairs of $ D3$-brane modes (in
  units of $\calr =1$).   The  brane   is located at 
vanishing azimuthal angles    $ \phi _1 =
  \phi _2 = \psi =0$. The first three line entries refer to the
  configurations of polar angles $ I.  \t _1 =0,\ \t _2 =0; \ II. \ \t
  _1 =0,\ \t _2 =\pi /2 ; III. \t _1 = \t _2 = 2 \pi / 5 , $ and the
  fourth line entry $ IV$ refers to smeared distribution of branes
  averaged over $ \t _1,\ \t _2 $.}
\begin{tabular}{|c|| ccccc ccccc| cccc|}
\hline & $h $ & $C_1 $ & $C_2 $ & $C_3 $ & $C_4 $ & $C_5 $ & $C_6 $ &
$C_7 $ & $C_8 $ & $C_9 $ & $h^{(1)} $ & $h^2 $ & $h_2 ^{(1)} $ & $h_2
^2 $ \\ \hline $ I$ & 0.707 & 0.00083 & 1.502 & 2.247 & 2.957 & 1.006
E-6 & 6.983 E-7 & 2.844 & 0.00135 & 4.06 & -0.707 & 0.707 & -0.000746
& 0.000727 \\ $II$ & 0.707 & 1.176 & 1.502 & 2.247 & 2.957 & 2.012 &
1.396 & 0.& 1.920 & 0.& -0.707 & 0.707 & -1.056 & 1.028 \\ $ III$ &
0.707 & 0.791 & 0.462 & 0.803 & 1.150 & 0.349 & 0.632 & 0.269 & 0.0489
& 0.447 & -0.707 & 0.707 & -0.710 & 0.692 \\ $ IV$ & 0.707 & 0.739 &
0.751 & 0.864 & 0.961 & 0.796 & 0.698 & 0.711 & 0.809 & 0.782 & -0.707
& 0.707 & -0.663 & 0.646 \\ \hline
\end{tabular} \end{center}   \end{table}    \vskip 0.3 cm

We      next  examine   how deformations     affect  the couplings. For a given 
configuration  of charged  modes disallowed  by the selection 
rules   in Eq.~(\ref{selecteq}),  it is always  possible to  select       some    harmonic components    of  $ \d (e ^{-4 A (r, \T )} )$   
that      neutralize the   overall charge from the interacting   modes.  For  instance,   including the   warp profile spurion   mode of   charge  $\bar C_ {\nu }  $ in the disallowed
cubic coupling $ hh C_\nu  $,   yields the       deformed coupling  
constant     
\bea &&  \hat \l _3  (hh C_{\nu  })  \equiv 
({164 \eta ^3 \over \sqrt 2 M_\star w
 })^{-1}  \l _3  = w \int dr r  \int d^5 \T \sqrt {\g
  _5}  (\Psi _h   \Psi _h  \Psi _{\nu }  )  \d (e ^{-4 A  (r, \T ) } ) 
, \cr &&   [\d (e ^{-4 A  (r, \T ) } ) \simeq 
-\ud \sum _\nu  \hat  c ^- _\nu (\Phi _-) w ^{-4 +\D _\nu  ^-  + Q
  _\nu  ^- }   ({r \over w  
\calr })^{\D _\nu  ^-  - 4 }    Y ^{\nu  ^\dagger  } (\T )  ]   . \eea  
The  deformed    coupling constants  of graviton  modes
depend strongly on the  scaling  dimension  of the  warp factor 
harmonic mode    but are     insensitive  to
other  parameters.  For the modes $C_i ^T,\ [i= [1, 2,  6, 7]]$,
the    cubic  couplings  are
$\l _3 (hh C_i ^{ T } )   \sim  \eta ^3  w ^{\a  ^-_{i}  }   / (w
M_\star ) ,  \ [\a  ^-_{_i  } = [1.5,\  2 ,\  3 ,\ 3.29 ]] $.
Note for  comparison    that the  small   variation  range       $ \a
^-_{\nu } \in [1.29    \  -\ 3.29 ]$   was  found    sufficient 
in~\cite{kofman08} to   satisfy    the      cosmological constraints 
  from CMBR    deformations, neutrino  emission and   diffuse $\g
  $-ray  background  due to late decays  of   charged warped  modes.    
The  cubic couplings   of gravitons   in   a  set of   forbidden 
configurations that   become allowed   through  deformation effects
are  given in Table~\ref{tab4}.  

The  couplings   to $D3$-branes  are  proportional
to  the warped modes     wave  functions.   We  here   consider
massive gravitons coupled to a $D3$-brane
localized  at  the  radial distance  $ r_\star $ and   base manifold  
angles $ \T = \T _\star $. (In the deformed conifold case, the base
manifold at the tip has a collapsing $S^2$ factor   and  the 
moduli space of  $ D3$-branes  is reduced to the  $ S^3$ factor of
finite radius.)  Due  to the    wide freedom of choice  and the weaker 
dependence on azimuthal angles,  we   choose to set $ \phi _1=
\phi _2= 0 , \psi = 0 $ and consider three discrete  choices  for the
polar angles,
$I :\ \t _{1}^\star = \t ^\star _{2} =0 , II :\ \t _{1}^\star =0,\ \t
^\star _{2} =\pi /2 , III :\ \t _{1}^\star = \t ^\star _{2} = 2\pi /5
$,   along with   the case $ IV$   obtained  taking by   averages 
over $ \t _{1}^\star $ and $ \t ^\star _{2} $   for  
uniform   distributions of smeared  
$D3$-branes.       The reduced dimensionless 
coupling constants   $\hat \l _{3, 4 }^B$ for cubic and quartic
order couplings of single and pair of graviton modes to pairs of
$D3$-brane modes,   \bea && \l _3 ^B = { 2 \sqrt 2 (2\pi
  \eta )^3 \over M_\star w \sqrt {V_5 /4 } } \hat \l _3 ^B \simeq 2
{164 \eta ^3 \over M_\star w } \hat \l _3 ^B ,\ \l _4 ^B  \simeq   4 
({ 164 \eta ^3 \over M_\star w })^2 \hat \l _4 ^B
,\ [\hat \l _3 ^B = {w \Psi _m ( y_\star ) }  ,\ \hat \l _4 ^B = \vert
  {w \Psi _m ( y_\star ) }\vert ^2 ]  \eea 
satisfy  the rule   $\hat \l _4 ^B  =
\vert \hat \l _3 ^B \vert ^2 .$
The couplings   to a $D3$-brane at the   conifold tip $r_\star = w$,  
with  the      angles  set   as in the four cases  above, are    given in
 Table~\ref{tab6}.  Since  they   track  the radial  wave
functions  in Eq.~(\ref{tiprwfeq1}),   the    couplings    
fall  off exponentially  for increasing radial  distances. 
The  variations between  modes of different   charges 
arise  from  the oscillatory  dependence on    polar angles. 
Only the modes of vanishing $r$-charge 
 couple to $D3$-branes located at  the north poles $\t _1= 0 $
 or $ \t _2= 0 $.     (The charged bulk modes  of $r \ne 0$  
could couple to such  $D3$-branes  only through 
the  admixture  of   modes $r = 0$  induced  by  
the warping   or  breathing fields   deformations.) 

We  finally wish      to  assess the  accuracy   of the  estimates 
for  the   bulk  coupling  constants  $ \l _n $  obtained  in the tip approximation    by  evaluating   the  overlap integrals  for
the radial    wave functions in Eq.~(\ref{tiprwfeq1}),
$R_m \simeq  q_m w^{-1} (r/w) ^{-(2+ \nu _m ) } $. The resulting
formula   for  the reduced  coupling  constants  
\bea  && \hat \l _n=   {w ^{n-2}\prod _{i=1} ^n q_{m_i} 
  \over  \sum _{i=1} ^n   \nu _i + 2(n-1) }     
\int d ^5 \T \sqrt {\tilde \g _5 } \prod _{i=1} ^n  Y^{\nu _i}(\T ),  \label{tipradcoup1} \eea 
yields   estimates for     cubic   couplings, 
$  \hat  \l _3  ( h C_i  \bar C_i  ) = (4.7, 14.6, 21.0 , 5.2) 
,\ i=0,1,2,3] $ 
an  order of magnitude    larger    than 
the  corresponding   values    in  the first array of
Table~\ref{tab2}.  The   overestimates   get  more amplified    for
quartic  and quintic order  couplings.
The   deformed   reduced   coupling constants, derived  from
the approximate formula  in Eq.~(\ref{eqdefcc})  using the
 tip approximation wave functions,  
\bea && \hat \l _n \simeq  (\ud c_\nu ^ - ) {w ^{\a _\nu } 
\prod _{i=1}^n q_{m_i } \over \vert  2 - S_n +  \D _\nu ^ - -4 \vert
}   A _{\nu , m }  ,\ [S_n = \sum _{i=1}^n (2  +\nu _i )] \eea 
gives  estimates for cubic   couplings,  
$ (-2 / \hat c _\nu ^-) \hat \l _3   ( C_0 C_0  C_{i}  ) =
(3.39, 4.21, 10.9, 27.7),\ [i= (1,2,3,4)] $
an order magnitude   larger  that  the
corresponding values quoted in
Table~\ref{tab4}.  Except  for the  warp factor power,     the   
dependence on other   parameters is     very      smooth. 

\section{Warped  modes  in early universe} 
\label{sec6}

The information gathered in the present work motivates us to reexamine
the impact of warped modes on the early universe cosmology.  We shall
develop the discussion in the context of the $D3 -\bar D3$-brane
inflation scenario~\cite{kklt03,kklmmt03}.  This  utilizes the same
background solution of type  $II \ b$ supergravity on
$M_4 \times \calc _6 $ spacetime     embedding 
a  fixed $\bar D3$-brane and a mobile $D3 $-brane  near the   tip and bulk  of  a warped conifold  throat.
The quasi-de Sitter expansion of $ M_4$,  due to
supersymmetry breaking from the $\bar D3$-brane,  occurs during the slow
roll motion of the $D3$-brane induced by the attractive inter-brane
radial potential $V_{I} (r) $.  The Hubble constant $ H_\star $ during
the $N_e$-fold expansion and the energy density perturbations $ \d _H
$, seeded (at horizon exit) by quantum fluctuations of the inflaton
field $r (\xi ) $, are related to the inflation  throat warp factor  $
w_A =M_{I}/ M_\star $ as~\cite{frey05}
\bea && 3 M_\star ^2 H_\star ^2 =
V_{I} (r) \equiv 2  N \mu _3 w^4_A (1- {4 \over 9 N} ({ w _A \calr \over r
})^4  ) + \d V_{I}(r) ,\ \ \d _H = C_1 N_e ^{5/6} w _A ^{4/3} ({\tau _3 \over    M_\star ^4})^{1/3} , \cr && [ \mu _3= {2 \pi  \over \hat l_s ^4 } = 
{g_s ^2 M_\star ^2 \over 8 \pi \calv  ^2} =
  {M_\star ^2 \over 8\pi  g_s \calv _e ^2 } ,\  g_s^{3/2} \calv _e =
  \calv = ({L \over l_s })^6 ]    \label{coulpot1}\eea 
where the correction $\d V_I(r) $ to  the Coulomb potential  
term in $ V_{I} (r)  $  includes the effect of compactification~\cite{baumann06,baumann08,baumann208} and the overall
factor $ N$   comes from the 5-flux sourcing the throat. The
experimental observations from the CMBR,  $ H_\star /M_\star
< (10^{-8} \ - \ 10^{-5}) ,\ \d _H = 1.91 \ 10^{-5} $,  combined   with the       choice  of parameter  values~\cite{kklt03},    $N_e \simeq
60$ and $ g_s \simeq 0.1,\ V_W ^{1/6} \equiv \calv ^{1/6} \hat l_s
\simeq 5 l_s $,   yield  the two values     for the  inflationary
$A$-throat warp   factor,       
\bea && \bullet \ w_A= ({12 \pi \over N g_s })^{1/4}  ({\calv _e  H_\star
  \over M_\star })^{1/2} = ({12 \pi \over N })^{1/4}  ({\calv  H_\star
  \over  g_s ^2 M_\star })^{1/2} 
\simeq 1.2 \ 10^{-4} N^{-1/4} ({0.1 \over g_s})
({\calv \over  (5/2\pi )^6 }  { H_\star / M_\star \over 10^{-8} } )^{1/2} . \cr && \bullet \ w_A= ({ 8 \pi
  \calv _e^2 \over N C_1 ^3 } )^{1/4} N_e ^{-5/8} \d _H ^{3/4} \simeq
2.86 \ 10^{-4} N^{-1/4} ({C_1 \over 0.39 }) ^{-3/4} ({ \d _H \over
  1.91 \ 10^{-5}} ) ^{3/4} ( {\calv _e \over 8 })^{1/2} ({N_e \over 60
} )^{-5/8}  . \eea
The numerical   values are obtained from~\cite{kklt03}. 
Both   estimates   are compatible with 
$ w_A   \approx 10^{-4}$~\cite{frey05}.   We thus conclude that 
the  Standard Model  may be   hosted
either  on    unification  scale $D$-branes   embedded   in the $A$-throat  
at the  radial distance $ r_\star  /\a ' \simeq
10^2 w_A M_\star \sim 10^{16} $\ GeV, based on   the  construction   of
orbifolds of the conifold~\cite{cascales04},   or  on  TeV scale $D$-branes    embedded  near the tip of another   throat  of warp factor $ w_S =
M_{SM} / M_\star $. As is known, the  statistical counting of
solutions for flux compactifications on manifolds with large numbers
of 3-cycles~\cite{dougkach05}  or   conifold
singularities~\cite{hebeckerrussell06}   
favours string vacua with multiple throats.

The  cosmic   evolution   following $ D-\bar D$-branes annihilation 
is discussed  in several studies~\cite{barnaby04,kofman05,chialva05,chen06}.   
It is  believed  that the  inflationary throat   is initially   
filled    with     excited closed strings  that decay  fast  into 
a gas   of    weakly  interacting  massive warped   modes
which    reheat  the universe and possibly   leave some  cold  thermal relics. Our   presentation will focus  mostly on the single inflationary
$A$-throat   case  with brief  digressions on   the option involving  
a  Standard Model   $S$-throat of  warp factor $ w_S $.   
We shall  settle on the value
 $w _A = 10^{-4}$ advocated in~\cite{kklt03,kklmmt03} and the value
$w _S = 2\ 10^{-14} $   advocated in~\cite{guirkshiuzurek07,shiuetal07},
which is  larger than that adopted  in warped models 
$ e ^{-k\pi r_C} \simeq  4.2\ 10^{-16} $.

\subsection{Cosmic evolution  following  $ D-\bar D $ inflation}  
\label{subsec6.1}

\subsubsection{Closed string phase}

In  the approach  of~\cite{sentachyon98,sentachyon99} that we follow,
the exit   from   inflation   is triggered  by open string tachyon mode(s) $T (x)$     that  appear  when the  $Dp-\bar Dp$-brane  radial distance 
reaches   a  value $ r = \sqrt {\vert y ^m \vert ^2  } = O(l_s) $ where the
tachyon   sits  at an  extremum of the potential.    Below this  critical distance  the tachyon   field   rolls downhill   before condensing
upon  sitting at   the potential minimum.        
The   collapsed $ D\bar D$ system   dynamic  was   initially described
in terms of the Born-Infeld  type $ U(1)$ gauge theory  of action
\bea && S = - 2\mu _p \int d ^{p+1} \xi  V(T)
(-Det (\eta _{\mu \nu } + \calf_{\mu \nu } ) ) ^{1/2} ,  \eea 
where the tachyon potential $ V(T)$   factor  acts as an
effective charge that vanishes at tachyon condensation.
Kinetic terms for the transversal coordinate
fields  can  also  be included inside the determinant. 
   The    instability   is interpreted classically
as a Higgs mechanism evolving    the   system of confined 
electric    fluxes~\cite{bergmann00}     to a true vacuum at $ V(T)\to 0$
involving free relativistic flux lines~\cite{gibbons00}.  The  decay
products   must then lie preferentially along the initial directions
of the  decaying branes~\cite{sentachyon02}. An alternative 
version  involves   a  self-coupled tachyon  of   action  
\bea &&  S_p=- 2 \mu _p \int d ^{p+1} \xi V(T) (- Det ( \eta _{\mu \nu
} + \dh _{\mu } T \dh _{\nu } T) )^{1/2} .\label{tacheq1} \eea  
The tachyon potential and kinetic energy terms
in both approaches  can be   evaluated 
from the string theory  functional integral on the
disc~\cite{kutasov00}.  Explicit calculations of these quantities    
are   reported  in~\cite{kraus00} for the $D \bar D
$-brane  system and for  the   formally
related  $D D $-brane  angled system   in~\cite{jonestye03}. 
For instance, the  ansatze    for the  potential $V(T) $ of form
$1 /(1+ T T^\dagger ) $ or $ e ^{- T T^\dagger } $ reproduce the 
tachyon masses in the bosonic and superstring theories, $ \a ' M^2_{T}
= -1 $ and $ -\ud $.   As one moves along the positive and negative
semi-axes for  $T$, away from the
unstable maximum at $ T=0$,  $ V(T)$  either decreases
monotonically out to $ \vert T \vert \to \infty $ or decreases  to a
minimum  before rising out forever.  The classical evolution
for a spatially homogeneous field,  $ T (\xi ^0) = T(t)$,  
involves a downhill rolling from the initial
values,  $ T(t_I) = 0 ,\ \dot T (t_I) =0 $, that  ends at large times  
at  the  true minimum   $ \dot T (t_F) =1 $   
associated  to  the closed string sector phase of condensed tachyons. Comparing the   hydrodynamical 
stress tensor,  $T_{\mu \nu } (t )\equiv p \eta _{\mu \nu } + (p +\rho )
u_{\mu } u_{\nu } $,  with the solution  implied by      the action
Eq.~(\ref{tacheq1}),   \bea && T_{00} = V (T) (1 - \dot
T ^2)^{-1/2} = \rho ,\ T_{ij} = - V (T) (1 - \dot T ^2)^{1/2}\d _{ij}
= p \d _{ij} , \label{eqhydroem}  \eea 
shows that the  classical system evolves from  an
arbitrary spatial distribution of the conserved energy density 
$ T_{00} (t_I) = \rho = - p = V(T) $  to  a  pressureless non-thermal
and  non-relativistic  tachyon matter  with  $T_{ij} (t_F)= p \d _{ij}
=0$.       The       quantum 
 description is   more interesting, however,   since the branes
act as  boundary  sources of  closed string states.  The decay
process  can  then be   described classically in terms of spacelike
$S$-branes or 
Euclidean $D$-branes~\cite{gutperlestro02,gutperlestro03}  of the open
string field theory or   the boundary string field
theory~\cite{sentachyon02}.  The time evolution in the latter approach
utilizes the Wick rotated (Euclidean) 2-d string theory for the
non-compact time component $X^0 (z) $,    subject to  a (kink-antikink)
time-dependent boundary perturbation.  The  resulting bosonic string
action     on the disc surface of coordinate $z $ and boundary of
coordinate $\tau $ (in units $\a ' =1$), in   a  boundary state 
for   the time coordinate of coupling constant $\tilde \l $, 
 \bea && S (X^0) = -{1\over 2\pi } \int d ^2 z \dh _z
X^0 (z,\bar z) \dh _{\bar z} X^0 (z,\bar z) + \tilde \l \int _{\s =0}
d \tau \cosh (X^0 (\tau ) ) ,\eea  is a solvable 2-d conformal
field theory whose solution   was obtained  in~\cite{sentachyon02} 
as a power expansion in  the  string
oscillator    modes $ a_s ^0 (z) $     along the time  direction, 
\bea && \vert B> _{X^0} =(f (X^0) + g
(X^0)  a ^0_{-1} a ^0_{-1} +\cdots ) \vert 0> ,\cr && [f ( X^0) =
  {1\over 1 + \hat \l e ^{X^0} } + {1\over 1+ \hat \l e ^{-X^0} } -1,
  \ g (X^0) = 2 (1 - \hat \l ^2) - \rho (X^0) ,\ \hat \l = \sin (\pi
  \tilde \l ) ] \eea 
where the     coefficient fields $ f (x^0) ,\ g (x^0) $
are related to the  effective field theory   stress tensor 
in Eq.~(\ref{eqhydroem})     as 
$ T_{00}= \mu _p (f (x^0) + g (x^0) ) ,\ T_{ij} = - 2 \mu _p
f (x^0) \d _{ij} $.  
The total boundary state includes  additional    contributions from   
the string time and spatial coordinates $\vec X (z) $  and  the   
 ghost  fields~\cite{lambert03,shelton04}, $  \vert B >
= N_p \vert B >_{X^0} \vert B >_{\vec X} \vert B >_{g} ,\ [ N_p = \pi
  ^{11/2} (2\pi )^{6-p} ] $. If the oscillator     excitations 
of  $\vec X (z) $ in the   boundary state   are   treated as external sources, the    corresponding  mode   operators   $a _s ^\dagger
,\ \tilde a _s ^\dagger $    are    replaced by
coherent state operators $ e ^{ \a _s a   _s^\dagger } $,
where $\a _s $ denote  the  one-point disc amplitudes of the closed 
string vertex operator $ V_s$  for modes  of energy $E_s$,
\bea \a _s = { \cala _s\over \sqrt { 2
    E_s} } = { N_p \over \sqrt { 2 E_s} } < V _s > _{disc} ,\ [< V _s
  > _{disc} \simeq <e ^{i E_s X^0} > _{X^0} < V _{s} >_{\vec X} ] \eea
where the left and right movers  must be identified since  one deals
with  states of   open strings   attached to  $Dp-\bar Dp$-branes. 
A similar  treatment applies to ghost  fields. 
  The average number $ \bar N $ and energy  $ \bar  E $ of
closed strings emitted in spacetime  can be evaluated  from the matrix
element of the closed string  propagator in the   boundary state    $
< B\vert \D \vert B>$.  
Integrating over the transverse momenta $\vec k _\perp $ and summing
over the oscillator levels $ N $  for closed string modes   with
identified   left-right movers of  same density of states $d_n$,
yields  the formulas~\cite{lambert03} 
\bea && {\bar N / V_p \choose \bar E
  / V_p} = \ud \sum _s {1/ E_s \choose 1} { \vert \cala _s \vert ^2 }
= { N_p ^2 \over 2} \sum _N d _N \int _{\vec k_\perp } {1/ E_{k , N}
  \choose 1} \vert I ( E_{k , N} ) \vert ^2 ,\cr && [\int _{\vec
    k_\perp } \equiv \int {d ^{25 -p} k _\perp \over (2\pi )^{25
      -p}},\  E_s = E_{k , N} = ( \vec k_\perp^2 + 4 N ) ^{1/2} \simeq 2
  \sqrt N + {\vec k_\perp^2 \over 4 \sqrt N } , \ d _N \sim N ^{-27/4}
  e ^{4\pi \sqrt N },\cr && I(E_s) = i \int _{-\infty } ^{\infty } d t
  f (t) e ^{iE_s t} = {\pi (e ^{-i E_s \log \hat \l } - e ^{i E _s
      \log \hat \l } ) \over \sinh \pi E_s } ] \eea 
where $ V_p$ is the spatial volume  occupied   by 
the $Dp$-brane  and the   formula   quoted for  the source function $ I(E _s)$
corresponds  to the so-called full-brane solution.  In the sum over
string oscillator  levels  $N$,
the exponentially rising   factor for  the  level number degeneracy   $ d
 _N $   is compensated by the falling source function factor  which behaves as 
$\vert I( E_{s} ) \vert ^2 \simeq (2\pi )^2 e ^{-2\pi E _s} $ at large $N$. The suppression of    transverse momenta $ k_\perp / m
\sim 1 /\sqrt N $ implies that the initially emitted closed strings
are mostly in bosonic states of high excitation energy and low velocity.  The
resulting  asymptotic   contributions  to  the   
average energy and number densities  are dominated
at large $N$  by   the terms    \bea && {\bar E
  \over V_p } \sim  \sum _N N ^{-p/4 -1 /2} ,\ {\bar N \over V_p } \sim \sum _N N ^{-p/4    -1} , \eea   
that  diverge    for $ p \leq 2$,  meaning
that the branes tension energy goes mostly into
closed strings.  Energy conservation requires, however, setting an
ultraviolet cutoff on the level parameter $N$.  This    can be 
determined       by  considering  a $ Dp-\bar Dp$-brane system wrapped
on a torus $ T^p(R_p)$ of large radius $R_p$ and going to the $\calT
$-dual    picture of   a torus  $ T^p(R_0)$ of small radius $
R_0$   and  a $D0 -\bar D0$-brane system~\cite{lambert03},    via the $\calT
$-duality  correspondence   for the  radius and string coupling constant
\bea &&  T^p(R_p)\to T^p(R_0) : \ R_p \to R_0= l_s^2 / R_p,\ 
g_s \to g_{s,0} = (R_p/l_s) ^p g_s  . \eea  
Assuming  that   the closed string energy
is bounded by the $D0-\bar D0$-brane  system mass,   $ E_s \simeq
m_s \sqrt N \leq 2 \tau_0 = 2 m_s / g_{s,0} $,   yields 
the  bound   on  the  string excitation   level   $ N\leq N_{max}
\sim 1/ g^2_{s,0} $.  Bounding also the closed string transversal
momentum  by that of open strings attached to $D0-\bar D0$-brane 
implies  $ k_\perp \sim N^{1/4} \leq 1 / \sqrt {g_{s,0}}$. 

The   total  $D - \bar D$-brane  pair annihilation rate  itself  
can   be independently accessed  from  the string theory 
vacuum amplitude for a world sheet bounded by  branes at fixed
positions    $0,\  y ^m$   along the transversal directions.  The 
cylinder amplitude  gives  the interbrane potential energy  
as a    sum over   open string  modes computed  from  the  closed string propagator  matrix element, 
$V(y) = <\bar Dp (y ) \vert \D \vert Dp (0) > $.  
Below    the interbrane radial distances   where the   Coulomb potential
tail    dominates,  the  contributions from radiative corrections
are  expected    to turn over the real part  of the
potential $ \Re ( V(\vert y \vert ) )$ over 
by  producing     a shallow local minimum~\cite{saratye03}.    
Meanwhile, the regularization of infrared (ultraviolet)
divergencies in the  open string (closed string)
channel~\cite{marcus89}  induces  an imaginary part  in 
the potential,    related  via the optical theorem to the
forward   scattering  amplitude.  
The  resulting $ Dp-\bar Dp $ annihilation
rate   from   tachyon exchange   at short  distances,  
\bea && \D \G _s  \equiv  \G (Dp/\bar Dp)
= 2 V_p \Im V(0) \simeq { V_p \over \hat l_s ^p} {m_s \over \G
  ({p+3\over 2}) 2 ^{p+1 \over 2} } \sim m_s ,\cr && [\Im V (\vert y
  \vert ) = {2\pi \over \G ({p+3\over 2}) } ({\vert M_{T} ^2 \over
    4\pi } ) ^{p+1\over 2} + O( \vert y \vert ^2) ,\ \a ' \vert m_{T}
  \vert ^2 ={\vert y \vert ^2 \over \hat l_s^2} -{1\over 2} ]   \eea
is  seen to be   proportional  to the    spatial volume  
occupied by the $ Dp-\bar Dp$-branes  whose 
size   is approximately set by the string scale, $V_p \simeq  \hat l_s ^p $.  

At this point, we     shall make  the bold assumption  that the above  flat spacetime  predictions can  be adapted to the $A$-throat background by simply trading the string  scale with the local warped scale,  $ m_s \to w _A
m_s$~\cite{barnaby04,chialva05}.  One then infers the order of
magnitude  estimate   for the  decay lifetime,   $\D t_s = 1/\D \G
_s \sim 1/ (w _A m_s ) $.  If the $D3-\bar D3$-brane energy
density $\rho _I \equiv \rho (t_{I}) = 2 \tau _3 w _A^4 $
is assumed to be  efficiently  transferred   to  excited closed strings,  one obtains
the   following estimates   for  the average energy, number density
and transversal momentum per  closed string mode~\cite{sentachyon02,kofman05} 
\bea   && E_S = {\rho _I (t_{I}) \over (w _A m_s)^3 } \sim {w _A m_s
  \over    g_s },\ n _{S} 
= {\rho _{I} \over E_{S} } \sim (w _A m_s ) ^3 ,
\ p^\perp _{S} \sim {w_A m_s \over \sqrt {g_s} } .  \eea 
The   estimate  predicted  for the   average  velocity,   $v^\perp _S =
p^\perp _{S} / E_S  \sim \sqrt{g_s}$,    indicates  
that the  produced closed strings   are non-relativistic.  
The  massive   strings   are expected  to decay
via 2-body processes $ S_N \to S_{N'} + S_{N_0} $  that reduce
their oscillator excitation levels in successive steps $ N \to N' \to
\cdots $ down to massless and Kaluza-Klein modes.  The 
description   using folded  classical string rings,   with or
without appearance of    cusps~\cite{iengo06},     predicts 
for modes of mass $M\sim m_s $  decay   rates     $ \G _{S}
(M) \sim g_s ^2 M ^{5- D} \sim g_s ^2 m_s $   of string scale size
for a spacetime   of dimension $D =4$.   It is useful to 
compare  this  estimate       with   the perturbative prediction     for
string theory~\cite{amati99,chialvarusso04}     compactified   on  an
internal manifold $ S^3$ and a radial $S^1$ of radius $R$.  The
integrated rates    for the  two-body decay reactions $ S_N \to
S_{N'} + h_m (E ,k _\perp ) $,   with emission of  low lying closed string modes
$h_m $ of 4-d energy $E$, transversal momentum $ k _\perp $ along $
S^3$ and longitudinal string momentum $ m/ R $ along $ S^1$,  are given
(in string units) by \bea && \G _S (N)  \simeq g_s ^2 \pi V(S^3)
\int d E {e ^{-2 a E} \over (1- e ^{-a E} )^2 } E^2 \sum _{m=1}
^{m_{max} } ( E^2 - ( {m \over \calr })^2)^{1/2} ,\ [E \simeq {N-N'
    \over 2\sqrt N } = {N_0 \over 2\sqrt N } ,\ a = 2\sqrt 2 \pi ]
. \eea      
The total rates  for the cascade
decays  were    estimated in~\cite{chialva05} by noting that the 
energy distribution  of $\G _S(N) $   in the released  energy $ N_0 $ 
is  strongly  peaked     at $ N_0 ^{max} \simeq
\ a_{KK } \sqrt  N $,  hence giving  $\G _{S} (N) \approx a _{KK } c_{ KK}   \sqrt N ,\ [c_{KK}     \simeq 2.  \ 10^{-4} ,\
  a _{KK }\simeq 0.65].  $    
Summing over  excited levels  up to the cutoff $ N<  N_{max} = 1/g_s ^2 $
and  rescaling  $  m_s \to w_A m_s$,  
yields the    approximate decay lifetime  of  massive closed strings   into warped modes \bea
&& \D t _{S} \equiv {1\over \G _{S} } = \sum _{N=1} ^{1/g_s ^{2} } \D
t _N \simeq {1 \over c _{KK} w _A m_s } \sum _{N} {1 \over \sqrt N }
\simeq { \sqrt { N_{max} } \over c ' _{KK} w _A m_s } \simeq (c '_{KK}
g_s w _A m_s )^{-1} ,\ [ c ' _{KK} \simeq  10^{-4} ] .  \eea 
Note that the formula   for the cascade  decay time   proposed in~\cite{kofman05}, 
$  \D  t _S \simeq N / \D \G ( S\to S' + K) \sim  1 / (g_s ^4  w _A
m_s )  $ yields an estimate of comparable  size
for $ g_s \approx 0.1$. The   decay lifetime   to massless gravitons 
is     found to  be of    comparable    size,   with $ c ' _{KK} \to 
c_g \simeq 1.8 \ 10^{-3} $,  but
 the warp factor enhancement is absent in this case.  
The resulting ratio of the final state  energies $ E_{KK} / E_{g} \sim
0.2 \ w _A^{-2} $  thus   disfavors the graviton component.  

We  note finally that  inflation   might also leave 
excited    open strings  supported  by  $D$-branes that  contribute to the cosmic bath.  The amplitudes      for       decay  processes of massive 
 open string  modes  into  pairs of lighter modes      was computed   
in terms of     two point  functions on  the annulus  string   world sheet 
in  backgrounds  with  $D9$-branes~\cite{mitchell89} or with
$Dp$-branes~\cite{balakleb96}. 
For   open strings  of squared  mass and angular momentum, 
$ M_l^2 =2l ,\ J = l+1 $,   the two-body decay rates  $\G _D$  for
transversally polarized (to branes) string  states  into  fixed  final
states  and the total two-body decay rates $\G _N$  
for  longitudinally  polarized string  states  are  
\bea &&  \G _D (l (M_l) \to (l-1) + 1) =
{ (\pi / 2 )  ^{p/2 +1 }   m_s \over \G (p/2) } 
(M_l/m_s)^{2 (2-p)}  ,\  \G _N   (l (M_l) \to all ) \simeq  \sqrt {2} \pi ^2 M_l
  ({\ln ( m_s / M_l ) \over \pi } )^{p-9 \over 2} . \eea    
We see that the  predicted  decay  rates  for  massive open strings    are  suppressed relative
    to  those   of closed  strings  by   power    factors    
for  $p > 2$ and    logarithmic  factors, respectively. 

\subsubsection{General discussion  of  warped  modes thermal evolution} 

After  closed   strings have decayed,  the $A$-throat  is     populated by     an  interacting   gas of
warped modes     sharing   the   energy  density
$\rho _I$  from  the branes tension  energy.
Alongside  with  bulk  modes, light open string   modes  
on embedded  branes     might   also  be
present. Since the kinetic energies of the various  modes   should
exceed     initially  their  rest masses,  it is natural to   
start from  a relativistic gas  of bulk modes in a   radiation dominated   (RD)   universe.   The initial 
 Hubble expansion  rate $H_I $,   cosmic temperature $  T_I  $ and
 cosmic time  $t_I$    are then  determined  from  the  
Friedman   equation  
\bea && H_I = H (t_ {I} ) \equiv ({\rho
  (t_{I} ) \over 3 M_\star ^2})^{1/2} = ({\pi ^2 g_\star (T_ {I})
  \over 90 M_\star ^2})^{1/2} T^2 _I \simeq {1\over 2 t_{I} } ,\ 
[\rho ( t_I) = 2 \tau _3 w _A^4 = {4\pi w_A^4 \over \hat l_s ^4
    g_s} ]   \eea 
where  $g_\star (T_ {I})$    counts  the number  of  spin   
degrees of freedom from   relativistic  bulk and brane  modes 
and  the initial energy density $\rho _I$ was  set  at the  $D-\bar D$-branes rest mass.  We record  in the table  below the resulting    formulas for 
 the  initial   data  for  
$\rho _{I}= \rho ( t_I)  ,\ t _I ,\   T_I $,  
expressed as a function of    the physical  parameters 
$M_\star ,\ w_A , \ z ,\ \eta $.    
\begin{center} \begin{tabular}{|c|c||c|}   \hline    
$ \rho _{I} $& $t _I $& $ T_I $ \\ \hline $ {(w _A m_s )^4 \over 4\pi
      ^3 g_s} $ &$ {\sqrt 3 M_\star \over 2 \sqrt { \rho _I } } $& ${w
      _A m_s \over (g_s g_\star )^{1/4} }$ \\ $ {M_\star ^4 w _A ^4
      \over 4 \pi ^{5/2} \eta ^3 z^3}$ & $ {(\eta z )^{3/2}\over w_A^2
      M_\star } $ &$ ({15 \over 2 \pi ^{9/2} g_\star })^{1/4} {w _A
      M_\star \over (\eta z) ^{3/4} } $\\ $ { 10^{-2} \ M_\star ^{4} w
      _A ^{4} \over (\eta z) ^{3} } $ &$ {5\ 10 ^{-24} (\eta z) ^{3/2}
      \over M_\star w _A^2} $ &$ {0.1 g_\star ^{- 1/4}M_\star w _A \over (\eta z)
      ^{3/4} } $ \\ $ {5. \ 10^{68} w^4_A \over ( z \eta / 10 )^{3} }
    \ GeV^4 $ &$ {6. \ 10^{-41} ( z \eta / 10 )^{3/2} \over w^{2}_A }
    \ s$ &$ {2.\ 10^{17} w_A g_\star ^{- 1/4}\over ( z \eta / 10
      )^{3/4} } \ GeV $ \\ \hline
\end{tabular}  \end{center}   \vskip 0.4 cm
For   $ w_A = 10^{-4}  $  and   $ \eta z = M_\star L  = O(10)  $,
the approximate  values   for   $t_I \sim 10^{-33 }  \ s $  and $
 T_I \sim  10^{13} \ GeV $    are   of  same  order of magnitude  as  
the    interaction time  and energy of  closed string modes, 
$ \D t_S \sim 10  ^{ -33}\  s $ and $ E_S   \sim 10
 ^{14} \ GeV $.  The      estimates for  the   average energy density, energy and number density    of  the  warped modes gas    \bea && \rho _K = \rho _I, \ E_K= {w_A   m_s } ,\ n_K = {\rho _K \over E_{K}} \sim {w_A ^4 m_s^4 g_s ^{-1}
  \over w_A m_s} \sim {(w_A m_s )^3 \over g_s } , \eea   
are seen to predict a   large     Lorentz factor   $ \g _K \equiv
{E_K    \over m_K } \sim m_s \calr _- \sim \l \hat \s g_s^{1/4} $     
consistent   with  the assumption that  the gas is     initially
relativistic.   Note that the conditions that   thermal        excitations
stay below  the   
string  scale  but above the Kaluza-Klein threshold   scale, 
$ m_K   \leq T_I \leq w m_s  $,     yield  the  respective lower and
upper      bounds    $ (g_s g _\star )^{1/4}  \geq 1 $
and $ z/\eta ^3  \leq 0.4\  g_\star .$  

To    facilitate   the   discussion,  we have  rewritten
the  formulas  for    the  thermal and  kinetic properties 
of   the closed  strings  and   warped    modes  phases  in Table~\ref{thermal123}  as a function   of the    physical parameters.  
Before  addressing  the dynamical properties in these  phases, 
we recall  that  thermalization occurs     through  scattering
reactions     between     bulk modes  while   reheating occurs
through    decays  of   bulk  modes   transferring  the  false  vacuum
energy     to  the lighter      LCKP   and brane  modes, including 
massless gravitons. 
We   shall restrict consideration    to the  decay    rates of  bulk
modes    to  pairs of $D3$-brane  scalar  modes,  
$ \D \G _{1\to 2 } (h\to  \varphi +\bar \varphi )  $,     
and    to the  interaction rates   of  pairs of bulk  modes, 
$\D \G _{2\to 2} (h +\bar h \to h+\bar h)$, 
using the   approximate  formulas borrowed  from~\cite{kofman05}, 
\bea && \D \G _{1\to 2 } = { w_A m_s\over 32 \pi } ({m_K\over
  w_A m_s })^9 ,\ \D \G _{2\to 2} = n _K  <\s _{2\to 2}  v_K> \simeq { (w_A
  m_s)^3 \over 4 \pi ^3 g_s} {g_s ^2 \over (w_A m_s)^2 } = {w_A m_s
  g_s \over 4 \pi ^3}  \cr &&  
\Longrightarrow  \D t_{1\to 2}   \simeq 10^{-40} ({w _A \over 10^{-4}})
^{-1} \eta ^{-9} ({\eta z \over 10}) ^{3} \ s ,\ \  
\D t_{2\to 2}   \simeq 10^{-35} ({w _A \over
  10^{-4}})  ^{-1}  ({\eta z \over 10}) ^{3/4} \ s ,\cr && \Longrightarrow  
 { T_{RH} \over M_\star }  \simeq 10 ^{-3} ({w\over
  10^{-4}})^{1/2}    ({\eta \over 10})^3 ({z \over 10^4})^{-3/2}
, \label{eqdecann1}    \eea
where we have    quoted in the last line  entry the reheating  temperature
evaluated   from the    decay rate  $\D \G _{1\to 2 } $  by  means of Eq.(\ref{RHtemp1}) below  and  have used the  definition $\D t = 1/\D \G $ for the interaction and decay  times. 
We see that   for   natural values for the   parameters, 
$\eta = O(1)  ,\ z = O(10)$, the      decay and
relaxation  times     lie  somewhat   below  the      
Hubble expansion  time,   $ t_I =(2 H(t_I) )^{-1} \sim 10^ {-33} (w _A/
10^{-4})^{-2}   (\eta z/10) ^{3/2} \ s  $.  
The       gap    between the predicted decay and interaction times  
in Eq.(\ref{eqdecann1}),   which  scale  as  $w ^{-1} $,  and  the expansion time,   which scales    
as     $ w^{-2}$,     gets    wider  for  longer  throats.    
To assess whether thermalization  does  occur, it
suffices      to   restrict to  short throats,  
$w_A = 10^{-4} $.     We see that while     the   interaction   time
is    short   enough to ensure  thermalization,  the      decay time
to   brane modes    is far   too  short to  prevent   an 
early  removal  of   warped modes.   Note that    
the above    lifetime  is      much  shorter       
that   the decay time  of  massive closed  strings $ \D t _S  = 1/ \D
\G _S \sim  2. \ 10^{-33} (w_A / 10 ^{-4} ) ^{-1}  \ s $  and
also   shorter  than   the 
decay time to  bulk  modes  $  \D t _{1\to 2}  
\sim 10^{-37} (w _A/ 10^{-4})^{-1} \eta ^{-9}  (\eta z/10) ^{3} \ s $, 
calculated from the formula 
$ \D  \G _{1\to 2 } = \calf ' {\l _3 ^2 m_K^3  / (960 \pi ) } $ 
in  Eq.(\ref{eqrelax1}) below.  The     inequality   
$\D t_{2\to 2}  >> \D t_{1\to 2} $     that we have found   
contradicts the claim  of~\cite{kofman05}      that     bulk modes
thermalize  before heating    the  Standard Model brane.   However,
given  the sensitive dependence on   parameters, it is   clear that 
one could   equally obtain the opposite hierarchy   $\D t_{2\to 2}  <<
\D t_{1\to   2} $   by  adopting, for instance,  the  admittedly  
unnatural    value $z >  O(10^4) $ 
or  $ \eta = O(10^{-1})  $,
or   by  localizing the brane    far from the tip  or in
the bulk, hence shutting off the decays to  brane modes.     

To  complete the picture, it is necessary to discuss   the
tunneling processes   of  massive warped modes from  
$A$-throat to  bulk  and from  $A$- to
$S$-throat.  The  predicted migration  rates 
out of the $A$-throat   are seen from 
Table~\ref{thermal123}  to scale    with     large
powers of    the   modes  mass, 
  $ \D \G _{tun} \sim   (m _K \calr  _A )^{5  \   \sim  \    9}    \sim
(x _k w  _A )  ^{5  \ \sim      \  9}   $. 
The  transitions from  a shorter throat
are   faster   and, due to the conservation mass,    populate    modes
of higher    excitations   in the   longer throat. 
For  modes  of finite angular
momentum $l$, the  centrifugal  barrier  suppresses 
rates       by   the extra   factor   $ w ^{4 l }$.  
The  disparity   between the    Randall-Sundrum~\cite{dimo01}
and Chen-Tye $D3$-brane~\cite{chen06}  predictions in Table~\ref{thermal123}      
partly  reflects  the fact that the    former 
refers   to   an    inter-throat  transition and  
the   latter    to     a  throat-bulk  transition.
The  bulk to throat  transition is given by  
the   throat absorption probability  per unit  bulk  volume  $L^6$,
and      is thus suppressed   at  large  volume.
If one argues that the  inter-throat  $ A\to  S$  rate    in the $D3$-brane model   is     typically   the   smallest of 
the throat to bulk  and  bulk to throat rates,    
\bea &&  \G _{A\to S} = min(\G_{A\to X} ,   \G _{X\to S} ) ,\    
[\G _{A\to X} ^ {D3} = {\pi    ^2 \over 2 ^8 \calr _A}  w_A (m _K
  \calr _A)^8 ,\      
\G _{X\to S} ^ {D3}  = {P_{X\to S} ^ {D3} \over  L^6} = {\pi ^4 \over 8\calr
  _S} ({\calr _S \over L })^6  (m_K \calr _S )^3  ]   \eea  
then for the  natural  choice $ L\simeq \calr _S \simeq  \calr _A $, 
 the  throat to bulk time  
$\D  t _{A \to X} $     dominates  over the bulk to throat time 
$\D  t _{ X \to S} \sim 10^{-33} (w_A / 10 ^{-4} ) ^{-3}  (z/10 )  \ s
$,     hence  yielding for   the    inter-throat   tunneling  time   
$\D t ^{D3} _{A\to X} \sim   1.6\  10^{-13 }  (w_A / 10 ^{-4} )
^{-9}  (z/10 ) \  s$.  In spite of the   large ratio 
between the predictions  in  Randall-Sundrum and $D3$-brane  models, 
$\D t ^{RS} _{A\to S} /   \D t ^{D3} _{A\to S}  \sim  10^{10} $, 
the tunneling  in both   models  is accomplished well before  the
cosmic nucleosynthesis  and recombination times, $ t _{Nucl} \sim 180
\ s ,\    t  _{Rec} \sim 1.2  \ 10^{13} \ s  $,    
and     lies   fortunately     far   below   the
 universe age    $ t_0 \simeq   6.5\  10^{17} s $.    
The inter-throat  energy transfer    and   the r\^ole of tachyonic
warped  modes  for tunneling  in  Randall-Sundrum   model    
are examined  in~\cite{harlinghn07,harling08}
and~\cite{langfelder06,halterRE09},  respectively. 

We are  now in position to  speculate about the   fate  of warped   modes. 
For the  case of a single $A$-throat of $ w_A \sim
10^{-4}$  that embeds    Standard Model  branes, the warped   modes 
are ultra-massive  and their  thermalization  
is possible  only  within a corner  of parameter  space  involving
large $z $ and  small $\eta $,    enough to   prevent their   rapid
disappearance into    gravitons  and   visible sector modes.  The 
upper limit  on the warp  factor deduced  from  Eq.(\ref{CDMabI}) below, 
$ w _A  \sim  10 ^{-13} \ \eta ^3  (\O _K ^I)^{1/2} $,  via
the universe overclosure bound $\O _K ^I < 1$~\cite{chen06},  
is far  smaller than the value     deduced     from 
inflation      unless   $\eta \sim 10 ^{3}  $. 
If  no Standard Model   branes are     present in throat $A$, one would  get  an   LCKP thermal relic     decoupled  from  the visible sector
while   only  gravitons  get   reheated.   An excessive  
fraction of gravitons today
is    unlikely, however,  due to  the   small production  rates of
gravitons, as discussed near   Eq.~(\ref{sect4.eq14})  below.  

Turning next  to the    double throat scenario, 
the starting point  is same  as  for  the
$A$-throat   case      except that   warped modes    of decay  lifetimes   
$  \D t _{1\to 2} >  \D t_{tun} $   can tunnel  out  to the $S$-throat  
    with probability $ P  _{ A\to S}   = 
{\pi ^2 \over 2 ^8} (m \calr _A) ^8 ,\ [m= x_m w _A /  \calr _A]$. 
The    reheating  of the $S$-throat and  Standard Model branes by
the fraction  $ P  _{ A\to S} $  of the  initilally produced  modes  takes   place     at     lower temperatures    
 $  (T_{RH})_S  \sim P _{X \to S} ^{1/4}   (T_{RH})_A.$    
The  abundance of  LCKP  thermal relics in  the   $S$-throat, $\O _K \sim  10^{-19}  \  (\eta  / 10 )^{-6}   (w _S / 10^{-14})^2   $,  is negligibly small  due to the  warp  factor. In contrast, the  abundance in the $A$-throat,  which is   slightly reduced  due to the shorter period  of matter  domination 
until   tunneling  ends in the $A$-throat,     is related  to the warp factor   by   the  formula~\cite{chen06}   $ w _A^{11/12}  \sim  10 ^{-4} \ \eta  z
^{-1/12} (\O _K ^{II} ) ^{1/6}   $,  deduced  from  Eq.(\ref{CDMabII})  
below. The resulting  upper limit  set  on $ w_A$ by  universe overclosure 
is   less  stringent  than that  deduced  above.  
    
Finally, we note that the  LKCP  modes $C_2 (100) =  S^- (\pi , \ b) $     re could  exhibit    late decays   to     
massless   gravitons or photons induced through tree
diagrams  involving multiple vertices  and modes   deformations 
$\d g _{ab} (110) $  and  $ \d C_{abcd }   (210) $ 
from    compactification  effects~\cite{berndsen07}.    
The   estimated  rates  for these decays (omitting the  coupling
constants),  $\D  \G  _{dec} (b\to h + h) \sim  w ^{3.4 }   (m_{3/2 } /   M_\star )^2 , \ \D   \G  _{dec} (b\to h + h + \g ) \sim     w ^{11.9} /  M_\star ^3  $,
indicate  that   the cosmic   temperatures 
at which  they  occur,   $ T _{b\to h h }   \sim    O (1\ - \ 0.1 ) \ TeV $ and $ T _{b\to h h    \g }  \sim   O(1 ) \   MeV $,     are close   to the   baryogenesis   and   nucleosynthesis era, respectively.

\begin{table} \centering
\caption{\it \label{thermal123} Kinetic and dynamic properties of the
two post-inflation  phases   for    non-relativistic  
closed strings (first 3 columns) and    relativistic  
warped modes (last 6 columns) with the average  energy, number density
and decay and interaction rates into massless modes   denoted 
$ E_S ,\ n_S,\ \D \G _S $ and $ E_K ,\ n_K,\ \D \G _{1\to 2}  ,\ \D \G
_{2\to 2}  $.    The  $A$-throat  tunneling rates 
$\D \G ^{RS ,\ D3} _{tun} $  refer to estimates using
Randall-Sundrum~\cite{dimo01,chialva05}  and 
$D3$-brane~\cite{chen06} models.   The  first line entry lists  
order of   magnitude estmates  as  a function of the string
  theory parameters. The  various  formulas   are expressed  in terms   of
  the physical    parameters $ M_\star ,\ \eta , \ z$ in the second
  line entry and   evaluated  (in GeV units) 
in the last line entry  using the input numerical values
  $M_\star= 2.43\ 10^{18} GeV, \ \eta =1, \ g_s=1/10 ,\ z=10 ,\ 
  c ' _{KK} = 10^{-4} $.  The conversion of   
decay and interaction     times $ \D t _{1\to 2} =  1/ \D \G _{1\to 2}
,\ \D t _{2\to 2} =  1/ \D \G _{2\to 2}  $   from      $ GeV ^{-1} \to
$     seconds  units is realized  by  the      rule
  $ 1 \  GeV ^{-1}= 6.57 \ 10   ^{-25} s $. In particular, 
the Planck time and universe age set at  the values 
 $ t _\star \equiv M_\star ^{-1}  = 3.\ 10^{-43} \ s ,\ t_0 \equiv
H_0^{-1} = 10^{42}   GeV ^{-1} \sim 6.5 \ 10^{17} \ s $.}
\tiny
\begin{tabular}{|ccc||cccccc|}   \hline  
& Closed   Strings  &  &    & & & KK Modes &&   \\ \hline
$ E _{S} $&$ n_{S} $& $ \D \G _{S} $ & $ E_K $&$ n_K $&$ \D \G _{1\to
    2} $& $ \D \G _{2\to 2} $& $ \D \G _{tun} ^{RS}$&$  \D \G _{tun}
  ^{CT}$ \\ \hline ${ 2 w_A m_s \over g_s}$&$(w_A m_s )^3 $& $ c '
  _{KK} w _A m_s g_s $& $ w_A m_s $ & $ { (w_A m_s) ^3 \over 4 \pi ^3
    g_s } $& $ {m_K ^9 \over 32 \pi (w _A m_s )^8 } $& $ {g_s w_A m_s
    \over 4 \pi ^3 } $& ${\pi ^2( m_K \calr _A )^4 w_A \over 2^6\calr _A  } $&$
  {\pi ^2 ( m_K \calr  _A )^8 w_A  \over 2^8  \calr _A} $
  \\ $ { 2\pi ^{1/8} w_A M_\star \over (\eta z )^{3/4} g_s^{3/4}} $& $
     { g_s ^{3/4} (w_A M_\star )^3 \over 8 \pi ^{21/8} (\eta z )
       ^{9/4} } $& $ {c'_{KK}  \pi ^{1/8} g_s^{5/4} M_\star w_A \over
       (\eta z)^{3/4} } $& ${\pi ^{1/8} g_s ^{1/4} M_\star w_A \over
       (\eta z) ^{3/4} } $ & $ { (w_A M_\star )^3 \over 4 \pi ^{21/8}
       g_s ^{1/4} (\eta z ) ^{9/4} } $& ${564.  \eta ^6 M_\star w _A
       \over g_s^2 z^3} $& $ {g_s ^{5/4}  M_\star w_A \over 4 \pi ^{23/8}
       (\eta  z)^{3/4} } $& $ {33\  M_\star w_A ^5 \over z} $&$
     {2.\ 10^3  M_\star w_A^9 \over z} $ \\
 $ 5.\ 10^{18}w_A $ &$9. \ 10^{49}w_A^3 $& $ 2.8\  10^{12} w_A $& $ 2.8\ 10^{17} w_A $& $    
     2.\ 10^{51} w_A^3 $& $ {1.4 \ 10^{20} w_A \eta ^6  \over
       ({z / 10})^{3}      ({g_s / 0.1})^{2} } $& $
 {2.3 \ 10^{14} w _A  ({g_s /  0.1})^{5/4}   \over ({\eta z / 10})^{3/4} }  
     $& $ 8. \ 10^{18} w_A^{5} $& $ 4. \ 10^{20} w_A^9 $ \\ \hline
\end{tabular} \end{table}    \vskip 0.4 cm
\normalsize


Before   delving  into  a semi-quantitative  treatment  of the   warped
modes  thermal  evolution,  we  review briefly three general 
constraints  on   the cosmology of relic particles set by 
compactification  effects~\cite{kolbky84,bailin87}. 
Recall    first that very  weakly  interacting  species are expected to
 decouple  early  as hot relics   with a present day (red shifted)
abundance $ \O _K h^2 = {g_K \over g_\star } {m_K \over 0.19 eV} $,   
yielding   the bound $ m_K < 100 $ eV.   For    weakly
interacting massive particles (WIMPs)  of      annihilation
cross section rate  $< \s _{ann} v >  \geq 10^{-36} \ cm ^3 /s $,
the unitarity  mass   bound   from universe overclosure  
is  $m_X < (10^6 \ GeV \ - \ 500 \ TeV
) $~\cite{griestkam90}.  
For    non-thermal ultramassive (wimpzilla)
relics, the mass  bounds     are $ m_X < ( 10^{10} \ GeV \ - \ 10^{13} \ GeV)
$~\cite{chung198,chung298}.      For Kaluza-Klein particles with annihilation   
rates $ < \s _{ann} v > = 1 / M_C ^2 $ to  radiation   set by the
compactification mass scale $ M_C$, the bound on
the ratio of number densities~\cite{kolbky84} $n_K / n_\g
\leq 10^4 T_{\g _0} / M _C < 10^{-14} \ \Longrightarrow \ M_C / M_\star  \leq
10^{-12}  $,     requires   a large hierarchy with Planck scale.
Note that  adapting   this  condition   to  a warped compactification yields 
the weaker bound $  M_C / M_\star \leq 10^{-12} / w $. 
We  also   mention  for  completeness 
the   proposal  in~\cite{dimo01},    using a   double throat  
Randall-Sundrum   model with Standard
Model and conformal  sectors        on  left and right hand sides of  the Planck
boundary,   where  thermalization is   realized  by a  population 
of metastable bulk scalar modes   on the left that   tunnel to the right
before  decaying  lately to a     conformal   dark matter  component.  

The second constraint     concerns         the   cosmic graviton
radiation    produced   through   warped modes  decays.   
The  graviton fraction  is suppressed  because   the decay  reactions
$ h \to ( g )^n $ are forbidden  by  the orthogonality of 
massless and massive  modes  wave functions  while  the
annihilation   reactions
$h +\bar h \to ( g )^n  $   have     small    cross sections  
 of  size $\s _{Kg} \simeq  \s ( h +\bar  h \to g ) \sim 1/ M_\star ^2 $. 
The  graviton  energy density and   abundance 
was estimated  in~\cite{chen06} by  converting  the  kinetic
equation $ d(\rho _g a^4 (t) ) / dt =  T(t) n_{K}^2 (t) a^4 (t) \s
_{Kg}  ,\ [ n_K(t) \sim g_\star T^3 (t) ] $ 
to a convolution integral. Noting  that the integral 
 is  dominated at the lower   bound $ t= t _I $,
and using   the scaling laws,  $ a(t)\sim 1/T,\ T\sim 1/\sqrt t
,\ \rho _g (t) \sim t^{-1/2} / a^4 (t)  \sim (t ^{-5/2}   \ - \  t ^{-2})
,\ T(t) \sim g_\star ^{-1/4} (M_\star / t )^{1/2} $,   yields 
the predicted  energy  density of gravitons and its 
contribution  to the ratio to  the universe 
critical    energy density  $\O _g $  at big bang nucleosynthesis (BBN) or radiation domination to matter domination (RDMD),
\bea && \rho _g (t) = \int _{t_{i} } ^ t dt ' T (t') n_K^2
(t') \s _{Kg} ({ a (t') \over  a (t) } )^4 \
\Longrightarrow \ \O _{g}\approx ({g_\star \over  g_s })^{1/4} {m_s   w_A \over M_\star } .\label{sect4.eq14} \eea  
If the  energy density  of gravitons  is expressed  in terms of 
the effective number of light cosmic neutrinos
today,   defined    by the ratio $ \rho _{tot} (t_0 ) / \rho _{\g_0}
 -1 = {7\over 8} ({4\over 11})^{4/3} N_{eff} $,   one   finds the
 contribution  $ \d  N_{eff} \sim w _A $,   
 which is  far below the     experimental bound   $ (N_{eff}  -3 ) \simeq [1.34 ,\ 0.9 ]$   set  by  CMBR (cosmic microwave background radiation)  and BBN (big bang nucleosynthesis)  constraints.
 We also  note that a large  initial  graviton   component    
would   get  reduced  relative to the  
visible radiation by a factor  $ g_\star (T_{RH} )/  g_\star
(T_{BBN} )$, due to the entropy  released at temperature $T_{RH}$ in
decays of light warped modes.  Additional dilutions   of the   graviton
fraction     by     factors $ T_{X} / m_X$  take place   
at   each     successive  
out of equilibrium   decay of  a   heavy mode  of mass $m_X$
   at  temperature $T_{X}  > T _{BBN} $~\cite{barnaby04}.  

The third constraint    concerns     the possibility  that 
the Hubble expansion in the $A$-throat back-reacts on the  
$S$-throat in case the  inter-throat   tunneling is completed
before the end of  inflation~\cite{frey05}. 
 If $ \G _{tun} (A\to S) > M_{SM} ^2/M_\star $,     the long open
 strings   produced in the  
Hagedorn regime would reheat  the $S$-throat   before the RD
phase takes over.  For the effective field theory
description  to remain valid, one must    impose 
  the upper limit $ w_S > w^2_A$,
inferred from the   bound  on the induced scalar curvature $\tilde R ^{(4)}
\sim H^2 w^2_S \sim M_\star ^2 \calv _E ^{-2} w_A^4 / w^2_S < 1 $,
or the upper limit  $ w_S >  (w_{A})_{min}$  set    on  the 3-flux parameter $N_S$ sourcing the $S$-throat,  $ N _S \tau _3 w_S^4 \geq H^4 _\star \ \Longrightarrow \ w _S \geq (w_{A})_{min}\equiv   (\pi \sqrt 2) ^{1/2} H_\star
\calr _S / \sqrt {N_S}$.   To avoid a     stringy
type scenario  in   case   the  bounds are  violated, it was proposed in~\cite{frey05}  to consider   a   time-dependent warp factor $ w_S
(t) \simeq \sqrt { C } (\calv _E / (g_s ^2 n_f ) ) H (t) / M_\star $
that relaxes to $ w_S$ at the end of inflation.  

\subsection{Warped modes  phase}  
\label{subsec6.2}

The thermal   evolution   of  warped   modes species should  be
ideally   discussed   within the    wider  framework using   the
coupled   Boltzmann equations   for the  number  densities   of
various species sourced  by the  relevant   annihilation   and decay 
reactions~\cite{kolbturner90,bertone05,aoki12}.
Since this program is technically   hard,  we  here  consider  a simplified  approach   ignoring    cascade reactions       and mass
thresholds  and  assuming  that
the   species evolve independently in a   RD universe
where the     various scattering and decay reactions    occur between 
neighboring   modes.  We first   examine   whether     
thermalization    does occur   and     next    analyze   the
constraints imposed  by  matching the  abundance   
of cold thermal relics to observations.  

\subsubsection{Relaxation to   thermal equilibrium}

Our starting assumption  is that the  warped throat is   
populated by  singlet and charged massive  modes that
interact   through  operators   of  
 dimension $5$ or higher    at   the effective   mass scale $ w M_\star $, 
\bea && L = \l _3 ^{(l, m, m')} H ^ {(l)} \dh H ^
{(m)}\dh H ^ {(m')} +\l _4 ^{(l , l' ; m, m ')} H ^ {(l)} H ^ {(l')}
\dh H ^ {(m)}\dh H ^ {(m ')} + \cdots ,\cr && [\l _n \simeq 2 ^{n/2
    -2} ({ 164 \eta ^3 \over M_\star w }) ^{n-2} \hat \l _n ,\ H^ {(m)} = [h^ {(0)} ,\ (h^ {(m)} ,\ \bar h^ {(m)}) , 
    (C^ {(m)} _i ,\ \bar C^ {(m)} _i ) ]
,\ n\geq   3,\ m \geq   0 ] .  \label{sub6eq1} \eea   
We restrict to graviton modes  although much of
the discussion  could apply to other  modes   of the supergravity  multiplet.  The number conserving $2\to 2 $ scattering reactions  
are  distinguished from the number 
changing $1\to 2 $ decay, $2\to 1 $ inverse decay and $2\to 2 $  pair
annihilation    reactions.   The channels  for    decay and 
annihilation reactions  $ h ^{(n)} \to h ^{(m)} + \bar h ^{(m)} $ and $
h ^{(n)} +\bar h ^{(n)} \to h ^{(m)} +\bar h ^{(m)} $ are open for $
E_n \geq 2 E_m $ and $E_n \geq E_m   ,  \ [E_m = x_m w /\calr ]$.
The tree level s- and
t-channel exchange Feynman diagrams for the processes $ h^{(n)} +
 \bar h^{(n)} \to h ^{(x)} \to h^{(m)} +  \bar  h^{(m)} $
contribute   an  effective  4-point coupling constant,
 $ \l ^{ex} _{4} \sim
\vert \l_3 \vert ^2 ( s/t + t/s ) \simeq \vert \l_3 \vert ^2 (E_m / E_x
)^2  \sim \l_3 ^2  $,    whose size is seen to  be  comparable  to that
of the point  coupling $ \l _4 .$   We   retain  the contributions from cubic
couplings  only, ignore various kinematical factors and   evaluate 
thermal   contributions  to the cross sections   of 
inverse decays and  annihilation   by  means of the  approximate
formulas,  $\s _{2\to 1 } \simeq \l _3 ^2
$ and $\s _{2\to 2} \simeq \l _4 ^2 E_m ^2 $,    deduced  via    
 dimensional analysis.    The rates for scattering and  decay 
reactions and  Hubble expansion are  computed     
by  means  of  the formulas 
\bea &&  \D \G _ {2\to 2} = n_K
^{R} <\s _{2\to 2 } v_K >  = {g_K \zeta (3) \calf \over \pi ^2} \l _4
^2 v_K  T^5 ,   \ \D  \G _{1\to 2 } = {\l _3 ^2 m_K^3 \calf ' \over
  960 \pi },\  H (T) = { \pi ({g_\star \over 90 }) ^{1/2} } {T^2 \over M_\star } 
, \cr && [n_K^R ={g_K  \zeta (3) \over \pi ^2 } T^3 , \ <\s _{2\to 2}
  v_K >  =\calf  \l _4 ^2 T^2   ,\ \l _4 = \vert \l _3 \vert ^2 =\vert
  {164 \eta ^3    \hat \l _3 \over     \sqrt 2 M_\star w_A } \vert ^2
]    \label{eqrelax1}      \eea 
where  $ n_K^R$   is the number density of relativistic modes, 
 $v_K \sim (T/m_K)^{1/2}  $ is the thermal  relative velocity    and 
 $ \calf ,\ \calf ' $ count the number of open
channels  in decays to  bulk and brane modes. 
 Including light scalars, Dirac
fermions and vector bosons in the bulk yields  $ \calf
\simeq ( n_S + 3 n_F + 6 n_V) $, while including the massless gauge
modes  
and $ N_g$ generations of quarks and leptons on a Standard Model brane
yields   $\calf ' \simeq 1 + 3 N_g (6+1 +1/2) + 72 $.  
The  reheating temperature from    decays to  massless modes  
is  derived   by matching   the   energy  density of warped modes  
to that of radiation,  in the approximation  $\D t _{1\to 2} >>t_I$,  
\bea && \rho _K = {3 M_\star ^2 \over 4 ( t_I +\D
  t _{1\to 2} )^2 } = {\pi ^2 g_\star \over 30} T_{RH} ^4
\ \Longrightarrow \ T_{RH} \sim ({90 \over 4\pi ^2 g_\star })^{1/4}
\sqrt { M_\star  \D \G _{1\to 2} } .  \label{RHtemp1} \eea  
Nucleosynthesis  and gravitino non-regeneration  impose
the lower and  upper bounds,  $ 1  MeV  \leq T_{RH} \leq 10^{9}  GeV$.
In order   that the throat      thermalization  be effective,
the ratios   of   the scattering and decay rates to the  Hubble
expansion rate,  $R  (T) $  
and $\D  (T) $,   must satisfy  the    inequalities  \bea &&
R (T) \equiv {\D \G _ {2\to 2} \over H (T) }  =   {g_K \calf  \zeta (3) \over \pi ^2 } \l _3 ^4   T^3   \geq 1 , \ \D (T) \equiv {\D \G _{1\to 2 } \over H(T)}  = {\calf '\l _3 ^2 m_K^3  \over 960 \pi } {M_\star \over 1.6 g_\star ^{1/2} T^2 }  \leq 1   .    \label{WTeq1} \eea 
Selecting  the following values  for     the  input data,   $  x_m = 3.83, \ 
\calf =10  ,\ \calf '=10 ,\ v_K =1,\   \zeta (3) =1.202  
\l _3 =  \xi  \eta ^3 \hat \l _3 / (M_\star w _A) ,\ 
[\xi =164 /\sqrt 2  ,\ \hat \l _3 = 0.481  ,\ m_K= x_m  w /\calr  ] $ 
yields the  following  estimates   for the    
thermalization, decay   and Hubble rates  and the reheating temperature, 
\bea &&  \D \G _{2\to 2}  =5.9  \  10 ^7  {T^5   \eta ^{12}
\over M_\star ^4   w ^4 }  ,\  \  
\D \G _{1\to 2}= 5.8  \  10 ^2  {M_\star w \eta ^6 \over 
z^3 } \cr && \Longrightarrow \ 
R(T)  =  10^9 \   \eta ^{12} T_K^3  w^{-1} z^{-3} ,\  
\D (T) =  12 \   \eta ^6 w^{-1} T_K^{-2} z ^{-1}  , \ 
T_{RH} / M_\star = 10^{-4}\ ({w\over 10^{-4}})^{1/2}    ({\eta \over
  10})^3 ({z \over 10^4})^{-3/2}  .\label{WTeqx1}   \eea 
To analyze the  implications of the  constraints  in 
Eq.(\ref{WTeq1}),  it  is   advantageous to  consider 
fixed values of  the    temperature $  T $ and 
infer   the   resulting  allowed  values  of the free parameters
$ z, \  \eta $ as a function of $T$.
The   inequalities     $R  (T )   \geq 1  $ and $\D  (T)   \leq   1 $  
yield     the respective  upper and  lower  limits on $z$,  
$  z_L \equiv  11.9 \   \eta ^6 w^{-1} T_K^{-2} \leq  z \leq 10^3 \eta
^{6} w^{-1/3} T_K \equiv z_U , \ [T_K= T / m_K]   $.   
Consistency with the condition  $\eta >1$  is achieved only
in   the  relativistic regime  for temperatures  
$ T_K= T / m_K > 1  $.   Increasing  the temperature is  seen to 
increase  and decrease  the upper and lower  bounds  $ z_{U} $,  
hence widening  the  admissible window  for  $z$, 
\bea &&  z_L \equiv 10^9\ ({\eta \over 10})^6 ({w\over 10^{-4}} )^{-1}({T_K\over
10})^{-2} \leq  z \leq  2.\ 10^9\  ({\eta \over 10})^4 ({w\over
10^{-4}})^{-1/3} ({T_K\over 10})\equiv  z_U    . \eea
  The window  closure   (for  $z_L= z_U$) is avoided  provided 
$ \eta  \simeq 12.\ ({w\over 10^{-4}})^{1/3}  ({T_K\over
  10})^{3/2} $, which  implies $ z   \simeq 5. 10^9 \ {w\over 10^{-4} }   
  ({T_K\over 10})^{7} .$   The  allowed parameter region     
is     thus   sizeable   for  short throats   and    
of  reasonable size   for long throats.
  
\subsubsection{Thermal  relics  from  massive metastable charged  modes} 

We  next   examine  whether  the   lightest   metastable charged
warped particles   might survive  as  cold dark matter   relics.
There are two cases of interest depending on whether 
the number changing  reactions 
involve    the  $2 \to 1 $  or  $2 \to 2 $
pair annihilation reactions   that     freeze  out  of equilibrium
in  relativistic and  non-relativistic regimes, respectively.   We
shall use the dimensional  
estimates for  the thermal annihilation rates in these two cases,  
\bea && \bullet \  \D \G _ {2\to 1} = n
_{K} ^R  (T) <\s _{2\to 1} v _K > = n _K^R (T) \l ^2 _3  ,\ [n_K ^{R} (T) =
  (N_B + {7 \over 8 } N_F) {g_K \zeta (3) \over \pi ^2} T^3 ] .\cr && 
\bullet \  \D \G _{ 2\to
  2} = n_{K}^{NR} (T) <\s_{ 2\to 2} v _K> = n _{K}^{NR} (T) \l_4 ^2 
m_K ^2  , \ [\ n_K ^{NR} (T) = g_K ({m_K T \over 2\pi })^{3/2} e
  ^{-m_K  / T} ] .  \label{sub6eq2} \eea 
  The  Hubble expansion rate  consists of additive
contributions from  relativistic and non-relativistic massive modes, 
\bea && H (T) ={1\over   \sqrt 3 M_\star } (\rho _{SM} +\rho _{K} )^{1/2} \simeq {1\over    \sqrt 3 M_\star } { ( {\pi ^2 g_\star \over 30 } T^4 + n^{NR}_K (T)
  m_K ) ^{1/2} } , \eea  where the spin degeneracy statistical factor
$ g_\star $   counts    the  number of  relativistic   degrees of
freedom  
 among   light bulk  and  brane modes.       The   second term from
 massive   bulk  modes   can be dropped since     it is
negligible relative    to the first term in all  cases of interest.
We shall  set  $ g_\star = 100 $   corresponding to the 
 typical contribution   from the Standard Model at large $T$. 

Let us first consider the relativistic case for $2 \to 1$  pair annihilations,  
ignoring the kinematic constraints  from energy conservation  and
setting  the    rate  from   the   above  formula for $ \D \G _{2 \to 1}$  with   the input  data  specified above  Eq.~(\ref{WTeqx1}).  
The freeze-out temperature and
equilibrium number density of left out particles (and antiparticles)
are  determined   by the onset of  decoupling   when  the annihilation
rate exceeds the   Hubble rate.    Repeating the  same analysis
as that of~\cite{chen06},  we    compute the decoupling  time  
$\D t^{dec} _{NR} $  at  which    the    species    
become non-relativistic   ($ T \leq  m_K $)  and    evaluate the    resulting 
number density    $ n_K ^{dec} $   from the 
Hubble   rate at that time   by means of the formulas 
\bea &&  \D t^{dec} _{NR}  = {1\over 2 H (t_{NR})}=  {\sqrt {90} g
  _\star ^{-1/2}  M    _\star \over 2 \pi m_K^2  } 
\simeq  {1.\ 10^{-2} z^2 \over M_\star w_A^2} 
\simeq 
{6.7 10^{-27} z^2 \over (M_\star  / GeV ) w_A^2} \ s  \simeq
2.8\ 10^{-35} ({z\over 10})^2 ({w_A \over 10^{-4} })^{-2} \ s  ,
\cr &&   n_K ^{dec}  \simeq  {H
  (t_{NR} ) \over < \s _{2\to 1}  v_K > } =  
{g _\star ^{1/2} m_K^2  \over < \s _{2\to 1}  v_K >  M_\star } .\eea  
Setting  $\rho _K  ^{dec}  = n_K  ^{dec} m_K ,\ \rho ^{dec} _{tot} =
g_\star m_K^4  $  and  including the suppression  factor  
$a(t  _{RDMD}) / a  (t _{NR} ) \simeq (\rho _{tot}/  \rho
_{RDMD} )^{1/4}\simeq g _\star ^{1/4} m_K /  \rho ^{1/4}_{RDMD}$,  
that accounts  for the different   expansion  rates  
of   the  warped and   massless modes  cosmic components 
up to   the  transition  from    radiation
domination to matter domination (RDMD),  yields   the  abundance   
\bea &&   \O ^I _{K} = {\rho _K  ^{dec}  \over   \rho ^{dec} _{tot} } 
( {a    (\D t ^{dec} _{NR} )  \over  a (\D t  _{RDMD} )} )^{-1} 
={ 1  \over  g _\star ^{1/4 }   M_\star  \rho
  ^{1/4}_{RDMD}    \l _3 ^2}  = 
{2 g _\star ^{-1/4 }  M_\star w_A ^2 \over \xi ^2 \eta ^6 \rho
  ^{1/4}_{RDMD}  }
\simeq   7.8 \  10^{23} g_\star ^{-1/4} w_A^2 \eta ^{-6} ,  \label{CDMabI} \eea
where  we have set $ \rho  ^{1/4} _{RDMD}  \simeq 10^{-9}  \ GeV $.
In the   double throat scenario,  with 
$A$  and  $S$-throats  hosting   hidden  and    visible  
sectors, respectively,   the  suppression 
factor  from expansion      is effective   out to  $\D t_{tun } $ only.
Using     the  $D3$-brane  model estimate for the tunneling  time, one finds   the     abundance   
\bea && \O^{II} _{K} \simeq \O^I _{K}  ({ \D t ^{dec} _{NR} \over  \D  t
_{tun } }) ^{1/2}   
= { g_\star ^{-1/4}M_\star ^{3/2} w _A^2 \D
  \G _{tun} ^{1/2} \over \xi ^2 \eta ^6 \rho 
  _{RDMD} ^{1/4}  m_K } = 1.4   \ 10^{6}    {(M_\star / GeV )  w_A ^{11/2} z ^{1/2}
  \over  \eta ^6  g_\star       ^{1/4} } . \label{CDMabII} \eea  
It is seen that   the upper bound  on the warp  factor  from
Eq.~(\ref{CDMabI})  in the single throat  case,
$ w _A <   (\rho _{RDMD} g _\star )^{1/8} \eta ^3  (\O  ^I_K  / M_\star
) ^{1/2} $,   relaxes   in the double  throat  case to 
$ w _A <   ( 10 ^{-27}   \eta ^6 z ^{-1/2}    \O^{II} _{K} )   ^{2/11}
$.    Matching   the   above  two   predictions   
to  the    observed abundance   yields  estimates    for  the
parameters, 
\bea && \bullet \   \O ^I _K \simeq 7.8  \  10^{23} g _\star ^{-1/4}   \eta ^{-6}
w_A ^2     \ \Longrightarrow  \ 
 \eta  \simeq  10^4 \   g_\star ^{-1/24} w_A^{1/3} (\O _K ^I)^{-1/6} \approx 5.\ 10^{2}  (w_A / 10^{-4})^{1/3}  (\O _K ^I)^{-1/6}   .\cr &&   
\bullet \ \O _K ^{II} \simeq 10^{25}  w_A ^{11/2} g_\star ^{-1/4}\eta ^{-6} (z/10)^{1/2}  \ \Longrightarrow  \ \eta  z ^{-1/12}  \simeq 1.2 \  10^4
g_\star ^{-1/24} w_A^{11/12} (\O _K ^{II})^{-1/6}  \approx 3.\ 10^{2}  (w_A / 10^{-4})^{11/12} (\O _K ^{II})^{-1/6}   . \eea
For $ \O ^I _K   \approx 0.11$ and $ w_A = 10^{-4} $, the  fitted  parameter  values    agree qualitatively  with the  findings in~\cite{chen06}

\begin{figure} [ht]  \centering
\begin{center} \vspace{-1cm} 
\epsffile{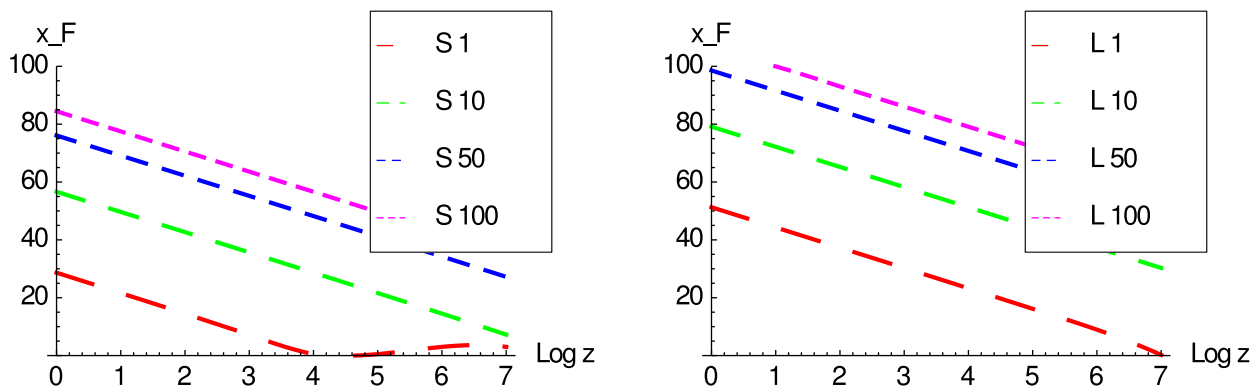}
\end{center}
\vspace{-3cm} 
\caption{\it \label{wf12}
  The  mass to temperature  ratio   $ x_F = m_K/T_F$
  at   decoupling (freeze-out), solution  of  the  equation 
  $ \G _{2\to 2} / H=1$      is   
plotted     as  a function of $ \log _{10}  z ,\  [z =
  M_\star  \calr ] $ at    fixed values of   $\eta = L/\calr $.    
The  (red, green, blue, pink)  dashed curves of decreasing  dashes length  from   bottom to top  refer to
the values  $\eta  = (1, 10, 50,   100)  $ as indicated in the legend. 
The left   and right hand panels correspond to 
 the  short and  long  throat cases  of   
 warp factors  $ w _S=  10^{-4} $ and  $w _L = 2\ 10^{-14}$.}  
\end{figure}

We now  proceed to  the non-relativistic  regime controlled    by $2 \to 2$ annihilation   reactions. The freeze-out
temperature $ T_F$   at which  one saturates 
the inequality $ \D \G (T) / H(T) \geq 1  ,\ [\D \G (T) = n^{NR} _K \l_3 ^4 m_K^2 ] $ is   first evaluated by   solving the equation 
\bea && 1= R (T) \equiv  {\D \G ( C + \bar C \to h +\bar h ) \over H (T) } 
= B {\sqrt{x _F} e ^{-x _F} \eta  ^{12} \over w z^3} ,\ [ x_F = {m_K \over T_F},\ B = 5.20 \ 10^7] \label{sub6eq6}  \eea 
where we have used the  parameter values  quoted  near 
Eq.~(\ref{WTeq1}).  We  show  in  Fig.~\ref{wf12}  plots of  
the numerical solution for $ x_F    $   as
a function of $z$  for a range of values of $\eta
$.   Both  parameters   have an exponential dependence on     $x_F$ 
with $ z$ (at fixed $\eta $) falling       very   rapidly
and $\eta   $  (at fixed $z $)  growing     slowly  
with  increasing  $x_F$.    The  variation  range $ x_F \in (1, 10) $ is admissible only in short throats     for the     parameter
values $\eta \in (1, 10) ,\ z \in (10^3  , 10) $, while
the variation  range   $ x_F \in (10, 50) $ 
is admissible in long  throats    for    parameters varying
inside the intervals   $\eta \in (1, 10) ,\ z \in (10^5  , 10^3). $
Therefore,  the  natural   values   $ \eta \sim 1 , \ z \sim  (10 \eta ) ^3  $ 
are favoured in short throats    for  $x_F  \sim 5$
and in long  throats   for $x_F  \sim 30 $.
    
\begin{figure} [ht]  \centering  
\begin{center}
\epsffile{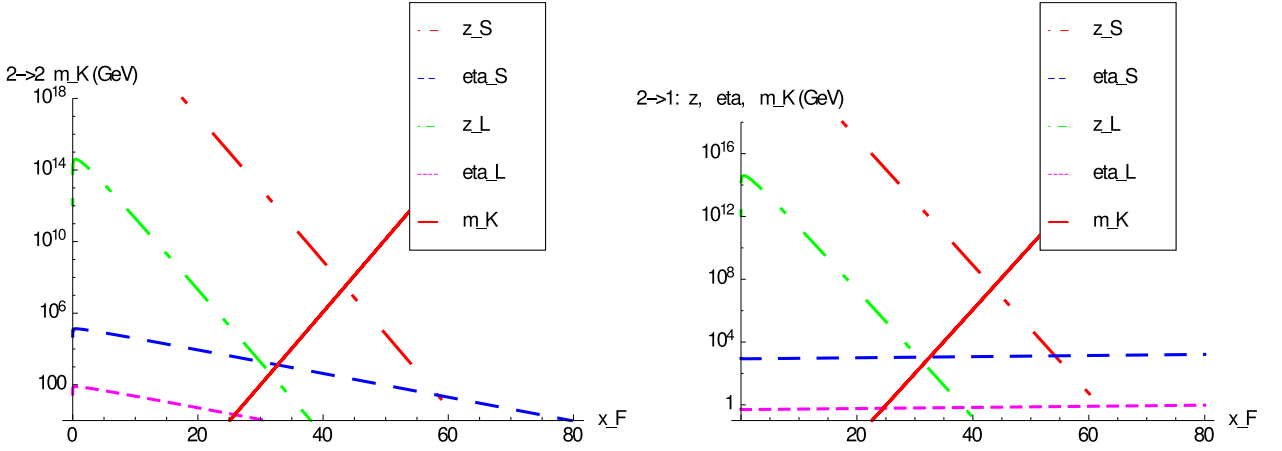} 
\end{center} 
\caption{\it \label{wf13}
The  solutions for the free parameters  $ z  $ and $\eta  $ 
fitted to the thermal relic abundance  and  decoupling
equations  $\O _K= 0.11 ,\  \G _{ann} / H =1$ are plotted  as a function of the     modes mass to temperature ratio $ x _F = m _K/T_F $. The      
short and long throats of warp   factors  
$ w _S=  10^{-4} $ and  $w _L = 2\ 10^{-14}$  are distinguished 
by   the  lower suffix labels $ S,\ L $.
The dot-dashed curves   (in red and green)     refer to $ z_S ,\ z_L$ and the  
dashed   curves (in blue and  pink)  refer to $ \eta _S ,\ \eta _L$ (from top to bottom), as also  indicated  in the legend.
The    fitted   mass of the   warped mode  in GeV units is
 given by the   full  line curve (in red).
The  left hand panel refers to  $ 2 \to 2$  annihilation rates and the
right hand panel refers to   $ 2 \to 1$  annihilation rates.}
\end{figure}

We  next compute the  abundance   by  making   
use of the approximate formula  for thermal relics abundance adapted
from~\cite{weinbergs08},  \bea &&
\O _K \equiv {\rho _K \over \rho _c} = {8\over 11} {6.1\   m_K B_W
  ^{-0.95} T_{\g 0} ^3 \over 3 M_{\star } ^2 H_0 ^2 } \simeq 
A {w ^{1.95} z^{1.85} \over M_\star (\eta ^{12} )^ {0.95} } \simeq
0.11 , \ [B_W = ({90
\over g_\star \pi ^2 } )^{1/2} M_{\star } m_K <\s _{ann} v _K > ,
\cr &&   H_0 = 1.44 \ 10^{-42} GeV ,\ T_{\g 0} = 2.34 \ 10^{-13} \ GeV
,\ \rho _c = (0.003 \ eV )^4 h^2 ,\ A = 5.57  \ 10 ^{38}] .\label{sub6eq2p} \eea
Matching   the  empirical value  $\O _K  \simeq 0.11 $  
while imposing the  decoupling condition  $ R (x_F) \equiv {\D \G
  _{2\to 2} \over H }  =  1$  in Eq.~(\ref{sub6eq6}),  yields two
equations that  we   may use  to find  a unique solution for the 
parameters $z = M_\star \calr ,\ \eta =L/\calr $ as a function
of $x_F= m_K /T_F  $.  In our  present procedure,  the  warped mode  mass, evaluated with the fitted  values of $ z,\ eta $,  turns out  to  be  independent of the warp factor,  
\bea && m_{K} = {x_m w M_\star \over z} = x_m {M_\star ^2 \O _K \ \D \G
  _{2\to 2} \over 
  H A B } ({e ^{x_F} \over \sqrt{x_F} }) ^{0.95} \simeq 2.1 \ 10^{-10}
x_m {\O _K\over 0.11 } {\D \G _{2\to 2}   
\over H } ({e ^{x_F} \over \sqrt{x_F} })^{0.95}  .\eea
The  solutions   for $ z$ and $ \eta $  are plotted    
as a function of the freeze-out ratio  $ x_F = m_K/T_F$ in
Fig.~\ref{wf13}.   As $ x_F$  decreases,      $ z  $ 
fastly  increase  while   $ \eta $  slowly  increases, as
indicate   by the approximate  formula $ z \eta ^{-4} \sim w ^{-1/3}
e ^{-x _F/3}$. The   slopes    of  $z ,\ \eta $ are
independent   of the warp factor
but their  absolute values   are  significantly  
larger for   short  throats.     
The  simultaneous    fit  to decoupling and abundance assigns 
wide  ranges    of values    to the parameters  that  extend  well
outside  the ranges   in    Fig.\ref{wf12}, hence     
ruling out  large portions of the parameter space  contemplated  there. 
For instance, $ x_F=20 $ selects $ \eta \sim 10^3 ,\ z \sim 10 ^{18} 
$ for short throats and $ \eta \sim 10 ,\ z \sim 10 ^8  
$ for  long throats.   The   LCKP mass    grows exponentially 
with $ x_F$    with typical  values   
$ m_K \in [10 ^{-2}    \ GeV   \ - \  10^3  \ GeV ]  $  over the interval     
$ x_F \in (10\ - \ 30)$.  The natural parameter  values     
favoured  in short throats  occur for   $ x_F \sim 50$   
with $  z_S \sim 10^5 ,\ \eta _S \sim 10^2 ,\  m_K  \sim 10^8 \ GeV$  
and  those favoured  in long throats occur for $ x_F \sim 20$  with 
$  z_L\sim 10^7  ,\ \eta _L \sim 10 ,\  m_K  \sim 1  \ GeV$. 
It is usefult to  compare our findings  with  those of~\cite{kofman08}
where a   different fitting   procedure to  the observed abundance
was used that    gives    mass  values  in short   throats  $ m_K
\sim  (10^{11} \ \ - \ 10^{12 } )  \ GeV $ and in long throats
$ m_K \sim (0.2 \ - \ 5    ) \ TeV $     at  fixed    
$L  /l_s \in  (1-5) ,\  \calr  m_s \in (0-30)$.   
Our  procedure  uses a floating   $ x_F$     and   covers a wider  region of the parameter space,    as seen from  the ratios of  their  free parameters
to ours,  $  \calr m_s  / z  = \sqrt {\pi / \calv _s } g_s,\ (L /l_s)  / \eta   = \l  g_s ^{1/4}    $. 
Finally, we   note that if the  pair  annihilation  rates  were  described in terms of  the $ 2\to 1 $  reactions instead,   solving  the resulting  equations 
\bea && R (T) \equiv {\D \G _{2\to 2 } \over H } = B {\sqrt{x
    _F} e ^{-x _F} \eta ^{6} \over w z} ,\ \O _K= A{w ^{1.95} \over
  M_\star (\eta ^{6} )^ {0.95} z ^{0.05} },\ [B  = 1.140 \ 10^3,\ A
  = 1.487 \ 10 ^{43} ]    \eea
for  the parameters $\eta ,\ z $   as  a function of $ x_F$  yields 
 the       plots  in  the  right hand   panel  of Fig.~\ref{wf13}.       
Similar features  to  those 
 found  in the $2\to 2 $ annihilation case  are      found  with 
slower slopes and    narrower ranges of variation for the parameters. 

\vskip 0.2 cm 
\vskip 0.2 cm 


\section{Summary and conclusions} 
\label{sec7}
The  purpose   of this work  was to    test      the viability 
of  a version of warped  models    inspired  by  string theory. Our
presentation  draws  heavily on previous published  studies.
We have   presented a    comprehensive review of 
Kaluza-Klein theory   for flux compactifications   of type $II \ b
$ supergravity   on   the  GKP   type  background  of a  warped
throat   region   glued to  a  conic  Calabi-Yau  orientifold. 
It is  important  to address    first  the objection that strong warping  is valid  only    in the regime of small compactification volumes,   which  is incompatible with   a  supergravity    description~\cite{blumen16}. This is answered by  noting that   one  need only   require  the inequality
$\calv \vert \e  \vert ^2 << 1$ on the
product of the volume  by the complex structure  modulus.
It  thus seems    consistent  to   work within supergravity provided    one 
stays  very close to the singularity,  at  small   values  of the
complex structure moduli.   

The   throat modelization by the warped conifold   benefits  from 
continuous symmetries  that are helpful in classifying warped modes  
and  identifying the leading contributions  to 
deformations of the  classical background by compactification
effects.    At  large radial distances away from
the conifold apex and mouth, the harmonic analysis applies 
for modes of the base manifold $T^{1,1}$ propagating in an
asymptotic $AdS_5 $ spacetime.  
We considered    the  extensively  documented  model of an $ AdS_5 $ slice truncated  by  hard walls where the
Sturm-Liouville eigenvalue problem  for    warped modes  is 
solved in terms of Bessel functions.
The  axial gauge  formalism~\cite{shiu08} provides a  convenient 
framework  to  discuss the Kaluza-Klein  theory for warped compactifications. 
We have  reviewed and hopefully clarified  several aspects of the    dynamics   but  by no means all.  In particular, the r\^ole  of
non-transversal and trace parts of the   metric
tensor  was  not  fully documented.  

The most robust  predictions  are those  made for
the metric tensor field along $ M_4$.  The  cubic and higher order
gravitons self-couplings  and   their cubic couplings to  bulk and  brane  modes  were  evaluated  approximately  by    ignoring the spacetime 
structure of interactions.   With  proper account  of the   wave
functions normalization,  the  cubic  couplings    between various
 charged and  radially excited  modes 
are of   nearly same size in all  cases and they  usually 
dominate over    the higher order couplings.     The effective
gravitational scale    in th Standard Model throat
lies around $ \L _K \sim 28 $ TeV.  The  modes of lowest mass  appear
in the  coupled system of scalar  fields,   $ \pi \sim \d g ^a_a \sim
\g ,\ b \sim C_{(4)} $  as   charged  states    
under the isometry group      with   quantum numbers $ (jlr )= (010)
+ (100) $.  The favoured candidates for lowest-lying charged particles (LCKPs)
might be produced   through   two-body decays  of the singlet graviton modes 
 $ h ^{(0)}  \to b + \bar b  $.
 The properties of    moduli and various  other warped modes
 descending  from the supergravity   multiplet     were  
also examined  but    less systematically. 
The  thermalization and    reheating
of warped  throats produced  after  brane inflation  
are more easily   accomplished  for   a gas of 
relativistic warped modes.  Our fitting procedure of the cosmological
constraints   on  a  cold   thermal relic  of  long lived 
non-relativistic  warped modes   yields   somewhat   large   
values  of the parameters  $ z = M_\star  \calr $ and  
$\eta  = V_W^{1/6} / (2\pi   \calr ) $. 
Since  this is     due     to  the 
exponential  sensitivity   on   the   value of the   temperature-mass
ratio $x_F = T_F / m_K$ at freeze-out, it is   hoped   that our
preliminary conclusions  could be  confirmed  through a
detailed  study   of  the thermal evolution  using   Boltzmann  equation. 

\begin{appendix}   
\appendix 
\section{Useful tools  for  flux compactifications  
for  type $II\ b $  supergravity theory}  
\label{app1} 

We  present in this Appendix   tools  of use  in developing 
the Kaluza-Klein theory for warped flux compactifications of type $II\ b $
string theory. The construction   relies   on the  differential
calculus  on Riemann manifolds  in the context of general relativity.  Our
presentation  draws    on existing 
reviews~\cite{vannieuw84,duffpope86,appel87} and lecture
notes~\cite{Popeweb,blau16} as well as 
recent physical~\cite{douglas09} and
mathematical~\cite{dobarro04}   discussions of warped Riemann
manifolds.

We shall provide subsets  of  useful  differential geometry formulas are provided  in
Subsection~\ref{subapp0} for Riemann manifolds and in
Subsection~\ref{subapp2} for warped Riemann manifolds.  Turning  
next   to  the bosonic field components of 
10-d supergravity theory, we present
in Subsection~\ref{subapp1}   the classical field equations and
Bianchi identities,   in Subsection~\ref{subapp3}   the classical field
equations adapted to warped spacetimes $ M_4 \times X_6$ 
and in  Subsection~\ref{subapp4}      
the linearized wave equations for  the   various  field fluctuations.    In
Subsection~\ref{subapp5}  we   present a useful   formula  
for the reduced  curvature action exhibiting the  
full spacetime structure of the    gravitational couplings of 
(massless and massive)  gravitons in an   arbitrary gauge.
In Subsection~\ref{subapp6}  we review the   construction  of the low energy
effective  bosonic and fermionic action  for $Dp$-branes. 

\subsection{Differential  calculus on  Riemann manifolds}  
\label{subapp0}

Riemann manifolds  are  differentiable  varieties  equipped with
a metric tensor  field $g$, a   volume   field $\e $ and   
derivative  operators. 
In  manifolds of dimension $D$,  parameterized by
real coordinates $ X^M ,\ [M=0, 1 ,\cdots , D-1] $ the   real
symmetric metric tensor  $g _{MN}$     is used to    construct the
differential distance squared    $ ds ^2 =  g _{MN}  d  X^M d  X^N$,  the 
rank $D$ totally antisymmetric   tensor field  $\e = \e  ^{(D)} $ of
components $\e _{M_1 \cdots M_D}  = \pm \sqrt {\vert g_D\vert } =\pm
Det ^{1/2}   (\vert g_D\vert )$   and  the  covariant derivative   $\nabla _M $.   Under the (diffeomorphism) group of differentiable coordinates 
transformations,   $ \d X^M = \xi ^M
(X) ,\ \d g _{MN} = 2 \nabla _{( M } \xi _{N )} ,\  \d \e _{M_1 \cdots
  M_D } =    0  $,  with  the transformation law for 
covariant quantities    being    independent    of  
the  variation  field $\xi (X)$  derivative.
 The  Minkowski and Euclide   type curved spacetimes  
of    metric tensor signatures $ (-+\cdots +)$ and $ (++\cdots +)$,
with   odd and even numbers of   negative  eigenvalues,   are   distinguished by the
parameter values $t=1,\ 0$.  The  standard  operators   of    differential calculus  are  the differential  $d$ and its
adjoint  $\d $, the wedge product $\wedge $, the Hodge star operation
$\star = \star _D$ and the Hodge-de Rham (or Laplace-Beltrami)
quadratic differential operators $\D _p $ acting on $p$-forms $
V_{(p)} $.   
We list below     a small set  of 
useful    definitions   and formulas illustrating 
the action of the various operators, including        
translation  formulas to  the  tensorial formalism.
\bea && \bullet \ V_{(p)} = {1\over p !}  V_{M_1 \cdots M_p } d
X^{M_1}\wedge \cdots d X^{M_p},\ ( d V_{(p)} ) _{M_1 \cdots M _{p+1}}
= (p+1) \dh _{[M_1 } V_{M_2 \cdots M _{p +1} ]} ,\ [d^2  V_{(p)} =0] . \cr &&
\bullet \ (V_{(p)} \wedge W_{(q)} ) _{ M_1 \cdots M_{p+q} } = {(p+q)! 
 \over  p!q!}   V_{ [M_1 \cdots M_{p} } W _{ M_{p+1} \cdots M_{p +q}
] } , \ d (V_{(p)} \wedge W_{(q)} ) = d V_{(p)} \wedge W_{(q)} +
(-1)^p V_{(p)} \wedge d W_{(q)} . \cr && \bullet \ (\star V_{(p)}
)_{M_{p+1} \cdots M_{D} } = {1\over p!}  \e ^{M_1 \cdots M_p}
_{\ \ \ \ \ \ \ \ M_{p+1} \cdots M_{D} } V _{M_1 \cdots M_p} ,
\ (\star V_{(p)} \wedge W_{(q)} ) = { vol( M_D) \over p!}  V_{M_1
  \cdots M _{p}} W ^{M_1 \cdots M _{p}} , \cr && \star \star V_{(p)}
= (-1) ^{p (D-p) +t} V_{(p)} ,\ \star 1 =  vol(
M_D) = {1\over D !} \e _{M_1 \cdots M_D} d X^{M_1}
  \wedge \cdots \wedge d X^{M_D} =\sqrt{ \vert g_D \vert} d X^{1}\cdots
  \wedge d X^{D}  , \cr && 
\star d \star d \phi = \sqrt {\vert g_D
  \vert } \nabla ^2 \phi =  \dh _M \sqrt{ \vert g_D \vert}
g ^{MN} \dh _N \phi ,\ d \phi \wedge \star d \phi = \sqrt {\vert
  g_D \vert } g^{MN} \dh _M \phi \dh _N \phi ,\cr && \star  d ( e ^{p \phi }
\star d C _0 ) = \sqrt {\vert g_D \vert } \nabla ^M (  e ^{p \phi } \dh _M C _0)
,\ F_{(p)} \wedge \star F_{(p)} = \sqrt {\vert g_D \vert }  (F_{(p)} \cdot
F_{(p)} )  = { \sqrt {\vert g_D \vert } \over p!} F_{M_1 \cdots M_p} F^{M_1
  \cdots M_p} ,\cr && [\e _{M_1 \cdots M_D} = \pm \sqrt {\vert g_D\vert } ,\ \e
  ^{M_1 \cdots M_D} = \pm {1\over \sqrt {\vert g_D\vert } } ,\ \e
  ^{M_1 \cdots M_p }
  _ {\ \ \ \ \ \ \ \ M_{p+1} \cdots M_{D} } \e _{N_1 \cdots N_p M_{p+1}
    \cdots M_{D} } = p! (D-p)! \d ^{[M_1} _{[N_1} \cdots \d ^{M_p]}
    _{N_p]} ] . \cr && \bullet \ \d V_{(p)} \equiv \d \cdot V_{(p)} =
(-1) ^{ D p +t} \ \star d \star V_{(p)} ,\ [d^2 = \d ^2 =0] ,\cr &&
(\d V_{(p)} )^{M_1 \cdots M_{p-1} } = - \nabla _M V ^{M M_1 \cdots
  M_{p-1} } = -{1\over \sqrt {\vert g_D \vert } } \dh _M \sqrt { \vert g_D
  \vert } V ^{M M_1 \cdots M_{p-1} } ,\cr && \ \D _{p} V_{(p)} = (d\d
+ \d d ) V_{(p)} = (d +\d )^2 V_{(p)} , \ [\star \D _p =\D _p \star ]
. \label{appeqm1} \eea 
The square brackets, parentheses and curly brackets
enclosing subsets of indices  denote antisymmetrization,
symmetrization and  symmetrization with trace subtraction, respectively.  
For instance, $ V _{[M } W _{N
]} $ and $ V _{(M } W _{N )} $ stand for $\ud ( V_M W_N \mp V_N W_M) $
and $ V_{\{MN\} } = V_{(MN)}  - (g_{MN}/ D)  V ^P_{\ P} . $ 
The Christoffel affine   connection  $ \G $    enters the 
definitions of  the Levi-Civita  covariant derivative   action  on
tensors,   the Riemann  curvature  4-tensor $R _{MPNQ}= -R _{PMNQ}
=- R _{MP QN}=  R _{NQ MP}  $  and the  Ricci 
curvature 2-tensor $R_{MN}= R_{NM}$ and scalar $R $    as  
 \bea && \bullet \ \nabla _M V_N =\dh _M V_N - \G _{MN} ^P V_P,\ \nabla
_M V^N =\dh _M V^N + \G _{MP} ^N V^P,   \ 
[\G ^P _{MN} = \G ^P _{NM} = \ud g^{PQ} ( \dh
  _M g_{NQ} + \dh _N g_{QN} - \dh _Q g_{MN} ) ] . 
\cr && \bullet \ [\nabla _M, \nabla _N]  V^P=  R ^{P} _{\ \  QMN} V^Q ,\  
[\nabla _M, \nabla _N]  V_P=  -  R ^{Q} _{\ \  PMN} V_Q ,\
[\nabla _M, \nabla _N]  W_{PQ}=  -  R ^{R} _{\ \  PMN} W_{RQ} 
-  R ^{R} _{\ \  QMN} W_{PR} ,\cr &&   
[R _{MNP}^{\ \ \ \ \ \ Q} = - R ^Q _{\ \ PMN} = - 2 \dh _{[ M} \G _{ N
  ] P } ^ Q + 2 \G _{ P [M} ^ {S} \G _{ N] S} ^ Q   ]. \cr && 
\bullet \ R_{MN} = R _{MPN}^{\ \ \ \ \ \ \ P} = R ^P _{\ \ MPN} 
= - \dh _M \G ^P_{NP} - \G ^Q_{NP}  \G ^P_{MQ} + 
\dh _P \G ^P_{NM} + \G ^Q_{NM}  \G ^P_{PQ} ,\ \ \  R =
g^{MN} R_{MN} = R_M^M   . \eea 

It is also  helpful to introduce the flat    coordinates $ X^A ,\ [A=1,\cdots , \  D] $  of the manifold tangent space of constant metric $ \eta _{AB}$ and to
consider the  spin connection (matrix) field  $\o _M ^{AB}  = - \o _M ^{BA} $
with    indices $ A, \ B $ transforming   under the adjoint representation of  the  Lorentz symmetry group $ SO(D)$. This is needed
to   construct   covariant  quantities  for spinor  fields. 
The curved and flat  space components of vector fields are related
through    linear transformations, $  V_M = e^A_M V_A  , \   V_A =
e_A^M V_M, \ [\eta _{AB} e^A_M e^B_M     = g_{MN}  ]   $     
involving the basis of frame fields  matrices $
 e^A_M$  and  the    dual  inverse  matrices  $e_A^M$. 
The Riemann  tensor  with 
 mixed flat-curved indices is    similarly defined  by
 $ R_{MNAB} =  \dh _M \o _{N  AB} - 
\dh _N \o _{M  AB} + [\o _{M} , \o _N ] _{AB} .$ 
The   differential     calculus  makes use of   vielbein and
spin connection 1-forms  $e^A ,\ \o ^{AB} $
and  the torsion and Riemann  curvature   2-forms  $T^A ,\ R  ^{AB}$,    defined through the   Cartan structure equations,  
\bea && T^A  = D  e ^A =  d  e ^A +  \o _B ^A  \wedge
e ^B  ,\  R _{AB}  = d \o _{AB}  +  \o _{AC}  \wedge \o ^{C}
_{\ B}   ,\cr && [e^A=e^A_M d X^M ,\ \o ^{AB} =   \o ^{AB}_M d X^M , \
T^{A} = \ud  T ^{A}_{\ \ \ MN}   d X^ \wedge d X^N ,\
R ^{AB} = \ud  R ^{AB}_{\ \ \ MN}   d X^ \wedge d X^N ,\
R _{MNPQ} =  R_{MN AB} e ^A _P e^B_Q ] . \eea
The   covariant differential $ D   = d +\o $
acts on vector,  2-tensor  and spinor fields
$V^A ,\  V_N^A  ,\  \psi $ as
\bea && D_M V^A =\dh _M V^A +  \o _M  ^{AB} V^B  ,\  
D_M V_N^A =\nabla  _M V_N^A +  \o _M  ^{AB} V^B_N = \dh _M V_N^A 
- \G _{MN} ^P V_P^A +  \o _M  ^{AB} V^B_N ,\cr && 
D_M \psi =  (\dh _M   +  {1\over 4} \o _M ^{AB} \G _{AB} ) \psi
,\  [D_M , D_N ] \psi = {1\over 4} R _{MNPQ } \G ^{PQ} \psi,\cr &&  
[\G _A  = e_A^M \G _M,\ 
  \{ \G _A  , \G _B \} = 2 \eta _{AB} ,\ \{ \G _M  , \G _N \} = 2  g_{MN}  ] \eea
where $ \G _A  $ are  (constant) Dirac gamma  matrices.
The consistency  between the  action on  flat and  curved
 components  requires   the  conditions  $  D_M e_N^A =0$,  implying
 that     the torsion  tensor $  T^A _{MN}  =   D_M e_N^A -D_N e_M^A $
vanishes.  The   corresponding   $D^3$  equations   are solved
for  the   spin  connection  as a function of    vielbein  fields, 
\bea &&  \o _M  ^{AB} = \ud [ e ^{ A N} (\dh _M e _N^B - \dh _N e
  _M^B) -  (A \leftrightarrow B) ]   -  
\ud   e ^{AP} e ^{BS} (\dh _P e _S^C - 
\dh _S e _P^C )  e _M ^C . \eea 

Given   a    pair of distinct metric  tensors $ g $ and $\tilde g$, 
then the  covariant derivatives 
$ \nabla $ and $\tilde \nabla $ and curvature tensors $
R$ and $\tilde R $ adapted to these metrics   are  related  by~\cite{wald84} 
\bea && \nabla _M V_N =\tilde \nabla
_M V_N - \tilde \G _{MN} ^P V_P,\ \nabla _M V^N = \tilde \nabla _M V^N
+ \tilde \G _{MP} ^N V^P, \  R _{MNP} ^{\ \ \ \ \ \ Q} =
\tilde R _{MNP} ^{\ \ \ \ \ \ Q} - 2 \tilde \nabla _{[ M} \tilde \G _{
    N ] P } ^ Q + 2 \tilde \G _{ P [M} ^ {S} \tilde \G _{ N] S } ^ Q 
, \cr &&    [\tilde  \G _{NP} ^ M = \ud g ^{MQ} ( \tilde \nabla _{ N } g _{ P
  Q} + \tilde \nabla  _{ P } g _{ N Q} - \tilde \nabla _Q g _{NP} )
  ,\ \tilde \nabla  _{M } \tilde  g _{ N P} = 0 ] . \eea
The   inverse  transformations  from  $ \tilde  g \to g$  take the  same
form  by  trading the quantities with  tilde  for  those without tilde.  
The conformal transformations      are  defined  as  Weyl rescaling  of the  metric  tensor 
$ g_{MN} = e^{2\O (X)}  \tilde  g_{MN} $  and,   
in a wider  sense,  as conformal  isometry transformations,    
corresponding  to  coordinates diffeomorphisms   that  map the metric
$ g_{MN}  \to    e^{ - 2\O }    g_{MN} =  \tilde g_{MN} $    for  some
function $\O (X) $.  The     metric  rescalings    from   unwarped
to warped   metric tensors   $\tilde g _{MN} (X) \to g _{MN}  (X) = 
e ^ {2 \O (X) } \tilde g _{MN} (X) $   induce  transformations of   the   corresponding  geometrical  quantities,        distinguished    by the  tilde
 symbol.     For instance, the  covariant derivatives action on vector and spinor fields, $  V^M,\ \psi $,   the Laplace
operator   action on scalar fields and  the   Ricci    curvature
tensor and scalar    adapted to the warped metric  $ g$ are
related  to those adapted to the   unwarped metric $ \tilde g$  
by the useful  formulas
\bea && \bullet \ 
\nabla _N V^M =\tilde  \nabla _N V^M  + V^P 
( 2 \d _{(N } ^M \tilde \nabla _{P )} - g_{NP}  g^{MQ} \tilde 
\nabla _{Q } )  \O   ,\ 
\nabla _M \psi  =  \tilde \nabla _M \e +\ud \tilde \G _{MN} e
^ {-\O } \tilde \nabla ^N (e ^{\O } ) \psi  (X)   ,\cr && 
\bullet \ \nabla ^2  \phi   = e ^{-2 \O } (\tilde
\nabla ^2 + (D-2) \tilde  \nabla _M \O \nabla ^ {\tilde M} ) \phi ,\ 
 \cr &&  \bullet \ R _{MN} =\tilde  R _{MN}  + (D-2) ( -\tilde
\nabla _M \tilde \nabla _N \O + \tilde \nabla _M \O  \tilde \nabla _N
\O  - \tilde g _{MN} (\tilde \nabla \O  )^2 ) - \tilde g _{MN} \tilde \nabla
^2  \O   ,\cr &&  R =   ^{-2 \O }  (  \tilde R -  2 (D-1)
\tilde \nabla ^2 \O - (D-1)(D-2) (\tilde \nabla \O )^2  ) , \cr && [\int
  _{M_D} d ^D X \sqrt {g } R = \int _{M_D} d ^D X \sqrt {\tilde g } e
  ^{ (D-2) \O } [ \tilde R -  (D-1)(D-2) (\tilde \nabla ^M \O ) (\tilde
    \nabla _ M \O ) ] ]       \eea
where   $  \tilde \G_{M N}   = \tilde \G _{[M } \tilde \G _{N]} ,\ [M\ne N]$. 
Note that  the  Laplacian  term $\tilde \nabla ^2  \O $ in  the
formula for $R$   in the last entry above  reduces   to    a  boundary
term   that  drops out    in the  volume    integral of $R$  
if  $ M_D$ is a closed (compact) manifold. 

  The  Hodge-de Rham operators $ \D _p$, 
generalizing  the scalar Laplace operator $ \D _0 = -\nabla ^2$, are
invariant quadratic differential  operators 
whose  action on (antisymmetric tensor) $p$-form fields
commutes  with the covariant derivative,
\bea && <V_{(p)} , \D _p V_{(p)} > = (
<d V_{(p)}, d V_{(p)} > + <\d V_{(p)}, \d V_{(p)} > ) \geq 0 ,\ \ \ \D \nabla _{ [M }   V_{ M_1 \cdots M_p]} = \nabla _{[M} \D V_{M_1 \cdots M_p ]} . \eea
In our present conventional  choice of  phase  in the definition of
$\d $, the $p$-form  eigenfunctions of $\D _p$  have non-negative
eigenvalues $\D_p V_{(p)} = \l V_{(p)} ,\ [\l \geq 0]$. 
The action of $\D _p$  decomposes into a  differential part
$ \nabla ^2 $  and a   algebraic  part involving the curvature tensors, 
as illustrated  by  the identities    for forms of order  $ p=1, 2,
3$~\cite{duffpope86} 
\bea && \D _1 V _M = - \nabla ^2 V _M + R _M ^{\ N} V _N ,\ \D _2 V _{MN} =
- \nabla ^2 V _{MN} + 2 R ^{P \ \ \ Q} _{ \ M N } V _{PQ} - 2 R ^ P
_{\ \ [M } V _{N ] P},\cr && \D _3 V _{MNP} = - \nabla ^2 V _{MN} - 6
R ^{ \ \ \ \ QR } _{ [M N } V _{P] QR } + 3 R _{ [M } ^{\ \ \ R} V _{N
    P ] R} ,\ [\nabla ^2 = g^{MN} \nabla _M \nabla
_N ].  \label{appeq17} \eea
Analogous  invariant  quadratic differential operators  
acting on  symmetric 2-tensor fields  are
defined   by   examining   the Ricci  tensor variation 
under  metric perturbations.  The transformation  of $ R_{MN}$ 
under  $\d g_{MN} =h_{MN} $  defines  the 
Lichnerowicz operator $\D _L $  and the 
related wave  operator $\D _K$ as~\cite{gandhi11}   
\bea && \d R _{MN} (g)
\equiv R _{MN} (g +h ) - R _{MN} (g) = E^R_{MN} = - \ud \D _K h _{MN}
\cr &&  =-\ud   (\nabla ^2 h _{MN} + \nabla _{(M } \nabla _{ N )}
  h ^P_P - 2 \nabla ^P \nabla _{ (M } h _{N) P} ) 
= \ud (\D _L h _{MN} - \nabla _{(M } \nabla _{ N )} h^P _P + 2 \nabla
_{(M } \nabla ^P h_{N ) P} ) , \cr && [\D _L h _{MN} = -\nabla ^2 h
  _{MN} + 2 R _{P (MN) Q} h ^{PQ} + 2 R _{(M } ^{\ \ \ \ P} h _{N ) P} ] .  \label{appeq16}
\eea 
For  metric tensor fields in the
transverse-traceless gauge, $ h = g^{MN} h _{MN} = h^M_M =0,\ h_N = \nabla ^M h
_{MN} =0$, the Ricci tensor variation  simplifies to $ \d R _{MN} (g) = \ud
\D _L h _{ M N } .$ 

Note  finally that the  Einstein (homogeneous) spaces are   
manifolds  of  maximal symmetry and constant scalar curvature $R$
with    Riemann and Ricci  tensors     of form
\bea && R_{MNRS } ={s \over \calr ^2 } (g_{MR } g_{NS } - g_{MS }
g_{NR } ) ,\ R_{MN } \equiv R_{MPNQ} g ^{PQ} = {s \over \calr ^2 }
(D-1) g_{MN } ,\ R = {s \over \calr ^2 } D (D-1) , \eea 
where $\calr $ sets the  overall scale  and 
$s = [-1, 1, 0] $  for  the  $[AdS_D ,\ dS _D ,\ M_D ]$
spacetimes.

\subsection{Warped  Riemann manifolds}  
\label{subapp2} 

We   here  focus  on  spacetimes $ M_D$   given   by  
direct products $ M_D= M_d \times X_k$  of 
 warped external  and  internal  spaces of  dimensions $ d,\ k ,\ [D = d + k ]$
 equipped   with metric tensors $ g_{\mu \nu } ,\ g_{mn}$ of Lorentzian 
 and  Euclidean  signatures  ($ t =1,\ 0$) and
 coordinates systems  $ x ^\mu \in  M_d $ and $ y^m\in  X _6 $.
The  properties of warped Riemann manifolds are   discussed  from a
general angle in the mathematical literature~\cite{dobarro04}.
For the special type of warp   profile $ A (y)$ of $ M_d $  and conformal profile $ B (y)$  of   $X_k$,    
 depending on the $y^m$-coordinates   only,  the   differential
 distance  elements    are   defined   as 
\bea && ds ^2 _D = g_{MN} dX^M dX^N  \equiv ds ^2 _d + ds ^2
_k  , \  \  \ [ds ^2 _d = e ^{2A (y)} d \tilde s ^2 _d =
  g_{\mu \nu } dx ^\mu dx ^\nu = e ^{2A (y)} \tilde g_{\mu \nu } dx
  ^\mu dx ^\nu ,\cr && ds ^2 _k = e ^{2 B (y)}d\tilde s ^2 _k = g_{mn
  } dy ^m dy ^n = e ^{2 B (y)} \tilde g_{mn } dy ^m dy ^n ]  \label{appeq5} \eea
where  the tilde symbol  is used to distinguish between 
geometrical quantities and index contractions associated to the
unwarped  $\tilde g_{\mu \nu } ,\ \tilde g_{mn } $  
and warped  $ g_{\mu \nu } ,\  g_{mn } $ metric tensors.  
The discussion in the text is mostly specialized
to the case $B (y) = -A (y)$.   
We quote below useful  relations between warped and
unwarped quantities    that  illustrate some
of the conventions adopted in the present work,  \bea
&& \bullet \ \tilde R^{(d)} = \tilde g^{\mu \nu } \tilde R_{\mu \nu
},\ \tilde R^{(k)} = \tilde g^{mn } \tilde   R_{mn }, \tilde \nabla _{k} ^2
\phi 
= \nabla _m \nabla ^{\tilde m} \phi = {1\over \sqrt {\tilde g_k} } \dh
_m \sqrt {\tilde g } \tilde g ^{mn} \dh _n \phi ,\ ( \nabla _k \phi )
^2 = e ^{2A} ( \tilde \nabla _{k} \phi ) ^2 .  \cr && \bullet \ G
_{mn} = e ^{-2A} \tilde g _{np} G _{m} ^{\ \ p} ,\ G _{ \mu m} = e ^{2
  A} \tilde g _{\mu \nu } G _{\ \ m} ^\nu ,\ G _{ \mu \nu } = e ^{2 A}
\tilde g _{\mu \l } G _{\mu } ^{\ \ \l } .  \cr && \bullet \ \vert
V_{(p)} \vert ^2 = e ^{2p A (y)} \vert \tilde V_{(p)} \vert ^2 =
{1\over p !} e ^{2p A (y)} V_{m_1 \cdot m_p} V^{ \tilde m_1 \cdot
  \tilde m_p} ,\ ( V_{(p)} \cdot W_{(p)} ) \equiv {1\over p!}  V_{(p)}
\cdot W_{(p)} ={1\over p!}  V_{m_1 \cdots m_p } W^{m_1 \cdots m_p } .
\cr && \bullet \ 3!  (G _3 \cdot \bar G _3) \equiv G _3 \cdot \bar G
_3 = G _{mnp} \bar G ^{mnp} = {e^{6 A} } G_{ mnp } \bar G ^{\tilde m
  \tilde n \tilde p } = e ^{6 A} G _3 \tilde \cdot \bar G_3
. \label{appeq8} \eea
No  confusion should hopefully  arise  from using  
the same symbol $G$  to denote the Einstein tensor
 $ G_{mn}$ and the complex 3-form field strength $ G_{mnp}$.     
The  Ricci curvature tensor   components along  $ M_D$  
split additively  into   diagonal and off-diagonal parts  involving 
diagonal and off-diagonal components 
along  the directions   $ M_d$ and $ X_k$
\bea && R^{(D)} _{MN} = R ^{(d)} _{MN} + R ^{(k)} _{MN} = R _{M\l N} ^{\ \ \ \ \ \ \l 
} + R _{M k N} ^{\ \ \ \ \ \ k} , \ R^{(D)} = g^{MN} R^{(D)} _{MN} = 
R ^{(d)} + R ^{(k)}  ,\ [(MN) = (\mu \nu ),\ (m n) ,\cr &&  R^{(D)} _{\mu \nu } =
  R _{\mu \l \nu } ^{\ \ \ \ \ \ \l  } + R _{\mu  p \nu } ^{\ \ \ \ \ \ p},\
R^{(D)} _{mn} = R _{m \l  n} ^{\ \ \ \ \ \ \l  } + R _{m  p n } ^{\ \ \ \ \ \ p},\  
R ^{(d)} = g^{\mu \nu } R^{(d)} _{\mu \nu } ,\   R ^{(k)} = g^{mn} R^{(k)}_{mn} ,\   
] . \label{appeq6} \eea 
For   the class of warp  profiles      varying  over $ X_k$ only, 
the total Ricci tensor $ R _{\mu \nu } (M_D)$ along $M_d$ is related to its counterpart $ R  ^{(d)} _{\mu \nu } (M_d)  $  for the isolated space $M_d $  by  means of  the formula~\cite{firouz06} 
\bea && R_{\mu \nu }  ^{(d)} (M_D)  = R ^{(d)}
_{\mu \nu } (M_d)  -\ud \nabla _{k}^2 g_{\mu \nu } - {1\over 4} g^{\rho \l }
\nabla _m g_{\mu \nu } \nabla ^m g_{\rho \l } +\ud g^{\rho \l } \nabla
_m g_{\nu \rho } \nabla ^m g_{\mu \l } .\label{appeq7} \eea 
We list  below useful    results, borrowed from~\cite{douglas09}, 
that link   components of the  various geometrical quantities   adapted to
the warped and unwarped metrics. 
The  formulas  for the  Christoffel symbols  are 
\bea && \G _{np} ^m - \tilde \G _{np} ^m= \d _n^m \dh _p B + \d _p^m \dh _n B - \tilde g_{np} \dh  ^m B ,\ \G _{\nu m} ^\mu - \tilde \G _{\nu m} ^\mu = \d _\nu ^\mu \dh _m A 
,\cr && \G _{\mu \nu } ^m - \tilde \G _{\mu \nu } ^m= - e ^{ 2A - 2B}
\tilde g _{\mu \nu } \dh ^m A 
,\  \G _{mn} ^\l - \tilde  \G _{mn} ^\l = - e ^{- 2A + 2B} \tilde g _{mn } \dh ^\l B  ,\ \G _{n \mu } ^m - \tilde  \G _{n \mu } ^m= \d _n^m \dh _\mu B   . \eea 
The  formulas  for the Riemann curvature tensor components
are \bea && \bullet \ R _{\mu \l \nu }
^{\ \ \ \ \ \rho } =\tilde R _{\mu \l \nu } ^{\ \ \ \ \ \rho } - 2 e
^{2(A-B)} \tilde g _{ \nu [ \mu } \d ^\rho _{ \l ] } (\tilde \nabla A
)^2 ,\cr && \bullet \ R _{\mu m \nu } ^{\ \ \ \ \ n} = e ^{2(A-B)}
\tilde g_{\mu \nu } ( - \tilde \nabla _m \tilde \nabla ^{\tilde n}  A - \tilde
\nabla _m (A-B) \tilde \nabla ^{\tilde n} (A-B) + \tilde \nabla _m B \tilde
\nabla ^n B - \d _m ^n (\tilde \nabla A \cdot \tilde \nabla B ) ) ,
\cr && \bullet \ R _{m \mu n } ^{\ \ \ \ \ \nu } = \d _\mu ^\nu (-
\tilde \nabla _m\tilde \nabla _n A - \tilde \nabla _m(A-B) \tilde
\nabla _n(A-B) + \tilde \nabla _m B \tilde \nabla _n B - \tilde g_{mn}
\tilde \nabla A \cdot \tilde \nabla B ) ,\cr && \bullet \ R _{m p n }
^{\ \ \ \ \ q} =\tilde R _{m p n } ^{\ \ \ \ \ q} + 2 \d ^{ q } _{[ m
  } \tilde \nabla _ {p ] } \tilde \nabla _ {n } B - \tilde g _{ n [ m
  } \tilde \nabla _ {p ]} \tilde \nabla ^ { q } B + \tilde \nabla _ {
  [m } B \d _{ p ] } ^ {q} \tilde \nabla _ { n} B - \tilde \nabla _ {
  [m} B \tilde g _{ p ] n} \tilde \nabla ^ { q } B -\tilde g _{n [ m}
  \d ^{ q } _{ p] } (\tilde \nabla B ) ^2 . \label{appeq9} \eea 
The components  $ R ^{(d)}_{\mu \nu }   (M_d) ,\   R ^{(k)}_{mn}   (X_k)  $
of the Ricci curvature tensor
$ R_{MN}  (M_D) = R _{MN} ^ {(D)} $  of $  M_D$  along
$M_d,\ X_k $,   as well as the corresponding components
$ R ^{(d)}   (M_d) ,\   R ^{(k)}   (X_k)  $ of the
Ricci curvature scalar  $ R(M_D)$   and $ \nabla _d^2 = (\nabla _D )_d ^2 ,\ \nabla _k^2  = (\nabla _D )_k ^2  $   of  the scalar
Laplacian $\nabla _D^2 ( M_D) $  are  given   by  the  useful  formulas   
\bea && \bullet \ R ^{(D)}_{\mu   \nu } \equiv R ^{(d)}_{\mu \nu } = \tilde R ^{(d)}_
{\mu \nu } + e ^{2(A-B)} \tilde g_ {\mu \nu } ( -d (\tilde \nabla A
\cdot \tilde \nabla A ) - \tilde \nabla  _k ^2 A - (k-2) (\tilde \nabla A
\cdot \tilde \nabla B ) ) ,\cr &&  R  ^{(D)}_{m n } \equiv
R ^{(k)}_{m n } = \tilde R ^{(k)}_{m n } + (k-2) ( -\tilde
\nabla _m \tilde \nabla _n B + \tilde \nabla _m B \tilde \nabla _n B -
\tilde g _{mn} (\tilde \nabla B )^2 ) - \tilde g _{mn} \tilde \nabla
^2 B   \cr && + d (- \tilde \nabla _m \tilde \nabla _n A - \tilde
\nabla _m (A-B) \tilde \nabla _n (A-B) + \tilde \nabla _m B \tilde
\nabla _n B - \tilde g _{mn} \tilde \nabla A \cdot \tilde \nabla B) )
.  \label{appeq10} \cr && \bullet \ R ^{(d)} = e ^{-2A} \tilde R
^{(d)} + e ^{-2B} d ( - \tilde \nabla ^2 A - (k-2) (\tilde \nabla B
\cdot \tilde \nabla A ) - d (\tilde \nabla A )^2 ) , 
\cr && R ^{(k)} = e ^{-2B} [\tilde R ^{(k)} - 2 (k-1) \tilde \nabla ^2 B -
  (k-2) (k-1) (\tilde \nabla B)^2 + d (-\tilde \nabla ^2 A - (k-2)
  (\tilde \nabla B \cdot \tilde \nabla A ) - (\tilde \nabla A )^2 )] .
\cr && \bullet \ \nabla ^2 (X_k) = e ^{-2 B}
(\tilde \nabla ^2  (X_k) + (k-2) \nabla _m B \nabla ^{\tilde m} )
,\ \nabla ^2 (M_d ) = e ^{-2 A} \tilde \nabla ^2 (M_d ) ,\cr && 
 \nabla _k^2  = ( \nabla _D^2 )_k =  e ^{-2 B}
(\tilde \nabla ^2 _k  + (d \nabla _m  A + (k-2)\nabla _m  B)  \nabla ^{\tilde m}
),\ \nabla _d^2  = ( \nabla _D^2 )_d =   e ^{-2 A} \tilde \nabla ^2 _d .  
\label{appeq11} \eea 
Note that the Riemann tensor components with mixed indices  are
 related as $R _{\mu m \nu } ^{\ \ \ \ \ n} = g ^{n n'} g _{\nu \nu '}
R _{m \mu n '} ^{\ \ \ \ \ \ \nu '} .$  
No confusion should hopefully arise between 
the scalar Laplacians  $ \nabla ^2 (M_d) ,\ \nabla ^2 (X_k) $  
of $ M_{d}$ and $X_{k} $  and  the  projections
$\nabla _{d,k} ^2$  in   Eq.~(\ref{appeq11})   of the  scalar
Laplacian of $ M_D= M_d \times X_k$ along $ M_d $ and $  X_k$, 
  which   satisfy  the relations  
\bea && \nabla ^2 _D  = \nabla ^2 (M_D ) \equiv 
(\nabla ^2 _{D})_d + (\nabla  ^2_{D} )_k  = \nabla ^2 _d + \nabla ^2 _k 
= e ^{-2A} (\tilde \nabla ^2
_{D} )_d + e ^{-2B} [(\tilde \nabla _{D}  ^2)_{k} + \nabla _m (d A
  + (k-2) B ) \nabla ^{\tilde m} ]  , \cr &&  
\   [\nabla ^2 _d = (\nabla ^2 _{D})_d  = {1\over \sqrt  {\vert
g_D\vert } } \dh _\mu    \sqrt  {\vert g_D\vert }  
g ^{\mu \nu } \dh _\nu ,\ \ \     \nabla ^2 _d = 
(\nabla ^2 _{D})_k = {1\over \sqrt  {\vert g_D\vert } } \dh _m 
\sqrt  {\vert g_D\vert }   g ^{mn} \dh _n  ] .\eea 
Clearly, $ \nabla ^2 _D  \ne \nabla ^2 (M_d) + \nabla ^2 (X_k).$   
The   operators  $\nabla _{d,k} ^2$  are used extensively  in our discussions,    near   Eq.~(\ref{sub2eq11}), for instance.  
To relate to the discussion in the text, 
we rewrite  some of the formulas above in the special case $d=4,\ k=6,\ B=-A$, 
\bea && \bullet \  R_{\mu \nu } =
\tilde R_{\mu \nu } ^{(4)} - e ^{4 A} \tilde g_{\mu \nu }\tilde \nabla
_{6} ^2 A ,\ R _{m n} =\tilde R _{m n}^{(6)} + 8 \tilde \nabla _m A
\tilde \nabla _n A + \tilde g_{m n} \tilde \nabla _{6} ^2 A ,\cr && R
^{(4)} = e^{-2 A} \tilde R ^{(4)} - 4 e^{2A} \tilde \nabla _{6} ^2 A
,\ R ^{(6)} = e^{2 A} ( \tilde R^{(6)} - 8 (\tilde \nabla _{6} A) ^2 +
6 \tilde \nabla ^2 _{6} A ) , \cr && 
\bullet \ \nabla ^2 _4 = (\nabla ^2 _{10})_4 = e ^{-2 A} \tilde
\nabla ^2 _4 , \ \nabla ^2 _{6} = (\nabla ^2 _{10})_6 
= e^{2 A}  \tilde \nabla ^2 _{6} ,\ 
e ^{\pm 4 A}  \tilde \nabla ^2 _{6} e ^{\mp 4 A} = \mp  4 \tilde \nabla
^2 _{6} A  + 16 (\nabla _m A) (\nabla ^{\tilde m } A )     ,\ 
\cr &&  \nabla ^2 _{10} = e
^{-2A}(\tilde \nabla ^2 _{10} )_4 + e ^{2A}(\tilde \nabla ^2 _{10} )_6= e
^{-2A}\tilde \nabla ^2_4 + e ^{2A}\tilde \nabla ^2_6 .
\cr &&  \bullet \  \nabla ^2 (M_4) = e^{-2A} \tilde
 \nabla ^2 (M_4) ,\    \nabla ^2 (X_6) = e^{2A} 
(\tilde \nabla ^2 (X_6)   - 4  (\nabla _m A) \nabla ^{\tilde m } )  .
\label{appeqp11} \eea 

We   next consider     the 
first order variations of  geometrical quantities under infinitesimal
perturbations of the unwarped metric tensor. Note  that 
the second order variation of   the curvature   action $\sqrt {g_D} R ^{(D)} $
are  computed   from    the  first order variation of  Einstein tensor 
$ \d G _{MN}  \propto \d (\sqrt {g_D} R ^{(D)} ) /\d  g^{MN}   $. 
The   first order variations   of the Euclidean manifold $X_k$  metric determinant  and   of  the    Ricci  and Einstein   tensors of  $ M_4$  and $
X_k$     are   evaluated  by    means of the identities
\bea && \bullet \ \d \sqrt {g _k} ={\sqrt {g _k} \over 2} g^{mn} \d g _{mn}
=  - {\sqrt {g _k} \over 2} g_{mn} \d g ^{mn}
,\ \d (\sqrt { g _k } R ) = - \sqrt {g_k} \d g ^{mn} (R _{mn} + \ud g_{mn} R
),  \cr &&  
\bullet \  \d  R _{\mu \nu } = \ud (-\nabla _D ^2  \d  g _{\mu \nu } - 
\nabla _{\mu }  \nabla _{\nu } \d g ^\l _\l   + 2  \nabla  ^\l  \nabla _{ (\mu }  
 \d  g _{\nu )  \l  } ) , \cr && 
\bullet \   \d R_{mn} \equiv E_{mn} ^R \equiv -\ud \D _K \d g_{mn} 
= \ud (- \nabla ^2 _k \d g_{mn} +  2   \nabla ^p \nabla _{(m } \d g
_{n ) p}  - \nabla _{m } \nabla _{ n } \d g ^p _p )   
\cr && = \ud ( \D _L \d g_{mn } -  \nabla
_{m } \nabla _{ n } \d g + 2 \nabla _{ (m } \nabla ^p \d g _{n) p} )   
, \cr && [\D _L \d g_{mn} = -\nabla ^2_k \d
  g_{mn} +2 R _{p (m n) q} \d g ^{pq} + 2 R _{(m } ^{\ \ \ \ p} \d g _{n
    ) p}   ,\ \d g = g^{pq} \d g_{pq},\  \d g _p= \nabla ^{q} \d g_{pq}]   
\cr && \bullet \  \d G_{mn} \equiv E_{mn} = \ud ( -
\nabla ^2 _k\d g_{mn } - \nabla _m \nabla _n \d g + 2 \nabla ^p \nabla
_{( m } \d g_{n ) p } -R ^{(6)} \d g _{m n } + g_{m n } (\nabla ^2_k \d
g - \nabla ^p \d g _p ) ) , \label{eqsubvarR} \eea 
where  $ E_{mn} ^R $   is the inverse   graviton propagator 
and $\D _L$  is the  Lichnerowicz  operator of $ X_k$,   
already defined   in  Eq.~(\ref{appeq16}).  
The relations   between the first order variation of  Einstein tensor 
 diagonal and off-diagonal  components along $ M_d $ and $ X_k$ 
in the  warped and  unwarped  metric of Eq.~(\ref{appeq5}) 
are  given by the useful  formulas
\bea && \bullet \ \d G _{\mu \nu } = e ^{2 A} \nabla
_\mu \ \nabla _\nu ( 4 \d A - \ud \d \tilde g _k ) +\d _K \tilde G
^{(4)} _{\mu \nu } - \ud e ^{4 A} \tilde \nabla ^2 _k ( \d _K \d g_{\mu \nu } - \tilde g_{\mu \nu } \d _K \tilde g _d) \cr && + e ^{2
  A} \tilde g_{\mu \nu } \d ( {3\over 4} e ^{-6A} (\nabla _m e ^{4A} )
(\nabla ^{\tilde m} e ^{4A} ) - {1\over 2} e ^{-2A} \tilde \nabla _k
^2 e ^{4A} - {1\over 2} e ^{2 A} \tilde R^{(k)} ) , \cr && [ \d \tilde
  G ^{(4)} _{\mu \nu } = \d ( \tilde R ^{(4)}_{\mu \nu } -\ud \tilde g
  _{\mu \nu } \tilde R ^{(4)} ) \equiv \tilde E_{\mu \nu } = \ud ( -
  \tilde \nabla ^2 _d h_{\mu \nu } - \tilde \nabla _\mu \tilde \nabla
  _\nu h + 2 \tilde \nabla ^\l \tilde \nabla _{ (\mu }  h_{\nu  ) \l } 
-\ud \tilde  R^{(4)} h_{\mu \nu }  +
  \tilde g _{\mu \nu } (\tilde \nabla _d ^2 h -\tilde \nabla ^\l h _\l )
  )  ,\cr && 
\bullet \ \d G_{m n} = \d \tilde G ^{(k)} _{mn} - \ud
e^{-4 A} \tilde \nabla _4 ^2 \d \tilde g _{mn} + {1\over 4} e^{-4 A}
\tilde g _{mn} \tilde \nabla _4 ^2 \d \tilde g - \ud e^{-4 A} \nabla
_{ (m } ( e^{4 A} \nabla _ { n ) } \d _K \tilde g _4 ) -\ud e^{-4 A}
\tilde g _{mn} \d _K \tilde  R ^{(4)} -\ud \d \tilde g _{mn}  
\tilde  R ^{(4)}   \cr && + \ud e^{-2 A} \tilde g _{mn}
\nabla ^{\tilde p} ( e^{2 A} \dh _p \d _K \tilde g _4 ) + \d [ {1\over
    4} e ^{-8 A} (\nabla _p e ^{4A} ) (\nabla ^{\tilde p} e ^{4A} )
  \tilde g_{mn} - {1\over 2}e ^{-8 A} (\nabla _m e ^{4A} ) (\nabla _n
  e ^{4A} )]  ,\cr && 
[ \d \tilde G ^{(k)} _{mn} = \d ( \tilde  R ^{(k)}_{mn}  -\ud  \tilde
 g_{mn}\tilde  R ^{(k)} ) \equiv \tilde E _{mn} =
  -\ud ( \D _K h_{mn} + h_{mn } R ^{(k)} + g_{mn } \d R ^{(k)} ) 
\cr && = \ud
(\D _{L} h_{mn} -  \tilde \nabla _m \tilde \nabla _n h +
2 \tilde \nabla _{ (m } \tilde \nabla  ^p h _{n) p} ) 
 ,\  \D _{L} h_{mn} = \ud ( - \tilde \nabla _k ^2 h_{mn} + 2 \tilde R
^{ k      \ \ \ l } _{\ mn } h _{\tilde k \tilde l } + 2 \tilde R ^{\ \ k}
  _{(m } h _{n ) \tilde k} ) ] \label{appeq12} \cr && \bullet \ \d G
_{ \mu m} = \ud \nabla _\mu ( \nabla ^ {\tilde n} \d \tilde g _{mn} -
\ud \nabla _m \d \tilde g _k - 4 \d \tilde g _{mn} \nabla ^ {\tilde n}
A )  ,  \label{appeq20} \eea 
where the  Einstein   tensor mixed  components $ G_{\mu m}$   arise
from  perturbations  of the    off-diagonal     term in the metric 
$ ds ^2 \supset 2 g _{\mu m}  d x ^\mu d y ^m $ and we used 
the notations $ \d _K 
\tilde g _{\mu \nu} = h_{\mu \nu } ,\ h _\mu = \nabla ^\nu h_{\mu \nu },\ h =
\tilde g ^{\mu \nu } h_{\mu \nu} ,\ h _{mn} =\d \tilde g _{mn} ,\
 $     with  the   indices raising    rule 
$\d G_\mu ^ \nu =  g ^{\nu \l } \d G_{\mu \l  } =  2
^{-2A} \d G_\mu ^ {\tilde \nu }.$ 
 In  the harmonic gauge for an  underlying   Ricci flat manifold, $ 2 E _{mn} =
\D _{L} h_{mn} = ( - \tilde \nabla ^2 _k h_{mn} + 2 \tilde R ^{ \tilde k \ \ \ \tilde l
} _{\ mn } h _{\tilde k \tilde l } ) .$ 
 The  inverse   graviton propagator $ \d G_{\mu \nu } = \tilde E_{\mu
   \nu} $   in $    
 M_4$, takes the following    forms  in the  4-d 
transverse-traceless   (TT)     and de Donder (H for harmonic) gauges,
ignoring  curvature terms, 
\bea && \bullet \ TT:\  h _\mu = \nabla ^\nu  h _{\mu \nu }  =0 ,\ 
h = \tilde  g^{\mu \nu } h _{\mu \nu }  =0 \ \Longrightarrow  \   
\tilde E ^R_{\mu \nu } = - \ud \tilde \nabla ^2 _4  h_{\mu \nu }  +  C _{\mu \nu } , \cr && 
 \bullet \  H:\   \nabla _\nu ( h ^{\mu \nu } -\ud \tilde  g ^{\mu \nu
 } h ) = (h ^\mu - \ud \nabla ^\mu  h )  =0 \ \Longrightarrow  \   
\tilde E ^R_{\mu \nu } =  - \ud \tilde \nabla _4 ^2  (h_{\mu \nu }  - \ud
\tilde g _{\mu \nu }   h )  +\ud \tilde \nabla _\mu\nabla _\nu 
h  +  C _{\mu \nu }  ,\cr && 
[C _{\mu \nu }= - \ud \tilde  R^{(4)} h_{\mu \nu } +     
\tilde R ^ {\l \ \ \ \ \rho } _{\ (\mu \nu ) }  h _{\l \rho }  +
\tilde R _{ (\mu } ^ {\ \ \l } h_{ \nu ) \l } ] . \label{appeq36}  \eea 
 
\subsection{Type $ II \ b$ supergravity}  
\label{subapp1} 

The     bosonic sector    of 10-d type $II\ b $   supergravity theory 
consists of the   metric tensor, 2-form and dilaton  field $g _{MN}, \ B_{MN} ,\ \phi  $    and the 0-, 2- and 4-form field  potentials
$ C _0,\ C_{MN} ,\ C_{MNPQ}$  governed  by  the  Einstein frame  action   in
Eq.~(\ref{sub2eq0}).  The classical  field equations  
and  Bianchi  identities (enclosed  inside square  brackets)    are    
\bea &&  \bullet \ \star d \star d \phi  = \ud ( 2 e^{2\phi }  F_1  \wedge \star
F_1   -     g_s  e^{-\phi }   H _3  \wedge \star  H _3   +  g_s
e^{\phi }   \tilde  F _3 \wedge \star   \tilde  F_3    )  , \cr &&    
 \bullet \  \star d (e^{2\phi  } \star  F _1)  = - g_s  e^{\phi }   H
 _3 \wedge \star \tilde  F_3   , \ [ d F_1= 0] \cr &&   
\bullet \ d (e^{\phi  } \star \tilde  F _3)  =   g_s F_5  \wedge
H_3  , \ \ [d F_3 = d H_3 =0,\
  d \tilde  F _3 = - d C_0   \wedge H_3  = - F_1  \wedge H_3  ]   \cr &&     
 \bullet \ d \star ( e^{-  \phi  }    H_3  - C_0  e^{\phi }  \tilde  F_3  )  = -
g_s  F _5 \wedge F _3  ,   \  \cr && 
 \bullet \   d \star \tilde  F_5  = 
- F_3 \wedge  H_3   + {2 \kappa ^2 \tau _3 \over g_s } \rho _3 ^{loc} , \  
[\star \tilde F _5 =\tilde F _5 ,\ d F_5 =0 ]   \cr &&  \bullet \ R_{MN} = 
\kappa ^2  ( T _{MN}  -{1\over 8}  T_P^{\ P })   
 = \ud   \dh _M \phi \dh _N \phi +  \ud
e^{2\phi } \dh _M  C_0 \dh  _N  C_0 + { g_s^2 \over 96}  \tilde  F_{MPQRS} F_{N}
^{ \ PQRS}  + { g_s \over 4 } ( e^{- \phi  }   H_{MPQ}  H_{N} ^{ \ PQ} \cr && +
e^{\phi  }  \tilde  F_{MPQ}    \tilde  F_{N} ^{ \ PQ}  )  -
{g_s \over 48}  g_{MN}  (  e^{- \phi  }   H_{PQR}  H^{PQR}   +  
e^{\phi  }  \tilde  F_{PQR}  \tilde F^{PQR} )  +  
\kappa ^2  ( T ^{loc} _{MN}  -{1\over 8}  (T^{loc }) _P ^P ) ,\cr &&  
[\rho  ^{loc} _3 (X)  =   {\d ^{( 6 )} (y - y_0) \over  \sqrt {g_{6} }}
 ,\ T_{MN} = - {2 \over \sqrt {-g} } {\d S \over \d g   ^{MN} } ]  .  
\label{appeq2}  \eea  
Using    the   identity  
$ \ud \d _\phi   ( d \phi \wedge \star  d \phi
) = -  vol  (M_D) \star  d   \star d \phi = - vol  (M_D) (-\D _0
\phi ) =  -   \sqrt { \vert g_D \vert }  \nabla _D ^2 \phi $, 
one   can rewrite       the  axio-dilaton   field    equation  
 in     tensorial   notation as 
\bea && 0= \nabla ^2 _{10} \tau -  {1\over i \tau _2 } 
\nabla _M \bar \tau  \nabla ^M  \tau  - {1\over 24 i } G_+\cdot
\bar G_-  \cr  && \Longrightarrow \ 
\nabla ^2 \phi = 
e^{2\phi }  (F_1 \cdot   F_1)  +  {g_s \over 2} (-e^{-\phi }  (H_3 \cdot
H_3)     + e^{\phi }  (\tilde F_3 \cdot  \tilde  F_3 ) ) , \ 
\nabla ^M   (e^{2\phi  }\dh _M C_0) = -g_s  e^{\phi }   (H_3 \cdot
\tilde  F_3) .  \eea 
The  energy-momentum stress tensor,   entering  the  trace reversed
Einstein field     
in the last entry  of  Eq.~(\ref{appeq2}),
acquires contributions  from  bulk fields and  localized   brane   sources. 
(Recall that the complex scalar field $\Phi $ of Lagrangian 
$L= - D _{M } \Phi ^\dagger  D^{M }  \Phi  - m^2   \Phi ^\dagger \Phi
$     contributes  $ T_{MN  }=  + g _{MN    } L   +   2 D _{ ( M } 
\Phi  D _{N  )} \Phi  ^\dagger $.)   The 
 contributions   from   the supergravity multiplet bulk fields  
are  of form 
\bea && T_{MN} = - {2 \over \sqrt {-g} } {\d S \over \d g
  ^{MN} } = {1\over 2 \kappa ^2 } [   \dh _M \phi \dh _N \phi
-\ud   g_{MN} \dh _P \phi \dh  ^P \phi + 
e^{2\phi } \dh _M  C_0 \dh  _N  C_0  -  \ud   g_{MN} \dh _P C_0 \dh^P
C_0     \cr && + {g_s ^2 \over 48}  (\tilde F _{M
  M_1 M_2 M_3 M_4 } \tilde   F _{N } ^{ \  M_1 M_2 M_3 M_4 }
- {1\over 10} g_{MN}  \tilde  F _{M_1 M_2 M_3 M_4 M_5}
\tilde  F ^{M_1 M_2 M_3 M_4 M_5} )\cr &&  +  {g_s e^{-\phi } \over  2 }  (H _{M
    M_1 M_2 }  H _{N } ^{ \ \   M_1 M_2 } - {1\over 6 } g_{MN} 
H _{ M_1     M_2 M_3}    ^2 )    
+ {g_s e^{\phi } \over  2 }  (\tilde F _{M M_1     M_2  } \tilde  F_{N } ^{ \ \  M_1
  M_2 }   - {1\over 6 } g_{MN} F_{ M_1     M_2 M_3}^2  )  ] .\label{appeq3}   \eea
For a $ Dp $-brane   of world volume  extended over $ M_4$
and wrapping  a $ (p-3)$-cycle of $ X_k$,  the    
energy-momentum stress tensor and RR  charge  density  are  
\bea &&  T ^{loc} _{\hat \mu \hat \nu  } (X) = -{2\over \sqrt {-
    g_{(p+1)} } }  {\d S ({Dp})  \over \d g   ^{ \hat \mu \hat \nu } }   
= -T _p g _{\hat \mu \hat \nu } {\d ^{(9-p)} (y  -y_0) \over
  \sqrt {g_{ 9-p } }} , 
\ \rho  ^{loc} _p (X)  = {\d ^{(9-p )} (y - y_0) \over
  \sqrt {g_{ 9-p } }}  ,  \cr &&  [ T_p = {\mu _p \over  g_s ^{(p+1)/4} }
 e ^{(p-3 ) \phi /4 },\ \mu _p = {2\pi \over \hat l_s ^p+1} ]  \eea    
where   $  x _{\hat \mu  }$ and $ y _m $  denote  the  $(p+1)$ and $
(9-p)$ coordinates   along   
longitudinal and transversal directions  to the brane world volume.
To  ease the comparison  with  studies  using the standard  Einstein
frame,   we   quote  below   some     of the substitutions
needed at the 10-d and 4-d levels 
to translate   from the   standard Einstein frame~\cite{gkp01}
to  the special Einstein frame  used in the present work,
\bea && \bullet \ g_{MN} \to  g_s ^{-1/2}  g_{MN},\ \sqrt {g _{10}}  
\to  g_s ^{-5/2} \sqrt {g _{10}},\  
\star _{10} \to   g_s ^{-5/2}\star _{10} , \ 
T_{MN} \to  g_s ^2 T_{MN}  , \ \rho ^{loc} _3 \to g_s ^{+3/2} \rho ^{loc} _3 
, \  R = g^{MN} R _{MN} \to g_s ^{1/2}  R ,  
  \cr && \bullet \  M_\star ^2 = {V_W \over \kappa ^2 _{10} } 
 \to  g_s ^{1/2} {V_W \over \kappa ^2 } =  g_s ^{1/2} M_\star ^2 , \ 
 V_W \to g_s ^{-3/2} V_W  ,\  V _{flux}  \to g_s  V  _{flux} , \eea
where     $T_{MN}  ,\ R_{MN} $ are  the stress  and  curvature tensors in
$ M_{10}$  and  $  V _{flux} $ denotes the potential  energy density
in $M_4$,  while   keeping unchanged the  parameters   $\kappa ^2 _{10},\ \mu _p $ and   the   quantities
$H_3,\  C_{(q)} ,\ R_{MN} ,\ G_{MN}= R_{MN} - \ud g_{MN} R,\  \  \int
\sqrt { -g _4}  V_{flux  } ,\   \int _{X_6}  \sqrt {g _6} \rho
^{loc} _3 ,\   d \star _{10} F_5  = H _3 \wedge F _3  + 2 \kappa ^2
_{10}  \mu _3  \star _{10}  (\rho_3 )  . $ 

\subsection{Classical field equations  for flux  compactifications}  
\label{subapp3}

The classical  ansatz for GKP  vacua of warped flux compactifications 
is specified    by   metric tensor  components   along  $
M_4 $ and  $ X_6$,   axio-dilaton field and   the 5- and 3-form
field  strengths    of form 
\bea &&  g_{\mu \nu } (x,y)  = e ^{2A(y) } \tilde g_{\mu \nu } (x),\ 
g_{mn } (y) =   e ^{-2A(y) } \tilde g_{mn } (y)   ,\cr &&   \tau  (y)
= C_0 + i e^{-\phi } ,\ g_s \tilde F_5  (y) = (1+\star _{10} ) d \a
(y) \wedge vol(M_4) , \  G_{mnp} (y)  =  F_{mnp} -\tau H_{mnp} . \label{app3EQ1}\eea 
The    field  equations and Bianchi identities adapted to this metric
 are  
\bea &&  \bullet \  \tilde \nabla _6  ^2 e^{4A} = \tilde R ^{(4)}   +  e^{2A} {1
  \over 12 \tau _2 } G_{mnp} \bar G^{mnp} + e^{-4 A} (\nabla _m \a
\nabla ^ {\tilde m }  \a
+ \nabla _m   e^{4A} \nabla ^ {\tilde m } e^{4A} ) +{ \kappa ^2  \over
  2 g_s} e^{2A} (T^m_m -T_\mu ^\mu ) ^{loc} ,    \cr && 
\bullet \  \tilde \nabla ^2 _6 \a (y) = { 1 \over  12 \tau _2 } e^{2A} G_{mnp}
     ( \star _6  \bar G^{mnp} )  + 2 e^{-4 A} \nabla _m \a \nabla
^{\tilde  m }  e^{4A} + {2\kappa ^2   \mu _3  \over g^2_s}   
e^{2A} \rho _3  ^{loc} , \cr &&  
\bullet \   e ^{-2A} (\tilde \nabla ^2 _{4} \tau 
- {1\over i \tau _2 }    \nabla _\mu \bar \tau   
 \nabla  ^{\tilde \mu } \tau
) + e ^{2A} ( \tilde  \nabla ^2 _{6} \tau - { 1 \over i \tau _2 }  
\nabla  _m  \bar \tau \nabla  ^{\tilde  m } \tau )-  {1\over 24 i } 
 e ^{6A} G_+ \tilde \cdot  \bar G_- =  \d _\tau  S (Dp),   
\cr &&    [ \d _\tau  S (Dp)  =   - {4 \kappa ^2 \tau _2 ^2
\over \sqrt {- g _{10} } }  {\d  S (Dp) \over \d  \bar \tau } 
= 4 i\kappa ^2   \mu _p e^ { (p +1) \phi  \over 4  }  \tau _2 ^2 
{\sqrt {- g _{p+1} }  \over \sqrt {- g _{10} } }   \d ^{ 9-p}  (y)] 
 \cr && 
\bullet \  d (e^{4A} \star G_3 -i \a 
 G_3 ) + {i\over \tau _2} d \tau \wedge \Re (e^{4A} \star G_3 -i
     \a G_3 ) =0   , \cr &&  
\bullet \   \tilde R_{mn } =  {1\over 4 \tau ^2_2 } ( \nabla  _m \tau   
\nabla _n \bar \tau  + \nabla  _n \tau \nabla  _m 
\bar \tau )  - {e ^{4 A} \over  16   \tau _2 }   (G_{ + (m }
^{\ \ \ \ \  \tilde p\tilde q }    \bar  G _{-
  n ) pq  }  +  G_{ -  (m } ^{\ \ \ \ \  \tilde p\tilde q}   \bar  G _{+   n )p
  q}    )    
 \cr &&  + \ud  e ^{-8A}  \nabla  _{ (m} \Phi _+  \nabla  _{ n ) } \Phi _-   
-  e ^{-4 A} \tilde g _{mn }  \tilde R ^{(4)} +  
 \kappa ^2   (T_{mn} ^{D7} - { g _{mn} \over 8} T ^{D7} ) . \label{appeq21}\eea 
The     formula  in the first  entry  for      the trace reversed  Einstein 
equation,  $ g ^{\mu  \nu } R_{\mu  \nu } = \kappa ^2  g ^{\mu
  \nu }   ( T_{\mu \nu }  - (g _{\mu   \nu } / 8 )  T ^L  _L )   $,  acquires  contributions from $ Dp$-branes   
and   that in the    last entry  for the trace reversed  Einstein 
equation  of   $ R_{mn } $, acquires
contributions  from   $ D7$-brane  sources. 
The  Bianchi  identities for  the 5-  and  3-form fields   are included   in  the equations  in the second and     fourth  line      entries.
Rewriting the  constraint equation in the first entry above
in terms of the inverse   warp  profile    
\bea && \tilde \nabla ^2 _6 (e^{-4 A} ) = - e^{-8A} \tilde R^{(4)}
- {1\over 12 \tau _2}  e ^{-6 A} G  \cdot \bar G - e^{-12 A} (\nabla _m \a  \nabla ^ {\tilde m }  \a  -  \nabla _m   e^{4A} \nabla ^ {\tilde m } e^{4A} )  -  {2 \kappa ^2 \over g_s}  e ^{-6 A} \hat T^{loc} , \eea    
shows that  in  GKP vacua,   with $ \a =e ^{4 A}$ and $ \tilde R^{(4)} =0$,
the dependence on the  warp and 4-form profiles   drop  outs     from
the  right hand side   since $ G  \cdot \bar G \sim  e ^{6 A}  ,\  \hat T^{loc} 
 \sim  e ^{6 A} .$   This  equation is  of use  in  discussing the
back-reaction of   branes  on    the embedding space geometry. 
For convenience, we    also  provide     below     the equivalent    
field equations  for the  set of independent  fields $ \Phi _\pm  (y) ,\     G_{\pm } (y)  ,\ \tau  (y) ,\  \L  (y)  $~\cite{mcguirk12,gandhi11},   
  \bea && \bullet \  0= \tilde \nabla ^2 _6 \Phi _\pm - { (\Phi _+ +
    \Phi _-)^2  
\over 96 \tau _2 } \vert \tilde   G_\pm \vert ^2 -  \tilde R ^{(4)} -
       {2 \over \Phi _+ + \Phi _- }  
\nabla _m  \Phi _\pm  \nabla ^{\tilde m}\Phi _\pm -
{ 2\kappa ^2 \over  g_s } e^{2A} 
( \hat T^{loc}  (y)  \pm {\tau _3 }  \rho ^{loc} _3 ), 
\cr && \bullet \  0= d \L + i {d\tau \over 2\tau _2 } \wedge (\L +\bar \L ) ,\ 
[d H_3 =0,\  0= d F _3 \equiv d (G _3 +  \tau H _3) ]  \cr && 
\bullet \  0= \tilde \nabla ^2 _6 \tau + {i\over \tau _2}
\nabla _{m} \tau   \nabla ^{ \tilde   m} \tau 
 + {i\over 48} (\Phi _+ + \Phi _-) G_+ \tilde \cdot
G_-  + source   
,\cr && \bullet \  0= \tilde  R_{mn} - {1\over 2 \tau _2 ^2 } \nabla _{(m}  \tau \nabla _{n)}  \bar \tau - {2\over (\Phi _+ + \Phi _- ) ^2 }
\nabla _{(m} \Phi _+ \nabla _{n)} \Phi _-
\cr && + {\Phi _+ + \Phi _- \over 32 \tau _2 } (G _{+ (m}  
^{\ \ \ \ \ \tilde   p\tilde q} \bar G _{- n) pq} + G _{- (m}
^{\ \ \  \ \   \tilde  p\tilde q} \bar G _{+n) pq} )  
 +  {1\over 2 (\Phi _+ + \Phi _-  )  } \tilde  g_{mn} \tilde R ^{(4)} 
 -\kappa ^2 ( T^{D7} _{mn} - {g_{mn} \over 8} T^{D7} )  ,\cr &&  
[\Phi _\pm = e^{4A} \pm \a ,\ G_{\pm } = (\star _6 \pm i ) G_3 ,\  
\L  (y)  =  (\Phi _+ G_- +   \Phi _- G_+ )=  2 (e ^{4 A} \star G - i \a G )
,\cr && G_3= F_3-\tau H_3 ,\ \tau = C _0 +i e ^{-\phi }  ] \label{appeq24}  \eea  
where we used the relations,
$ \vert  \tilde G_\pm \vert ^2 = G_{\pm  } \tilde \cdot \bar
G_\pm  = G_{\pm  ,  mnp } G_{\pm  } ^{\tilde  m\tilde n \tilde  p },\
G _+\cdot  \bar G_-    = 2  G \cdot \bar  G ,\
\dh _\tau \dh _{\bar \tau  } ( {G\cdot \bar G / (24 \tau _2 )}) =
( {G\cdot \bar G / (48 \tau _2 ^3 )})$.  The 3-forms,
$(G_+  + G_-) = 2 \star _6 G,\  (G_+   -  G_-) = 2 i G  $, satisfy  also  the useful  relations,  $(\star _6 \pm i ) G_\mp = 0,\
(\star _6 \pm i ) G_\pm = \pm 2i G_\pm .$ 
We     also    quote below  for   convenience      the equivalent form of the 
constraint   equations   for the   inverse  warp  profile  fields,
\bea && 0=\tilde \nabla _6 ^2 \Phi ^{-1} _\pm  +{(\Phi _+ + \Phi _-)^2 
\over 96 \tau _2 \Phi
_\pm ^2} G _\pm  \tilde \cdot \bar G _\pm + {\tilde R ^{(4)} \over \Phi _\pm
^2} + {2\over 
\Phi _\pm ^2} ({1\over \Phi _+ + \Phi _- } -{1\over \Phi _\pm } ) 
\nabla _m  \Phi _\pm  \nabla ^{ \tilde m }  \Phi  _\pm 
 + { 2  \kappa ^2 \over g_s  \Phi_\pm ^2 }    e ^{2A(y)}  
(\hat T ^{loc}  (y)  \pm    \tau _3 \tilde  \rho _3
^{loc}  (y)   ) .   \label{appeq25}  \eea  
For GKP  type  vacum backgrounds    with    
$\Phi _{\mp }=0  ,\ G_{\mp } =0,\  \tilde R ^{(4)} =0$,  respectively,
(ISD/IASD   3-fluxes  and    flat   spacetime $M_4$) 
the     inverse warp profiles  satisfy  Poisson  type equations 
sourced    by the charge  densities    of localized  $ D3/\bar D3$-branes,  
\bea && \tilde \nabla ^2 _6  h  (y; y_\star )  = - {2\kappa ^2  \over
  g_s}    e ^{-6   A(y)}  \hat T ^{loc}  (y) 
= -  {2\kappa ^2  \mu _3  \over   g_s^2}    e ^{-6   A(y)} \rho _3 ^{loc}  (y)  
= - {\hat l_s ^4  }  {\d ^{(6)} (y-y_\star ) \over
\sqrt {\tilde g_6 } } ,\cr &&   [{2\kappa     ^2   \mu _3 \over   g_s
    ^2} = 2\kappa _{10}^2   \mu _3 
= {\hat l_s ^4} ,\ \sqrt { g_6 } = e ^{-6A} \sqrt {\tilde g_6 } ,\ 
2 \Phi  _{\pm }  ^{-1} (y)  =  h (y; y_\star ) = e ^{-4
  A (y; y_\star ) }  ] .    \eea 
The   inverse  warp  profile $ h  (y; y_\star ) $ describe   the  back reaction  on the   throat geometry  at $y$
of    the $ D3$-brane located  at $ y_\star $  and   are seen from  
the above  simplified    eqaution to  identify to    
the Green's functions  $ h  (y; y_\star )  $  for  the    scalar
Laplacian  of $X_6$.  The    harmonic   decomposition  of the solution   
for a $ D3 $-brane    located at a point
$y_\star  = ( r_\star , \T _\star )   $  of
the deformed  conifold     is discussed in~\cite{pufu10}. 
If one  focuses on      large   radial  distances,
where the  conifold asymptotes  the  undeformed conifold,
the decomposition of  $ h  (y; y_\star ) $  on scalar
harmonics    of the  compact  base  $ T^{1,1}$
yields the  explicit solution for the perturbed   part of  the warp  profile 
\bea && \d h  (y; y_\star ) \equiv   h  (y; y_\star ) - ({\calr \over
  r} )^4  = \sum _ {\nu , M}
{\hat l_s ^4  \over 2 ( \D _\nu -2 )}(  \t (r_\star   -r) 
{r ^{\D _\nu - 4} \over r _\star  ^{ \D
_\nu } } + \t (r-r_\star ) {r _\star ^{ \D _\nu - 4} \over r ^{\D _\nu
}} ) Y ^{\nu  , M \star } (\T ) Y  ^{\nu , M} (\T _\star ) , \eea  
where $\D _\nu = 2 + \sqrt { H_0 ^\nu +4}  $.  
If the $ D3$-brane   sits at the ultraviolet boundary 
$ r _\star = r_{uv}  $,  the  solution  simplifies  to 
 \bea && \d h (y; y _\star ) = 
\sum _{\nu , M}  {(\hat l_s /  r ) ^4 \over 2 \sqrt {H_0^\nu
    +4 } } ({r\over r_{uv} })^{\D _\nu } Y  ^{\nu  , M \star } (\T ) Y
^{\nu , M} (\T  _\star ) . \eea   
One can use  the above  result  to obtain  the $D3/\bar D3 $ interaction  potential  felt by a mobile $ D3 (r_\star = r) $ due to  the background  deformation induced by a   $p\  \bar D3 (r = r_0 ) $  stack  near the tip.
With the  $ D3$-brane classical  potential from the Born-Infeld  
action given by  $\tau _3 h^{-1} (r)$, the total perturbed potential  evaluates  to   
\bea &&  
V (D3 (r) /\bar D3 (r_0)) = 2 p\tau _3 h ^{-1} (r) (1 + {\d h (r; r_\star )\over  h
  (r) } ) ^{-1}  = 2p \tau
_3 ({r_0\over \calr })^4 (1 - {1\over N }({r _0 \over  r })^{\D _{\nu } }) 
\simeq  2p \tau _3 ({r_0 \over  \calr })^4 (1 - {1\over N } ({r _0
  \over  r } )^{4 }) .  \eea   
The  attractive   Coulomb potential tail $ -r^{-4}$   corresponds to 
the  contribution  from the  flavour    singlet   operator  of lowest
dimension  $ O^V _{(000)} ,\ \D _{\nu } = 2 + \sqrt { H_0 +4} = 4 $
of the dual gauge    theory.  The resulting formula matches with 
the  standard  one quoted in Eq.(\ref{coulpot1})  up to constant
factors   that can always be absorbed   by redefining $  r $.

\subsection{Linear wave equations  for fields fluctuations}  
\label{subapp4} 

The   linearized  wave  equations  for fields
fluctuations       can be derived  from the  second order
variation    of the    action  or,  for  a  fixed   background
solution,   from the  first order  variation    of the 
field  equations.  Let us first  consider  
the   gravitational  wave equations   for  the  
metric  tensor   components   
$  \d G_{MN} -\kappa ^2 \d T _{MN} =0$, adapted  
to  the warped  metric ansatz  in  Eq.~(\ref{app3EQ1})    
for warp  profiles  $ B (y) = -A  (y) $.    
Combining  the  contributions to the 
 stress energy-momentum  tensor   from bulk   fields 
\bea &&  \bullet  \ \kappa ^2 T _{\mu  } ^{\nu } =  - {1\over 4} e
^{2A} \d _\mu ^\nu  
(\dh _m \phi  \dh ^{\tilde m } \phi   + {1\over  \tau _2 ^2}  
\dh _m  C_0  \dh ^{\tilde m }  C_0 )  - { g_s \over 24  \tau _2 } \d
_\mu ^\nu  ( G\cdot \bar G )   
+  {  g_s ^2 \over 4} e ^{-6 A}  \d _\mu ^\nu (\dh _m \a \dh
^{\tilde m } \a  )  , \cr &&  
\bullet  \ \kappa ^2 T _{m  } ^{n } =   \ud e ^{2A} \bigg  ( \dh _m
\phi  \dh ^{n  } \phi - \ud \d _m ^n (\dh _p \phi \dh ^{\tilde  p} \phi  )  
+ {1\over \tau _2 ^2}   ( \dh _m C_0  \dh ^{n }  C_0
- \ud \d _m ^n (\dh _p  C_0 \dh ^{\tilde  p}  C_0  ) ) \bigg )
\cr && + {g_s \over  4\tau _2 } ( G_{mpq} \bar G ^{n p  q} +  \bar  
G_{mpq}  G ^{npq} - {1\over  3} \d _m ^n   ( G\cdot \bar G ) )   
+ {  g_s ^2 \over 2} e ^{-6 A} [ 
\dh  _m \a    \dh ^{n } \a -\ud   \d _m ^n 
(\dh _p \a \dh ^{\tilde  p} \a  ) ] , \label{appeq22} \eea 
and the contributions  from localized sources 
with  Eqs~(\ref{appeq20})   for the  variations  of Einstein tensor
components, yields  after several cancellations
to  the     diagonal  and off-diagonal projections of the wave equations  
 along   $ M_4$ and $ X_6 $~\cite{shiu08}     
\bea &&   \bullet  \  0= \d ( G _{\mu \nu }   -\kappa ^2  T ^{loc} _{\mu \nu } )  =     \tilde \nabla _\mu \tilde \nabla _\nu ( 4 \d A  - \ud \d \tilde g _6 )  
+\d _K \tilde  G  ^{(4)} _{\mu \nu } 
 - \ud e ^{4 A} \tilde \nabla ^2 _6 ( \d _K \tilde g_{\mu \nu }    -  
\tilde g_{\mu \nu }  \d _K \tilde g _4)    \cr &&  -  \ud e ^{4A}  \tilde g_{\mu
  \nu } \d \tilde R^{(6)} - \kappa ^2 ( g_{\mu \nu } \tau _3 \d \rho
^{loc} _3   + \d T ^{loc} _{\mu \nu }  ) ,\cr &&    
[ \d _K   \tilde G  ^{(4)} _{\mu \nu } 
= \d (\tilde  R ^{(4)}_{\mu \nu } -\ud
\tilde g_{\mu \nu } \tilde  R ^{(4)}) =   \tilde   E_{\mu     \nu } =  
\ud ( - \tilde \nabla ^2 h_{\mu \nu } - \tilde \nabla _\mu \tilde \nabla
_\nu h  +2 \nabla ^{\tilde \l } \nabla _{\tilde (\mu } h _{\nu ) \l }
) - \ud  \d  \tilde  g_{\mu \nu } \tilde  R ^{(4)} + \ud  \tilde  g_{\mu \nu }
(\tilde   \nabla _6^2      \d g -  \nabla ^{\tilde \l }  \d  \tilde g
_{\l } ) ] .     \label{appeq30} 
\cr  &&  \bullet \  0= \d  ( G_{mn}   -\kappa ^2  T^{loc} _{m   n}) 
= \d  \tilde G ^{(6)} _{mn} - \ud  e^{-4 A}  \tilde
\nabla _4 ^2  \d  \tilde g _{mn}   + {1\over 4} e^{-4 A} \tilde g _{mn} \tilde
\nabla _4   ^2  \d  \tilde g - \ud   e^{-4 A} \nabla  _{(m} (  e^{4 A}
\nabla  _ {n)}  \d _K \tilde g _4)  -\ud e ^{-4A} \tilde g_{mn} \d \tilde R ^{(4)}    \cr && 
 - \ud  \d \tilde g _{mn} \tilde R ^{(4)} +
\ud   e^{-2 A} \tilde g _{mn} \nabla
^{\tilde     p} ( e^{2 A} \tilde\nabla _p \d _K \tilde g _4 )  -  \d [
  {1\over 4\tau _2 }  
(G_{npq} \bar G _m ^{\ \ pq} - {1\over 6} g _{mn} G \cdot \bar G ) 
- \kappa ^2 T ^{loc} _{mn}  ] ,  
\cr  &&   [\d  \tilde G ^{(6)} _{mn} = \ud \D _{L} \d  \tilde g _{mn} 
+ \nabla _{(m } \nabla ^p \d  \tilde g _{n) p} -  \nabla _{(m }
\nabla _{ n) }\d  \tilde g 
,\  \D _{L} \d  \tilde g _{mn}  = - \tilde  \nabla _6 ^2   \d  \tilde g _{mn}   + 2 \tilde R ^{\tilde k \ \ \ \ \tilde l }   _{\ mn } \d  \tilde g _{\tilde k \tilde l } 
+ 2 \tilde R ^{\ \ \ k} _{(m }  \d g _{n )   \tilde k}] .\cr
&& \bullet \ 0=  \d G _{ \mu  m}   =  \ud  \tilde\nabla  _\mu  
[ \nabla   ^ {\tilde n}  ( \d \tilde  g _{mn}  - \ud \tilde  g _{mn}  \d
\tilde  g _6 ) - 4 \tilde  \nabla ^n A (y) \d \tilde  g _{mn} ] .    
\label{appeq31} \eea
We have separated out from $ \d T_{\mu \nu } ^{loc}  $
in the first equation above   the contribution from
$ D3$-branes, $   g_{\mu \nu }  \tau _3 \d rho _3 ^{loc} $. 
The    cancellation  of the  longitudinal   part  of 
$ \d G _{\mu \nu } $ and  of  $\d G _{ \mu m} $  yields  the
(non-dynamical)      constraint  equations 
\bea &&  0= 8 \d  A (y, u ) -\d \tilde  g _6 ,\ \ 
0= \nabla   ^ {\tilde n}  (\d \tilde  g _{mn}  - \ud  \tilde g _{mn}
\d \tilde  g _6    - 4  \tilde \d g _{mn}  \nabla ^ {\tilde
  n} A  )   .  \label{appeqx31}   \eea 
The wave  equations  for the fields fluctuations  
$\d  \tilde g_{\mu \nu}, \ \d  \tilde g_{mn} ,\  \d \tau $   are 
encoded  within the   effective action reduced on $ M_4 \times X_6$~\cite{shiu08}   
\bea &&  S ^{(2)} =  - {1\over 4 \kappa ^2 } \int  d^{4}x 
 \sqrt {-\tilde  g_{4} }  \int  d^{6}  y  \sqrt {\tilde  g_{6} }    
\bigg [  \d _K \tilde g ^{\mu \nu }   [  e ^{-4A}\d _K   G ^ {(4)} _{\mu
    \nu } 
\cr &&  -\ud  \tilde \nabla _6 ^2 (\d _K \tilde g _{\mu \nu } - \tilde g
_{\mu \nu }  \d _K \tilde g _4 ) ] - \ud \d _K \tilde g _4 \d \tilde R_6 
  + \d \tilde g ^{mn} [ \d \tilde  G  ^{(6)} _{mn}   - \ud e ^{-4A}   
    \tilde \nabla _4 ^2    \d  \tilde g _{mn} ]
  + {3\over 8}\d \tilde g _4 \tilde \nabla ^2 \d \tilde g _6
\cr && + \ud e ^{-2 A} [ - e ^{-2 A}  \d \tilde g _{mn}
    \nabla ^{\tilde m} (e ^{4A} 
 \nabla ^{\tilde n}\d \tilde g _4 ) -\ud  \d \tilde g _6 \nabla ^{\tilde p} (e ^{2A}  \nabla _{p}  \tilde g _4 ) + {1\over 4} e ^{-2 A}  \d \tilde g _6 
  \nabla ^{\tilde m} ( e ^{4 A}   \nabla _m \d \tilde g _4 ) ] 
  \cr && +  e ^{-2A} \vert \d _K \tau \vert ^2 \dh _\tau \dh _{\bar \tau }  ( {
    G\cdot \bar G \over 12 \tau _2 } )   
-  e ^{4A}  \d \tilde g _m ^{\tilde n}  \d _\a [ {1\over  4 \tau _2 } 
( G_{npq} \bar G^{ \tilde  m\tilde  p \tilde q} - {1\over 6} \d _n ^m
G \tilde \cdot       \bar G   ) ]  - {1\over  2 \tau _2 ^2 } e ^{-4A}   
\d _K \bar \tau ( \tilde \nabla _4 ^2 +   e ^{4A}  \tilde \nabla _6
^2 )    \d _K  \tau
\cr &&  - \kappa ^2 e ^{-2A}   [\mu _3 \d _K \tilde g _4 \d \rho _3 ^{loc} + 
\d  \tilde g _{\mu }  ^{\tilde \nu }   \d (T ^{loc} ) ^{\mu  } _\nu   
+ \d \tilde g _{m} ^{\tilde n }    \d (T ^{loc} ) ^{m } _n 
- {1\over 4} \d \tilde  g \d (T ^{loc} )^{m } _{m }]   \bigg ]
,  \label{appeq34}  \eea  
where  $\d _K \tilde g _4  =\tilde g ^{\mu
    \nu }  \d _K \tilde g _{\mu \nu } ,\ \d  \tilde g _6 =\tilde g
  ^{mn }  \d  \tilde g _{m n } $    and $ \d
_K g _\mu ^{\tilde \nu } \d (T ^{loc} )^\mu _\nu =  \d _K g _{\mu \nu } \d
(T ^{loc} )^{\mu \nu }  ,\   d g _m ^{\tilde n } \d (T ^{loc} )^m  _{n} =  \d
g _{mn  } \d (T ^{loc} )^{mn }  .$  
Imposing the     conditions  in Eqs~(\ref{appeqx31})
substantially simplifies    the     wave equations in
Eqs~(\ref{appeq31})  and the   reduced effective action.      
For instance, the  three terms inside
brackets in the  third  line entry of Eq.~(\ref{appeq34}) 
cancel out   when  one imposes  the constraint    equation
$\d G _{m \mu } =0$.    The     classical background
 for GKP (BPS supersymmetric)   vacua~\cite{gkp01},  
$ \Phi _- =0 ,\ G_-=0 ,\ \hat T ^{loc } -\mu _3 \rho  _3^{loc } =0$, 
yields  the simplified     wave equations,  
\bea &&    \bullet \ 0= \d (G_\nu ^\mu  - \kappa ^2 T _\nu ^\mu  )  = 
e ^{-2 A} \d _K  \tilde  G ^{(4)  \tilde \mu }_ \nu  - {1\over 2}  e ^{2 A} 
\tilde \nabla _6 ^2  (  \d _K \tilde g ^{\tilde \mu }_\nu -\d ^{ \mu }_\nu 
 \d _K  g )   - {1\over 2} \d ^{ \mu }_\nu e ^{2 A} \d  \tilde R
 ^{(6)}   ,\cr &&   
 \bullet \  0= \d (G_n ^m  -\kappa ^2 T_n ^m )= e ^{2 A} \d  \tilde G ^{(6)
   \tilde m }_n   
-\ud e ^{-2 A} ( \tilde  \nabla _4 ^2 \d  \tilde   g_n ^ {\tilde m} -
{1\over 2}    \d _n ^ {m}  \tilde  \nabla _4 ^2 \d  \tilde   g ) 
- \ud  e ^{-2 A}  \tilde  \nabla ^{\tilde m} ( e ^{4 A} \nabla  _n \d _K g )
\cr &&  + \ud \d _n ^ {m}  \tilde  \nabla ^{\tilde p} ( e ^{2 A} \dh
_p \d _K g ) - \ud e ^{-2 A} \d _n ^ {m}  \d _K R ^{(4) \tilde \mu } _\mu   
- \d [ {1\over 4 \tau _2 } ( 2 G _{npq}
\bar G ^{mpq}  - {1\over 6} G\cdot \bar G ) ]  .\label{appeq18}   \eea
The linear wave equations   for the  axio-dilaton, the  2- and 4-form
potentials  and the  internal  metric tensor  are  
\bea && \bullet \  0= (\nabla ^2 _{10} -   {\tau _2 ^2 \over  6 }
\dh _{\bar \tau  }\dh _{\tau  } ({ G \cdot \bar G  \over \tau _2 }) )      
\d _K  \tau    \equiv  
(e ^{-2A} \tilde \nabla _4^2  +   e ^{2A}  \tilde \nabla _6^2 
-  {e^{4A} \over 12 \tau _2 } G \tilde \cdot \bar G   )    \d _K  \tau  
. \cr && \bullet \
  d ( (\Phi _+ \d G_- + \d \Phi _-  G_+)  + (\d \Phi _+ G_- + \Phi _-  \d G_+) )  + i \d ( {d \tau \over \tau _2} \wedge  (e^{4A} \star _6  (G+\bar G) -
i \a   (G-\bar G) ) ) =0.   \cr && 
\bullet \  \d  (\star _{10} \tilde F_5 )  =  \d \tilde F_5 ,\ [
d \d \star _{10} \tilde F_5 = \d ( H_3 \wedge F_3) + 
{2 \kappa ^2  \tau _3 \over g_s}   \d \rho _3 ^{loc}  ] . 
\cr && \bullet \    
-\ud \D  _K  \d \tilde  g_{mn}     =  {2  \over  \Phi  _+ ^2 } \tilde
\nabla  _{( m} \Phi _+ \tilde \nabla  _{ n )}    \d  \Phi _- - {\Phi
  _+ \over 32 \tau _2  } ( G _{+( m}  ^{\ \ \ \ \  \tilde  p\tilde
  q} \d G_ {-n) p  q}  +   \d \bar G _{-( m}  ^{\ \ \ \  \ \tilde p\tilde  q}  \bar  G_ {+n)p  q}
) - {e ^{-4 A} \over 4} (\d \tilde  g_{mn}    
\tilde \nabla _4^2 \d \tilde  g_6 - 2 \tilde \nabla _4^2 \d \tilde
g_{mn}  )   ,\cr &&
[-\ud \D  _K \d \tilde  g_{mn} \equiv  -\ud (
\tilde \nabla  _6 ^2 \d \tilde  g_{mn} +\tilde  \nabla _ { (m } \tilde  \nabla
_{n ) }  \d g - 2 \tilde \nabla ^p \tilde  \nabla _{(m} \d \tilde
g_{n)p}  ) ]  
.\label{appeq33}   \eea
Note the useful relations  for the 3-form field,
$[ \d ( (\star _6   \pm  i )   G_\mp  )  =0,\
  \d ( (\star _6   \pm  i )   G_\pm  )  =  \pm 2i \d G_\pm  ,\   
  d (\d G_+ - \d G_-) = -2i (d \d \tau  ) \wedge  H_3 ] $, and the
simplified  wave equation in GKP vacua,
$d  (\Phi _+ \d G_- + \d \Phi _-  G_+) =0 ,\
[\Phi _- = 0,\ G_- =0,\ d \tau =0 ]$.  
It is also  useful  to record the  related     equations
for vacuum field   deformations  of use in   describing the  back-reaction on the geometry  from  compactification   or  localized sources effects. 
Specializing   to fluctuations   constant in
$ M_4$,   one obtains  from Eqs.~(\ref{appeq33}) and (\ref{appeq24})  the linearized   equations   for  the warp profiles,   axio-dilaton,  2- and 4-forms and   the internal space metric  around  the  GKP  vacua~\cite{gandhi11} 
\bea  && \bullet \   \d  (\tilde \nabla ^2 \Phi _- )  = 
\tilde \nabla ^2   \d \Phi _- =   0. \cr   &&  
  \bullet \  - \d ( \tilde \nabla _6^2  \Phi _+ ^{-1} )  =   
{g_s \over 96} G _+  \tilde \cdot  \tilde G _-  
(-g_s Im (\d \tau ) + {2  \over  \Phi  _+ }\d \Phi
_- ) - { 2   \over  \Phi  _+ ^4} \nabla _m \Phi  _+  \nabla ^{\tilde
  m}  \Phi  _+  \d  \Phi _- \cr && + {g_s \over 96} ( \d G_+ \tilde
\cdot \bar G_+ +    
G_+ \tilde \cdot  \d \bar G_+  + 3 \d \tilde g ^{pq} G_{+ mnp} \bar G _{+
  \ \  q } ^{\ \ \tilde m \tilde n} ) .
\cr  &&    \bullet \ \tilde \nabla  ^2 \d \tau = -{1\over 8} \Phi _+
(G_+\cdot \d    G_-).     \cr &&
\bullet \  d (\Phi _+ \d G_- + \d \Phi _- G_+  ) = 0,  
\  [\d   ( (\star _6 \pm i )  G_\mp ) =0 ]  .\cr &&  
\bullet \ -\ud \D  _K \d \tilde  g_{mn} 
= {2\over \Phi _+^2} (\tilde \nabla _{(m} \Phi _+ \tilde \nabla _{n)} \d \Phi
_- - {\Phi _+ \over 32} (G _{+ (m} ^{\ \ \  \  \  \tilde p  \tilde  q } \d \bar G _{- n) pq} + \d G
_{- (m} ^{\ \ \  \  \  \tilde p\tilde  q} \bar G _{+n) pq} ) .\label{appeq35} \eea 
 
\subsection{Local self couplings  of  graviton modes} 
\label{subapp5}

The self couplings of the  4-d    metric field   $g _{\mu \nu } $ in the background of  metric $ d s ^2 _{10}= e ^{2 A(y) } d\tilde s ^2 _4 + e
^{-2 A(y) } d\tilde s ^2 _6  $ produce a  variety of gravitational  interactions   between the 4-d  graviton  harmonic modes.
These   couplings   arise  
from    the perturbed  10-d Einstein-Hilbert action by expanding  the  Ricci curvature  tensor components  $R  ^{(d)} _{\mu \nu } =  R _{\mu  \l \nu } ^ {\ \ \ \ \ \l } + R _{\mu  m  \nu } ^ {\ \ \ \ \  m}  $ in powers of linear fluctuations of the  metric tensor  $ g _{\mu \nu }  +\d g _{\mu \nu },\ 
[\d g_{\mu \nu } = e^{2A} h_{\mu \nu } ].$ The contributions  in the
first  term come  from the  external space $  R   _{\mu \nu } (M_4) $   only while  those  in the second  part     involve  covariant  derivatives $\nabla _m $     of the internal  space $X_6$    acting on  $h_{\mu \nu } (x, y)$ and    the   warp  profiles,  as      illustrated    by the  identity in  Eq.~(\ref{appeq7}).   We shall   consider here only the former contributions 
 with the view to provide   an expanded    formula  for the perturbed   curvature tensor $  \tilde R ^{(4)}  (\tilde  g_{4}  +h )$    encoding the   all order  expansion in powers  of linear fluctuations of the  metric tensor  $h_{\mu \nu } (X)$.  The relevant  part  of the 10-d    curvature action,
after  extracting  out the overall  dependence on the warp profile, is given by 
\bea && S _{grav}  (g +\d g ) = {1\over V_W} \int _{X_6} d^6 y \sqrt {\tilde
  g_6} e ^{-4 A (y)} S_{4} (\tilde g _4 +h) ,\ \ [ S_{4} (\tilde g _4 +h) = 
{1\over 2 \kappa _4^2 } \int  d^4 x  \sqrt {\tilde  g_{4} +h } \tilde R ^{(4)}  
(\tilde  g_{4}  +h ) ] .   \eea   
The  non-linear dependence  of    Ricci   curvature scalar  $ \tilde R ^{(4)}  =  \tilde g ^{\mu \nu }  R _{\mu  \l  \nu } ^{\ \ \ \ \l } $ on the metric   
yields self couplings  among  the   $ h_{\mu \nu }$ of arbitrarily high   orders.  It is tacitly understood that the  metric tensor  fields  $h_{\mu \nu } (x, y)$  in each term  of the expansion   are decomposed   into
Kaluza-Klein  modes   with  the products  of   graviton modes wave
functions    transferred under   the integral  over  the  internal  manifold  volume.       The    spacetime  structure  of  the metric tensor fields   couplings  is fully  encoded  in  the   following closed  formula  for the perturbed Ricci tensor of $ M_4$, obtained   via   the   Mathematica package `Ricci'~\cite{lee92},
\bea &&  \tilde R ^{(4)} (\tilde  g +h)  = {1\over 4 } \bigg [ 2  \tilde R ^{(4)}    + 2 \tensor R{\down
\rho \down \lambda } \tensor G {\up \rho \up \lambda } + \tensor G{\up
\nu \up \rho } ( 2 \tensor h{\down {\mu \nu ; } \up \mu \down \rho }
-2 \tensor h{\down \mu \up {\mu } \down {;\nu } \down \rho } - 2
\tensor h{\down \nu \down {\rho ;} \up \mu \down \mu } +2 \tensor
h{\up \mu \down \nu \down {;\rho } \down \mu } ) + 2 \tensor h{\up \nu
\up \sigma } \tensor R{\down \nu \down \rho \down \lambda \down \sigma
} \tensor G {\up \rho \up \lambda } \cr && + \tensor G {\up \nu \up
\sigma } \tensor G {\up \rho \up \lambda } ( \tensor h{\down \mu \down
{\nu ;} \up \mu } \tensor h{\down \rho \down {\lambda ;} \down \sigma
} - \tensor h {\down \mu \up \mu \down {;\nu } } \tensor h{\down \rho
\down {\lambda ;} \down \sigma } -2 \tensor h{\down \mu \down {\nu ;}
\down \rho } \tensor h{\up \mu \down {\lambda ;} \down \sigma } -
\tensor h{\down \nu \down {\rho ;}\down \mu } \tensor h{\up \mu \down
{\lambda ;}\down \sigma } ) \cr && + \tensor G {\up \nu \up \lambda }
\tensor G {\up \rho \up \sigma } ( \tensor h{\down \nu \down {\rho
;}\down \mu } \tensor h{\down \lambda \down {m;} \up \mu } +\tensor
h{\down \lambda \down {\sigma ;}\down \mu } \tensor h{\up \mu \down
{\nu ;}\down \rho } +2 \tensor h{\down \mu \down {\nu ;}\down \rho }
\tensor h{\up \mu \down {\lambda ;}\down \sigma } ) \cr && + \tensor G
{\up \nu \up \rho } \tensor G {\up \lambda \up \sigma } ( -2 \tensor
h{\down \mu \down {\nu ;} \up \mu } \tensor h{\down \rho \down
{\lambda ;}\down \sigma } +2 \tensor h{\down \mu \up \mu \down {;\nu }
} \tensor h{\down \rho \down {\lambda ;}\down \sigma } -2 \tensor
h{\down \rho \down {\lambda ;}\down \sigma } \tensor h{\up \mu \down
{\nu ;}\down \mu } + \tensor h{\down \nu \down {\rho ;}\down \lambda }
\tensor h{\up \mu \down { \sigma ;} \down \mu } ) \bigg ] ,\ [ G ^{\mu
    \nu } =( (\tilde g + h) ^{-1} ) _{\mu \nu }] . \label{appeq37}  \eea
(We mention incidentally that symbolic   calculations for non-linear  fluctuations   might also be performed   with the  Mathematica  packages `xAct'  and `xPert'~\cite{brizuela08}.)
The  background   curvature scalar is denoted 
$\tilde  R  ( \tilde g) =  \tilde g ^{\mu \nu }  R _{\mu \nu } $, 
semi-colons   refer  to    covariant  derivatives   
$ \tensor h{\down {\mu \nu ; } \down \rho \up \mu  }
=    \nabla ^\mu   \nabla _\rho  \tensor h{\down {\mu \nu } }  = 
\nabla ^\mu  (\dh _\rho  h  _{\mu \nu } + \G ^\l 
_{\rho \mu } h   _{\l  \nu }  +   \G ^\l  _{\rho \nu }   h _{\l  \mu
}  )  $ and  the   inverse  metric and integration  measure
determinant    are  computed  from      the  power expansions 
\bea &&  G ^{\mu \nu }  = 
(\tilde g  ^{-1} ( 1 - \tilde g   ^{-1} h + ( \tilde g  ^{-1} h )^2
+\cdots ) )_{\mu \nu }   ,\cr &&   
\sqrt { \tilde  g+h } \equiv   Det ^{1/2} (\tilde g+h) = 
e^{\ud Tr \ln  (\tilde g+h)}  =  \sqrt {\tilde  g}  + \ud Tr (\tilde g^{-1} h )
 - {1 \over 4} Tr (\tilde g^{-1} h \tilde g^{-1} h ) +
{1 \over 8}  (Tr (\tilde g^{-1} h) )^2 +\cdots  .\label{appeq39}  \eea 
In flat spacetime,    $ \tilde  g _{\mu \nu  } = \tilde  g ^{\mu \nu  }  
= \eta _{\mu \nu } = diag (-1, 1,1,1) $  and covariant derivatives   
reduce to  ordinary derivatives.
The   $ O( h^0) $ (zeroth order)  terms     in Eq.~(\ref{appeq37})   
reproduce     the unperturbed  scalar  curvature $R=  R( \tilde  g) $, 
the   $O(h)$ terms  reproduce    Einstein  equation ($ R _{\mu \nu }
-\ud  \tilde g_{\mu \nu } R  =0$)  in the form    
\bea &&   \d \tilde R  ^{(4)} ( \tilde g) =  
\tilde  R^{(4)} ( \tilde g+h) - \tilde  R^{(4)}(g)  = \ud \tilde  g ^{\mu \nu }
(h _{\l \mu ; \ \nu } ^{\ \    \ \l }  +  h _{\ \  \mu ; \nu  \l } ^{\l } +
h _{\l \ \ ; \mu   \nu } ^ {\ \l }    - h _{\mu \nu ; \ \  \l }
^{\ \ \ \    \l } ) +{1\over 2}  h  ^{\mu \nu }  \tilde  R ^{(4)} _{\mu \nu }  ,\label{appeq40}  \eea 
while the quadratic  and higher   order  terms reproduce 
the  kinetic energy (inverse propagator)     and  couplings.    
The   effective action for  the 4-d   massless and massive graviton
fields $  h ^{(m)} _{\mu   \nu } (x)  $ from the    decomposition
$ h_{\mu   \nu } (X)  = \sum _m h ^{(m)} _{\mu   \nu } (x)  \Psi _m
(y)   $  is   built   by  integrating the products of participant 
wave functions  over  $X_6$,  as in Eq.~(\ref{sub4eq3}). 

 We   mention in conclusion two    general   warnings  concerning the 
implementation   of a  perturbative   gravitational  
theory  in a spacetime with  extra dimensions.  
Recall  first that the standard  formulation of  
massive gravity theories  suffers from ghosts  and strong
coupling problems   for which    possibly   cures using 
improved formalisms~\cite{derham14} do exist.  
Observe  in second that   
the  ultraviolet truncation  of Kaluza-Klein  theories  to  finite 
sets of  massive spin 2 modes
breaks the  general coordinate invariance  explicitly 
and   is  inconsistent with string    theory~\cite{duffpopestelle89,duffrapope90}. 
If a consistent gauge invariant  truncation could  be defined, 
the structure   of couplings  would     simplify considerably  
by    choosing a  suitable gauge.   One could then develop the  
powerful string theory  methods expressing      the 
 scattering  amplitudes   for massive spin 2    modes 
as  products of left  and right   scattering amplitudes  of spin 1 
modes.   Compare, for instance,  with the simplified  form  of 
the  massless gravitons  action   in
Feynman gauge  up to  cubic order~\cite{berngrant99}, 
$  L= \ud h _{\mu \nu } \dh ^2 h ^{\mu \nu }- 
\kappa _4 (h ^{\mu \nu } h _{\l \s   , \mu \nu }  h ^{\l \s } +
h ^{\mu \nu } h _{\l  \mu    , \s }  h ^{\l \s } _{\ \  , \nu }  +    
h ^{\mu \nu } h _{\s   \nu    , \l }  h ^{\l \s } _{\ \  , \mu } ) .$  

\subsection{Low energy action  of $Dp$-branes}
\label{subapp6}
  
The  geometrical description of   $Dp$-branes  dynamic utilizes a  non-linear    sigma model for the open string  theory  fields
with   the  background  fields  associated   to  closed   strings   and 
interactions  determined   by  matching to  string  scattering
amplitudes.   The  embedding  of the  brane  world volume $ M_{p+1}$ 
in super-spacetime  $ M_{10} $ is  described  by     bosonic   fields $ X^M (\xi  ) $  and   fermionic  Majorana-Weyl (MW) bi-spinor
fields $ \T  _{ \a } (\xi )= ( \T  _{i  \a } (\xi )  ) ,\ [i=1,2] $, 
functions of  $ (p+1)$   coordinate  variables  $\xi  = (\xi ^{0},\cdots , \xi ^p  )   $,    in  irreducible  representations of   the tangent spacetime
group $SO(10)$,    
\bea &&  X^M (\xi )= (X^\a ( \xi ) , X^m ( \xi ) )\sim 10  ,\ \T
( \xi ) = { \t _1 ( \xi )  \choose \t_2 ( \xi ) } \sim (16 \oplus 16 )
,\cr &&    [ X^\a   = \s  A^\a ,\ \s = 2\pi \a ',\    
\G _{(10)} \t _i = \t _i ,\  \t _i = B^\star  \t _i ^\star ,\  
i=1,2] \label{sub4eq6}   \eea 
where   $\G ^M,\  [\{ \G ^M ,\G ^N\} = 2 g ^{MN} ]  $ 
  are    Dirac matrices of   $SO(10)$
with $\G _{(10)} ,\  B,\ [B B ^\star   =1,\ B \G ^M B^\star  = (\G ^M
  )^\star ] $   denoting  the  chirality and charge conjugation
matrices.   The    curved  spacetime     components  of
the vector  fields  $ X^M,\  \G ^M$ are related  via the frame  fields $ e  ^A _M ,\ [ e  ^A _M  e  ^B _N  \eta _{AB} = g _{MN} ] $ to the 
(flat) tangent   spacetime     components,     $ X^A =  e  ^A _M  X^M\,  \G ^A =  e  ^A _M  \G^M  .$       In the presence  of a $ Dp$-brane,  the flat and curved   spacetime  indices
$M ,\ A \in (0,\cdots , D) $   are split into  pairs of indices 
$  M=  ( \a , m)   ,\   A=  ( \underline{\a } ,\underline{m} ) $   
  referring  to  the   brane world volume longitudinal and transversal
  directions.  The frame fields of $ M_{p+1}$  are given by  $ e ^M_\a = \dh _\a  X^M ,\ [ \dh _\a = \dh /\dh _{xi ^\a } ].$
  For a $ Dp$-brane  wrapped on a $\S _{p-3}$-cycle  
of  $ X_6$, the  longitudinal directions are labelled  by the indices 
$ \a = (\mu , a ) \in M_4 \times \S _{p-3}$.    
The low energy  field content  consists 
of  massless gauge  bosons and  matter moduli  $ X^{\a } ,\ X^{m} $
 and their  superpartners,
identified to the brane  embedding  in the  super-spacetime  for $
M_{10}$   with coupling constants   identified  to the classical 
bulk supergravity    fields. 
For instance, the   couplings to graviton modes are obtained  by 
replacing   $  \tilde  g_{\mu \nu }  (X) \to \tilde   g_{\mu \nu } +
h_{\mu \nu } ,\   \tilde  g^{\mu \nu }  (X) \to  \tilde   g^{\mu \nu }
- h^{\mu \nu }  $.             
The scalar   fields  and   products  with fully saturated
spacetime   indices      have   similar  forms  in terms of spacetime
and brane  coordinates,  $ \G ^\a D_\a =  \G ^\mu D_\mu +  
\G ^a D_a $     while   the  components of tensorial
quantities   are related as    $ V _\a = \dh _\a X^M V_M= \dh _\a X^M  e ^A_M V_A .$ 

The   $D$-brane action  inherits from string theory  several 
symmetries comprising    diffeomorphisms of the variables  $ \xi ^\mu $, Lorentz-Poincar\'e, gauge   and supersymmetry transformations   of the   coordinate  fields  $ X^M ,\  \T  $  and   discrete duality symmetries. 
The extremal $Dp$-branes of  type $ II\ b$ theory have
odd-$p$  and preserve half of the 32  bulk   supercharges.
For convenience, one often  selects   the static gauge choice 
fixing  the  longitudinal coordinates  as  
$X^\a ( \xi ) = \xi ^\a  \ \Longrightarrow  \   \dh _\a  X^M ( \xi )
=  \d ^M _\a .$   
The  axio-dilaton, metric  tensor  and  $p$-form fields 
are then replaced  by their  pull-back  transforms on the brane world   volume~\cite{grana03}   
\bea &&  P( \tau  (\xi )  )  = \tau  ( X(\xi )) ,\  
P( g_{\a  \b  } )  = \dh _{ \a  }  X^M \dh _{\b } X^N  g _{MN}  =
g_{\a  \b  } + \dh _{ \a  }  X^m  g _{m\b  }  +  \dh _{ \b  }  X^n  g _{\a   n}  + 
\dh _{ \a  }  X^m \dh _{\b } X^n  g _{mn } ,\cr && 
  ( P (C ^{(q)} ) ) _{\a  _1 \cdots \a  _q} = C_{\a  _1 \cdots \a  _q}
-q \dh _{\a  _1}  X^m  C_{m \a  _2\cdots \a  _q}  + {q (q-1) \over 2} \dh
_{\a  _1}  X^m  \dh _{\a  _2}  X^n  C_{m  n \a  _3 \cdots \a  _q} +
\cdots  . \label{sub4eq7}   \eea   
The  pull-back  transforms      of the   metric,  Kahler form and
3-form  fields   $\tilde g_{mn} ,\    \o _{mn}  ,\  G_- ,\  \O $ along
the    brane transversal directions are evaluated at  the brane location,  with the      coordinates $ X^m (\xi ) $ 
interpreted   as  fluctuations  of the  brane world  volume 
 about the  equilibrium positions     which are usually  
set  at $X^m _\star = X^m (y_{\star } ) =0$.  
The fields   functionals $ F (X)$  depending on   transversal
coordinates,  are  evaluated    from   the Taylor expansions
$  e ^{ X^m  (\xi ) \nabla  _m }  F (X (y_{\star } ) )  $.  

The    $Dp$-brane action of lowest  order    in
bosonic  fields  derivatives~\cite{myers03}   and  in fermionic 
fields~\cite{aganagic96,cederwall96,grisaru97,marolf03,martucci05,grana02,trivedi05}    
is conveniently    constructed with the help   of $T$-duality,  
moving  the      brane dimension   up  from $ p=2$    for 
the  $M2$-brane or  down    from $ p=9$    for the $ D9$-brane.    
The   bosonic  action   includes 
the  Dirac-Born-Infeld (DBI)  part  depending  on $U(1)$ 
gauge  boson  and matter brane fields $ A^\a  ( \xi  ) ,\  X ^m (\xi )$
and   bulk fields $ g _{\a  \b } ( \xi ) ,\ B_{\a \b } ( \xi )$  
and the  topological  Chern-Simons or Wess-Zumino (CS or WZ)   
action  depending on the gauge   field strength
$ F_2 = d A $ and the  pull-backs of  the  antisymmetric       
bulk  potentials  $ B_2 = B ^{(2)}  $ and $C_q  =  C ^{(q)}  $.    
The bosonic   action    of $ Dp / \bar
Dp$-branes is given in    Einstein frame  by  
\bea && S_{B} (Dp /\bar Dp) = - \tau _p   \int _{M_{p+1}}  d ^{p+1}
\xi e ^{(p-3 )  \phi /4 } (- Det (g  + g_s ^{1/2} e^{- \phi /2 } \calf  ) )^{1/2}  \pm  \mu _p \int _{M_{p+1}} [\sum _q C ^{(q)}
  \wedge  e ^{\calf _2} ]_{p+1}    ,  
\cr &&  [\tau _p = { \mu _p  \over g _s
    ^{(p+1) / 4}  } ,\  \mu _p   = {2\pi \over \hat l _s ^{p+1}
} ,\  \calf _2 = B_2 +\s F_2 = B_2 + {\hat l_s ^2 \over 2\pi }  F_2
]  \label{app4EQ1}  \eea    where $\tau _p ,\ \mu _p$  denote  the
tension and charge parameters and the determinant  
is   conveniently    evaluated  by means of the   expansion 
\bea &&  Det ^{1/2} ( 1 + M)\simeq  1 + \ud Tr (M) -
{1\over 4} Tr (M^2)  + {1\over 8} (Tr M)^2 + {1\over 6} Tr (M ^3) -
{1\over 8}  Tr (M) Tr (M^2) +  {1\over 48} (Tr M)^3  +  \cdots
\label{subexpdet}. \eea
Although we shall  restrict   in the sequel  to single $Dp$-branes, 
note    that for  $N Dp$-brane stacks     the     fields 
$ A_\a ,\  X^m $         become non-commutative 
 matrices   in the  adjoint representation  of  the gauge group
 $U(N)$~\cite{myers03}.   The   $U(1)$ covariant  derivative  $ D_\mu =
 \dh _\mu   + i A  _\mu  $   for  the single  brane case $ N=1$, 
becomes $  D_\mu =    \dh _\mu   + i [A  _\mu ,\  ] $  and 
the    Taylor expansion    of  the   topological  action   
involves    the    interior products $i_X $  along the $ X^m$
directions,   which  act  as  
anti-derivative operators     reducing   the order of field  forms by
single units,    
\bea &&  e ^{i_X i_X}  C ^{(q)}   = 
C ^{(q)}_{\a  _1 \cdots \a  _q} +{ q!  \over 2 (q-2)!}   
[\dh _{\a  _1}  X^m ,\dh _{\a  _2}  X^n]   
C ^{(q)}_{nm\a  _3 \cdots \a  _q} +\cdots . \eea    
The   extra terms introduced  by the pull-backs are 
generally    higher order in the string scale,  since the canonically normalized
fields $\varphi ^m $ are   defined      by replacing $ X^m \to \s  
\varphi ^m  ,\ [\s = 2\pi \a ' ] $.   While  we   mostly restrict     to    the
leading     order   terms  in $\a ' $,  it is  useful to discuss
briefly  the contributions  to the effective  $Dp$-brane  action  
of higher orders in $ g_s $ and $\a '$ arising from 
string loop and sigma-model perturbative corrections.
The parity even    part    of $O(\a ^{'2} )$  was    derived
in~\cite{bachasgreen99} by matching to  the
2-graviton  disc  string   scattering amplitude on $ Dp$-branes. 
The typical  contributions  involve    the  quadratic order combinations  
$[ (R_{T, 4}  ^2 - R_{T, 2} ^2 ) - (R_{N, 4}  ^2 - R_{N, 2}
  ^2 ) ] $     of   the    Riemann and Ricci curvature  4- and 2-tensors 
$ R ^{T,N} _{4, 2} $  for the tangent and normal bundles  $ T$ and
$N$     over the  brane world volume $ M_{p+1}$.  
The      lowest order radiative  correction to the     action  string
frame   is given in the simplest case  by     
\bea && \d S (Dp)  = {\hat l_s ^4 \over 4 !\  32 \pi ^2} \mu _p 
\int d ^{p+1}\xi  \sqrt {g_ {p+1}}     e ^{-\phi } 
 [( R_{\a  \b  \g  \d  }  R^{\a  \b  \g  \d  } -2 
R_{\a \b  } R^{\a \b } )  -( R_{\a \b  mn }  R^{\a \b  mn } -2 
R_{mn } R^{mn } ) ] .\eea     
The  quartic order  actions for orientifolds  and   compactified
theories  are  discussed in~\cite{junghausshiu14}. 
The string  one-loop  effects  also induce    parity  odd
topological couplings in  the RR  potentials  $C_{(q)}$  of higher
order   in   the gauge and   gravity field  strengths  $ F,\ R$. 
The derivation,   based on    the  dimensional descent
approach to   Wess-Zumino   type actions~\cite{harvey99},
employs   the  formula for the anomaly in 
the variation of the $Dp$-brane generating functional  under 
coordinate and  gauge   transformations  
 $\d \o = d \t + [\t ,   \o  ],\  \d  A = d \l + [\l ,  A  ] $, which 
is given  (for $\a ' =1$)  by   
\bea &&  \d _\L \log Z (A, \o ) = 2\pi  \mu _p \int _{M_{p+1} }
     [Tr _\rho (e ^{\calf_2 } ) \wedge   \hat    \cala ( \hat l_s ^2 R )] 
\vert ^{(1)} _{(p+1)}        ,\cr &&  
[\hat \cala (R) = 1 - {p_1 (R)    \over 24 } + \cdots
= 1 -  {R^2  \over 16 \pi ^2 }  + \cdots , \ \calf _2 = B_2 + \s
F_2,\   F _2 = d A + A^2,
\ R _2  = d \o + \o ^2 ]    \label{app4EQ2}   \eea    
where  in the integrand  term inside  brackets 
the   first exponential    factor  is   
the polynomial   for Chern characteristic classes in the  
chiral fermion representation $\rho $ of the gauge group    
and the   second  factor  is  the  Dirac operator $A$-genus polynomial
for  Pontrjagin   classes $p_n (R) $.   The 
lower  suffix in $[ \cdots ] ^{(1)} _{(p+1)} $  
selects  the  net $(p+1)$-form  and the upper suffix  selects the 
linear order  terms in   the  variation fields $\L = (\t , \ \l )  $. 
The anomalous   action  is   then  defined as the counterterm action
that  must  be  added in order to cancel  the   anomalies in   case 
the  $Dp$-brane  is   embedded in  the  background 
of   some  non-anomalous   theory.   In  the complete theory, chiral  fermions      can arise   from open string  sectors  of
intersecting branes or branes at  singularities. 
The anomaly  cancellation is  thus realized  as  
an RR charge inflow  mechanism   from   the bulk.
The resulting   anomalous   $Dp$-brane action is given by 
the  generalized   Chern-Simons   action~\cite{harvey99}   
\bea &&  S_{CS} (Dp)  = \mu _p \int _{M_{p+1}}  [\sum _q 
C ^{(q)}  \wedge Tr_\rho    (e ^{\calf _2 } ) \wedge 
\sqrt {\hat A ( \hat l_s ^2 R _T ) \over \hat A ( \hat l_s ^2 R _N) }
] _{p+1} . \label{app4EQ3}   \eea 
The  $Op$-planes  also  acquire  the  anomalous   
gravitational      action, $ S_{CS} (Op)  = - 2 ^{p-4} \mu
_p \int \sum _q C ^{(q)}  \wedge \sqrt {\hat L ( \hat l_s ^2 R /4 ) }
$, involving    the Hirzebruch signature  $L$-genus.  
The  anomalous  brane action     can  also be  
derived   by matching  to one-loop   string amplitudes $ \g g^2,\  \g g^4 $ 
for gauge bosons on branes coupled to   massless gravitons~\cite{morales98}.

The  fermionic action    of  extremal (BPS)  $Dp$-branes
is  usually derived   as a supersymmetric completion of the    bosonic      action  by invoking the  fermionic  $\kappa $-symmetry
utilized     introduced initially in    the   superstring and
super-membrane theories. The  construction   was developed  for   flat background  
spacetimes  in~\cite{bergshoe96,bergshoe97,aganagic96,cederwall96}  and  was generalized   
to   general     bosonic    backgrounds  with curved spacetimes in two different  approaches~\cite{grisaru97,marolf03},   limited so far   
to    quadratic orders     in  spinor fields.   
We     here follow the     discussions  in~\cite{bergshoe97,marolf03}
in   the updated  presentation  of~\cite{martucci05}. 

The   bi-spinor  fermion    field $\T ( \xi )$
the    corresponding bulk    field $\T (X) $, while the bi-spinor  fermion field $\e  $  generating  supersymmetry transformations is  constant. 
The  $\kappa $-symmetry acts by local  transformations  
$\d   \T (\xi )  =(1 +\G ) \kappa (\xi )  $,
implying the gauge  equivalence $ \T (\xi ) \sim \T +(1 +\G )   \kappa (\xi )  $ under shifts involving arbitrary  
brane  bi-spinor  fields $\kappa (\xi )  $, with     $\G $   some
functional of the brane   fields and $\G _\a $  matrices. The construction   assumes   that the combined      $\kappa $-symmetry  and  
supersymmetry   must be  realized   to leading order of $\T $
by  $ \d \T  (\xi ) = ( 1 + \G ) \kappa (\xi ) + \e  + O(\T ) ,\ [  \G  ^\dagger = \G ,\ \G ^2 =1 ]$.
Imposing  the gauge  fixing     $\calp \T =0 $    via the field dependent  projection  $\calp ^2 = \calp $,   one finds that  the compatibility    with      the  above   transformation requires    $ \calp \d \T =
(1- \calp  ) \d \T =0 \ \Longrightarrow \ 0= \d \T = ( 1 + \G ) \kappa
(\xi ) + \e   $.  The    selected residual   supersymmetry is then
generated  by  bi-spinor  variation fields  satisfying the  complementary projection $ ( 1 -\G ) \e  =0 $. The decoupling of half of the  unphysical     fermionic  degrees of freedom    has removed half of  the bulk  supercharges. 
 The   $ \kappa $-symmetry projection 
operators    $ P ^{Dp}  _\mp $ for $ Dp/\bar Dp$-branes, including
the dependence on the  NSNS  and   $ U(1)$ gauge fields 
in  the  2-form  field $\calf _2 $ to all  orders      are given by~\cite{bergshoe97}   
\bea &&  P_\pm  ^{Dp}  = {1 \pm \G ^{Dp} \over 2} =    
{1\over 2} \pmatrix {  1 &  \pm  \tilde \G _{Dp} ^{-1} \cr \pm
  \tilde \G _{Dp} & 1}  
,\  \tilde  \G _{Dp}  =  \G ^{Dp} _{(0)}  \  \L ( e ^{-\phi ' /2 } \calf )   
,\cr &&   [\L (\calf  )=  {\sqrt  {g   _{p+1} }  \over 
\sqrt  {g   _{p+1}  + \calf } }   \sum _{q\geq 0} { 1 \over 2 ^q q!} 
\G ^{\a _1 \b _1 \cdots   \a _q \b _q} \calf _{\a _1 \b _1
  } \cdots    \calf _{\a _q \b _q} \G ^{' Dp} _{(0)}  (-\s _3  )^q    
=  \hat e ^{-a/2}   \G ^{'Dp} _{(0)} \hat e ^{a/2} ,\cr && 
a= \ud  Y_{\a \b }  \G^{\a \b } \s_3 , \ \G ^{' Dp} _{(0)}  =\G ^{Dp} _{(0)}  ((\s _3 )^{p-3 \over 2}      i \s _2 )
,\   \G ^{Dp} _{(0)}  = {1\over (p+1)! \sqrt  {g   _{p+1} } }  \e
_{\mu _1 \cdots \mu _{p+1} }   \G ^{\mu _1 \cdots \mu _{p+1} }  ]  \label{app4EQ5}  \eea   
where $ (\G ^{Dp} _{(0)} )^2 = (-1)^{p(p+1) /  2}  $  and
$\L (\calf )  = 1 + O(\calf ) $ includes  the dependence on   the Abelian
2-form   field strength  $\calf _{\a \b  }  $. 
    The notation $ \hat e $ refers to  the antisymmetrization of  
distinct  pairs of  indices  $ \a _i \b _i $ in the   polynomial expansion 
of the exponential  $ \hat e ^{\mp a /2} $     while 
$ Y_{\a \b }$  and  $ \calf _{\a \b }$  are  related  by  the  formal  relation 
$  \calf  \simeq  \tan Y $ or $   \tanh Y $    applying  to  the 
diagonal  basis   of the corresponding matrices.
The Hermiticity   and trace  free properties 
$ (\G ^{Dp})^2 =1,\ Tr (\G ^{Dp}   )=0 $  ensure  the
requisite  conditions for projection operators,     
$ (P ^{Dp}  _\mp )^2 =P ^{Dp}  _\mp ,\ Tr (P ^{Dp}  _\mp ) = 16. $ 
The $ Dp/\bar Dp$-brane fermionic action    of   quadratic order 
in  $\T ( \xi )$   is  then given  in   the string frame   by  the  formula  
\bea && S_F ^s ({Dp/\bar Dp})  =  i  \mu _p \int d^{p+1} \xi e
^{ -\phi } \sqrt { - Det (g +  \calf  ) } \bar \T P_\mp 
^{Dp}   ( ( \calm ^{-1}  ) ^{\a \b } \G  _\b D_\a  - {O \over 2}) \T , 
\cr &&  [D_\a = \nabla _\a   + {1\over 4} \Hslash _{3 \a } \s _3 
+ {e ^{\phi } \over 8} ( \Fslash _1 i\s _2    + 
\Fslash _3 \s _1   +  \Fslash _5  i \s _2   ) \G _{\a  } ,\ 
O= \dslash   \phi +  {1\over 2} \Hslash _{3 } \s _3  
- e ^{\phi } ( \Fslash _1 i\s _2    +  {1\over 2} \Fslash _3 \s _1)
,\cr &&  \calm _{\a \b }  =   g_{\a \b }   + \calf _{\a \b }   \G _{(10) }
\otimes \s _3,\ \calf  _{2} =B_{2} + \s  F _{2}  =  
B_{2} + \hat l_s ^2 f _{2}   ,\ \Fslash _p = {1 \over p !} F_{M_1 \cdot M_p}
\G ^{M_1 \cdot M_p},\ \bar \T = \T ^\dagger \G^0 \s _3   ]    \label{app4EQ4}    \eea  
where   $ D_\mu $  and $ O$    are  the   brane induced   operators  that    realize the  supersymmetry  transformations   of the  gravitino  and    dilatino pull-back  fields   in the bulk,    
$\d \psi _\mu   =  D_\mu \e ( \xi ),\ \d \l  = O\e ( \xi ) $.
(The  brane Killing (variation) spinors   $ \e ( \xi )$   satisfy the  equations  $ 0 =  D_\mu  \e ( \xi )  = O\e ( \xi ) $.) 
We  use    the  notational conventions   as~\cite{marche08}, which are same as
those  of~\cite{mcguirk12}  except   for  an  opposite sign  convention
of the  antisymmetric symbol.  The  Pauli  matrices  $\s _a ,\  [a=1,2,3]$ act  
within    the  $ SU(2)$ group space  of the 
bi-spinors $\T  = ( \t _1,\ \t _2 )   $, 
and  the  Dirac matrices $\G _\a    $     acting in  the
representation   space   of the   spinors  $ \t _{1,2} \in 16 $    
are  given by  the   brane pull-backs  
$ \G _\a  =   \dh _\a X^M \G _M = \dh _\a  X^M  e _M ^A \G _A  $, 
where $ e _M ^A  $  are   the   spacetime  frame vectors
transforming    curved  indices  $M$ to flat indices $A$. 
The  rescaling   from string to our special Einstein  frame
for the metric  tensor
and  Dirac matrices must  be accompanied by redefinitions  of  the 
superspace  bi-spinor $\T (\xi )$  and the     supersymmetry generators  
$D _\a  $ and $ O $~\cite{marche08,marche10,mcguirk09,mcguirk12},      
\bea && g_{MN} \to  g_s ^{-1/2} e ^{\phi /2 }g_{MN} ,\   \T \to
 g_s ^{-1/8} e ^{\phi /8 } \T   , \ \G _\a  \to g_s ^{1/4} e ^{+\phi  /4}  \G _\a  , 
\ D_\a \to  g_s ^{-1/8}  e ^{\phi /8}  (D _\a + {1\over 8}
\G _\a O )   ,\  O\to  g_s ^{1/8}  e ^{-\phi /8}  O  , \label{app4EQ6} \eea   
where  $ \phi '  = \phi -  <\phi > \ \Longrightarrow \ e ^{\phi '} =
g_s ^{-1} e ^{\phi }$ and   the tacit understanding  that  omitted
quantities remain unchanged.   One can write these  relations
explicitly by  assigning the suffix  label  $ E$  to     the Einstein  frame quantities,
$  \T = e ^{\phi '/8 } \T
^E ,\    \bar \T =   \bar \T  ^E ,\ \e  =e ^{\phi ' /8 } \e ^E  , \
D_\a = e ^{\phi ' /8} ( D_\a ^E + {1\over 8} \G _\a ^E O^E ),\    
O  = e ^{-\phi ' /8}  O ^E $,  yielding  the
reverse  relation $ D_\a ^E = e ^{-\phi ' /8}( D_\a  - {1\over 8 }
\G _\a O ) $. Substituting in Eq.~(\ref{app4EQ4})   while
suppressing the suffix label $ E$,  for convenience,  gives
the  fermionic    action  in Einstein  frame   
\bea && S_{F} ({Dp / \bar Dp }) =  i   \tau _p \int d^{p+1} \xi e
^{(p-3)\phi /4}   ( - Det (g + g_s ^{1/2} e^{-\phi '/2} \calf _2 )
^{1/2}  \bar \T P_\mp ^{Dp}  (  ( \calm ^{-1} )^{ \a \b } 
 \G  _\b (D_\a    + {1\over 8} \G  _\a O ) -\ud O    ) \T , 
\cr &&  [D_\a    = \nabla _\a  + {1\over 8} \Fslash _5 \G _\a    (i \s
_2)  +{1\over 8} e^{\phi /2} (\calg ^+ _3 \G _\a  + \ud \G _\a    \calg ^+
  _3 )    
+ {1\over 4} e^{\phi }   F_{1 \a   } (i \s _2 ) ,\  O  ^E = 
\G ^ \a   \nabla _\a    \phi  - \ud e^{\phi '/2} \calg ^- _3 -
e^{\phi '} \Fslash _1 i\s _2 ,\cr &&  \calg ^\pm _3 = \Fslash _3 
\s _1 \pm e^{- \phi '} \Hslash _3 \s _3 ,\ 
\G ^M  \nabla  _M \phi = (\G ^\mu   \dh _\mu  + \G ^m   \dh _m  ) \phi
,\  F_{1 M } = \dh _M C_0 ,\  \Fslash _p = {1\over p!} F_{M_1 \cdots
  M_p} \G ^{M_1 \cdots  M_p}] .\label{sub4eq8}    \eea 
The calculations are  greatly simplified by fixing   the $\kappa $-symmetry.
For instance, the    $\kappa$-gauge   projection  choice,  defined by 
$  P ^{Dp}  _{-}  \T  = (1 -\G  \otimes \s _2 ) / 2 = 0 $,  
is solved  by    $\T = (\t , \ i  \G   \t ) $. 
For   bi-spinors of form $ \T = (a \t , \ b \t ) ^T ,\ [a^2 + b^2 =1]$   
one has   $\bar \T \T  = \bar \t  \t   , \ 
\bar \T  \s _1 \T  = 2 ab \bar \t  \t ,\ \bar \T  \s _2 \T    = 0  
  ,\    \bar \T  \s _3 \T  = (a^2 -b^2) \bar \t  \t .$ 
We    shall   specialize  below  to  the  $\kappa $-symmetry gauge
 $ a=1 ,\ b=0 \ \Longrightarrow  \ \t _1 = \t ,\   \t _2 =0.$  

We  now   focus on      the  case of space filling  $D3 $-branes   
embedded at   points of $X_6$  of the 
warped  spacetime  $ M_{10} =  M_4 \times X_6$  of  metric $  ds ^2 _{10} = h^{-1/2}  (y)  d \tilde s ^2 _{4} + h^{1/2}
(y)   d  \tilde s ^2 _{6}, $ as in Eq.~(\ref{sub2eq1}).   
Ignoring the gauge field   and setting the   Dirac matrices and $\kappa $-projector as $ \G _{D3} =  \G _{(4)} \otimes \s _2 ,\  \tilde \G _{D3}^{(0)} =  i \G _{(4)} = - \G ^{0123} ,\ P ^{D3}  _\pm =  (1 \pm \G _{(4)} \otimes \s _2 ) /2  ,$  yields,  after rescaling $ \T  \to  \sqrt
{i}\  \hat l_s ^2  \T $,   the   fermionic  Lagrangian  
\bea && L_F(D3/\bar D3 ) =  {i \tau _3 \hat l_s ^4 \over 2} 
\bar \T ( 1 \mp \G _{(4)} \otimes \s _2 ) (\calm ^{-1} )^{\mu \nu }   
\G _\mu D _\nu \T ,\cr &&  [D _\mu  = \nabla _\mu  -   {1\over 16} 
e ^{\phi '/2} \G _\mu (\Fslash _3 \s _1 + e ^{-\phi  ' } \Hslash _3     
\s _3 ) +  ( {1\over 8} \Fslash _5  \G _\mu  + {1\over 4}  
e ^{\phi  ' }   F _{1\mu } ) i \s_2 ,\  e ^{\phi '  / 2} =  g_s ^{-1/2} e ^{\phi
  / 2}   ]   \label{app4EQ9} \eea
where the   translation   from  warped to unwarped   metrics 
for Dirac matrices is realized    by  the substitutions  
$\G _\mu \to   h ^{-1/4}  \G _\mu  +  h ^{1/4} \nabla _\mu
X^{\tilde m}  \G _ {m} ,\       \G _m   \to  h ^{1/4}   \G _m   ,\
\G ^\mu  D _\mu \to h ^{1/4}   \G ^{\tilde \mu }  D _{\mu } 
+ h ^{3/4}   \G ^{\tilde \mu } \dh _{\mu }X^{\tilde m }  D
_{m}  $.   We  have relabelled  the   $ X_6$ indices $m \in (4,\cdots , 9 )$ 
to $m \in (1,\cdots , 6).$ 
The  covariant derivatives, including the    contributions from  the 
classical 5-form  profile are  defined  by 
\bea && D _\mu  = \nabla _\mu +{1\over 8 } \Fslash _5 ^{int}
\G _\mu \otimes i \s _2 =  \nabla _\mu -{1\over 4} \G _\mu \dslash 
(\ln h (y) )  P_+^{O3} , \  
[\nabla _\mu = \dh _\mu + {1\over 4} \o _{\mu }^{ab} \G_{ab} ,\ 
P_\pm ^{O3} = \ud (1  \pm  \G _{(6)} \otimes \s _2 ) ]  \cr && 
D_m = \nabla _m  +{1\over 8 } \Fslash _5 ^{int} \G _m \otimes i \s _2 =
 \nabla _m + {1\over 8} \dh _m   \ln (h (y) )
- {1\over 4} \dslash  \ln (h (y) ) \G _m   P_+^{O3},\ 
[\nabla _m =\dh _m + {1\over 4} \o _{m }^{ab} \G_{ab} ]  \label{app4EQ7}   \eea 
where  $\o _\mu ,\   \o _m $  are  the spin connections of  $ M_4 $
and $ X_6$ and $  \G _{(6)} = -i \G ^{1\cdots 6} .$ 
The  splitting   of  spacetime  dimensions  $10 \to 4+6$, breaking   
the  tangent spacetime symmetry    $SO(9,1) \to SO(3,1)\times
SO(6) $, separates the Dirac matrices  into  4- and 6-d
Clifford algebras  for which we choose the following   representations 
\bea && \bullet \   \G ^\mu = \g ^\mu \otimes I, \  
\{ \g ^\mu , \g ^\nu  \} = 2 \tilde g ^ {\mu \nu  } ,\ [\G _{(4)} = \g
_{(4)}  \otimes I, \ \g _{(4)} =  i \g ^{0123} , \ \G ^2_{(4)} = 1] .\cr && 
\bullet \   \G ^m = \g _{(4)} \otimes \g ^{m } ,\ 
\{ \g ^m , \g ^n  \}   = 2 \tilde  g ^{m  n } ,\ [\G  _{(6)}  = 
\g _{(4)} \otimes  \g  _{(6)} ,\ 
\g  _{(6)}    = - {i  \over 6!}  \e _{ijklmn} \g ^{ijklmn} 
=  -   i \g ^{1 \cdots  6}   ,\ \g _{(6)} ^2=1  ] .\label{app4EQ8} \eea  
The resulting       Lagrangian 
in the    $\kappa $-symmetry gauge $\t _1 =\t,\ \t_2 =0$, 
\bea && S_F(D3/\bar D3 ) =  {i \tau _3 \hat l_s ^4 \over 2} 
\int d^4 x \sqrt {- \tilde  g_4 } \bar \t [e ^{3A} \g ^\mu \nabla _\mu +
\ud e ^{A} \g _{(4)} \g ^m \nabla _m  \phi  
 + {1\over 16} e ^{7A +\phi /2 } ( ( 1 \mp \g _{(4)} ) \Gslash _{\mp } 
- ( 1 \pm \g _{(4)} )  \bar  \Gslash _{\mp } ) ] \t  ,  \label{subeqa0}\eea
 shows  that   in  the GKP  type     vacua  (with
$  G_\mp =0,\ \Phi _\mp  =0,\   \dh _y \tau =0$),    
all  but the kinetic energy    terms cancel out  in the  
respective  $ D3/\bar D3 $-brane      Lagrangian. 

It is useful at this point    to widen the discussion to 
flux  compactifications  on general  backgrounds 
with or without   supersymmetry, hence preserving  4-d 
$\caln =  1 $ or $0 $   supersymmetry. 
Useful   mathematical   and physical discussions are  provided
 in~\cite{grana03,granarev05,lusttsimpis04,mcguirk12}       and  especially for
 Calabi-Yau  orientifolds  in~\cite{figuero97}.  
A    convenient   generalization is    provided by 
complex manifolds $X_6$   of  $ SU(3)$  structure  
group   in which the  transition functions between patches
belong to the subgroup  $ SU(3) \subset SO(6) $. 
Such  manifolds  admit   at least a  single   globally defined 
spinor   $ \eta _-  (y) $ of fixed (negative) chirality  $ \g _{(6)}
\eta _-  = -\eta _-  $  and  constant  
normalization   $ \eta _- ^\dagger  \eta _- $, along with  the
complex conjugate   spinor of    opposite  (positive) chirality,
$\eta _+  (y) \sim \eta _- ^\star (y)$.
Given   $ \eta _-  (y)  $ and  $\eta _+  (y)  $,  one can construct on  $X_6$ 
the  globally   defined   almost-complex  structure  matrix $J_m ^n$ and the pre-symplectic structure   3-form $\O  _{mnp} $ in terms of the   bilinear matrix elements of Dirac-Clifford  matrix  algebra,    
\bea &&  \O _{mnp} = \eta _- ^\dagger \g _{mnp}   \eta _+ ,\ 
\bar \O _{mnp} = - \eta _+ ^\dagger \g _{mnp}   \eta _-  ,\  
 \o  _{mn} =  +    i \eta _+ ^\dagger  \g _{mn}   \eta _+   = 
-     i \eta _-  ^\dagger  \g _{mn}   \eta _-   , \  
\cr  &&    [\eta  _\mp ^\dagger \eta  _\mp  =1,\ \g _{(6)}  \eta _\mp  =  \mp
  \eta _\mp ,\ \eta _+  (y) = B _6 ^\star \eta _- ^\star  ,\ B_6  B_6 ^\star
=1,\  B_6 ^\star \g   ^m B_6  =(\g ^{m } )^\star  ,\ B_{10} = B_4 B_6]  
\label{app4EQ10}  \eea 
where  the real 2-form $ \o _{mn} =  \tilde g_{mp} J^p _n  ,\ [(J^2) _m^n=  J_m   ^p J_p^n  = - \d _{mn} ] $     and the imaginary complex 3-form  $\O _{mnp} $   obey,  via the Fierz identities, $ \G \wedge \O  =0,  \ \O \wedge \bar \O = (4i/3)   \o  ^3, \ [ \star _6 \O = i\O ,\   \star _6 \bar \O = -i \bar \O ] 
$ and   the operators  $ (\Pi ^\mp  ) ^n_m =    {\d ^n _m \mp  i   J^n _m \over 2}  $ project  onto the holomorphic/anti-holomorphic  parts 
of $\O  $, $ (\Pi ^-  ) ^n_m \O _{npq} =\O _{mpq}  ,\
(\Pi ^+  ) ^n_m  \bar \O _{npq} = \bar \O _{mpq}  .$
The  various spinor  matrix  elements encountered  in
computations   involve products  of    $\O   _{mnp} $   with  Dirac
operators $\g ^m $ and projection    operators  $\Pi ^\mp  $
that   can be evaluated  by   means of the useful
identities~\cite{lusttsimpis04,mcguirk12}   
\bea && \bullet \    (\Pi ^-)_m^n  \bar  \O _{npq} =0, \ (\Pi ^+)_m^n
\bar  \O _{npq} = \bar  \O  _{mpq} , \ 
(\Pi ^-)_m^n    \O _{npq} =  \O  _{mpq} , \  (\Pi ^+)_m^n
\O _{npq} =  0  .    \cr && 
 \bullet \    \eta ^\dagger _- \g ^{pq} \g ^n \eta _+ \bar \O _{mpq} =
 2 (\Pi ^+   
)^ n _m \vert \O   \vert ^2 ,\ \tilde g ^{qt} \O ^{prs} \bar \O _{mpq}
 \bar \O _{nrs}     = 2 \vert \O   \vert ^2 \bar \O _{mn} ^{\ \ \ t} ,
 \cr &&    
\bullet \    \bar  \O _{mpq}  \O _{nrs } \eta _- ^\dagger \g ^{pq} 
\g ^{rs}  \eta _- =  2  \vert \O \vert ^2  \Pi ^{-}  _{mn } ,\
 \bar \O _{mpq}  \bar   \O _{nrs }  \eta _- ^\dagger  
\g ^{pq} \g _l\g ^{rs} \eta _+ = 
{ (-k +2 ) \over 3} \vert \O \vert ^2    \bar   \O _{mnl}    ,  
\cr &&  [\vert \O \vert ^2  =  {1\over  3! } \bar \O _{mpq} \O
  ^{mpq}    = 8,\   (\Pi ^{\pm } )^n _{m} =    {\d ^n _m \pm  i   J^n
    _m \over 2} ,   \ \Pi ^{\pm }  _{mn } =  {\tilde g _{mn} \pm  i  \o _{mn } \over 2} ]. \label{ideqs1}  \eea 
In order to  comply with  the   notational conventions  of~\cite{mcguirk12}, we  have been led    to use  in Eq.~(\ref{app4EQ10})   the  opposite sign   convention for $ \o _{mn}$  relative to that of~\cite{lusttsimpis04},  implying   that our $ \Pi ^\pm _{mn}  $ identify to  their $ \Pi ^\mp _{mn}  $.  In  Calabi-Yau manifolds, the  $SU(3)$   structure  group    coincides with the holonomy group, and  the   complex conjugate  pair of
opposite  (negative/positive) chirality spinor $\eta _-,\  \eta _+  (y) \sim \eta _- ^\star (y)$,    are  covariantly  constant,
$ \tilde \nabla _m \eta _\mp   (y)   = 0$, and similarly for   $ J $ and $ \O $.  

The  dimensional reduction $ 10\to 4+6$  is  realized  by decomposing the   spinor  fields   $\t  (X) \sim 16 $    on  bases of  
chiral spinor  fields  of $ M_4 $ and $X_6$.  
A convenient  choice  for the  internal     space spinors  
$\eta _\mp  (y) $  involves the basis  of fermionic modulino  
and  gaugino Weyl spinor   fields 
$ \eta ^{g, m} (y) ,\ [m=1,\cdots , 6]$ in   the  representations 
$  (3 + 1 )   + (\bar 3 +1)  $    of  $SU(3)  $, 
with  the coefficient  fields $ \psi ^{g, m} (x) $  corresponding 
to  superpartners of the  bosonic fields  $ A_\mu ,\ y^m  $. 
Upon   matching   the  pull-backs of the 10-d supersymmetry  transformations 
to  the  standard  form   in  4-d supersymmetry  field
theories~\cite{grana02,grana03,marche08,mcguirk12}, one obtains 
the  decomposition  in the chiral basis  
representation of  the   $ SO(3,1)$  Dirac matrices 
\bea && \t  (X )   =  e ^{ - 3 A /2  -  \phi /2 }
( {0\choose \psi^{g }   } \otimes \eta _{- }  -  { i \bar \psi ^g 
\choose      0 } \otimes \eta _+ )   + 
{e ^{ - 3 A  /2   } \over 4 \vert \O \vert }    ( {0\choose
\psi ^m } \otimes \O _{mnp} \g ^{np} \eta _{-}  - 
 {i\bar \psi ^{m  } \choose 0 } \otimes \bar \O  
_{mnp} \g ^{np} \eta _+    ) ,  \cr &&   
[\g ^\mu = \pmatrix {0 &   - \bar \s ^\mu   \cr \s ^\mu  & 0},\  
\g ^0 =\pmatrix {0 & -1 \cr 1 & 0},\  
\g ^{k } =\pmatrix {0 &  \s ^i  \cr \s ^i  & 0},\  
\g _{(4)} = \pmatrix {1 & 0 \cr 0  & -1} ] . \label{sub4eq13}  \eea 
Substituting these   formulas   for $\t (X)$ 
into the   brane action  
and rescaling  the      fermionic and bosonic  moduli  fields 
$\psi ^{m ,g } \to  \sqrt {i} \hat l_s ^2 \psi ^{m
  ,g }  ,\   y^m  \to \hat l_s ^2 \varphi ^m  $  and  the  Abelian 
$U(1)$ gauge theory  field strength  $F_{\mu \nu } \to 2\pi  f_{\mu
  \nu }  $, yields  the   single $D3/\bar D3$-brane Lagrangian~\cite{mcguirk12} 
\bea &&   L (D3/ \bar D3) = [ -\tau _3  (h^{-1} - \a )   
- {1 \over 4 g^2} e ^{-\phi }   f _{\mu \nu } f ^{\mu \nu } \mp 
{\t \over 16  \pi ^2 } f _{\mu \nu } \tilde f ^{\mu \nu } 
-  {\pi \over  g_s} \tilde   g^{\mu \nu }  \tilde g_{mn} \dh _\mu 
\varphi ^m \dh  _\nu \varphi ^n 
\cr && -  {\pi \over  g_s}  (i   (\tilde  g_{mn}  - i \o _{mn} )
\tilde g^{\mu \nu } \bar \psi ^m  \bar \s _\mu \dh _\nu \psi ^n  
+  2  i   e ^{-\phi }  \tilde g^{\mu \nu } \bar \psi ^g   \bar \s _\mu \dh _\nu
\psi ^g  ) - (m _{g}   \psi ^g  \psi  ^g + \ud m^F _{m}
\psi  ^g \psi  ^m +  m^F _{mn} \psi  ^m \psi  ^n   + H.\ c. ) ] 
\cr &&  - i (h _{mn}  \psi ^g \psi  ^m \varphi  ^n  + h _{mnp} \psi  ^m \psi  ^n
\phi ^p + H.\ c. )   +\cdots  ,\cr &&  
[h _{mn}  = {2\pi i \vert \O \vert \over g_s}
  \o_{mn} ,\   h _{mnp} = {\pi e^{\phi /2 } \over g_s }  \O _{mnp},\ 
{1 \over g^{2} } = \tau _3 \hat l_s^4  e ^{-\phi }  = {2  \pi \over g_s
}   e ^{-\phi }  ,\  \t = 2\pi C_0 , \cr && 
\pmatrix{ m _{g} \cr  m^F _{m} \cr m^F _{mn} } _{ D3}   = 
+ {\pi \over 8 g_s} e ^{4 A}  
\pmatrix{e ^{-\phi /2}   G_{-  }   \cdot \bar \O   \cr  
- {i \over 4} \vert   \O  \vert (\tilde  g_{ml} + i \o _{ml}) 
  (G_{-} )_{\ \ pq } ^{l }  \o ^{pq}    \cr   
\ud e  ^{\phi /2} ( \tilde  g_{l (m  } - i \o _{l (m   } )     
\O _{ n ) pq} )   G _- ^{lpq}  }   ] .    \label{sub4eq9}   \eea
The  3-fluxes    contribute the fermion number  non-conserving 
mass terms   for spinors  with  mass parameters 
$m_{g },\ m_F^m,\ m_F^{mn}$.
 We have quoted above the mass parameters for $ D3$-branes  only but  note
 that  those  for $\bar D3$-branes are obtained by replacing  
$G_{-  } \to  - \bar G_{+} $.    
We have also  quoted the kinetic terms and  Yukawa   cubic couplings 
of scalar and spinor fields,  but     omitted   
the    scalar  fields  couplings    of $ O(\phi ^3 ) ,\  O(\phi ^4) $. 

In  supersymmetric vacua    with     $ G_-  =0 ,\ 
\Phi _- =0 $,       the   fermionic mass    terms,  with the  exception    
of the  classical  and   kinetic energy     parts,   cancel   out 
in the $D3$-brane  action and are  doubled in the $\bar D3$-brane
action.   (An analogous   cancellation  holds  in the  
negative flux conjugate   vacua  with $ G_+  =0$.)   
If the complex structure   of the underlying   internal 
manifold  $X_6$    is  not  deformed by  the pull-back
transformations to      $ D3/\bar D3$-branes, 
the    spatial  coordinate  indices   $m ,\ n \in (1,\cdots , 6) $
are simply   expressed   into   linear      combinations of      
holomorphic  and  anti-holomorphic coordinate indices  
$ (i,\ \bar i )  \in (1, 2 , 3) $     of the  familiar   
diagonal basis    for   the complex
structure matrix  $  J ^i _j =  i  \d _{ij} ,\  J ^{\bar i} _{\bar j}
=  i  \d _{ij}  $.   The  cubic and quartic order  terms in scalar  fields  
are deduced   from      the  cubic superpotential 
$ W _{\caln =4}  $    inherited from  the underlying  $\caln =4$
supersymmetric    theory    and from D-terms. 
Part of the    bilinear  terms $ O( (\psi ^m )^2 )$ in the   fermion
triplet fields      are  also   absorbed inside  the  total  superpotential 
$  W  =  W _{\caln =4} + \ud  m^F_{ij}   \Phi ^i    \Phi ^j ,\ 
[W _{\caln =4}= - i {g \over 3!} \O _{ijk}  Tr (\Phi ^i    \Phi ^j
\Phi ^k ) ] . $  There also  occur non-holomorphic     cubic  scalar
couplings   $ c_{ij\bar k} Tr  (\varphi ^i \varphi ^j \bar  \varphi
^k)  $  which    turn out to   have  vanishing contributions~\cite{mcguirk12}. 

If  the       modification of  the pulled-back   
supercharges  $  Q_\a ^{GKP} $      to the   brane is
ignored,  one can classify the brane soft  supersymmetry  breaking couplings  induced  by 3-fluxes    
by the  Hodge  type   with respect to the    bulk   manifold complex
structure.   Recall that the  3-fluxes  preserving  the bulk 
$\caln =1 $ supersymmetry  belong   to the 
 ISD Hodge 3-forms $ G_{(0,3)} ,\ G^P _{(2,1)} ,\ G^{NP} _{(1,2)} ,\ [ G_-
  =0]$   while  those breaking it   belong   to the Hodge  IASD   3-forms  
 $ G_{(3,0)},\ G_{(1,2)}^P ,\  G^{NP} _{(2,1)}
,\ [ G_+  =0]$, where  the primitivity  property  of    mixed 
type 3-forms, signalled   by the   suffix $P$, 
means    a   vanishing   contraction with  the  Kahler  2-form, namely, $  G ^{P \ j} _{ij }  \equiv  
G ^P_{ij \bar k} \o ^{j\bar k} = 0.$   It follows 
  that      the  gaugino,   gaugino-matter  and pure matter
fermion masses    in $D3$-branes  originate from the following three
types of 3-fluxes   
\bea &&  m_g \propto G_{3,0} ,\ m^F _i   \sim \o ^{j\bar k} G
_{ij\bar k}  \propto G_{2,1} ^{NP}   ,\  m^F _{ij}    \propto
G_{1,2} ^{P}  .\eea    
For  $ G_-=0$,    all the fermion mass terms    vanish and     
half of the preserved $\caln = 1 $    bulk supersymmetry  generators   are
realized  linearly   by  half of the    fermions      left out  after     
$\kappa $-gauge fixing. 
The      contributions from   3-fluxes   to the soft supersymmetry
breaking  terms  on  $D3$- and  $D7$-branes    in toroidal orientifold
compactifications    with  $O3/O7$--planes    
are reviewed   within the 10-d  approach 
in~\cite{camara03,camara04}  and within  the  4-d  effective string
theory effective action    approach  in~\cite{blumber06}.    

The  bilinear  fermionic  mass terms   in  $\bar D3$-branes
originate from  the 3-flux types, 
$ m_g  \propto  G_{0,3 } ,\  m^F_i \sim
G_{1,2}^{NP} ,\ m^F _{ij}    \propto   G_{2,1} ^{P}  $. 
In  GKP    vacua  with    $ G^P _{(2,1)}
\ne 0 ,\  G_{0,3 } =0  $,     the introduction of a 
$\bar D3$-brane,      preserving the      orthogonal combination to  
$ Q_\a ^{GKP} $,    leaves no     invariant supercharges 
and, from  $ m_g =0, \ m_i^F  = 0  ,\ m^F _{ij}  \ne 0   $, 
enforces  the presence of  massless   gauginos    
and   massive (decoupled)  matter  fermions. 
This  means  that the    supersymmetry breaking on  $ \bar D3$  
is  spontaneous   and      realized non-linearly 
by a  nilpotent gaugino-Goldstino  multiplet superfield 
described (in terms of a redefined $\psi ^g $)   by   a 
Volkov-Akulov  type action.  The    
spontaneous supersymmetry breakings    that arise     on  $D$-branes 
for various    $ \kappa $-gauge   choices 
compatible      with   different $\caln = 1 $ orientifold
projections   are  discussed in~\cite{bergkalyen15,kalverwrase16}. 

The      fermion       fields on $D3$-branes    also    possess
couplings   of linear order  in the $U(1)$  gauge  field
strength. These      are  produced  through         the   
effective metric  tensor       
$\calm _{\a   \b }  = g_{\a   \b } +  g_s ^{1/2} e ^{-\phi /2 } 
f_{\a   \b } \G _{(10)}  \otimes \s _3$ and    the  background
axio-dilaton  field term.  The  resulting    magnetic dipole
couplings    of dimension $d= 6$    
contribute in the $D3$-brane Lagrangian  as 
\bea &&  L (D3) =- i {2 \pi   \over g_s} \hat l_s ^2    (f +i\star_4
f )^{\a \b }  [ \ud   e ^{ - 2 A }  (e ^{-\phi }
  \bar \psi ^g \bar \s _\b  \dh _\a   \psi ^g  +
  e ^{-\phi /2}  \Pi ^- _{nm}   \bar \psi ^m \bar \s _\b  \dh  _\a  \psi ^n   )  
\cr && +   {1  \over 32} e ^{\phi /2}{ \dh _p  \bar \tau }
 (\bar \O _{mn} ^{\ \ \  p}  \bar \psi ^m \bar \s _\b  \s _\a   
\bar \psi ^n   + \vert  \O \vert  (\Pi ^{+} )^{p} _m  
\bar \psi ^m \bar \s _\b  \s _\a \bar  \psi ^g  )  ]  + H.\ c.
,  \label{app4EQ13}    \eea 
where we made    use of  the identities   in Eq.~(\ref{ideqs1}).
The  first pair of   couplings   inside  square brackets  is enhanced by the   warp profile    but  suppressed by the fermion  field derivative, while   
the second pair of   couplings,  considered initially  in~\cite{grana02},  
are independent of   warping   but require an  axio-dilaton 
of  non-constant  profile. The   dependence on the linear
combination  $(f  \pm i \star _4  f ) $ is   expected  from
the $D3$-brane  action invariance under  electromagnetic
duality transformations.

We     note      finally  that the  Weyl spinor  formalism  can be 
translated  to   a Dirac  spinor formalism
 by  means  of the identities in  the   chiral basis  of 
Dirac  matrices,  
\bea &&   { (- i \bar \psi ^{m } \bar
\s ^\mu \dh _\mu \psi ^n   - i  \psi ^{m } \s ^\mu \dh _\mu  \bar \psi
^n  ) \choose   - (\psi ^m \psi ^n + \bar \psi ^m \bar \psi ^n)  } 
=  {-i\bar \Psi  ^{m} \g ^\mu   \dh _\mu \Psi ^n \choose 
- \bar \Psi ^m \Psi ^n  } ,\  
[\Psi = {i  (\bar \psi ) ^{\dot \a }  \choose \psi _{\a } },\ 
\bar \Psi = { \bar \psi ^{\dot \a } \choose \  i \psi _{\a } } ]      \eea  
where we       use  same   notations  as~\cite{wessbagger92}.   
The   kinetic  Lagrangian for   modulinos becomes  
\bea &&   L _{kin} (D3) = - {i\over 2}  (\tilde \calk _{mn} \bar \psi ^{m } \bar
\s ^\mu \dh _\mu \psi ^n +   \tilde \calk ^\star _{mn}   
\psi ^{m } \bar  \s ^\mu \dh _\mu \bar \psi ^n )  \to {i\over 2} 
\bar \Psi ^m \g ^\mu \dh _\mu \Psi ^n  ,\cr &&   
[  ( \Psi ^m  )^T  = ({i   \calk _{mn} (\bar \psi  ^ n ) ^{\dot \a }  
,  \ \psi  ^m _{\a } } )  ,\ \tilde \calk _{mn} = 
{\pi \over g_s} ( \tilde g _{mn} -  i \o _{mn} )  ] 
, \label{app4EQ11} \eea 
with a corresponding   formula   holding for  gauginos  in terms 
of   the  4-component Majorana  spinor field 
$\Psi  ^g = (i \bar \psi  ^g , \ \psi ^g)^T .$ 
  It is important  to distinguish the  present notation 
for Dirac  spinors     involving  complex conjugate pairs of  Weyl
spinors,   from  the  standard  notation    for quark and lepton
Dirac     spinor fields   built  from pairs of  (left-right) quark and lepton  Weyl  spinors of  opposite $ \mp 1 $  chiralities
  $ \Psi =  (i \psi _R^{\dot \a }  , \ \psi _{L   \a }  )^T  ,\   \bar \Psi = 
  ( \bar \psi _L^{\dot \a }  , \  i \psi _{R   \a }  )$
that belong    typically to  different  gauge  group   representations. 

\section{General  properties of warped  conifold} 
\label{app2} 

We  provide   in this appendix   a brief  review of the  
deformed conifold.   The  discussion    starts  with   some 
general algebraic  properties,  proceeds  with   the 
harmonic   analysis on the conifold   base $ T^{1,1}$   and
ends with   the dimensional reduction of the 10-d 
supergravity  multiplet fields. 

\subsection{Algebraic  description} 
\label{suc1app2} 
 
The deformed conifold $\calc _6$~\cite{candelas90}  is a special case  
of  the so-called Stenzel spaces $\calc _{2d -2 }  $~\cite{cveticpope00}, 
defined   algebraically    as  the loci of    the 
quadratic equation,   
$ \sum _{a=1} ^{d} w _a ^2 =\e ^2  ,\ [w_a \in C,\  \e ^2 \in
R ^+]   $.  The  isometric   mappings  for these  manifolds     
conist of  real   rotations of  the complex   variables $ w_a$
and    sign   reversals  $ w_a \to \pm  w_a  $,
generating the  symmetry group   
$G = SO(d) \times Z _2$. (At $\e =0$,  the $Z _2$  is enhanced 
to  an      $ U(1)$ symmetry  acting as $ w ^a\to e ^{i\a } w^a $.)  
The  spaces $\calc _{2d -2 }  $ of   complex dimension  $(d-1)$   
are  parameterized    by  a   real radial  coordinate 
$\tau \in [0,\infty ] $,     accompanied   by either  $d$  pairs of  complex  
conjugate  coordinates $   (y_a  ,\ \bar y_a )\in C^{2d} $  
or   $d$  pairs of  real coordinates   $ (u_a, \ v_a ) \in R
^{2d}   $,  
\bea &&  w_a = {\e \over \sqrt 2 }  ( e^{\tau / 2} y_a  +  e^{-\tau / 2}
\bar y_a  )  = \e (u_a \cosh (\tau /2) + i v_a \sinh (\tau /2)
) ,\   [y _a = {u_a +i v_a  \over \sqrt 2} ] \label{app2eq1}   \eea  
subject to the   respective constraint  equations 
\bea && \sum _a y_a ^2 = 0,\ \sum _a \vert y_a \vert  ^2 =
1 \ \Longrightarrow  \  \sum _a u_a ^2 = \sum _a v_a ^2=1, \ \sum _a u_a v_a =0
.  \label{app2eqI1} \eea 
The     spaces  $\calc _{2d -2 }  $  are  non-compact   Kahler manifolds 
whose  metrics    are  derived  from    
radially symmetric  Kahler  potentials  $ \calf (\tau )$,   
\bea && ds ^2  _{2d -2 }  = g_{a\bar b}
d w^a  d \bar w ^{b} ,\ [g_{a\bar b} = { \dh ^2  \calf
(\tau )  \over   \dh w_a \dh \bar w_{\bar b}  } ] .  \label{app2eq2}
\eea   
The Ricci  flatness  condition   is then  expressed  by a solvable second-order 
differential  equation  for  $  \calf (\tau ) $. 
The  fixed $\tau $ sections  (bases)  of $\calc _{2d -2 }  $, 
spanned  by  the  complex variables $ y_a$  subject to the
conditions  in Eq.~(\ref{app2eqI1}), 
are  homogeneous Kahler (Einstein-Sasaki)  manifolds given by 
coset spaces  $G /H = SO(d) / SO(d-2)$, 
corresponding to the so-called    Stiefel manifolds  $ V_{d,2}$. 
The   explicit construction of   the  cohomology 2- and   3-forms 
on $\calc _{2d-2}$ is   presented in~\cite{cveticpope00}. 
 
We  focus hereafter      on the  deformed conifold case, $ d=4 $. 
The  geometry  of $\calc _6$   is that of a real  cone with  
radial variable $ \tau \in [0, \infty ]$  and  base manifold   
$V_{4,2}= T^{1,1} =   G/H =  SO(4) / SO(2) = SU(2) _L\times SU(2)_R /
U(1)_H $.  The  Ricci  flatness condition $ R_ {a\bar b} =0 $   is
solved   by   a radial  Kahler potential      whose 
derivative  is  given by 
$ \calf '(\tau )  = \e ^{4/3} (\cosh \tau  \sinh \tau -\tau ) ^{1/3} .   $
The   algebraic equation   of the base manifold $ V_{4,2} $ 
is   expressed in matrix form by  grouping the $ w_a $ into  a  $ 2
\times 2 $ complex matrix $W$   satisfying  the two  conditions 
\bea && \sum _{a=1}^4 w_a ^2 =  -  Det (W) = \e ^2 ,\ 
\sum _{a=1}^4 \vert w_a \vert ^2 = \ud Tr (W^\dagger W )  = 
\rho ^2 \equiv ({2 \over 3})^{3/2} r^{3} \equiv  \e  ^2 \cosh  \tau  ,
\cr &&  [W= \pmatrix{ w^3 + i w^4 & w^1-i w^2 \cr
w^1+i w^2 & -w^3 + i w^4 } ] .    \label{app2eq6}  \eea 
The  metric at large radial distances  $\tau \to \infty $ 
takes  the  same   standard conic form   $ ds ^2 (\calc _6 ) = dr^2 +
r^2 ds ^2 ( T^{1,1})  $  as in the   undeformed     
conifold limit $\e =0$,  where  the conic radial  variable is given by 
   $ r ^2 =  {3 \over  2 ^{5/3} }   \e ^{4/3} e ^{2\tau /3 } $.  
The isometry  group  $SO(4) \sim  SU(2) _L\times SU(2)_R$  of $\calc
_6$   is  realized  by    the left and right   unitary
transformations $ W\to g_L W g_R ^\dagger ,\ [g_{L,R} \in SU(2)
  _{L,R}]$    that leave the determinant and trace in
Eq.~(\ref{app2eq6}) invariant.   The   coset space   structure
$ T^{1,1} \sim S^3 \times S^3 / U(1)_H  \sim SU(2)  \times SU(2) /
U(1)_H  $ of  the fixed $\tau $ slices    follows  from 
the   multiplicative  decomposition of  $W$~\cite{herzog01} 
 \bea && W= L_1 W_0 L_2  ^\dagger  :\  
[W_0 =\pmatrix{ 0 & \e e ^{\tau /2} \cr \e e ^{-\tau /2} & 0 } ,\ 
L_i = \pmatrix{\cos (\t _i /2) e ^{i (\varphi _i + \phi
_i ) / 2} & - \sin (\t _i /2) e ^{-i (\varphi _i - \phi _i ) / 2} \cr
\sin (\t _i /2) e ^{i (\varphi _i - \phi _i ) / 2} & \cos (\t _i /2) e
^{- i (\varphi _i + \phi _i ) / 2} } ] \label{app2eq8} \eea 
where    the  $SU(2)$   group  matrices  $ L_i$   
are  parameterized by  the pair of three  $ S^3_i $  Euler angles 
$ \t _i ,\ \phi _i,\ \varphi _i ,\ [i=1,2] $  while  the   
$U(1)_H$  quotient  imposing    the equivalence $ \varphi _1 \sim \varphi _2
$  arises from  the   fact that   $ W$   depends  only 
on $\varphi _1 + \varphi _2 = \psi $.  
The gauge choice $ \varphi _1 = \varphi _2 = \psi / 2$    
reduces   the   coordinate   variables to    the 5  angles  
$ [ \t _i \in [0, \pi ],\ \phi _i\in [0, 2\pi ],\ \psi
\in [0, 4\pi ]  ]$  for $ i=1,2 $ with the   isometry group factors  
$ SU(2)_1 \times SU(2)_2 \supset  SO(4)   $ rotating  the  spherical  angles  
$( \t _1, \phi _1 )$ and $( \t _2, \phi _2 )$
and  the     $Z_2 $   factor  rotating   $ \psi \to  \psi +2\pi   $.   
 The  metric  of the  compact  base manifold    is    expressed in  
diagonal  form in  the linearly related bases of   invariant  1-forms
$ e^ {(a)}  $   and $ g ^ {(a)}  ,\ [a=1,\cdots , 5]$  
\bea &&  ds ^2 ( T^{1,1} ) = 
 {1\over 9} e^{(5)2} + {1\over 6} \sum _{i=1} ^4 e^{(i) 2}  
= {1\over 9} g^{(5)2} + {1\over 6} \sum _{i=1} ^4 g^{(i) 2} 
\ = {1\over 9} (d\psi + \sum _{a=1}^2\cos \t _i d\phi _i )^2 + 
{1\over 6} \sum _{a=1}^2 (d\t ^2_i + \sin ^2\t _i d\phi _i ^2 ) ,       
\cr &&  [g^{(1,3)} = {1\over \sqrt 2} (e^{(1)}\mp e^{(3)} ),\ 
g^{(2,4)} = {1\over \sqrt 2} (e^{(2)}\mp e^{(4)} ),\ g^{(5)} = e^{(5)} 
,\cr &&  {e^{(1)} \choose e^{(2)} }  = { - \sin \t _1 d \phi _1
  \choose  d\t _1 },\  
{e^{(3)}  \choose e^{(4)} } = \pmatrix{ \cos \psi  & -\sin \psi  \cr 
\sin \psi & \cos \psi   }  {\sin \t _2 d \phi _2  \choose d\t _2 } ,\ 
  e^{(5)} =  d\psi + \sum _{i=1}^2\cos \t _i d\phi _i
] .  \label{app2eq9}  \eea  
The    cohomology 2- and   3-forms  of  $\calc _6 $  
are    discussed in~\cite{papado00}, the supersymmetry  preserving 
embeddings of $3$-branes  in $\calc _6 $ are discussed in~\cite{minasian99} 
and  the  scalar Laplace operator  is discussed in~\cite{krishtein08,pufu10}.
The moduli   $\e $ and $U$,  describing  the 
complex structure and Kahler    deformations 
of the supergravity background  for   the (deformed and resolved)
conifold~\cite{candelas90},  correspond   
to    the  gaugino condensate and  baryonic   operator   VEV
parameters  $ <\l \l > $  and $\calu $, 
parameterizing the  broken chiral symmetry $ U(1)_r $  and  baryon symmetry $ U(1)_b$    (mesonic  and  baryonic)
branches  of the  dual (Klebanov-Witten)  gauge theory moduli space. 
The  correspondence   relations  are~\cite{dymarberg05} 
$ \e ^{2/3} \sim \a ' S ^{1/3} \sim \a '(<\l \l > / N)^{1/3} $  
and   $  \calu \propto   Tr   (\vert A _i \vert ^2 - \vert B _i\vert
^2) $. 

\subsection{Harmonic   analysis  on singular conifold} 
\label{suc2app2}
 
 Harmonic analysis deals with  the formal properties of  Hilbert
 spaces   of  square integrable  (normalizable)   functions on
 manifolds.   
The  aim  is    to   represent  smooth classical
fields  on  a   given  curved   manifold   in terms of series
decompositions  on orthonormal bases of    a vector space  of
functions  living on the manifold.  
For  scalar,  tensorial or  spinorial
fields  on curved manifolds, the  functional   vector spaces   are spanned   by  eigenfunctions of the  diffeomorphism invariant Laplace-Beltrami (or Hodge-de Rham), Lichnerowicz and 
Dirac operators,   generalizing    the   scalar Laplacian.     
These   are  built from    quadratic  products  of  covariant
derivatives that  reduce in special  cases
to squares  of the  differential operator $d  $ and  its adjoint $\d $.
The   construction    for group   manifolds  is   detailed   in the 
monograph~\cite{vilenkinharm68}   and  that for   
coset spaces  in~\cite{pilch84,caste99}. 
The explicit decomposition   on Hyperspherical harmonics   of  $S^N$
is presented in~\cite{higuchi87}   and that for $ S^3$ is detailed  in~\cite{gerlachgupta73}. 

We start with a    brief review  of  coset spaces with  application
to the  undeformed conifold in  mind.   Recall  that
the elements  of the  coset space $ K= G/H$,  quotient of 
the groups $G,\ H \subset G$,  are arranged into  
equivalence classes represented by elements   $ L(\T ) $,
embedded  in $G $, that transform     under  left   multiplication by
$G $  as     
\bea && g  L(\T ) = L(\T' ) h (\T , g),\ [g\in G,\   h\in H    ,\ 
L(\T ) =  e ^{ \T ^a  T _a } \in G ,\ a=1,\cdots , d _K 
\equiv dim (K) = d_G - d_H ]   \eea  
where  $\T   =  \{\T ^a \} $  denote   local  coordinates of $K$ 
and  $ T_a ,\ [a =1,\cdots , d_K= d_G - d_H] $ denote the subset 
of  the Lie algebra generators 
$ T_A,\ [A =1,\cdots , d_G \equiv dim   (G) ]$  of the group $G$
that   are not  part  of  the Lie algebra generators $ T_H ,\ [H =1,\cdots , d_H \equiv dim   (H) ]  $ of its subgroup $H$.
The geometrical description is based on   the decomposition
of the left-invariant   1-form  $L^{-1} d L $ on a  basis of
invariant 1-forms $ \o ^A $  on the   $G$ group space satisfying   the  Maurer-Cartan equations,
\bea && d \o ^A + \ud C^A _{BC} \o ^B \o ^C =0,\
     [L^{-1} d L = \o ^A  T_A = \o ^a  T_a + \o^H T_H  ,\
[T_B, T_C]= C ^A _{BC} T_A ] \eea 
where  $ C^A _{BC} $ are the   group  $G$ structure constants, the  1-forms  
$  \o ^a  $      identify   to  the reference frame  of
$K$  and    the 1-form $ \o ^H =  \o ^H _a T^a $ to  the    spin connection    of $H$.  The (torsion free) spin  connection and  the curvature
1- and 2-form  fields  $ B  _{ab} $
and $  R _{ab}$     of  $K$  are   defined  via the  Cartan 
structure  equations,   
$0= \calT ^a \equiv d \o ^a - B ^{ab} \o _b , \  R_{ab}= d   B _{ab} -
B ^a_c B^{cb}$  with the covariant  derivative  $ D= d +  B $.    
The basic (scalar) harmonics  $Y^{G} _H \sim
(L^{-1} (\T ))^ {G } _H    $    are   built  from  the 
linear (matrix) representations $ D^ {(\nu )}  (g)  $   of $G$,     
$ Y^{ (\nu )   M } _ { (h) , q  } (\T )
= ( D ^ {(\nu )} _{(h)} (L^{-1} (\T ) ) ) ^ M _{q}  , $
where  $M$  label   the magnetic quantum numbers     of     
the representation $\nu $ of $G$   and  $q$
label the components of the   irreducible representations  $h $ of 
$H$ into  which  $\nu  $  splits   via  the symmetry breaking $ G\to
H$.   The   ranges  of variation     of  $q $  are  determined  by  the
embedding of  $H$  in     the Lie  algebra of   
the   tangent space group $ SO(d_K)$ of $K$. 
The     set of   harmonic functions  $Y^{\nu  , M }  $ for variable
 $\nu , \  M$     constitute   a basis for  the vector space of
 smooth square integrable  functions on  $ G/H$.  
 The orthonormalization and completeness relations
 are defined in terms of the invariant scalar product in $G/H$  as 
 \bea &&
\int d^5 \T \sqrt {\tilde g_{X_5}} Y^{\nu  , M \star } (\T ) Y^{\nu ' , M' }
 (\T ) = \d _{\nu \nu '} \d _{MM'} ,\ \sum _{\nu ,M} Y^{\nu , M  \star
} (\T ) Y^{\nu  ,   M } (\T ') = {\d ^{(5)}(\T -\T ') \over \sqrt
{\tilde g_{X_5}} } . \label{app2eq11} \eea 
The harmonic functions   are   eigenfunctions of
the   Laplace  (differential) operator on  $G/H$ 
whose eigenvalues  are     given  by   the 
difference    of quadratic  Casimir operators~\cite{pilch84},  
$ \nabla ^2  _{G/H} = C_2 ^G (\nu ) - C_2 ^H (  h ).$

The parameterization   of 
the  undeformed  conifold ($\e =0 $)    to which we specialize
hereafter,   is  realized   in terms of  a  radial variable $r$  that  
separates from  the  angle variables $\T ^a$  of the 
5-d compact base  manifold,   given by  the  coset space  
$T^{1,1} = SU(2) \times SU(2) / U(1)_H \equiv G/H $.  
The   harmonic    analysis on $T^{1,1}$  is 
developed by means of  either   differential~\cite{gubser98}  or group
theoretical  methods~\cite{ceresole99,ceresoII99}. 
The    Lie algebra generators of   $G:\ 
T_A = (T_i,\   \tilde T_i) \in g = su(2) \oplus su(2)  ,\ [i=1,2, 3]$
split into  the coset  algebra generators 
$ T_a = [T_{i } ,\tilde T_{s} ,  \ T_5  = T_3 -\tilde T _3 ]
,\ [a= 1,\cdots , 5, \ i=1,2,\ s= 1,2] $  
 and   the      generator  of  $
U(1)_H:    T_H = T_3 +\tilde  T_3 $.
The reference  frame of $K$ and the spin connection of $H$
are  defined   through the   Lie algebra decomposition  of the  left
invariant  1-form  of the coset space   
\bea &&    L^{-1} d L = \o ^i T_i +
\tilde \o ^s \tilde T_s + \o ^5 T_5 +  \o ^H T_H  , \ 
[L(\T ) = e ^{\T ^i T _i}   e ^{\T ^s T _s}    e ^{\T ^5 T _5}  ,\ 
T_H = C_H^{ab} T_{ab},\  \o ^H =  \o ^H _a T_a  ] \eea  
where $ T_{ab} ,\ [a<    b =1,\cdots , 5] $ denote the Lie algebra
generators   of the tangent   space  group $SO(5)$ of $K$ and   
$T_H \in U(1)_H$  is  represented above by  its   embedding in  $SO(5)$. 
It is important to       distinguish the  generators
$T_{ab}  $,   acting on tensors of  $SO(5)$  as 
matrix  transformations,    $ V_{c d \cdots }  \to (T_{ab} ) _{c} ^{c'}
V_{c' d   \cdots } + \cdots  $,   from   the  operators
$ T_a $   
 acting on the coset space
$ K= SU(2) _1\times SU(2)_2  / U_H(1) $. 
The covariant derivative $ D   $ in $K$   splits  additively into the 
$U(1)_H$ spin    connection and $SO(5)$ group  parts,
\bea && D   \equiv d + B^{ab}   \hat  T_{ab}  = d +  \o ^H T_H + M^{ab}   \hat  T_{ab}  \equiv D^H  +  M^{ab}   \hat  T_{ab} ,\ [D^H =  d + \o ^H  _a  T^a ,\
  a= ( i , s , 5) ] \eea
where the $H$ covariant derivative $ D^H  $  is  also found 
(just as the  $ SO(5)$   generators)      to act  algebraically on
harmonic   fields~\cite{ceresole99,ceresoII99},   
$ D Y ^{(\nu  ) } _{h, q }=  (- \o ^a   (T_a)_q^{q '}    +
M^{ab} (T_{ab} )_q^{q '} )    Y ^{(\nu  ) } _{h, q' } .$     
The covariant   derivative components  along the   
basis of rescaled 1-forms,  $e^a = r(a) \o ^H_a,\  [r(i)= r(s)= \sqrt 6 ,\  r (5) = 3/2]   $ are given     by the explicit formulas
\bea && D_i= - r(i)  T_i - 2 \e _i^j
T_{5j} ,\ D_s=-  r(s)   \tilde T_s + 2 \e _s^t T_{5t} , \ 
D_5 = - r (5)  T_5 -(T_{12} - \tilde T_{34} )  ,\ [D =   e^a D_a  =
  r(a) \o ^a   ]  \eea 
where $ \e _{ij},\ \e _{st},\ $   are the antisymmetric tensors of $SU(2) _{1,2}$.
The  irreducible representations  of $G$   consist   of   2 conserved
angular momenta    
 $ j , \ l $   with the  corresponding  magnetic   quantum 
numbers  $ (j_3, \ l _3 ) = (m, \ n)$. 
The  conserved charge $r$  (that identifies to  the R-symmetry  charge 
of  the   $\caln = 1$ dual gauge  theory)  is associated to   the subgroup  $ U(1)_R
: \ r = T_5 = T_3 - \tilde T_3 = j_3-l_3  $,     orthogonal to 
the    non-conserved     charge $q$  from 
$ H  = U(1) _H:\ q = T_H = T_3 +\bar T_3 = j_3+l_3  $. 
Since the  charges $ (q, \ r )$ are in 1-to-1 correspondence 
with the magnetic   quantum  numbers,   $(m, \ n)  =  (q \pm r )/2 $ 
or $ (q, \ r ) =  (m\pm n)  $, one can trade  the quantum
numbers  $ (m, n)$   for the   $(q,r)$.  
With $ 2q = 2 j_3 +2 l_3 \in (2 Z)$, it
follows that $2j +2 l \in (2 Z)$ and hence $ j,\ l $ are both even or
odd   integers  while   $ r = j_3 -l_3 $ is
correspondingly  an integer or half-integer.  The    basic  harmonics 
 of $T^{1,1} $    thus constitute    a subset of
the hyperspherical harmonics of  $ S^6 $   of isometry group  $SO(5)$
which can be   expressed by the non-redundant  notation 
\bea &&  Y ^{jl, M} = Y^{jlr} _{q} \equiv Y ^\nu _{q} , \ 
[\vert q +r \vert \leq 2 j ,\   \vert q -r \vert \leq 2 l ]  . \eea  
 The charge $q$ is not conserved
while the charge  $r$ is conserved only for scalar fields. 
The   coset  projection   for scalar   fields singlets of $U(1)_H $
 imposes the    equivalence relation   $q=  j_3  + l _3  \equiv m +n
 =0$, with the restriction $ \vert r \vert \leq 2 \tilde j ,\ [\tilde
   j =  min (j,l) ] .$  
The    basic harmonics  $ Y ^{j l , M } (\T )  $ 
can be constructed  algebraically~\cite{ceresole99,ceresoII99} 
in  terms of  the representation  space of  
the  completely symmetric    $ (2j)$-   and $ (2l)$-tensors 
$   (\ud )^{2j} $ and $   (\ud )^{2l} $    with respect to
the direct product   groups     
$ (SU(2)_{1} ) ^{2j} $  and   $ (SU(2)_{2} ) ^{2l} $. 
The Lie algebra   generators  are  then   given     by  direct sums 
of the respective  Pauli  matrices  
$\tau  _i $ and $\tilde \tau   _i $,          
\bea &&   T_i = - {i\over 2}   \sum _{p=1}^{2j}  \tau _i ^{(p)}  ,\  
\tilde T_s =- {i\over 2} \sum _{p=1}^{2 l} \tilde  \tau  _
s ^{(p)}   , \ [i=1,2,3,\  s =1,2,3 ]\eea    
where the  irreducible representations  of totally   symmetric tensors
are associated   to single row  Young tableaux  of $m = m_{2} + m_{1} $  boxes 
filled   with  $ m_{2,1} $ states of components $ \tau _3 ^{(p)}=  \pm
1/2  $   and  $n = n_{2} + n_{1} $ boxes filled   with  
$ n_{2,1} $ states of components $\tilde \tau _3 ^{(p)} = \pm 1/2 $. 
The    angular   momenta   and  magnetic quantum numbers $ (j,l )
,\  M= (m , n)$  and the parameters $ m_{2,1}  ,\ n_{2,1} $ 
are   connected  by the   linear relations 
\bea && \bullet \ j = (m_2+m_1)/2,\ j_3 = m =
(m_2-m_1)/2 = (q+r)/2 \in (-j , -j +1, \cdots ,j ) , 
\cr && \bullet \ l = (n_2+n_1)/2,\ l_3 =n = (n_2-n_1)/2 =
(q-r)/2 \in  (-l , -l +1, \cdots ,l ) .  \label{app2eq12}     \eea   
The  action  of  Lie   algebra generators    on  harmonic functions  
takes the simple form 
\bea && (T _1 \pm i  T _2) Y^{jl r } _ q = -i  ( j \mp {q+r \over 2})
Y^{j l (r \pm 1) } _ {q \pm 1} ,\ (\tilde T _1 \pm i \tilde T _2) 
Y^{jl r } _ q =-  i( j \mp {q-r \over 2}) Y^{j l (r \mp 1) } _ {q \pm 1} ,\   
(T _3 \pm \tilde T _3) Y^{jl r } _ q  =  i {q \choose r} Y^{jl r } _ q .  \eea

The      integer eigenvalues    $ q_a$ of $ T_H $  
in the    first  five low  dimensional  representations 
$ (1, 4, 5  , 10, 14) $  of
$SO(5)$ range   over the    respective intervals,     
$q(1)= 0,\ q(4) = (0^2, \pm 1),\ q(5) = (\pm 1 ^2 , 0),\  
q(10)= (\pm 2, \pm 1 ^2 , 0^4) ,\ q(14) =
(\pm 2^3, \pm 1 ^2 , 0^4)$.  

In the parameterization   of  $ T ^{1,1}\sim S^3 _1\times S^3_2 / U(1)_H$   
using the  5 angle   variables  $  \T ^a = [(\t_1 ,\ \phi _1 )\in S^2 _{1}  ,\ 
(\t _2 ,\ \phi _2)\in S^2 _{2}  ,\ \psi =  \varphi _1  +
\varphi _2  ] $,   the scalar    Laplace operator   action on harmonic
functions  $Y^{(\nu )} _{q=0} $   has the   separable structure~\cite{gubser98} 
 \bea &&   (  \nabla ^2 _{T^{1,1}}  +  H_0 ^\nu  ) Y ^{jlr } _{q=0} = 0
,\ [\nabla ^2  = D^a D_a  =  6 T_a T_a = 6 ( - \nabla ^2 _{S^2_1} -\nabla ^2  _{S^2_2}    +  { r^2  \over 8}     \dh _\psi ^2) ,\   
  H_0  ^{(jlr)}   = 6 (j(j+1) + l(l+1) - {r^2  \over 8} )  ]. \eea    
The  harmonics of  $ T^{1,1}$ are  trigonometric  functions 
of the angles $\T ^a$,    with  the   dependence on polar angles
involving    Hypergeometric functions  $_2F_1 (a, b, c , z)
$~\cite{gubser98,baumann06}   
\bea && Y^{jl ,  (mn) } (\T  _a)=  J_{j,m ,r} (\t
_1) J_{l,n ,r} (\t _2) e ^{im\phi _1 + in\phi _2+ i r \psi /2 } , \ [
m= j_3  = (q+r )/2 , \ n= l_3 = (q-r )/2   , \cr && 
J_{j,m ,r} (\t ) =  (\sin \t )^m (\cot
\t /2)^ {r/2} { {_2F_1 }(-l +m, 1+l+m, 1 + m -r /2 , \sin ^2 \t /2 )
\choose {_2F_1 } (-l +r/2 , 1+l+r/2, 1 - m + r /2, \sin ^2 \t /2 ) } ]
. \label{app2eq14} \eea 
The upper and lower entries  above  are  reserved to  the  intervals 
$ m \geq r/2 $ and $ m\leq r/2 $, respectively, and    the normalization
constants    are omitted.   Upon   imposing 
the  projection condition $ q=0$,  
the   various factors in $ J _{j,m, r} (\t )$    reduce   to    
polynomials of the  trigonometric   functions of $\t $.  
 The    angle dependent  factors obey the  useful  relation  
$\lim _{\t  \to 0}   J _{j,m,r} (\t )  \propto  \d _{r,0} $.  
For  concreteness, we  quote below the  explicit formulas   for    
 factors  of the  harmonic functions     of low  orders   
 \bea && J_{\ud , \pm \ud , 1 } (\t )= ( \cot {\t \over 2}  )  ^{\pm
   1/2} \sin  ^{1/2} \t  , \ J_ {1,0, 0} (\t ) = \cos \t ,\ J_{2,0,0}
 (\t ) = -1 + 3 \cos ^2 \t , \cr && J_{3,0,0} (\t ) = -3 \cos \t + 5
 \cos ^3 \t ,\ 
J_{2, 1, 2} (\t )= -1 + 2 \cos \t \cot {\t \over 2} \sin \t  ,\ J_{2,
-1, 2} (\t )= \sin  \t  \tan {\t \over 2} ,\cr && J_{1, 1, 2} (\t )=
\sin \t  \tan {\t \over 2}   ,\ J_{1, \pm 1 , 2} (\t )= (\tan 
{\t \over 2} ) ^{\mp 1}  \sin \t ,\ J_{{3\over 2}, \ud , 1} (\t )= -1
+ 3 \cos \t \ 
(\tan {\t \over 2}   \sin \t  ) ^{-1/2} .\label{app2eq15}  \eea
We  close  with  two  general  remarks concerning  
extensions of the harmonic  analysis  on  $T^{1,1}$.     
Observe first   that   $ T^{1,1}$     is part of    a   wider 
family  of compact     manifolds  $ T^{p_1,p_2} $,  
labeled  by pairs of relative prime  integers  $ (p_1, \ p_2 ) $, 
whose metrics  are related in a simple   way to  that of $T^{1,1}$. 
The geometries   of  $ T^{p_1,p_2} $  correspond  to  $ S^2 _1\times S^2_2
\times S^1 _r  $    spaces    with   magnetic monopoles 
of charges $\ud  p _{1} r ,\  \ud  p _{2} r  $       at the  centers of 
the $R^3 _{1,2} $  spaces embedding    the $ S_{1,2}^2$~\cite{jatkar99}.  
The   harmonic    analysis   is then      developed in terms of 
Fourier  series  in $ e ^{ ir \psi / 4} ,\ [\psi \in
S^1]$    multiplied  by  pairs of  monopole  harmonic  functions  of the
$S_{i}^2$, given by     the irreducible representation  matrices  of
$SU(2):\  \caly ^j _m = D ^ {j \star }  _{m , p_i } (\t _i ,\phi _i
)$~\cite{coleman81}.  
We note in second    the   useful connection between    
harmonic  analyses  in  the undeformed  and deformed conifolds.
Since the         fixed-$\tau $  sections  of  the  deformed conifold   are
isomorphic  to  $ T^{1,1}$, the angular momenta $ j,\ l$    are still 
preserved       but    only the parity  of the  $r$-charge   is conserved.  
The  differential operator  for the scalar 
Laplacian  of $\calc _6 $ was   initially examined  in~\cite{krishtein08}.
Although  the  dependence on the  radial  and angular   coordinate
variables  is    no longer  separable,   a useful   group theory  approach
was   developed~\cite{pufu10}    where     the scalar harmonic
functions  are expressed  as  finite  sums  of the  $ T^{1,1}$  harmonics
$  \caly ^{j l } =  \sum _{r  =- \tilde j} ^{ \tilde j}  f _r
(\tau ) Y ^{j l r} (\T  ) ,\  [\tilde j  =   \min  (j , l )] .$
The      coefficient  functions $ f_r  (\tau ) $  are defined    as solutions of systems  of    $ 2 \tilde j +1 $    coupled     second-order differential equations in   the radial  variable $ \tau $. 

\subsection{Harmonic   decomposition of  supergravity  multiplet  bosonic  fields} 
\label{suc3app2} 

The harmonic decomposition on $ T^{1,1}$ of tensor (or spinor) fields
$\phi _{R (ab \cdots ) }$,  in    representations $R$ of 
the tangent group  $ SO(5) $,  
utilizes solid harmonics $ Y^{jl, M}_{R (ab \cdots ) }$,  
given by     direct products   of the basic  scalar harmonics of 
angular momenta $ (j, l) $    by   the  representation  spaces  $R$   of 
$SO(5) \supset  SO(4) \sim SU(2)_j \times SU(2)_l$. The components  of
irreducible  solid harmonics are labeled by the
 magnetic quantum numbers $ M = (j_3, \ l_3)   \equiv (m, n)$    and 
sets of   vector and spinor  indices  $ R(a b
\cdots ),\ R (\a b \cdots ),\ [a \in (1, \cdots , 5) ,\ \a \in (1,
  \cdots , 4) ]   $      of appropriate symmetry   and trace  properties  
to build  the irreducible representations $ R
  = (4, \ 5,\ 10,\ 15, \cdots ) $ of $SO(5)$. 
The   basic  scalar harmonic  fields $
Y^{j,l,r} _q $     are  labeled  by the  conserved charges $ (j , l, r
)$  which characterize  the states  uniquely,  
since the projection $m  +  n = q=0
$  in the coset space $ T^{1,1}$  fixes the magnetic quantum numbers 
to  $ m=-n = r$.  Instead of  representing  solid  harmonics 
as  direct products of the basic scalar  harmonics $ Y ^{jl M} =  Y ^{jl r} _0 $
by    linear  representations $  R (a\cdots )$ of the   
tangent group $ SO(5)$, one can equivalently
consider  bases  of column vectors   of 
scalar  harmonic components    $ \caly ^{jlr } =  
\{ Y_{ q_i } ^{ (j, l, r_i ) } (\T ) \}  ,\ [i=1,\cdots ,  dim (R)]$. 
The    sets  of charges  $ (q_i,\ r_i) $    are  determined  by  the
closure   under the action of  $ SO(5)$ generators    or 
by  matching the eigenvalues of $T_H $   in the given  representation $  R $.   
Unlike the conserved quantum numbers $j, \ l $,     the 
charge $r$     labels  states  but  is not conserved. 
  For instance,  the solid vector and  spinor harmonics  
$ Y ^{jl, M} _a $ and $Y ^{jl, M} _\a  $    are   represented
by the   bases of  column vectors~\cite{ceresoII99}  
\bea && \bullet \ \caly ^{j l  r  } _a (\T ) = ( Y ^{jl
  (r+1) } _{+1},\ Y ^{jl (r-1) } _{-1},\ Y ^{jl (r-1) } _{+1},\ Y ^{jl
  (r+1) } _{-1},\ Y ^{jl r } _{0} )^T, \ [a =1, \cdots , 5]. 
\cr && \bullet \ \caly ^{j l   r  } _{\a }
= ( Y ^{jl (r-1) } _{0},\ Y ^{jl (r+1) } _{0},\ Y ^{jl r } _{-1},\ Y
^{jl r } _{+1})^T, \ [\a =1, \cdots , 4] 
. \label{app2eq17} \eea   
 Note  that the  magnetic quantum numbers $ m\equiv (q+r)/2 ,\ n \equiv (q-r)/2
$  of the various  components are   equal, modulo  constant shifts by
 $ \pm 1 $.  In the solid vector  harmonics,   $m = r / 2 +1 ,\ n = -r
 /2\ mod  (\pm 1) 
$ and in the solid  spinor  harmonics, $m = r/2 -1/2,\ n = 1/2 -r /2  
\ mod (\pm 1) $.  One advantage  in using   the  above matrix
formalism   for   
the harmonic  modes of   $ AdS_5$     fields $\Phi _{R(a\cdots )}$    
in        representations  of $SO(5)$  
is  to  enable      calculating   the   action of
Hodge-de Rham $ - \D   _1 $  or Dirac   $\Dslash $   operators 
by  algebraic  means.

\end{appendix}
  
\end{document}